\documentclass[11pt]{article}

\usepackage{bbm}
\usepackage{graphicx}
\usepackage{natbib}
\usepackage{amsmath}
\usepackage{amsfonts}
\usepackage{url}
\usepackage[a4paper]{geometry}
\usepackage{verbatim}
\usepackage[super]{nth}
\usepackage[utf8]{inputenc}
\usepackage{titling}
\usepackage{subfigure}
\usepackage[justification=centering]{caption}

\DeclareMathOperator*{\argmax}{arg\,max} 

\numberwithin{equation}{section}   

\newenvironment{proof*}{\begin{trivlist} \item[]
{}}{\nolinebreak
\hfill \rule{2mm}{2mm} \end{trivlist}}

\newcounter{ctr}

\newcounter{ctr1}

\newcounter{ctr2}

\newcounter{ctr3}

\newtheorem{definition}{Definition}				
\newtheorem{theorem}{Theorem}
\newenvironment{theorem*}[1]{{\bf Theorem #1} \begin{itshape}}{\end{itshape}}
\newtheorem{corollary}{Corollary}
\newenvironment{corollary*}[1]{{\bf Corollary #1} \begin{itshape}}{\end{itshape}}
\newtheorem{proposition}[definition]{Proposition}
\newenvironment{proposition*}[1]{{\bf Proposition #1} \begin{itshape}}{\end{itshape}}

\newcommand{\NN}{\mathbb{N}}

\newcommand{\RR}{\mathbb{R}}

\newcommand{\ud}{\, {\rm d} \kern-.015em }

\newcommand{\modulus}[1]{\left| \kern.05em #1 \kern.05em \right|}
\newcommand{\norm}[1]{\left\| \kern.05em #1 \kern.05em \right\|}
\newcommand{\inner}[1]{\left\langle \kern.05em #1 \kern.05em \right\rangle }

\newcommand{\bm}[1]{\mbox{\protect\boldmath $ #1 $}}

\newcommand{\pick}[2]{\renewcommand{\arraystretch}{0.6}
\left( \kern-.4em \begin{array}{c} #1 \\ #2 \end{array} \kern-.4em \right) }

\newcommand{\vartext}[1]{\, {\rm Var}( #1 ) }
\newcommand{\PP}[1]{\mathbb{P}\left( #1 \right)}
\newcommand{\EE}[2][]{\mathbb{E}_{#1} \left[ #2 \right]}

\newcommand\indpt{\protect\mathpalette{\protect\independenT}{\perp}}\def\independenT#1#2{\mathrel{\rlap{$#1#2$}\mkern2mu{#1#2}}}

\newcommand{\ie}{\textit{i}.\textit{e}.}
\newcommand{\eg}{\textit{e}.\textit{g}.}

\newcommand{\captionit}[1]{\caption{\small\textit{#1}}}

\makeatletter
\newcommand*{\defeq}{\mathrel{\rlap{%
                     \raisebox{0.3ex}{$\m@th\cdot$}}%
                     \raisebox{-0.3ex}{$\m@th\cdot$}}%
                     =}
\makeatother

\usepackage{multirow}
\usepackage{hhline}
\usepackage{array}

\makeatletter
\newcommand{\thickhline}{%
    \noalign {\ifnum 0=`}\fi \hrule height 1pt
    \futurelet \reserved@a \@xhline
}
\newcolumntype{"}{@{\hskip\tabcolsep\vrule width 1pt\hskip\tabcolsep}}
\makeatother

\newcommand{\thetahat}{\widehat{\theta}}
\newcommand{\betahat}{\widehat{\beta}}
\newcommand{\relb}{\mbox{RelBias}}
\newcommand{\rmse}{\mbox{RRMSE}}
\newcommand{\calS}{\mathcal{S}}

\usepackage{xr}
\usepackage{xcolor}
\newcommand{\overbar}[1]{\mkern 1.5mu\overline{\mkern-1.5mu#1\mkern-1.5mu}\mkern 1.5mu}

\graphicspath{{./}}

\newcommand{\xbar}{\overbar{x}}
\newcommand{\Xbar}{\overbar{X}}
\newcommand{\xhat}{\hat{x}}

\newcommand{\method}{\mathsf{M}}
\newcommand{\xhist}{\tilde{x}}

\title{Inference for extreme values under threshold-based stopping rules}
\author{Anna Maria Barlow, Chris Sherlock, Jonathan Tawn}
\date{}

\begin{document}
\maketitle
\vspace{-40pt}
\begin{center}
\emph{STOR-i Centre for Doctoral Training, Department of Mathematics and Statistics, Lancaster University, Lancaster, LA1 4YF, U.K.}
\end{center}

\begin{abstract}
There is a propensity for an extreme value analyses to be conducted as
a consequence of the occurrence of a large flooding event. This timing
of the analysis introduces bias and poor coverage probabilities into the associated risk assessments and leads subsequently to inefficient flood protection schemes. We explore these problems through studying stochastic stopping criteria and propose new likelihood-based inferences 
 that mitigate against these difficulties. Our methods are illustrated through the analysis of the river Lune, following it experiencing the UK’s largest ever measured flow event in 2015. We show that without accounting for this stopping feature there would be substantial over-design in response to the event. 
\end{abstract}

Keywords: Extreme value theory, flooding, stopping rules

\section{Introduction} \label{sec:intro}

The UK currently spends £400-500M per year on coastal and river flood
defence infrastructure, with 2 million properties exposed to the risk
of flooding. The agencies responsible for this spend monitor the
effectiveness of their investment at giving the level of protection
expected. After major flooding events renewed analysis is performed to
assess both existing flood defences and the cost benefit of potential new schemes, proposed in response to the flooding. 

Statistical extreme value methods, with likelihood-based inference,
have proved a core component of the required analysis in terms of
minimising the costs without jeopardising the level of accepted risk,
and hence have financial and societal benefits. However, there is a problem with using these methods when the statistical analysis has been prompted by the occurrence of a recent large event, since in this case the data-set size itself is also random. This can lead to substantially biased inference
and poor coverage properties and so result in inefficient
flood-defence designs. Omitting the new extreme data value from the data
set also seems unsuitable, as intuition suggests that flood risk will
then be underestimated; moreover it would appear perverse to flood management agencies to ignore events of the type most relevant to the design specification.

This paper aims to identify the extent of the inference problems when
an analysis has been triggered by a large event and to develop new conditional-likelihood methods which appear to overcome these problems. We do not \emph{suggest} when the timing of the data analysis should take place but study the analysis given that its timing has been determined by a data-dependent decision making process.

We consider modelling the extreme events of a time series of independent and identically distributed (iid) random variables $X_1,X_2,\hdots$. The classical approach to do this is to split the time series into blocks of equal size (often a year) and to model the maxima of these blocks. Linear normalisation is necessary since as the block size tends to infinity the distribution of the maxima degenerates to a point mass at the upper end point of the distribution of $X$. The generalised extreme value (GEV) distribution \citep{Coles} is the only non-degenerate limiting distribution of the normalised maxima as the block size tends to infinity.  The GEV has distribution function:
	\begin{align}
			G(x) = \exp\left(-\left[1+\xi\left(\frac{x - \mu}{\sigma}\right)\right]_+^{-\frac{1}{\xi}}\right) \, ,
	\label{eqn:GEV} 
	\end{align}

where $\mu$, $\sigma>0$ and $\xi$ are the location, scale and shape parameters respectively and $[z]_+ = \max(z,0)$. The shape parameter determines the behaviour of the upper tail of the distribution: for $\xi<0$ the distribution has an upper end point, for $\xi=0$ the tail is exponential and for $\xi>0$ the distribution has a power-decaying tail. There is particular interest in the occurrence of extreme events and so an important part of the analysis is the estimation of return-levels (quantiles). Under stationarity, the $y$-year return-level, $x_y$, is the value which is exceeded on average once every $y$ years. For the GEV distribution this can be calculated as:
\begin{align}\label{eqn:GEVret}
	x_y = G^{-1}\left(1-\frac{1}{y}\right) = \left\{ \begin{array}{ll} 
		\mu - \frac{\sigma}{\xi}\left\{1-[-\log{\left(1-1/y\right)}]^{-\xi} \right\} \qquad \qquad &\xi \neq 0 \\
		\mu - \sigma\log{\left[ -\log{\left(1-1/y\right)} \right]} &\xi = 0.
	\end{array}  \right. 
\end{align}

\begin{figure}
	\begin{center}
		\subfigure{\includegraphics[scale=0.28]{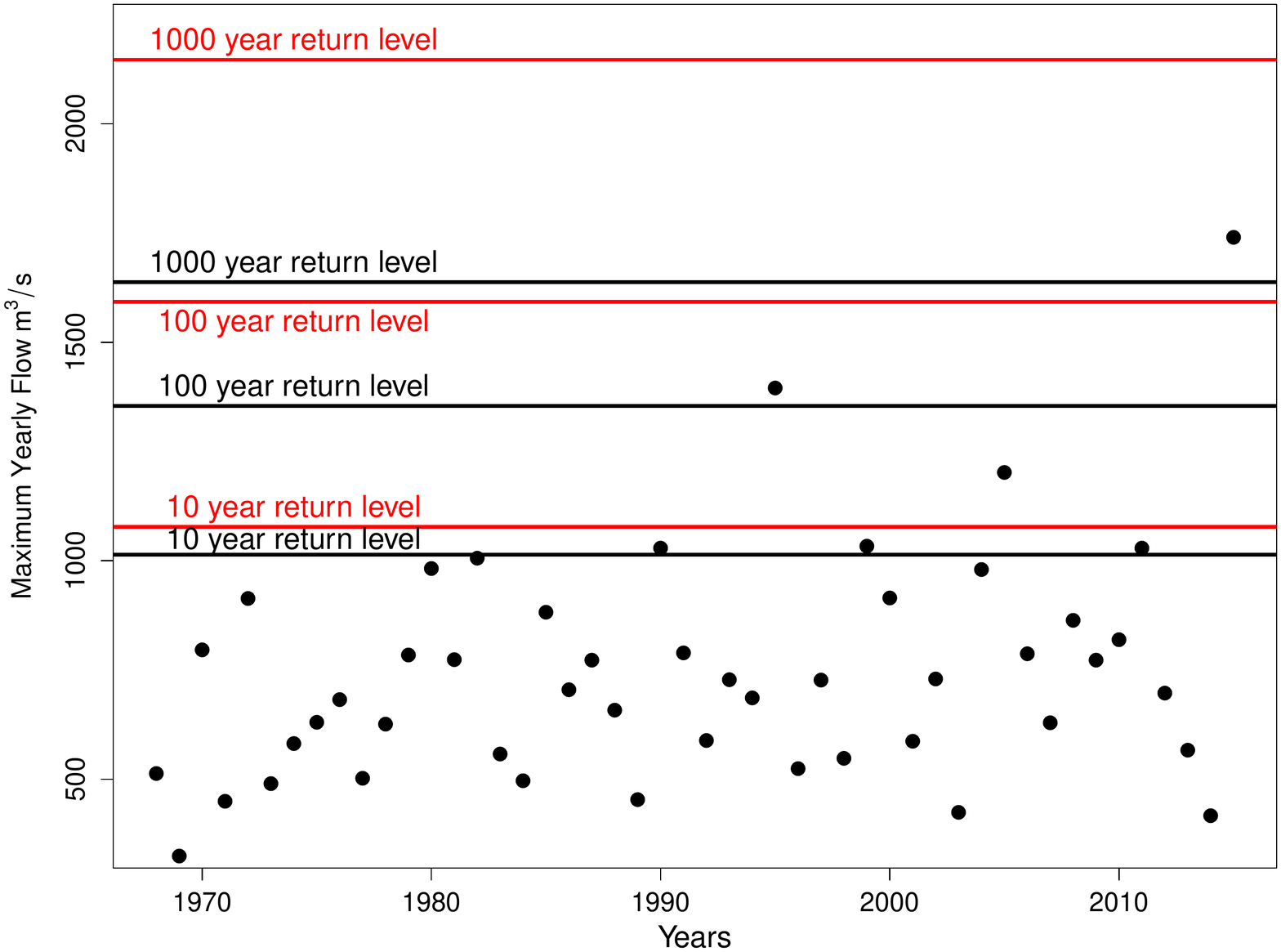}} 
		\subfigure{\includegraphics[scale=0.28]{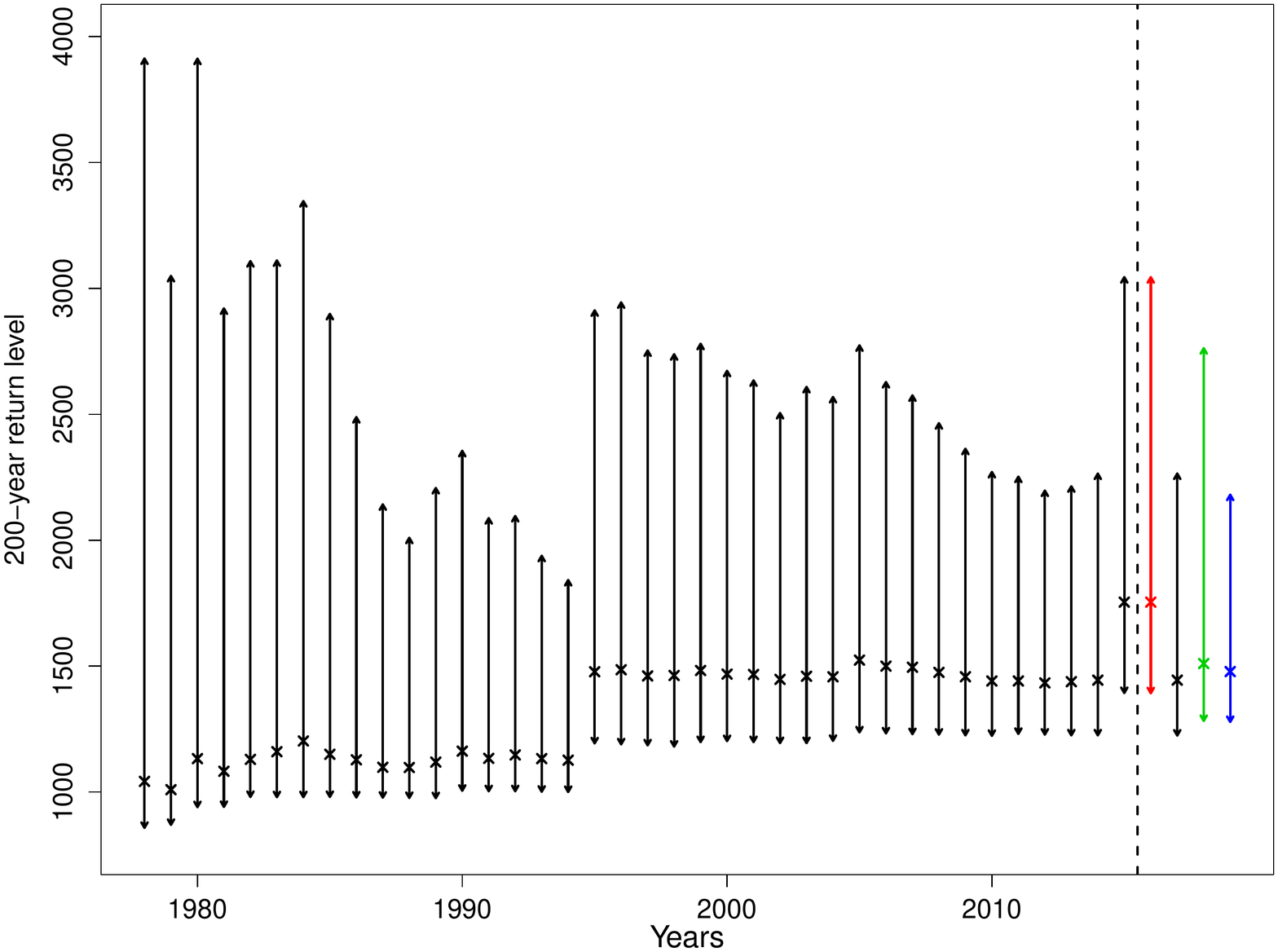}}
		\captionit{Left: The annual maxima of daily peak river flow data for the Lune at Caton with return level estimates before (black) and after (red) the 2015 flood. Right: 200-year return-level estimates based on all the data up to and including the current year for the Lune at Caton with 95\% profile likelihood-based confidence intervals. The four 200-year return-level estimates and associated 95\% confidence intervals to the right of the vertical dotted line are our new estimates that aim to address a fixed-threshold stopping rule of $c_k=1568$ based on the all the data up to and including 2015: standard likelihood (red), excluding the final observation (black), full conditioning (green) and partial conditioning (blue).}
		\label{fig:Lune}
	\end{center}
\end{figure}

One can also consider modelling daily observations above some high threshold (rather than just modelling the block maxima) by the asymptotically justified generalised Pareto distribution (GPD) \citep{DandS}. Threshold methods typically benefit from using more extreme-value data and hence are more efficient in their inferences than block maxima methods \citep{Coles}. We focus most of our analysis and developments on the GEV case, as similar benefits are found for both GEV and GPD inference, but with the GPD also sensitive to threshold choice. We illustrate some GPD results in \S\ref{sec:GPD} of the supplementary material.

We consider the analysis of annual
maxima of daily peak river flow data obtained from \cite{NRFA} for the Lune at Caton,
just outside Lancaster, from 1968 to 2015 (Figure~\ref{fig:Lune}, left panel) and illustrate the inference issues due to the timing of analysis being determined by the occurrence of a flood event. Under the assumption that the
annual maxima are independent and identically distributed (i.i.d.) we can fit the GEV distribution to the annual
maxima using likelihood-based inference (the likelihood is
$\prod_{i=1}^{n} g(x_i;(\mu,\sigma,\xi))$ where $n$ is the sample size
and $g$ is the density, $g=dG/dx$), and estimate return-levels
using \eqref{eqn:GEVret}. The estimated 10, 100 and 1000-year return
levels are shown in Figure~\ref{fig:Lune} (left panel) for the data up to 2014 (black) and including 2015
(red). The December 2015 floods resulted in the river Lune recording
the highest peak river flow (1740 $m^3/s$) of all UK rivers over all
years of records. This value is higher than the 1000-year return-level
estimate based on the observations up to 2014. However, once the 2015
event is included in the analysis the return-level estimates become
much higher. If we were to take these 2015 point estimates as the truth we
would expect to observe an event as extreme as that in 2015
approximately once every 200 years.
~For design purposes this level of sensitivity is highly undesirable, as the costs for flood protection would change dramatically.

Figure~\ref{fig:Lune} (right panel) shows a reanalysis of all data available at each year between $1978$ and $2015$. It provides the point estimate and profile likelihood-based $95\%$ confidence interval of the 200-year return-level, as it would have been produced in that year. The four additional point estimates and confidence intervals to the right of the vertical dotted line correspond to estimators introduced in \S\ref{sec:like} and their corresponding profile likelihood-based confidence intervals. At the beginning of the data collection the return-level estimates vary considerably, but they become more stable as the number of years increases, with the width of the confidence intervals generally decreasing over time. However, even after many years of data collection, the largest events can be seen to cause sharp increases in the estimates and their associated uncertainty. For example, the return-level estimate following the January 1995 floods and the 2015 floods are larger than those of previous years. 

This illustrative example is typical of when an analysis is performed immediately after a large event. Unless further analysis is undertaken it is unclear whether by analysing the data with the final extreme event we are introducing a positive bias into the inference. For example, the lower bound of the 95\% confidence interval of the 200 year return-level after the 1995 event is larger than almost all previous point estimates - directly after the event (without the knowledge of later years) this could have been seen as an indication of positive bias in the standard estimator. However, after the 1995 event the return-level estimates were fairly stable and larger than those before 1995, so it would seem the standard estimator for 1995 may not have been overestimating and before the event the shape parameter was estimated too low.

An alternative approach is to simply ignore the most recent year of
data when an analysis has been requested because we have large data in
that year, in which case the return-level estimate is lower and the
confidence interval is narrower - in particular the upper bound is
lower. However, we speculate (see also \S \ref{sec:expbias}) that this estimator is now negatively
biased due to the loss of information about the extreme
event. Moreover the estimator is inefficient since the larger data values are the most informative about
the upper tail \citep{DandS}. Finally, it would be hard to convince practitioners to exclude the largest events; for example, an event may be observed which is larger than the upper end point estimated from previous data, in which case it would be perverse not to make some update to the previously estimated return-levels.

The key issue that the Lune example illustrates is that when meeting
the flood management agencies’ needs, the time to undertake the
extreme value analysis is stochastic and triggered by a large
event. Thus, there is effectively some form of unwritten stopping rule, determined by the flood management agencies, which
determines the timing of the analysis. In contrast with a standard iid
sample of fixed size, when we use a stopping rule the time at which we stop (the sample size) is variable, we denote this by $N$.

One can attempt to mathematically formulate the characteristics of the stopping rule, though in reality a precise mathematical rule does not exist. The stopping decision may depend on (i) some absolute threshold, such as the height of existing flood defences or a critical level which when exceeded leads to severe flooding, or (ii) an assessment, based upon all observations to date, of what might constitute an ‘exceptional’ event. We consider two simple stopping rules based on a series of iid random variables, $X_1,X_2,\hdots$ which, in a sense, bracket this range of possibilities and we discuss other possibilities in \S\ref{sec:discussion}.

\begin{enumerate}
	\item \textbf{Fixed-threshold stopping rule} \\
	Stop when an observation exceeds a specified value, $c_k$, \ie :
	\vspace{0mm}
	\begin{align}\label{eqn:rule1}
		N = \inf\{n \in \NN : X_n > c_k\} \,,
	\end{align}		
	where $k$ is the true (but unknown) return period of $c_k$.
	\item \textbf{Variable-threshold stopping rule} \\
	Stop when an observation exceeds the return-level estimate, $\hat{x}_k$, corresponding to the fixed return period of $k$ years, calculated using previous observed values, \ie , when:
	\begin{align}\label{eqn:rule2}
		N = \inf\{ n \in \NN : X_{n} > \hat{x}_k(X_1,\hdots,X_{n-1})\} \, .
	\end{align}
\end{enumerate}

We do not suggest the stopping rule to use but study the analysis given that its timing has been determined by stopping rule. As far as we are aware, there has been no study of stopping rules and their effects on likelihood estimation in the extreme-value setting. 

Using a stopping rule to determine the sample size can lead to estimators, such as the maximum likelihood estimator (MLE), having different sampling properties to the fixed-sample case. To illustrate this feature consider an iid sequence of Bernoulli random variables, $Y_1,Y_2,\hdots$, each with probability of success of $\theta$. If one fixes the number of trials, $n$, the number of successes, $R$, in these trials is binomially distributed and $\hat{\theta} = R/n =: \hat{\theta}_1$; whereas if the number of successes is fixed as $r$, the number of trials, $N$, is negative-binomially distributed and $\hat{\theta} = r/N =: \hat{\theta}_2$. In both cases the MLE of the probability of success is the proportion of successes, however, $\EE{\hat{\theta}_1} = \theta$ whereas $\EE{\hat{\theta}_2} = r \EE{1/N} \geq r/\EE{N} = \theta$ by Jensen's inequality. The presence of a stopping rule affects the performance of the estimator which motivates an investigation into the performance of return-level estimation under stopping rules.

Testing the data against some `stopping criterion' at regular
intervals falls into the setting of sequential analysis, which has a
rich literature covering applications from quality control \citep{Wald}, to
clinical trials \citep{Todd} and abundance modelling \citep{Barry}. Many studies have considered
the influence of such stopping rules on likelihood inference, \eg
, \cite{Barndorff-Nielsen} consider the distribution of the likelihood-ratio statistic  under different stopping rules for systems with Brownian motion and Poisson processes, and in the clinical trial setting \cite{Whitehead86} derives an expression for the bias of the MLE of the treatment effect tested under a sequential probability ratio test. Some papers compare the bias under different experimental designs or stopping criteria (\eg , \cite{Bauer}). \cite{Cox}, \cite{Whitehead86} and \cite{Stallard} propose bias-reduced estimators by approximating the bias and subtracting this from the usual estimate. One such approach uses an iterative method corresponding to a bootstrap bias correction \citep{Efron}.

\cite{Kenward} consider iid sampling from a Normal distribution using a deterministic stopping rule and study the estimation of the mean parameter of this Normal distribution. \cite{Molenberghs} extend this setting to the use of a probabilistic stopping rule. They note that an unbiased estimator of the mean parameter can be obtained from the conditional likelihood (we derive such estimators in \S \ref{sec:like}) however, at the cost of an increased mean squared error (MSE) in comparison to the MSE of the sample average (the standard estimator if the sample size was fixed). The increased variance of a bias-reduced estimator appears to be an issue for many of the proposed bias reduction methods. For example, bias reduction using Rao-Blackwellisation \citep{Bowden} 
and shrinkage estimators \citep{Carreras} often have a worse MSE than the standard MLEs.

In \S\ref{sec:setup} we introduce the notation used throughout the
paper and discuss likelihood inference under stopping
rules. In \S\ref{sec:expbias} and \S\ref{sec:gamma} we discuss the
bias under the fixed-threshold and variable-threshold stopping rules
respectively and derive expressions for the bias when sampling from
some simple distributions. We introduce two conditioning-based
likelihood estimators in \S\ref{sec:like}. In \S\ref{sec:GEV} we perform a simulation study for sampling from the GEV distribution using the two stopping rules and discuss the properties of the estimators in this setting. 
We apply our estimators to the Lune river flow data in \S\ref{sec:Lune} and discuss our conclusions, the practical usage of the methods and extensions in \S\ref{sec:discussion}.

\section{Inference under stopping rules} \label{sec:setup}

\subsection{Introduction} \label{sec:intro2}

Throughout this paper we restrict our attention to sequences of
iid observations arising from some
distribution with a density of $f(x;\theta)$, where $\theta$ is the
parameter vector for the distribution. We sample consecutively until
some stopping criterion is met and denote the (random) time at which
the stopping criterion is met by $N$. We write $x_n$ for the $n$th
observation and $\bm{x}_{1:n}$ for the vector of observations
$(x_1,\hdots,x_n)$. We define the log-likelihoods of the sample, both
including ($\ell_{std}$) and excluding ($\ell_{ex}$) the final, large
observation as follows: 
\begin{eqnarray}
	\ell_{std}(\theta;n,\bm{x}_{1:n}) &=& \sum_{i=1}^{n} \log{f(x_i;\theta)} \label{eqn:likefull} \\
	\ell_{ex}(\theta;n,\bm{x}_{1:n}) &=& \ell_{std}(\theta;n,\bm{x}_{1:n}) - \log f(x_n;\theta) \, . \label{eqn:likeex}
\end{eqnarray}

Given data $(n,\bm{x}_{1:n})$ an estimate of the parameter vector is
obtained by maximising the log likelihood:
$\hat{\theta}(n,\bm{x}_{1:n}) = \argmax_\theta \ell(\theta)$. When the
nature of the data is clear we abbreviate this to $\thetahat$, and
depending on the likelihood used we have estimators
$\hat{\theta}_{std}$ or $\hat{\theta}_{ex}$.

In practice we would not consider estimating return-levels
(particularly for large return periods) from a sample of only a very
small number of observations.
 However, the fixed-threshold stopping
rules can result in samples of size 1, and this can lead to
parameter identifiability issues for data sets simulated from the
hypothesised data-generating mechanism. In reality,
if an analysis has been requested then sufficient information would be available to derive a meaningful estimate. This information could be historical information, hydrological knowledge, data from other sites,
or data at the current site collected before the instigation of a stopping rule. We call this the
\emph{historical data} and, for simplicity in this article, code the
historical data as some number, $n_0$ of data values collected before the
stopping rule could be invoked. Real decisions will incorporate this information, and our analysis should allow for this. 


We are interested in a set of $y$-year return-levels, $x_y(\theta)~(y\in \mathcal{Y})$, for some set $\mathcal{Y}$, such as $\{50,200,1000\}$. In particular, we wish to understand the behaviour of the estimators $x_y(\thetahat(N,\bm{X}_{1:N}))$ (with $x_y$ given by expression \eqref{eqn:GEVret} for GEV sampling) when the dataset arises from a stopping rule. In this section we focus on the relative bias, and in \S \ref{sec:GEV} we look at other properties including the relative root-mean-squared error, given respectively by:
\begin{align}
  \label{eq.relb}
  \relb(\xhat_y)&=\frac{1}{x_y(\theta)}\EE{x_y(\thetahat(N,\bm{X}_{1:N}))} - 1\\
  \label{eq.rmse}
  \rmse(\xhat_y)&=
  \frac{1}{x_y(\theta)}\sqrt{\EE{\{x_y(\thetahat(N,\bm{X}_{1:N}))-x_y(\theta)\}^2}} \, ,
\end{align}
where $x_y(\theta)$ is the true $y$-year return-level.

In \S\ref{sec:full} we detail a well-known result that the likelihood for the data $(n,\bm{x}_{1:n})$ with a random stopping time is the same as for data $\bm{x}_{1:n}$ with $n$ fixed. However, the properties of the estimator, such as its bias and variance as well as the coverage of any confidence interval, may be influenced by the different data-generating mechanism. 

The properties of likelihood-based estimators of tail quantiles under our stopping rules are intractable for data arising from the GEV or GPD distributions. However, for a particular special case of the GPD, the exponential distribution, certain properties are tractable and this provides insight into the behaviour observed in the simulation studies of \S \ref{sec:GEV} for the GEV. 
Specifically, in \S\ref{sec:expbias} we derive the bias in quantile estimates for exponential data under the fixed-threshold stopping rule, and in \S\ref{sec:gamma} show that, under the variable-threshold stopping rule, quantile estimates for gamma data (including the exponential as a special case) with a known shape parameter are unbiased.


\subsection{Likelihood in presence of a stopping rule} \label{sec:full}

In practice it is usual to assume the sample size, $n$, is fixed in which case the likelihood for data $\bm{x}_{1:n}$ is $L_{fixed}(\theta;\bm{x}_{1:n}) \defeq \prod_{i=1}^n f(x_i;\theta)$. Now, following \cite{Pawitan}, we derive the true likelihood for the data sampled using a general stopping rule which is a function of the data and not the unknown
parameter vector. We define a stopping region
$\mathcal{S}_n = \mathcal{S}_n(\bm{x}_{1:n-1})$ such that we stop
sampling if $X_n\in\mathcal{S}_n$ and continue to sample otherwise. We abbreviate $\PP{X_i\in \calS_i}$
by $p_i$ and we let $f_{X_i|\calS_i}$ and $f_{X_i|\calS_i^c}$ denote the densities of $X_i$ conditional on $X_i\in \calS_i$ and $X_i \in \calS_i^c$. The likelihood for the full data is
\begin{align*}
	L_{std}(\theta; n, \bm{x}_{1:n}) &= \PP{N=n, \bm{X}_{1:n}=\bm{x}_{1:n} | \theta} \\
	&= \PP{N=n}\PP{\bm{X}_{1:n}=\bm{x}_{1:n}|N=n,\theta} \\
	&= p_n \mathbbm{1}_{\mathcal{S}_n}(x_n) \prod_{i=1}^{n-1} (1-p_i) \mathbbm{1}_{\mathcal{S}_i^c}(x_i) \times f_{X_n|\mathcal{S}_n}(x_n|\calS_n) \prod_{i=1}^{n-1} f_{X_i|\mathcal{S}_i^c}(x_i|\calS_i) \\
	&= L_{fixed}(\theta;\bm{x}_{1:n}) \times \mathbbm{1}_{\mathcal{S}_n}(x_n) \prod_{i=1}^{n-1} \mathbbm{1}_{\mathcal{S}_i^c}(x_i) \\
	&\propto L_{fixed}(\theta;\bm{x}_{1:n}) \, .
\end{align*}

The logic here is that to have a sample of size $n$ the final observation must be in the stopping region and all other observations outside their respective stopping regions, hence $\PP{N=n}$ includes indicator functions of the observations being in the correct sets. The last step follows since the indicator functions do not depend on the unknown parameter, $\theta$, and so are absorbed into the proportionality constant. Thus inference purely from the likelihood leads to the same
conclusions whether we have a random sample size according to some
stopping rule or a fixed sample size. Thus, the MLE, $\thetahat$, and the observed Fisher information are the same in both cases. However, the properties of the estimators are different since the distribution of $\{N,X_1,\dots,X_N\}$ is different to the distribution of $\{X_1,\dots, X_n\}$ for some fixed $n$. In particular, estimators obtained from $L_{std}$ can be biased even when estimators from $L_{fixed}$ are unbiased, as seen in \S\ref{sec:intro} for Bernoulli sampling. 
                 

\subsection{Fixed-threshold stopping rule with exponential observations} \label{sec:expbias}

Let $X_i$ have an exponential distribution with an unknown rate parameter of $\beta$, which is a special case of the GPD used to model the tails of a distribution and is given by expression \eqref{supp:eqn:GPD} with $\xi=0$ and $\sigma=\beta^{-1}$.
The $y$-observation return-level is $x_y = (\log y)/\beta$ and, since this is proportional to $1/\beta$, the relative bias is $\beta/\betahat - 1$ whatever the value of $y$. The MLE of $\beta^{-1}$ for a sample of size $n$, whether
fixed or random is simply $\xbar$, where $\xbar$ is the
sample mean. When $n$ is fixed, the MLE, $(\betahat_{fixed})^{-1}=\Xbar_n$, is an unbiased estimator of $1/\beta$; however with 
the fixed-threshold stopping rule $N$ follows a geometric distribution where $1/k$ is the probability of a `success' \ie , an exceedance. The geometric distribution is a special case of the negative-binomial distribution and so we know the estimator of the  probability of exceedance of a fixed threshold is positively biased (\S\ref{sec:intro} under \eqref{eqn:rule2}). Now $(\betahat_{std})^{-1} = \Xbar_N$ and, similarly, the MLE when excluding the final observation is $(\betahat_{ex})^{-1} = \bar{X}_{N-1} = \sum_{i=1}^{N-1}X_i /(N-1)$. It is straightforward to show (see Appendix \ref{App:Exp}) the following.

\begin{proposition}
\label{prop.expfixed}
	Let $X_1,X_2,\hdots$ be a sequence of iid random
	variables with $X_i \sim \mathsf{Exp}(\beta)$. Let $N$
	arise from the fixed-threshold stopping rule \eqref{eqn:rule1} giving data $(N;\bm{X}_{1:N})$. Let $\hat{x}^{std}_y = x_y(\betahat_{std}(N;\bm{X}_{1:N}))$ be the estimator of the $\left(1-\frac{1}{y}\right)$th quantile (equivalently the $y$-observation return-level) obtained from the MLE for $\beta$ from the full likelihood and let $\hat{x}_y^{ex} = x_y(\betahat_{ex}(N;\bm{X}_{1:N}))$ be the estimator from the likelihood excluding the final observation.
	
	Then the relative biases of the return-level estimators are	
	\begin{align*}
	\frac{1}{x_y}\EE{\hat{x}^{std}_y} - 1 &= \frac{\beta c_k}{e^{\beta c_k} - 1} \left(\frac{\beta c_k}{1 - e^{-\beta c_k}} - 1 \right) \\
	\frac{1}{x_y}\EE{\hat{x}_y^{ex}|N>1} - 1 &= - \frac{\beta c_k}{e^{\beta c_k} - 1} \, .
\end{align*}
	
\end{proposition}

In Proposition \ref{prop.expfixed}, when excluding both the final
observation (and the fact that it \textit{is} the final observation),
when $N=1$ the MLE is undefined since there are no data; $x_1$ is unknown
and the fact that $N$ would be greater than zero was known before the
data-collection process began; we therefore condition on $N>1$.  

Proposition \ref{prop.expfixed} shows that the estimator of any
return-level using the full likelihood is always positively biased,
whereas if the final observation is omitted the estimator of any return-level is always negatively biased. The final data observation is the largest and has been shown by \cite{DandS} to be the most influential on the MLE fit so when this value, together with the information that it exceeded the threshold, is omitted from the dataset this changes the bias and, potentially, also the variance of the return-level estimator and risks being inefficient. Nevertheless, for thresholds with only a small chance of exceedance, \ie , large values of $\beta c_k$,
$\relb(\hat{x}_y^{std})\sim (\beta c_k)^2 \exp(-\beta c_k)$, whereas
$\relb(\hat{x}_y^{ex})\sim -\beta c_k \exp(-\beta c_k)$, that is, the
bias is a factor $(\beta c_k)^{-1}$ smaller for estimates where the final observation is ignored. The higher the threshold, the larger the typical data set that is generated before the stopping criterion is met and the less biased the estimate of any return-level.
                                
\begin{figure}
	\begin{center}
		\includegraphics[scale=0.3]{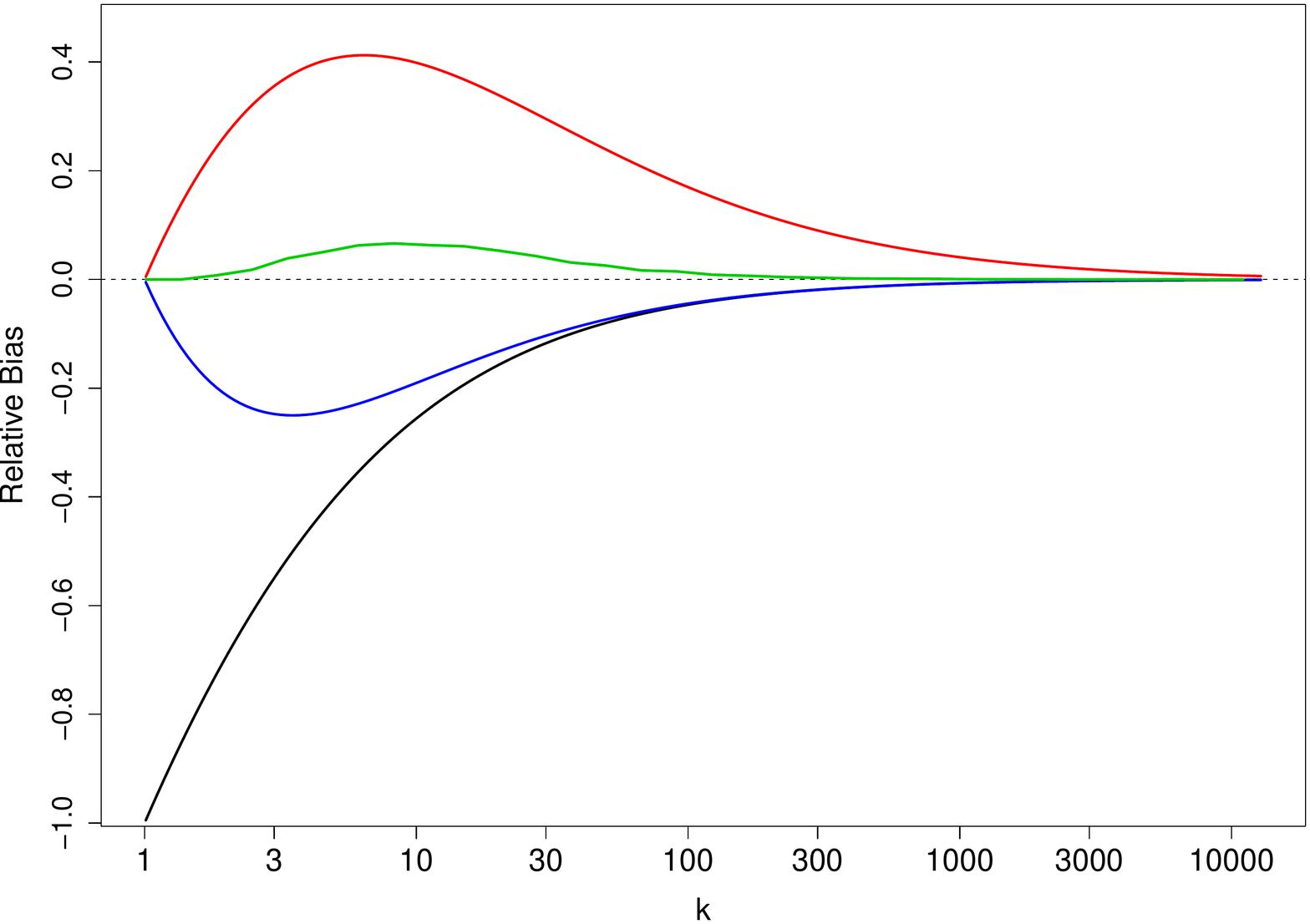}
		\caption{\small\textit{Relative bias of the return-level estimates against the return period, $k$, of the fixed-threshold stopping rule $c_k$ when sampling from the Exponential distribution with the fixed threshold stopping rule using: standard likelihood (red), excluding the final observation (black), full conditioning (green) and partial conditioning (blue).}}
		\label{fig:expanalytical}
	\end{center}
\end{figure}

In Figure~\ref{fig:expanalytical} we compare the relative bias of the
estimates of $\beta^{-1}$ (and hence also for the return-level estimates)
both when including and excluding the final observation
when varying $k$, the true return period of the stopping threshold,
$c_k$. The two additional curves correspond to estimators that will be
introduced in \S\ref{sec:like}. The maximum relative bias in the standard return-level estimator is $0.4$;
\ie , the estimator is around $1.4$ times the true value. This occurs
for a threshold corresponding to $k\approx 7$, \ie , when we stop
sampling if an observation exceeds the 7-observation return-level. Clearly
this will generally result in a very small sample so we would expect
return-level estimates to also be highly variable in this case. 
         
\subsection{Variable-threshold stopping rule with gamma observations} \label{sec:gamma}

The positive bias in return-level estimates that arises from the fixed
stopping rule is partly a result of the geometric distribution of $N$
(see \S \ref{sec:expbias}). For the variable-threshold rule $N$ no longer has a geometric distribution and we find empirically for the GEV (see \S\ref{sec:varthres}) that the bias is typically reduced; as we now show, at least for one parametric family of distributions, the bias disappears entirely.

Let $X_i\sim \mathsf{Gamma}(\alpha,\beta)$, where the shape parameter, $\alpha > 0$, is known but the rate parameter, $\beta > 0$, must be estimated. Notice when $\alpha=1, ~ X_i \sim \mathsf{Exp}(\beta)$, \ie , $X_i$ are GPD with shape parameter 0. In \S \ref{sec:intro2} we noted the need for a historical sample in practice; here, to reflect this, we suppose that the stopping rule is only implemented after an initial sample of independent $\mathsf{Gamma}(\alpha,\beta)$ variables, $X_{-n_0},\dots,X_{-1}$, whose mean is denoted by $\bar{X}_0$, with $\bar{X}_0 \sim \mathsf{Gamma}(n_0\alpha,n_0\beta)$.

As with the exponential distribution, the return-levels of the gamma distribution are proportional to $\beta^{-1}$, with the constant of proportionality depending on $\alpha$. Furthermore the MLE from the full likelihood satisfies $\betahat^{-1}=\xbar/\alpha$. Thus, for some constant of proportionality $\gamma$ (depending on $\alpha$ and the return period, $y$), the variable-threshold stopping rule is equivalent to
\begin{equation}
\label{eqn.stopvargam}
N= \inf\{n\ge 1: X_n>\gamma \Xbar_{n-1}\},
\end{equation}
where
\begin{align}
\label{eqn.meanwithinit}
    \Xbar_{k} = \frac{n_0 \Xbar_0 + X_1 + \hdots + X_k}{n_0 + k} \qquad k \geq 1 \, .
\end{align}

\begin{theorem} \label{thm}

With $N$, $\bm{X}_{1:N}$ and $\Xbar_0$ as defined in \eqref{eqn.stopvargam} and \eqref{eqn.meanwithinit},  
	for all $n\in\mathbb{N}$:
              \begin{align*}
	\Xbar_N|N=n \quad \,{\buildrel d \over =}\,  \quad \Xbar_n \sim \mathsf{Gamma}(\alpha (n+n_0), \beta (n+n_0)) \, .
\end{align*}

\end{theorem}

A proof for this theorem is provided in the Appendix \ref{proof}. From Theorem \ref{thm} we see that $\EE{\Xbar_n|N=n}=\alpha/\beta$, and hence:

\begin{corollary}
\label{cor.conseq}
For a sample obtained as in Theorem \ref{thm}, the sample mean and $y$-year return-level estimate are unbiased:
\begin{align*}
                \EE{\Xbar_N} &= \frac{\alpha}{\beta}.\\
    	\EE{x_y(\alpha, \betahat^{std}(N,\bm{X}_{1:N}))} &= x_y(\alpha, \beta).
\end{align*}
\end{corollary}

Contrasting Corollary \ref{cor.conseq} with
Proposition \ref{prop.expfixed}, both of which apply to the
exponential distribution, we see that the standard
estimator can be unbiased for the variable-threshold stopping rule
even though it is strongly positively biased for the fixed-threshold stopping
rule.

\section{Alternative Methods for Parameter Inference} \label{sec:like}

\subsection{New conditional likelihoods} \label{NewLike}



Motivated by the lack of bias in the stopping rule of \cite{Molenberghs}, we propose a similar estimator for our scenarios by conditioning on the fact that only the final
observation met the stopping criterion. The likelihood, therefore,
consists of the conditional
densities of the data values given that each of the first $n-1$ is
outside its stopping region and the $n$th is inside its stopping
region. The log likelihood, $\ell_{fc}$, is as given in \eqref{eqn:likeall} and the corresponding estimate is denoted $\hat{\theta}_{fc}$.
For the fixed-threshold stopping rule this effectively
conditions out the geometric distribution for $N$ (\S\ref{sec:expbias}); it might be hoped,
therefore, that it might remove that part of the positive bias that is
due to the randomness of $N$. 


By conditioning on the final observation exceeding its stopping threshold and all other observations not exceeding theirs we are effectively losing all of this information which will lead to larger uncertainty in
 the estimates, \eg , giving wider
 confidence intervals. Hence, we  consider a further likelihood which
 conditions \textit{only} on the fact that the final observation
 exceeds its threshold
. Like full conditioning, this results in the stochasticity of $N$ being less influential. We refer to this method as partial conditioning with log likelihood, denoted by $\ell_{pc}$, given in \eqref{eqn:likefinal}. The corresponding estimate is denoted by $\hat{\theta}_{pc}$.

In summary, the two new log likelihoods we consider are:
\begin{eqnarray}
	\ell_{fc}(\theta;n,\bm{x}_{1:n}) &=& \ell_{std}(\theta;n,\bm{x}_{1:n}) - \log \bar{F}(s_{k,n};\theta) - \sum_{i=1}^{n-1} \log F(s_{k,i};\theta) \label{eqn:likeall} \\
	\ell_{pc}(\theta;n,\bm{x}_{1:n}) &=& \ell_{std}(\theta;n,\bm{x}_{1:n}) - \log \bar{F}(s_{k,n};\theta) \label{eqn:likefinal} 
\end{eqnarray}

where $s_{k,i}$ is the lower boundary of the stopping set for the $i$th observation; then for the variable-threshold stopping rule $s_{k,i} = \hat{x}_k^{std}(\bm{x}_{1:i-1})$, \ie , it is the standard estimate of the $k$-year return-level using all the
data up to and including the previous observation, and for the fixed-threshold stopping rule $s_{k,i} = c_{k}$ for all $i$.

The examples of \S 2 have exponential tails with an unknown scale parameter. When data values are modelled using the GEV, uncertainty in the shape parameter, $\xi$, has a much larger impact on estimates of high quantiles than the uncertainty in the other two parameters, $\mu$ and $\sigma$ \citep{Coles}. So we now consider estimation of the shape parameter and high quantiles using the standard and partial conditioning likelihoods. For simplicity we focus on an idealised scenario where we take $\mu=0$ and $\sigma=1$ as known, 
so $X$ has a distribution function of $F(x;\xi)=\exp\left(-[1+\xi x]_+^{-1/\xi}\right)$ and a survivor function of $\bar{F}=1-F$. 

For quantile estimation the standard likelihood estimator of $\xi$, \ie , $\hat{\xi}_{std}$, leads to a positive bias for high quantiles. This can be seen as follows. The $y$-year return level can be written as
$\bar{F}^{-1}(1/y;\xi)=[\exp(a_y\xi)-1]/\xi$, with $a_y = -\log[-\log(1-1/y)]$, where $a_y\ge 0$ provided
$y\ge e/(e-1)\approx 1.6$. Return levels as low as 1.6 years are of no practical interest in our setting. When $a_y>0$, $\bar{F}^{-1}$ is an increasing, convex function of $\xi \in \RR$, so, whatever the likelihood, Jensen's inequality gives
$\mathbb{E}_{\xi}\left[\bar{F}^{-1}(1/y;\xi)\right]
\ge
\bar{F}^{-1}(1/y;\mathbb{E}_{\xi}\left[\xi\right])$.
The monotonicity of $\bar{F}^{-1}$ implies that even if $\hat{\xi}_{std}$ is unbiased, we should expect a positive bias in all quantile estimates, and this will only be exaggerated if (as we find in our stopping-rule simulations) $\xi$ is positively biased.

This bias in the estimator for high quantiles is guaranteed to be less positive when using the partial conditioning likelihood rather than the standard likelihood. To see this first note that $\ell_{pc}(\xi) = \ell_{std}(\xi) - \log[\bar{F}(c;\xi)]$ where $c$, the stopping threshold, has been standardised. The resulting MLE, $\hat{\xi}_{pc}$, satisfies $\ell(\hat{\xi}_{pc}) - \log[\bar{F}(c;\hat{\xi}_{pc})] > \ell(\xi) - \log[\bar{F}(c;\xi)] ~ \forall ~ \xi$. Also, $\bar{F}$ is an increasing function of $\xi$ since
\[
\frac{\partial}{\partial \xi}\log\left\{-\log F(x;\xi)\right\}
=
\frac{1}{\xi^2}\left\{\log[1+\xi x] - \frac{\xi x}{1+\xi x}\right\}\ge 0, \qquad \forall \xi x > -1 \, .
\]
So, as $\ell(\hat{\xi}_{std}) > \ell(\xi) ~ \forall ~ \xi$, it follows that
\begin{align*}
\ell(\hat{\xi}_{std}) - \log[\bar{F}(c;\hat{\xi}_{pc})] &> \ell(\hat{\xi}_{pc}) - \log[\bar{F}(c;\hat{\xi}_{pc})] > \ell(\hat{\xi}_{std}) - \log[\bar{F}(c;\hat{\xi}_{std})] \\
\Rightarrow - \log[\bar{F}(c;\hat{\xi}_{pc})] &> -\log[\bar{F}(c;\hat{\xi}_{std})] \\
\Rightarrow \hat{\xi}_{pc}<\hat{\xi}_{std}.
\end{align*}


Thus, if the standard estimator is positively biased the partial conditioning method will be less positively biased. Given that this effect is magnified for return levels, as shown above, we should expect improvements in return level estimates using the partial conditioning likelihood. An analogous argument to the above also applies to data modelled using the GPD, except that there is no restriction on $y$, and $a_y=\log y$.

\subsection{Application to exponential observations}

Consider iid sampling from the exponential distribution with rate parameter, $\beta$, using the fixed-threshold stopping rule. The relative bias for return-level estimators using the log-likelihoods \eqref{eqn:likefull} and \eqref{eqn:likeex}, are detailed in Proposition \ref{prop.expfixed} and
plotted in
Figure~\ref{fig:expanalytical}. Figure~\ref{fig:expanalytical} also
plots the relative bias for the likelihoods in \eqref{eqn:likeall} and \eqref{eqn:likefinal}, the latter has the form 
\begin{align}
\relb(\hat{x}_y^{pc}) = \relb(\hat{x}_y^{std}) - \frac{\beta^2 c_k^2}{e^{\beta c_k} - 1} = \frac{\beta c_k}{e^{\beta c_k} -1} \left[ \frac{\beta c_k}{1 - e^{-\beta c_k}} - 1 - \beta c_k \right] \, .
\end{align}

Estimator $\hat{x}_y^{pc}$ is
negatively biased but the bias is smaller than that for $\hat{x}_y^{ex}$. The bias of $\hat{x}_y^{pc}$ tends to $0$ as $c_k$ tends to infinity at the same fast rate as for $\hat{x}_y^{ex}$ (\S\ref{sec:expbias}).

We were unable to obtain a tractable expression for the bias of the full-conditional estimator. In Figure~\ref{fig:expanalytical} this bias was found using Monte Carlo methods.
The bias is very
low and tends towards 0 much faster than any of the other estimators
considered. 
This finding is similar to that of \cite{Molenberghs} 
for the mean of normally distributed observations with a probabilistic stopping rule;
however, the MSE of the unbiased estimator was found to be poor
compared to that of the standard estimator. In \S\ref{sec:GEV} we show that in our `extremes' setting, the full-conditional MSE for a return-level is often lower relative to the MSE of the standard estimator since the high variance of
return-level estimators using the standard likelihood is in part due to
the final observation being large. Furthermore in \S\ref{sec:GEV} we show that the partial conditioning approach results in estimators with much reduced variance and that this leads to lower MSE compared to the standard likelihood approach.

\section{Simulation results} \label{sec:GEV}

In this section we focus on the return-level inference when sampling from the GEV distribution with the two stopping rules of \S\ref{sec:intro}. In \S\ref{sec:fixed} we calculate the fixed stopping threshold, $c_k$, for a range of return periods, $k$, using \eqref{eqn:GEVret} and our knowledge of the true parameters $\mu, \sigma$ and $\xi$. In \S\ref{sec:varthres} we consider the variable threshold stopping rule over a range of $k$. Similar simulation results are given in the supplementary material for the GPD.  

\subsection{Simulation design}
We investigate true return-periods, $k$, between $20$ and $2000$. 
When generating the data, for each $k$, for the fixed-threshold stopping rule, we set $c_k$ to be the true $(1-1/k)$th quantile of the data-generating distribution (\ie , the $k$-yr return-level) whereas for the variable-threshold rule the threshold is the estimated $(1-1/k)$th quantile; with both rules we stop at the first exceedance. In the simulation study, for each combination of $\theta$, stopping
rule and $k$, a large number of data sets were simulated to evaluate the RMSE, bias and variance of the estimators. Given the likelihood $\ell_{\method}$ for $\method\in\{\mbox{std},\mbox{ex},\mbox{fc},\mbox{pc}\}$, detailed in equations \eqref{eqn:likefull}, \eqref{eqn:likeex}, \eqref{eqn:likeall} and \eqref{eqn:likefinal}, profile likelihood confidence intervals for a return-level are studied in terms of their coverage and width. 


One major issue with simulating data sets with stopping rules is parameter identifiability. For observation $i$, the stopping decision of the variable-threshold rule is based on the parameter MLEs using observations $1,\dots,i-1$. 
However, with $N\leq 2$ observations contributing to a likelihood the GEV parameters are strictly not identifiable, and for larger but low values of $N$ the parameters are still not practically estimable. As discussed in \S 2.2, in practice there is typically additional information which is incorporated into decisions, and our analysis should allow for this also. Such historical information is treated as {\it fixed} and introduces a fixed extra penalty term, $P_{hist}(\theta)$, into the log-likelihood; in a Bayesian analysis it would constitute prior information about the parameter vector. As our simulation studies are conducted without such evidence we treat the first $n_0$ simulated values as providing historical information on $\theta$; we call  $\xhist:=(x_1,\dots,x_{n_0})$ the \emph{historical data}. Thus each simulated data set has the penalty contribution to the likelihood: 


\begin{equation}
  \label{eqn.Penalty}
  P_{hist}(\theta)=\sum_{j=1}^{n_0}\ell(\theta;x_{j}),
\end{equation}
a contribution that does not depend on the stopping rule since we imagine that these data were available \emph{before} decisions to stop and analyse the data were being made.



We fix the historical data, 
$\xhist$, using an even spread of values:
\begin{align} \label{eqn:initial}
	x_{j} = G^{-1}(j/(n_0 + 1);\theta) \qquad \mbox{for} \quad  j = 1,...,n_0
\end{align}
where $G$ is the distribution function of the data-generating GEV
distribution. In addition to providing a natural spread of values and
stabilising the likelihood, for the fixed-threshold rule, provided $c_k$ is greater than the $1/(n_0+1)$ return-level, no historical value 
exceeds the stopping threshold. The stopping threshold, $\xhat_k(\bm{x}_{1:i-1})$ is
now, implicitly, also a function of $\xhist$. We take $n_0=10$, the smallest value that gave reliable numerical estimates for $\xhat_k(\bm{x}_{1:i})$ with $i \geq n_0$, across the set of different true values for $\theta$ that were used in the simulation study. 


\subsection{Fixed threshold stopping rule} \label{sec:GEVret}\label{sec:fixed}

\begin{figure}[t]
\centering
		\subfigure{\includegraphics[width=0.4\textwidth]{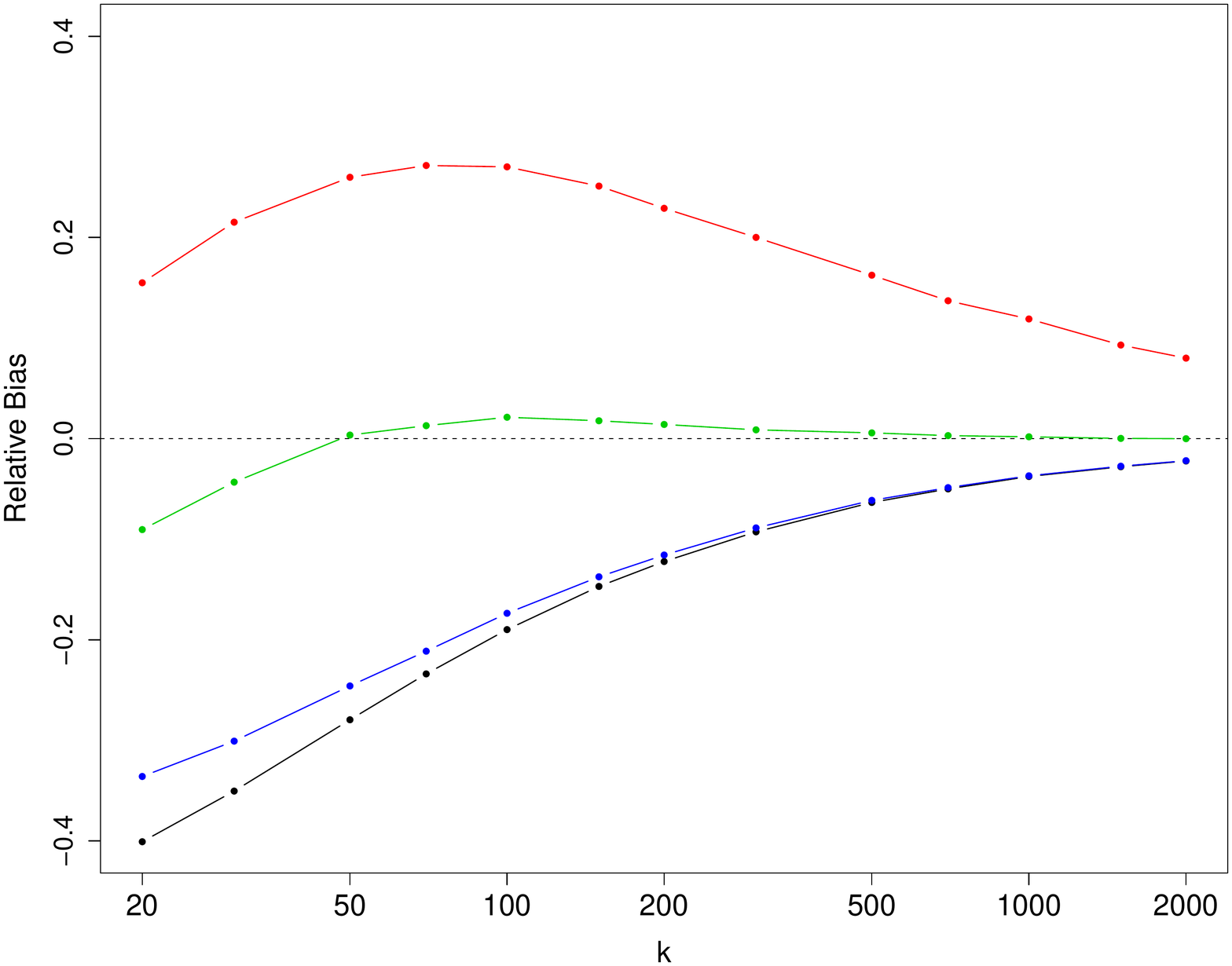}} \hspace*{3em}
		\subfigure{\includegraphics[width=0.4\textwidth]{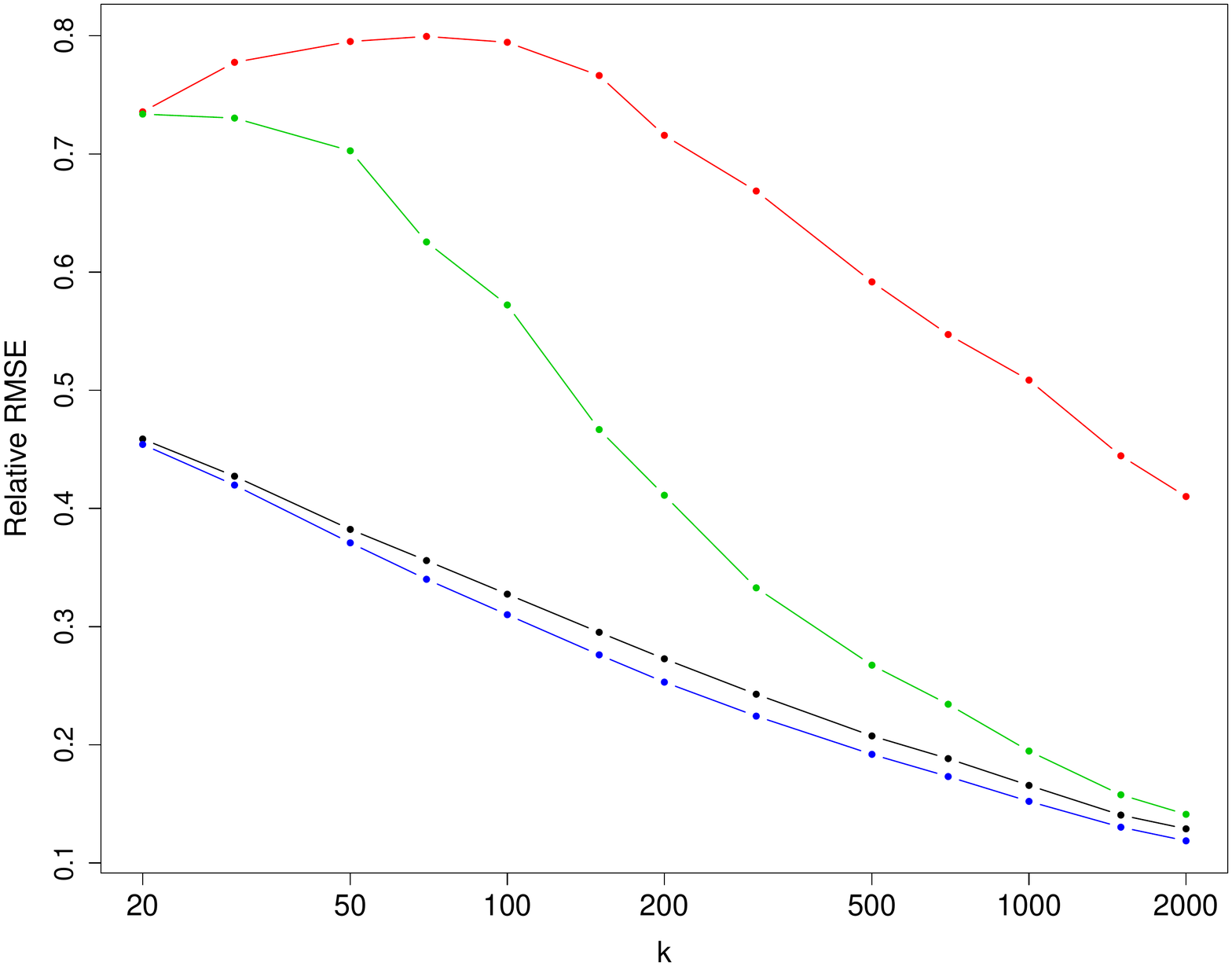}}
\hspace*{2em}
		\subfigure{\includegraphics[width=0.4\textwidth]{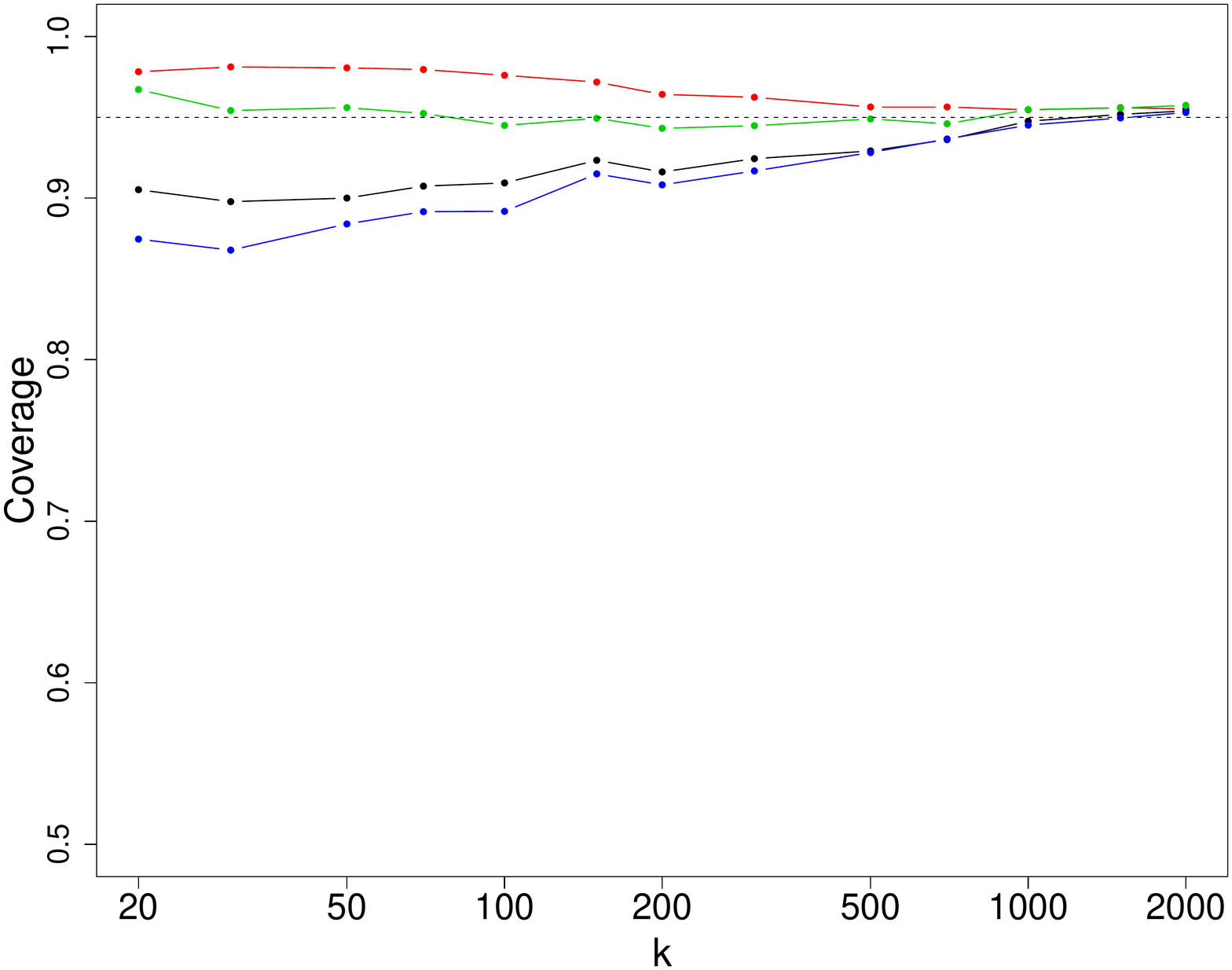}} \hspace*{3em}
		\subfigure{\includegraphics[width=0.4\textwidth]{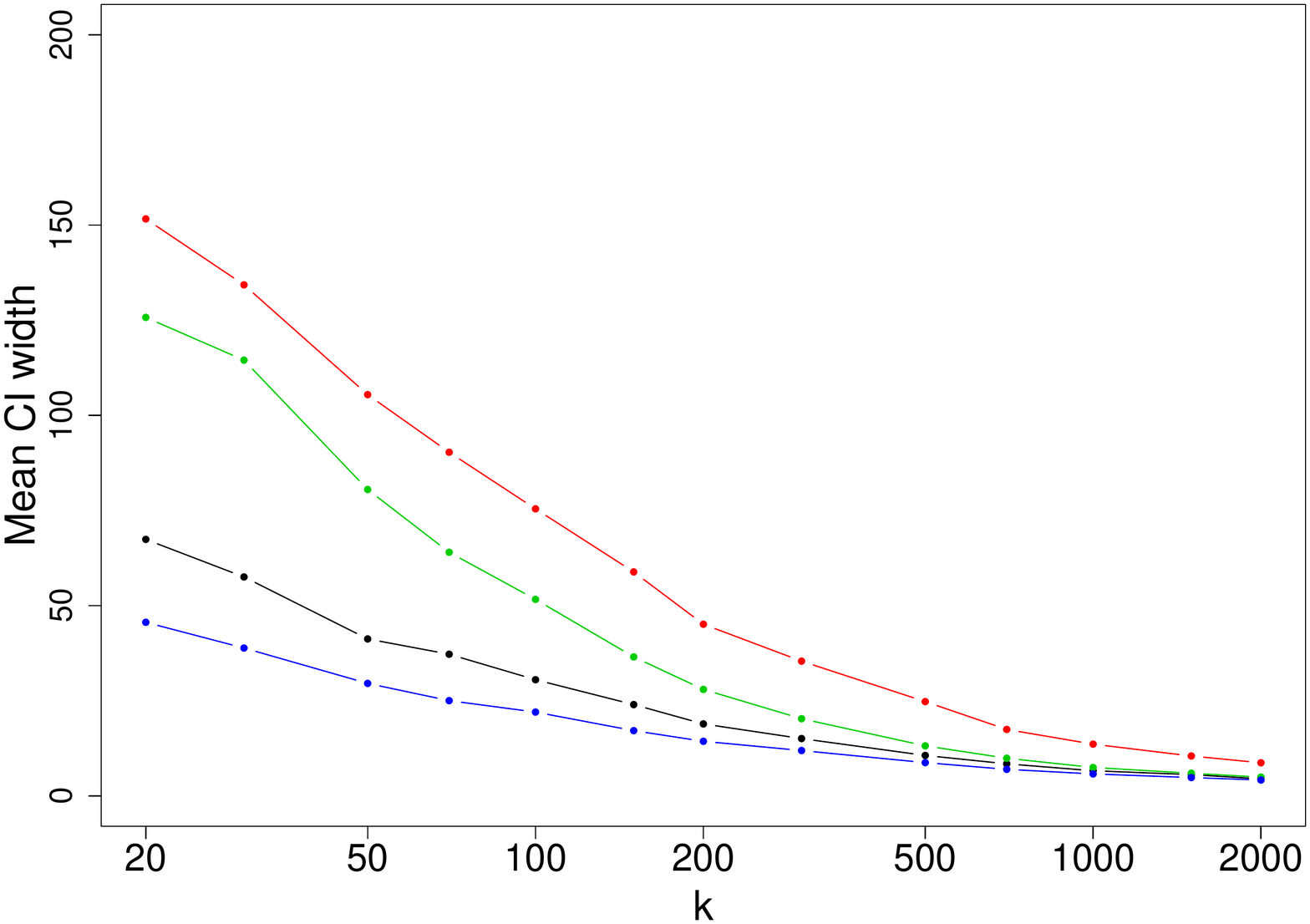}}
\hspace*{2em}
		\captionit{200 year return-level estimates when sampling from the GEV distribution with $(\mu, \sigma, \xi)=(0, 1, 0.2)$ using the fixed-threshold stopping rule over a range of thresholds. From left to right. Top: relative bias and relative RMSE. Bottom: coverage and average CI width. Colour scheme is the same as in Figure \ref{fig:shape}. Based on $10^5$ replicated samples with the historical data created using approach~\eqref{eqn:initial}. Coverage is based on 5000 replicated samples.}
		\label{fig:200ret}
\end{figure}

In the appendix \S\ref{sec:shape} we describe in detail the behaviour of the shape parameter estimates in our simulations. In particular we found the standard shape parameter estimator has both large positive bias and large variance. The formulae in \S\ref{NewLike} show that return level estimates are exponential in the shape parameter. For high return periods moderately large $\xi$ estimates can lead to unrealistically high return-level estimates which exert unwarranted influence on statistics based on empirical averages, such as estimated bias. Hence, we use trimmed averages here.

Figure~\ref{fig:200ret} shows the relative bias, variance and RMSE of the 200 year return-level estimators when sampling using the fixed-threshold stopping rule from the GEV distribution with $\xi=0.2$. Similar sets of plots for the 50 and 1000 year return-level estimator and $\xi=-0.2$ can be found in the supplementary material. The main driver of RRMSE in all cases is found to be the variance of the estimators, so changes in bias are not too important in this regard. Overall, the return-level estimator which results in the lowest RRMSE most consistently is $\hat{x}_{y}^{pc}$, mostly due to the low variance of these estimates whereas $x_y^{fc}$ has the lowest bias. Both conditioning estimators, $\hat{x}_{y}^{pc}$ and $\hat{x}_{y}^{fc}$, improve upon the $\hat{x}_{y}^{std}$ especially when we are estimating very high return-levels (\ie , for larger $y$) and/or the underlying distribution is heavy tailed. Although $\hat{x}_{200}^{ex}$ has somewhat similar properties to $\hat{x}_{200}^{pc}$ for $\xi=0.2$ it has larger RRMSE for $\xi=-0.2$. The fitted distribution  using $\ell_{ex}$ typically has a lighter tail and can even have an upper end point which is less than the excluded observation. 

The coverage for all likelihoods gets closer to the correct value (here 95\%) as $k$ increases for any return period, $y$. 
For $\ell_{std}$ we have overcoverage 
and the widest confidence intervals on average and using $\ell_{fc}$ we have good coverage, particularly when the distribution is heavy tailed. 
For the other likelihoods there is mostly undercoverage (coverage ranging from 80-95\%) due to upper bounds being too low. 
The exclusion of upper tail information results in relatively narrow confidence intervals from $\ell_{ex}$. 
In contrast, $\ell_{fc}$ produces a higher upper confidence limit and hence a wider confidence interval than $\ell_{pc}$ and $\ell_{ex}$ because the likelihood essentially neglects the distribution of $N$, \ie , the threshold exceedance counts, which contain some information about the upper tail of the distribution. The confidence intervals produced using $\ell_{fc}$ vary greatly in width across our simulations with a larger median width than those using $\ell_{std}$. 


\subsection{Variable threshold stopping rule} \label{sec:varthres}

\begin{figure}
\centering
		\subfigure{\includegraphics[width=0.4\textwidth]{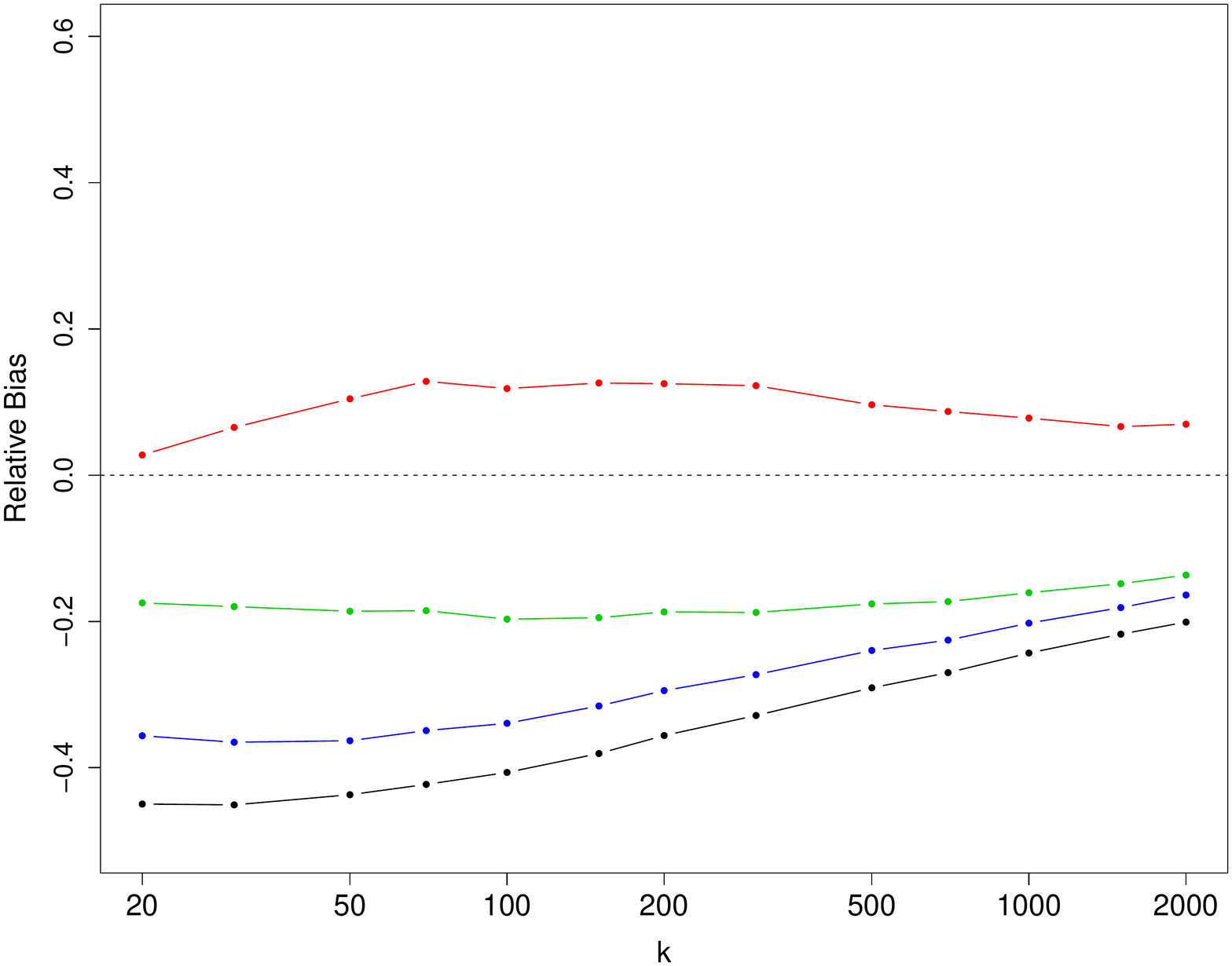}} \hspace*{2em}
		\subfigure{\includegraphics[width=0.4\textwidth]{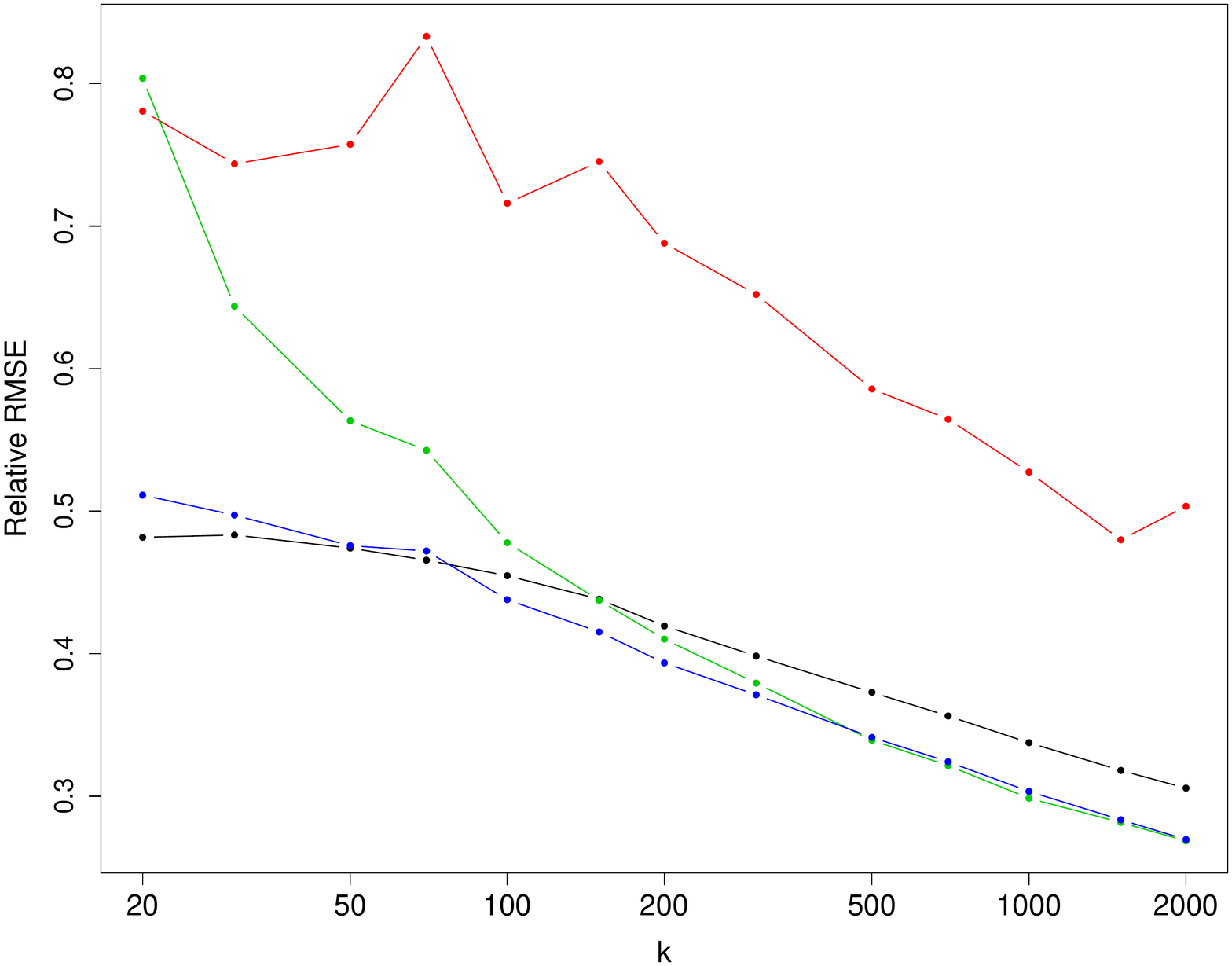}}
\hspace*{2em}
		\subfigure{\includegraphics[width=0.4\textwidth]{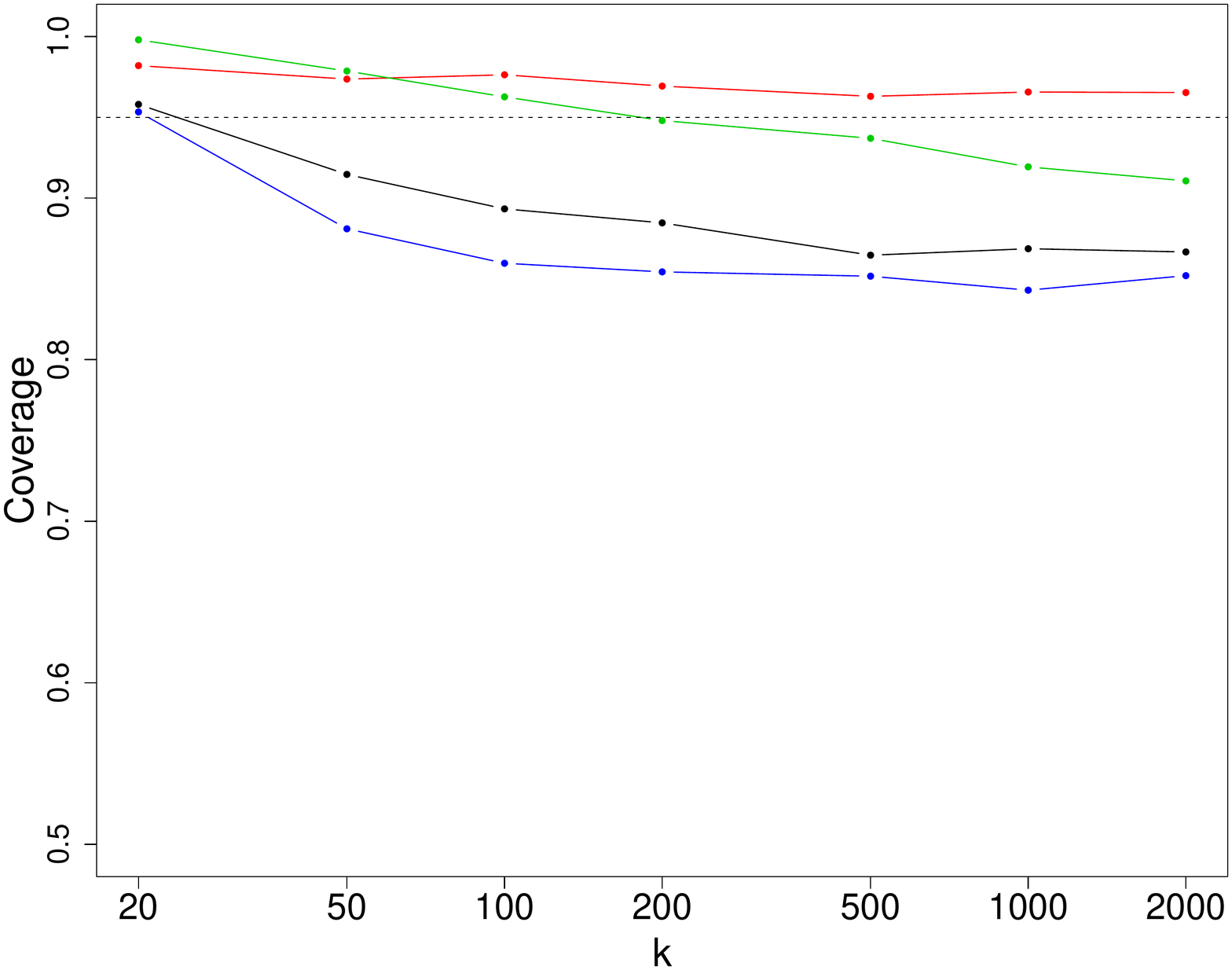}} \hspace*{2em}
		\subfigure{\includegraphics[width=0.4\textwidth]{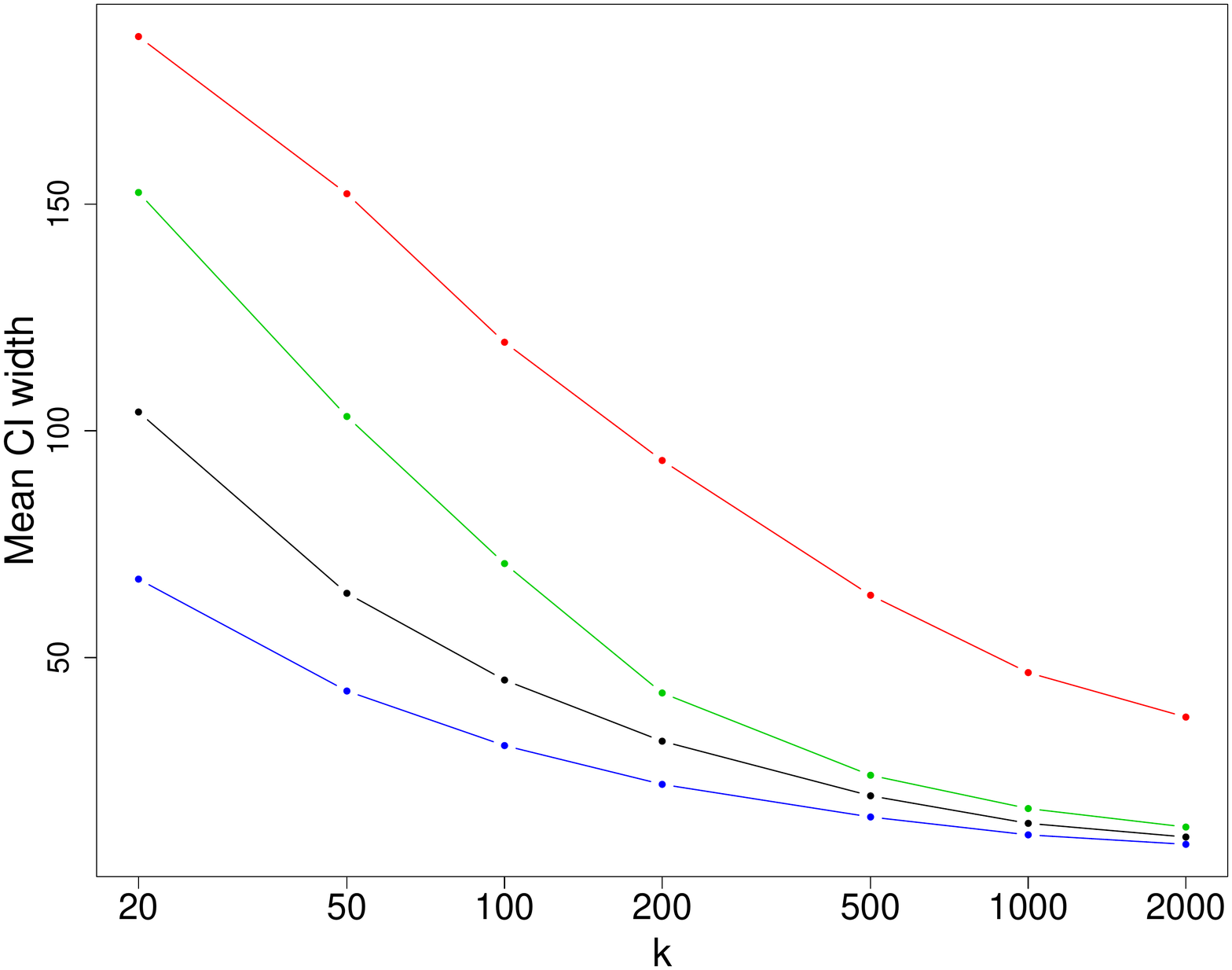}}
\hspace*{2em}
		\captionit{200 year return-level estimates when sampling from the GEV distribution with $(\mu, \sigma, \xi)=(0, 1, 0.2)$ using the variable-threshold stopping rule over a range of $k$. See Figure~\ref{fig:200ret} for associated detail. Based on $10000$ replicated samples with the historical data created using approach~\eqref{eqn:initial}. Coverage is based on 3000 replicated samples.}
		\label{fig:200retvar}
\end{figure}

Within the samples simulated we find the stopping thresholds, $\hat{x}_k(x_{1:m})$, over $m<N$ are generally less than the true $k$-year return level, $x_k$. As a result the samples are both smaller in size and consist of smaller values than when using the fixed-threshold stopping rule. So return level estimates calculated using $\ell_{std}$ have a small positive bias and those calculated using the other three likelihoods have a larger negative bias than observed for the fixed-threshold stopping rule. 

The properties of the 200 year return-level estimators for $\xi=0.2$ are shown in Figure~\ref{fig:200retvar}, the 50 and 1000 year return-levels and $\xi=-0.2$ are considered in the supplementary material (Figures~\ref{supp:fig:50retvar}-\ref{supp:fig:1000retvarneg}). For $\xi=0.2$ the conditioning methods provide the best return-level estimators in terms of RMSE despite the estimators having a larger squared bias than $\hat{x}_{y}^{std}$. The reason for this is that for heavy tailed distributions the variances of return-level estimators are generally larger than the bias. 
However, for lighter tailed distributions the bias plays a larger role as the relative variances of the different estimators are much closer together. As a result $\hat{x}_{200}^{pc}$ and $\hat{x}_{200}^{fc}$ can perform marginally worse than $\hat{x}_{200}^{std}$ in terms of RRMSE when the distribution has a light tail.


The coverage for $\ell_{std}$ is high (96-98\%), decreasing only slightly as $k$ increases but it has the widest confidence intervals generally. Using either $\ell_{pc}$ or $\ell_{ex}$ leads to undercoverage, 
as $k$ increases ranging from approximately 95\% to 83-87\% for $\ell_{ex}$ and from 93\% to 78-85\% for $\ell_{pc}$ with coverage higher when the distribution is heavy tailed.
The coverage for $\ell_{fc}$ also reduces with increasing $k$ from $99-100\%$ for $k=20$ to approximately $90\%$ for larger $k$. On average the confidence intervals using $\ell_{fc}$ are narrower than using $\ell_{std}$ but generally wider than those using $\ell_{ex}$ or $\ell_{pc}$. 

Overall, $\ell_{fc}$ provides the `best' results when using the variable-threshold stopping rule. The RRMSE of $\hat{x}_{200}^{fc}$ is generally lower than that of $\hat{x}_{200}^{std}$, coverage is above $90\%$ and the confidence intervals are narrower on average than those using the $\ell_{std}$. Although $\ell_{pc}$ provides estimators with a lower RRMSE than $\ell_{fc}$, particularly when the distribution is heavy tailed, it has more severe undercoverage. 


\subsection{Use in Practice}

In practice, for the analysis of data that we believe has been obtained by the flood management agencies using a the fixed-threshold rule we must set a threshold, $c$, and if they use a variable-threshold rule we must set a return period, $k$, neither of which may be known. This is important since the behaviour of the estimators can vary depending on the return period, $k$, associated with the stopping threshold (as we have seen in \S\ref{sec:fixed} and \ref{sec:varthres}). For the fixed-threshold rule, $c$ should lie between $\max_{i<n}x_i$ and $x_n$. For the variable-threshold rule $k$ should be such that $x_i\le \xhat_k(x_{1:i-1})$ for all $i<n$, but $x_n> \xhat_k(x_{1:n-1})$. To use the simulation study results to understand the properties of the estimators it is useful to narrow down a range of feasible $k$. For the variable rule, a range of possible $k$ can be determined from the data. However, for the fixed-threshold stopping rule $k$ is unknown. Nevertheless, we are likely to have some idea of the range of $k$ which corresponds to $c$, \ie , we have a prior belief for $k$.

The `historical data' also needs to be determined, maybe incorporating prior knowledge in some way. The simplest approach is to start using the stopping rule after the first $n_0$ observations of the data set and use these $n_0$ values as the historical data. 
The choice of $n_0$ only affects the point estimates and confidence intervals using $\ell_{fc}$. However $n_0$ and the historical data itself can have a large impact on the properties of the estimators, particularly when the sample size is small. In the simulation study, out of necessity, we have restricted ourselves to a particular fixed historical sample, so for low $k$ the properties of the estimators will differ slightly in practice. 



\section{Case Study - Lune at Caton} \label{sec:Lune}

We now consider the analysis of the 48 annual maximum river flow observations from the Lune at Caton introduced in \S\ref{sec:intro}. 
Figure~\ref{fig:Lune}, right panel, shows the inference for the 200 year return-level of the data, at yearly intervals as new data are observed, with the analysis not accounting for any stopping rule. We now estimate this return-level using the four inference methods (standard, exclude, and our full-  and 
partial-conditional) for both fixed- and variable-threshold stopping rules for a range of levels ($c$ and $k$ respectively), where we drop the subscript of $c$ as the return period of the stopping threshold is unknown.  The following discussion assumes that the sampling procedure is well approximated by these respective stopping rules for the selected $c$ and $k$. In all cases we take the historical data to be the first $n_0=10$ observations as in practice no estimates of long period return levels would be attempted from smaller samples. 
We also consider the implications if a trend in the annual maxima is also simultaneously estimated.

\subsection{Fixed-threshold stopping rule} \label{sec:CS}

First we discuss the inference using the fixed-threshold stopping rule with $c=1568m^3/s$, where, for illustration purposes, $c$ is taken to be the mid-point between the 1995 and 2015 levels and the realised value of $N$ is 38, \ie , we stop after 2015. Figure~\ref{fig:Lune}, right panel, to the right of the vertical dotted line, shows the estimates and the associated 95\% confidence intervals for the four inference methods. The estimates $\hat{x}_{200}^{std}$ and $\hat{x}_{200}^{ex}$ are identical to the estimates in the right panel of the figure for years 2015 and 2014 respectively. 
Both $\hat{x}_{200}^{fc}$ and $\hat{x}_{200}^{pc}$ (evaluated at 2015) are only slightly larger than the $\hat{x}_{200}^{std}$ estimates for the years before 2015 and $\hat{x}_{200}^{ex}$, despite the inclusion of the 2015 value. From \S\ref{sec:GEVret} we know that, when employing the fixed-threshold stopping rule, $\hat{x}_{200}^{std}$ is positively biased, $\hat{x}_{200}^{fc}$ is close to being unbiased and both $\hat{x}_{200}^{ex}$ and $\hat{x}_{200}^{pc}$ have some negative bias, therefore it is reassuring to see that $\hat{x}_{200}^{std} >> \hat{x}_{200}^{fc}>\hat{x}_{200}^{pc}>\hat{x}_{200}^{ex}$. 

The confidence interval for 2015 using $\ell_{std}$ is wider than the intervals of the previous 15 years, especially the 2014 interval (\ie ,~using $\ell_{ex}$), and both the lower and upper bounds are much larger. In this case study, the confidence interval of $x_{200}$ using $\ell_{pc}$ is similar but slightly narrower than when using $\ell_{ex}$. However using $\ell_{fc}$ the interval is wider (since the upper bound increases) than if we just ignored the 2015 event (using $\ell_{ex}$) so we are capturing some of the increased uncertainty in the heaviness of the tail that this event has caused. 
Nevertheless, the upper confidence bound of $x_{200}$ is lower than that using $\ell_{std}$.

The behaviour of the confidence intervals of these methods appears to be in line with our coverage and width results in \S\ref{sec:GEVret}. Indeed, here the shape parameter estimates, $(\hat{\xi}_{std},\hat{\xi}_{ex},\hat{\xi}_{fc},\hat{\xi}_{pc})$, are $(0.04, -0.07, -0.04, -0.05)$ so we expect coverage to be between the coverage values found in the simulation study for $\xi=0.2$ and $\xi=-0.2$. In the study we found that using $\ell_{std}$ with the fixed-threshold stopping rule leads to overcoverage (95-98\% for $\xi=0.2$, 97.5-99.5\% for $\xi=-0.2$) and the upper bound of the confidence interval found using $\ell_{std}$ is lower than $x_{200}$ only 1-2\% of the time, so it is likely that for the Lune data the upper bound of the confidence interval using $\ell_{std}$ is too high. This is further emphasised for the Lune estimates by the upper bound for 2015 exceeding the associated values for the previous 30 years (Figure~\ref{fig:Lune}). In \S\ref{sec:GEVret} we found that $\hat{x}_{200}^{ex}$ and $\hat{x}_{200}^{pc}$ exhibited narrow confidence intervals which together with their negative bias led to undercoverage, with the upper bounds being too low, especially when $\xi=-0.2$ and $k$ is low. For our chosen $c=1568$ we can obtain estimates of the corresponding return period, $k$, of $c$; in particular $\hat{k}_{std}=90$ and $\hat{k}_{ex}=550$ and we expect $k$ to lie between these two values. Thus, using the simulation study results, we expect that the coverage of the $\ell_{pc}$ and $\ell_{ex}$ confidence intervals to lie between 85 and 95\%. However the lower bounds of these confidence intervals were found to be less than $x_{200}$ for almost $100\%$ of simulated samples so it is highly likely that the true 200-year return level for the Lune data is above the lower bounds given by the $\ell_{pc}$ and $\ell_{ex}$ confidence intervals. For $\ell_{fc}$ and $90<k<550$, the coverage is 94-95\% with the percentage of upper bounds too low being 3-6\% suggesting that with the Lune data the upper bound of the $\ell_{fc}$ confidence interval is likely to be higher than the true $200$-year return level, $x_{200}$.

\begin{figure}[t]
	\begin{center}
		\subfigure{\includegraphics[width=0.45\textwidth]{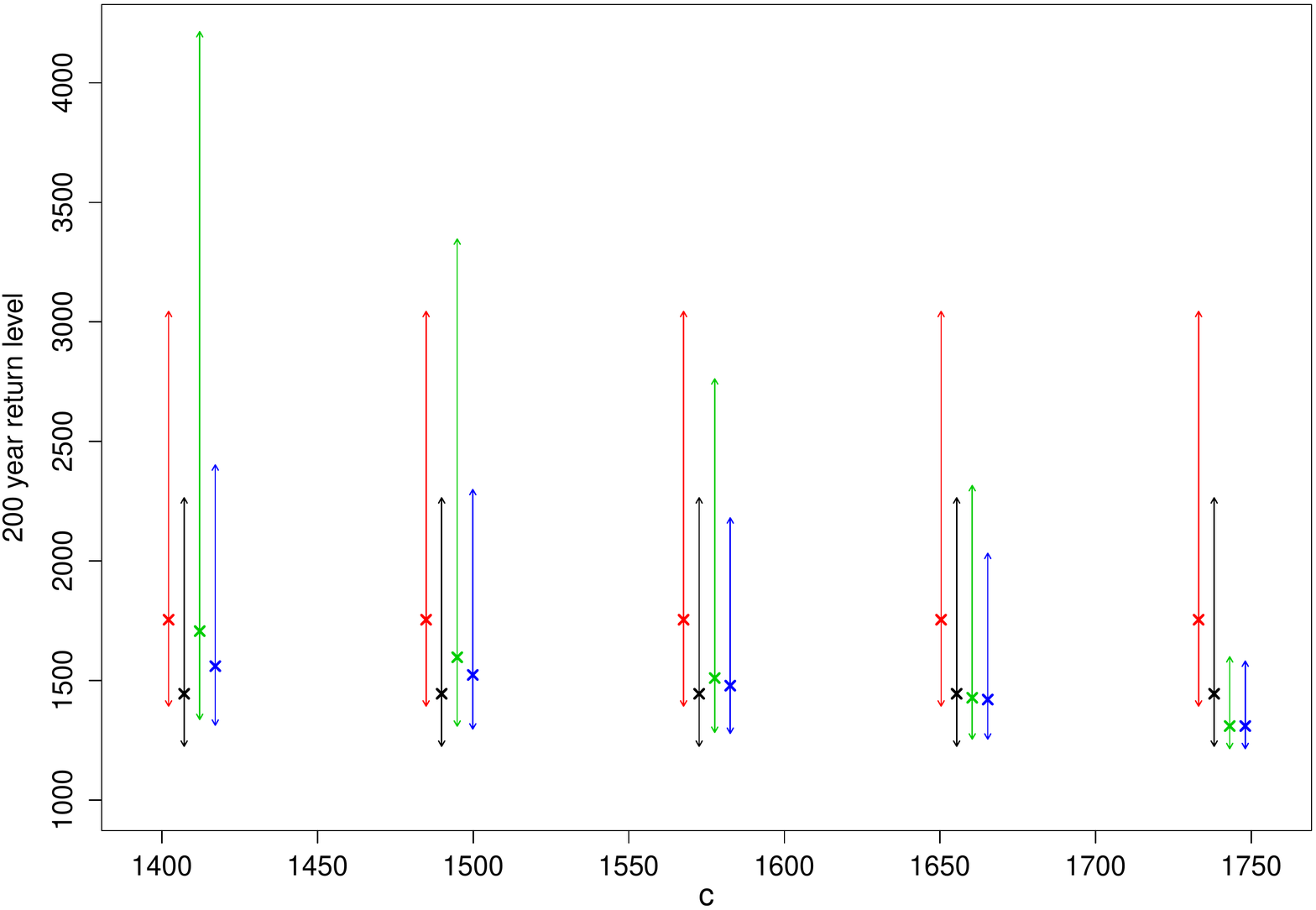}}
		\subfigure{\includegraphics[width=0.45\textwidth]{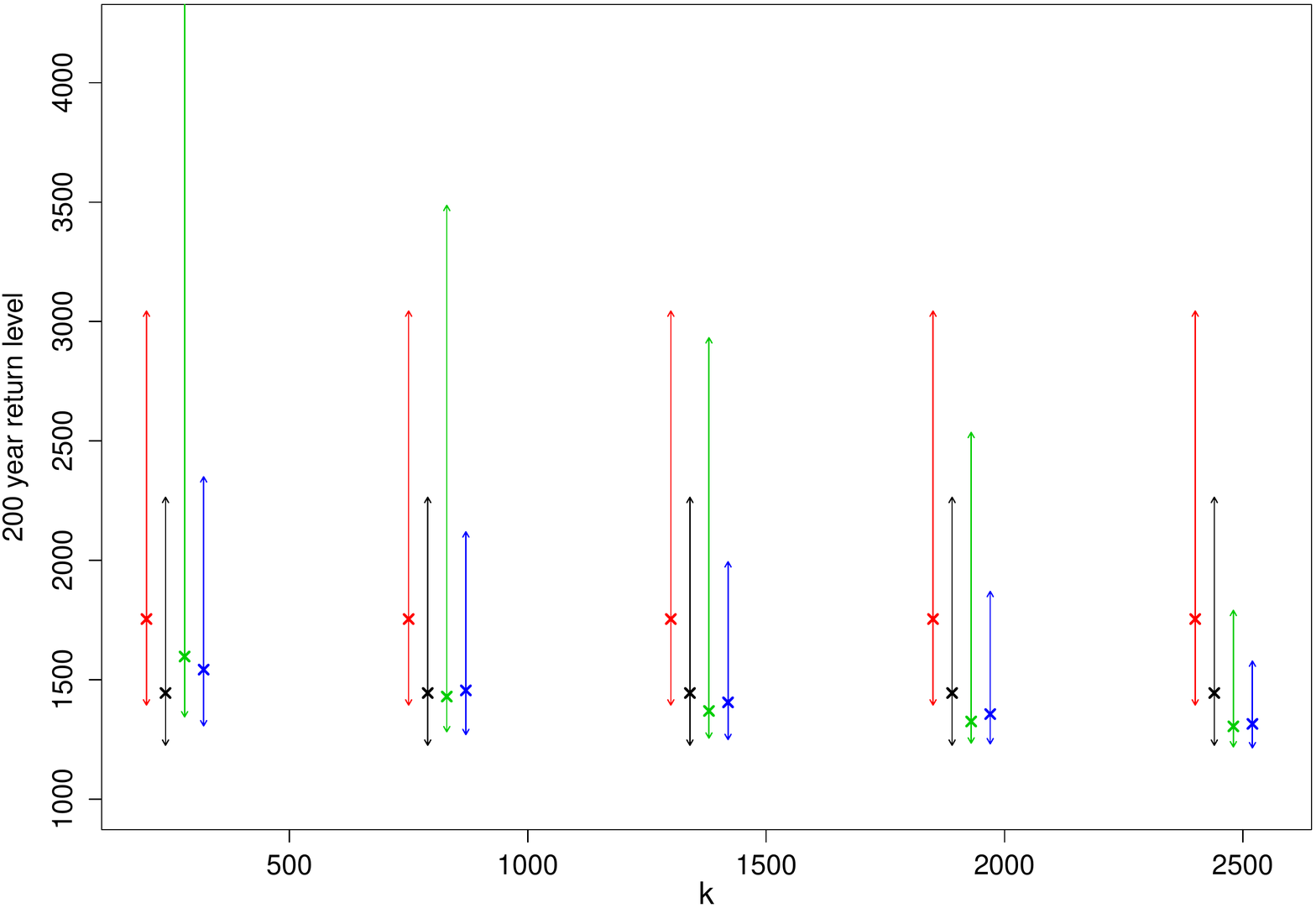}}
		\captionit{200-year return-level estimates based on all the data up to and including 2015 for the Lune at Caton with 95\% profile likelihood-based confidence intervals: left with the fixed-threshold stopping rule over a range of $c$ and right with the variable-threshold stopping rule over a range of $k$: standard likelihood (red), excluding the final observation (black), full conditioning (green) and partial conditioning (blue). Each group of 4 estimates applies for the same $c$/$k$ as for the standard estimate in each group and have been horizontally shifted for clarity.}
		\label{fig:changingc}
	\end{center}
\end{figure}

The above discussion assumed that $c$ was known. In some cases this may be true as $c$ could represent a known physical limit linked to flooding. This is not the case for the Lune at Caton, with our value chosen subjectively for illustrative purposes although it could be argued that lower $c$ values in this range would be more reasonable since the 1995 river flow observation was considered high as it led to flooding. To assess the impact of $c$ we consider a range of values for $c$ between the 1995 and 2015 observations, with the inference for the four methods presented in Figure~\ref{fig:changingc}, left panel (when $c=1568$ the estimates are those shown in Figure~\ref{fig:Lune} right panel). 
Now, $\hat{x}_{200}^{std}$ and $\hat{x}_{200}^{ex}$ and the corresponding confidence intervals are invariant to $c$ but as $c$ increases $\hat{x}_{200}^{fc}$ and $\hat{x}_{200}^{pc}$ both decrease. 
As noted earlier, $\hat{x}_{200}^{fc}>\hat{x}_{200}^{pc}$ but they become closer as $c$ tends to the 2015 event level because the information that $\ell_{fc}$ discards, \ie , the probability of the event that $c$ was not exceeded on the first $n-1$ observations, becomes less informative. 
The confidence intervals using the conditioning likelihoods notably narrow with increasing $c$; the lower bounds slightly decrease but the largest reduction is in the upper bounds. For lower $c$ values the $\ell_{fc}$ intervals are wider than for $\ell_{std}$, in contrast for the largest possible $c$ values the interval is very narrow (a reduction in size of factor 14 over the range of $c$ possible). For $\ell_{pc}$ the upper bounds are smaller than those using the $\ell_{std}$ for all values of $c$ and are slightly larger than those for $\ell_{ex}$ for low $c$. However, for large $c$ the upper bounds of both conditioning confidence intervals are much lower than that using $\ell_{ex}$ since the information that $c$ was exceeded on this observation becomes more informative about the tail of the distribution as $c$ approaches the 2015 observation. Thus if we stop after the first minor exceedance of $c$ we can be reasonably sure the tail is short. This is an unexpected but helpful finding. 
Further investigation into the confidence intervals 
can be found in \cite{thesis}.


\subsection{Variable-threshold stopping rule}

Now we consider the variable-threshold stopping rule and first determine a range of $k$ from the data. In the Lune data the maximum river level in $2015$ corresponds to
$\hat{k}=2561$ given the data up to $2015$ and to $\hat{k}=188$ using all the data. However, the river level in $1995$ corresponds to $\hat{k}=\infty$ (\ie , it is larger than the point estimate of the upper end point of the GEV fitted to the data up to 1995) so the variable threshold rule as given in \eqref{eqn:rule2} cannot have been applied for any $k<\infty$. Furthermore, the river level in $1980$ corresponds to $\hat{k}=111$. If the variable-threshold stopping rule had motivated a request for an analysis of the data up to and including $2015$, the request must have been triggered by the second such exceedance. In our analysis we explore values of $k$ between $200$ and $2500$ and simply amend $\ell_{fc}$ slightly by replacing the $\ell_{fc}$ contribution of the 1995 observation ($i=28$), $g(x_{28};\theta)/G(\hat{x}_k^{std}(\bm{x}_{1:27});\theta)$, by $g(x_{28};\theta)$. 

Figure~\ref{fig:changingc}, right panel, shows the same inferences as the left panel, but for the variable-threshold over a range of return periods $k\in [200,2500]$. Given the rarity of all events in this range we would expect a `true' $k$ to be towards the lower end of this range. 
The estimates $\hat{x}_{200}^{std}$ and $\hat{x}_{200}^{ex}$ and the corresponding confidence intervals are invariant to $k$ (and independent of the stopping rule used) but as $k$ increases $\hat{x}_{200}^{fc}$ and $\hat{x}_{200}^{pc}$ both decrease. For small $k$, $\hat{x}_{200}^{fc}>\hat{x}_{200}^{pc}$, as we would expect from our bias results in the simulation study. However, the inequality reverses for large $k$ perhaps as a result of there being more than one exceedance of the threshold. This is hinted at by the bias results and also since if one omits the 1995 observation from the data set then $\hat{x}_{200}^{fc}>\hat{x}_{200}^{pc}$ for all $k$. 
More investigation into the estimators when there are multiple exceedances would be useful.


The intervals using the conditioning likelihoods and variable-threshold stopping rule behave similarly to those using the fixed-threshold stopping rule. Again the $\ell_{fc}$ intervals are highly influenced by the `extremeness' of the stopping threshold. With the lowest possible $k$ for this data set (ignoring the 1995 exceedance) the $\ell_{fc}$ interval is almost double the width of the confidence interval using $\ell_{std}$ whereas for a large $k$ value it is less than half the width. 
The $\ell_{pc}$ confidence intervals also reduce in width with increasing $k$ but not as dramatically. 



\subsection{Non-stationarity}

\begin{figure}
       \subfigure{\includegraphics[width=0.5\textwidth]{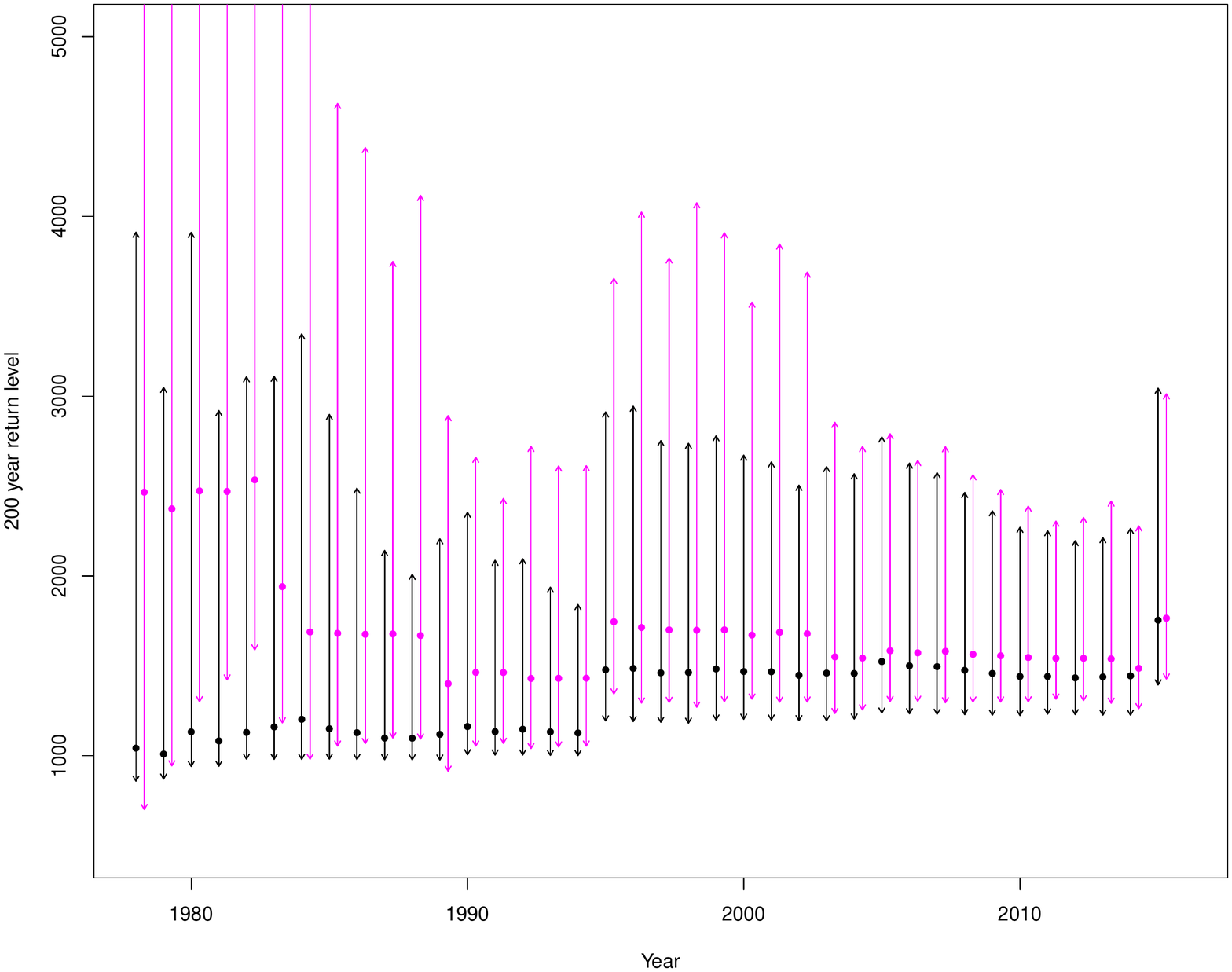}}
       \subfigure{\includegraphics[width=0.5\textwidth]{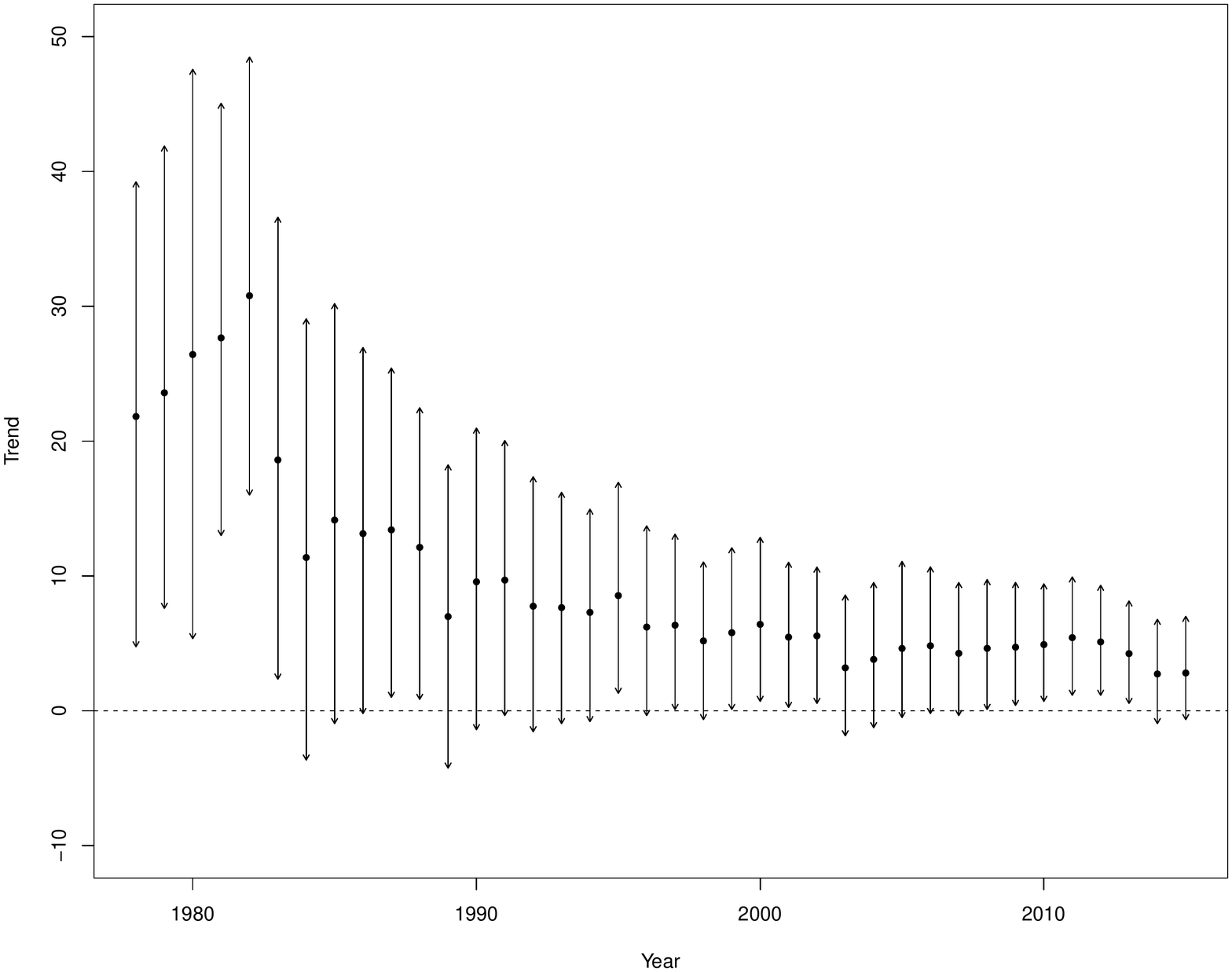}}
        \captionit{Fitting a GEV to all the data up to and including the current year for the Lune at Caton. Left: 200yr return-level estimates for $2015$ using progressively more data over the years with and without a trend in the location parameter (pink and black respectively). Each group of 2 estimates applies for the same year and have been horizontally shifted for clarity. Right: Slope parameter, $\hat{\beta}$, and it's 95\% confidence interval.}
        \label{fig:trends}
\end{figure}

The implications of using stopping rules on the estimation of trends in extreme levels is also a concern, as stopping with the final observation being large is likely to have a similar biasing effect as found  in \S\ref{sec:setup} and \S\ref{sec:GEV} for return-levels. This is particularly important given the interest in whether trends in extreme values differ from trends in mean levels \citep{Eastoe,Hannaford}. In Figure~\ref{fig:trends}
we illustrate the analysis of the Lune data with a GEV distribution including a linear trend $\mu_t=\alpha_0+\beta t$, showing both the resulting estimates of
the 200 year return-level for 2015, \ie , the estimates of the $0.995$ quantile of the annual maximum in 2015, and the associated  trend estimate $\hat{\beta}$ using progressively more data over time. With few data used the trend is estimated to be unrealistically large, with huge uncertainty,  and this results in very different point estimates of return-levels relative to the analysis with no trend. As more data are observed we can see that the trend estimates generally decrease, with reduced uncertainty, with positive jumps in $\hat{\beta}$ estimates after the large 1995 and 2015 events. Although the 2015 river flow is more extreme than that of 1995 its impact on $\hat{\beta}$ is much less. Furthermore, we see that $\hat{\beta}$ is not statistically significantly different from $\beta=0$ at the 2.5\% level. Thus here the effect of including the estimated trend is small on the 200 year return-level estimate and the stopping rule seems to have almost no effect on the trend estimate.
 
\section{Discussion} \label{sec:discussion}

In this paper and the associated supplementary material we have shown that return-level estimators based on the standard likelihood are positively biased when sampling from the GEV or GP distributions using certain stopping rules. 
The extent of the stopping bias 
is lower for lighter tailed distributions and when estimating low return-levels. We have proposed conditioning upon the stopping threshold in the likelihood. In most cases we have found that conditioning on the final observation exceeding the stopping threshold (partial conditioning) results in return-level estimates with the lowest RMSE despite the estimator being negatively biased. 


A balance must be struck between low RMSE and good coverage, however. Partial conditioning results in undercoverage despite the low RMSE of $\hat{x}_{y}^{pc}$. 
The full-conditional likelihood, which also conditions on the non-exceedance of all previous observations, gives the closest to 95\% coverage 
and though the intervals are wide, they are typically narrower than the confidence intervals obtained from the standard likelihood. The interval widths using the full and partial conditional likelihoods are smaller the closer the stopping threshold is to the final observation as the occurrence of the final exceedance becomes more informative on the tail of the distribution (see \S\ref{sec:CS}). 

Overall, the conditioning estimators presented here outperform the standard estimator when the decision to analyse data at a particular time was triggered by what was perceived to be a large observation. For the fixed-threshold stopping rule, partial conditioning has the best combination of RMSE and coverage for a range of $\xi$ with moderate $k$ and particularly when the distribution is heavy tailed, as is the case for most UK rivers \citep{FEH}. For the variable-threshold stopping rule, full conditioning provides the best balance of coverage and low RMSE. To apply the conditioning estimators in practice if the rule of the flood management agency is unknown the statistician needs to choose a suitable stopping threshold, $c$, for the fixed-threshold stopping rule and a suitable stopping `period', $k$, for the variable-threshold stopping rule if the values are unknown. A range of $c$ and $k$ can be considered provided that the observed data are below the resulting stopping threshold(s) up to the final observation. 


The decision to analyse data will likely be based on a confluence of many factors. 
Our work attempts to simplify the true decision making procedure by using stopping rules based on the occurrence of a \textit{single} large observation exceeding some threshold. An analysis may instead be prompted by a prolonged period of quite large (but not necessarily `extreme') observations or the observation of large values at many locations simultaneously, 
requiring more complex multivariate analysis since the observations at nearby locations will be dependent in some way \citep{Keef,Asadi}.


In practice if the stopping rule is unknown and the analysis is triggered by a large event, we suggest using the full conditional return-level estimator. However if $k$ is thought to be less than 50, or the full-conditional estimate and/or confidence interval are clearly too large then partial conditioning should be used instead.
We argue that the decision to `stop' and analyse data would in part be based on both past return-level estimates and thresholds set due to current infrastructure and so the `true' stopping rule is a mixture of the two rules considered here. Hence the `true' bias, RMSE and coverage of the estimators can be expected to lie between those which we found under the two stopping rules. It should be noted that this work does not address the
question of \textit{when} the data should be analysed, but rather how we can reduce the bias given the use of a particular stopping rule.
Nevertheless, if we are at a point in time where a stopping criterion has been met and triggered an analysis, this study
can give guidance on the behaviour of return-level estimators calculated at the current time whether based on the full likelihood, partial or full conditioning, or even excluding the most recent, `triggering' event.

In our theoretical and simulation studies we have not accounted for the possibility of a trend in the data, such as river flows gradually increasing over the years. We saw in \S\ref{sec:Lune} that the Lune data has a slight positive trend in the location parameter and fitting such a model at an earlier point in time resulted in a very large positive trend. This could cause problems for the fixed-threshold stopping rule, in particular it might become necessary to change $c$ after a certain number of years. Nonetheless, doing this is probably not too unrealistic since, for example, the height of a flood defence might be increased if there has been evidence of higher flow in recent years. On the other hand the variable-threshold stopping rule is more robust to data with an underlying trend as it is directly a function of the observed data.



\subsection*{Acknowledgements}

Barlow gratefully acknowledges funding of the EPSRC funded STOR-i centre for doctoral training (grant number EP/L015692/1). This research was also financially supported by JBA. The authors also thank NRFA for the Lune gauge data.

\bibliographystyle{apalike}
\nocite{*}
\bibliography{paper}

\section*{Appendix}
\appendix
\Alph{section}
\renewcommand{\thesection}{\Alph{section}}
\renewcommand\thefigure{A.\arabic{figure}}    
\setcounter{figure}{0} 
\renewcommand\theequation{A.\arabic{equation}}    
\setcounter{equation}{0} 
\section{Proof of results from \S\ref{sec:setup}} \label{App:Exp}

\subsection{Proof of Proposition 1}

For simplicity we denote $c_k$ by $c$.
Sampling from some general distribution with the first stopping rule, we have:
\begin{align}
	\EE{\frac{1}{N}} = \sum_{n=1}^\infty{\bar{F}(c)\frac{F(c)^{n-1}}{n}} = -\frac{\bar{F}(c)}{F(c)}\log(\bar{F}(c)) \, , 
\end{align}

where $F(x)$ and $\bar{F}(x)=1-F(x)$ are the CDF and survival function of the distribution of $X_i, ~i=1,\dots,n$. Thus,
\begin{align}
	\EE{\Xbar_N} &= \EE{\frac{1}{N} \EE{\sum_{i=1}^N{X_i} | N=n}} \nonumber \\
	&= \EE{\frac{1}{N} \left( (N-1)\EE{X|X \leq c} + \EE{X|X > c} \right)} \nonumber \\
	&= \EE{X|X \leq c} + \EE{\frac{1}{N} \left( \EE{X|X>c} - \EE{X|X \leq c} \right)} \nonumber \\
	&= \EE{X|X \leq c} +\EE{\frac{1}{N}} (\EE{X|X>c} - \EE{X|X \leq c}) \, .
\end{align}

Specifically for sampling from the exponential distribution:
\begin{align*}
	\EE{\frac{1}{N}} = \frac{\beta c}{e^{\beta c} - 1}.
\end{align*}

By the memoryless property of the exponential distribution:
\begin{align*}  
	\EE{X|X>c} = \EE{X} + c = \frac{1}{\beta} +  c \, ,
\end{align*}

and rearranging $\EE{X} = F(c)\EE{X | X \leq c} + \bar{F}(c) \EE{X|X>c}$ gives

\begin{align}
	\EE{X|X \leq c} = \frac{1}{F(c)} \left[ \frac{1}{\beta} -\bar{F}(c)\left( c + \frac{1}{\beta} \right) \right] = \frac{1}{\beta} - c \frac{\bar{F}(c)}{F(c)}\, .
	\label{eqn:explessc}
\end{align}
Therefore, for the exponential distribution,
\begin{align}
	\EE{\Xbar_N} = \frac{1}{\beta} + \frac{c}{e^{\beta c} - 1} \left(\frac{\beta c}{1 - e^{-\beta c}} - 1 \right) .
	\label{eqn:XN}
\end{align}

\vspace{2mm}

For the standard estimator based on the full sample we have
$1/\hat{\beta}_{std} = \Xbar_N$, the sample mean. The first part of
Proposition \ref{prop.expfixed} then follows from \eqref{eqn:XN}.

If the final data point is excluded from the sample then all included
samples are from the distribution truncated at $c$, so, from \eqref{eqn:explessc},
\begin{align*}
	\EE{\frac{1}{\hat{\beta}_{ex}}\Big|N>1} = \EE{X | X \leq c} = \frac{1}{\beta} - \frac{c}{e^{\beta c} - 1} \, ,
\end{align*}
leading to the expression in the second part of Proposition \ref{prop.expfixed}.

\subsection{Proof of Theorem \ref{thm}} \label{proof}

\begin{proof*} 
  We start by defining the following key quantities for each $k\geq 1$,
  \begin{align}
    S_{k}&:= (n_0+k) \Xbar_k = n_0 \Xbar_0 + \sum_{j=1}^{k} X_j,\\
    V_k&:=\frac{X_k}{S_k}.
  \end{align}
  Marginally $S_k\sim \mathsf{Gamma}((n_0+k)\alpha,\beta)$
  and $V_k\sim \mathsf{Beta}(\alpha,(n_0+k)\alpha)$; we denote their
  marginal densities as:
    \begin{align*}
    	f_{S_{k}}(s_{k}) &\propto s_{k}^{(n_0+k-1)\alpha - 1} e^{-\beta s_{k}} \\
    	f_{V_{k}}(v_k) &\propto v_{k}^{\alpha-1} (1-v_k)^{(n_0+k-1)\alpha - 1} \, .
    \end{align*}
  
The stopping time, $N$, is $n$ if $X_n > \gamma \Xbar_{n-1}$ and
$X_i < \gamma \Xbar_{i-1}$ for $1 \leq i < n$. However, $X_n>\gamma \Xbar_{n-1}$
	\begin{align*}
		\Leftrightarrow X_n &> \frac{\gamma}{n+n_0-1}(S_n-X_n)\\
		\Leftrightarrow \left(1 + \frac{\gamma}{n + n_0 - 1}\right) X_n &> \gamma \frac{1}{n + n_0 - 1} S_n \\
		\Leftrightarrow \left(1 + \frac{\gamma}{n + n_0 - 1}\right) V_n &> \frac{\gamma}{n + n_0 - 1} \\
		\Leftrightarrow \qquad V_n &> \frac{\gamma}{n + \gamma + n_0 - 1} \, .
	\end{align*}
	
    So the stopping rule can be written purely as function of the $V$s. Explicitly, we stop at time $n$ if $V_n > \frac{\gamma} {n + \gamma + n_0 - 1}$ and $V_i < \frac{\gamma}{i + \gamma + n_0 - 1}$ for $1 \leq i < n$. 
       
    We define the statement $\mathcal{A}_n \defeq
    ``V_1,\hdots,V_n,S_n$ are mutually independent".
    Below, we will show by induction that $\mathcal{A}_n$ holds for
    all $n\ge 1$.
Thus $\Xbar_n \indpt V_i \quad \forall i \leq n$; the distribution
of $\Xbar_n$ is independent of whether or not the stopping rule has
been triggered. Therefore, $\Xbar_N$ conditioned on $N=n$ is
equivalent to the mean of $n$ i.i.d. $\mathsf{Gamma}(\alpha,\beta)$
random variables, as stated in the theorem. 

    \underline{$\mathcal{A}_{n-1}\Rightarrow\mathcal{A}_n$}: 
    If $\mathcal{A}_{n-1}$ holds then the joint pdf of $V_1,\hdots,V_{n-1},S_{n-1}$ can be factorised:
   
	\begin{align*}
		f_{n-1}(v_1,\hdots,v_{n-1},s_{n-1}) = f_{S_{n-1}}(s_{n-1}) \prod_{i=1}^{n-1} f_{V_i}(v_i) \, ,
	\end{align*}	       
	Consider the change of variables
        $(V_1,\hdots,V_{n-1},S_{n-1},X_{n}) \rightarrow
        (V_{1},\hdots,V_{n-1},V_{n},S_{n})$,
        where $X_n=S_nV_n$ and $S_{n-1}=S_n(1-V_n)$.
        The Jacobian for this transformation is: 
	
	    \begin{align*}
            |J| = \left| \frac{\partial(v_{1:{n-1}},s_{n-1},x_n)}{\partial(v_{1:n},s_n)} \right| = \begin{vmatrix}
                    I_{n-1} & 0 \\
                    0 & A
            \end{vmatrix}
            = s_n \, .
    \end{align*}
    
    where $I_{n-1}$ is the $(n-1)\times(n-1)$ identity matrix and 
    
    \begin{align*}
            A = \frac{\partial(s_{n-1},x_n)}{\partial(v_n,s_n)} = \begin{bmatrix}
                    -s_n & 1 - v_n \\
                    s_n & v_n
            \end{bmatrix}.
    \end{align*}
    
	So,     since $S_{n-1}$ and $V_{1},\dots,V_{n-1}$ are independent of $X_{n}$,
	
	\begin{align*}
		f_{n}(v_{1:n},s_n) &= f_{n-1}(v_{1:n-1},s_{n-1}(s_n,v_n))f_X(x(s_n,v_n)) |J| \\
		&\propto \left(\prod_{i=1}^{n-1} f_{V_i}(v_i)\right) (s_n(1-v_n))^{(n_0+n-1)\alpha - 1} e^{-\beta s_n (1-v_n)} (s_n v_n)^{\alpha - 1} e^{-\beta s_n v_n} s_n \\
		&= \left(\prod_{i=1}^{n-1} f_{V_i}(v_i)\right)
                 s_n^{(n+n_0)\alpha - 1} e^{-\beta s_n} v_n^{\alpha - 1} (1-v_n)^{(n_0 + n - 1)\alpha - 1} \\
		&\propto \prod_{i=1}^{n} f_{V_i}(v_i) f_{S_n}(s_n) \, .
	\end{align*}	    
So $\mathcal{A}_n$ holds.

\underline{$\mathcal{A}_1$ holds}: We must show that $V_1$ and $S_1$ are independent. We do this by using the change of variables $(\Xbar_0,X_1) \rightarrow (V_1,S_1)$ to show that the joint pdf of $V_1$ and $S_1$ factorises.
    
    We have
		\begin{align*}
			f_{\Xbar_0,X_1}(\xbar_0,x_1) \propto  \xbar_0^{n_0 \alpha - 1}e^{-n_0 \beta \xbar_0} x_1^{\alpha - 1} e^{-\beta x_1}
		\end{align*}
		
	        and $X_1 = S_1V_1$ and
                $\Xbar_0 = \frac{1}{n_0} S_1(1 - V_1)$.
So Jacobian for the transformation is:

\begin{align*}
	\left| \frac{\partial(\Xbar_0,X_1)}{\partial(V_1,S_1)} \right| &= \begin{vmatrix}
		-\frac{s_1}{n_0} & s_1 \\
		\frac{1}{n_0}(1-v_1) & v_1
	\end{vmatrix} \\
	&= \frac{s_1}{n_0}
\end{align*}

Thus the joint pdf of $V_1,S_1$ is:
\begin{align*}
	f_{V_1,S_1}(v_1,s_1) &\propto s_1
        (s_1(1-v_1))^{n_0 \alpha-1}e^{-n_0 \beta (s_1(1-v_1)/n_0}
        \\
        &\qquad \qquad \left(s_1 v_1)\right)^{\alpha-1}e^{-\beta s_1 v_1}
         \\
	&=  s_1^{(n_0+1)\alpha - 1} e^{-\beta s_1} v_1^{\alpha-1}
        (1-v_1)^{n_0 \alpha - 1} \\
        &\propto f_{S_1}(s_1)f_{V_1}(v_1),
\end{align*}  
as required.
\end{proof*}

%
%
%
%
%

\section{Properties of the GEV shape parameter} \label{sec:shape}

\subsection{Fixed-threshold stopping rule}

\begin{figure}[t]
	\begin{center}
		\subfigure{\includegraphics[width=0.32\textwidth]{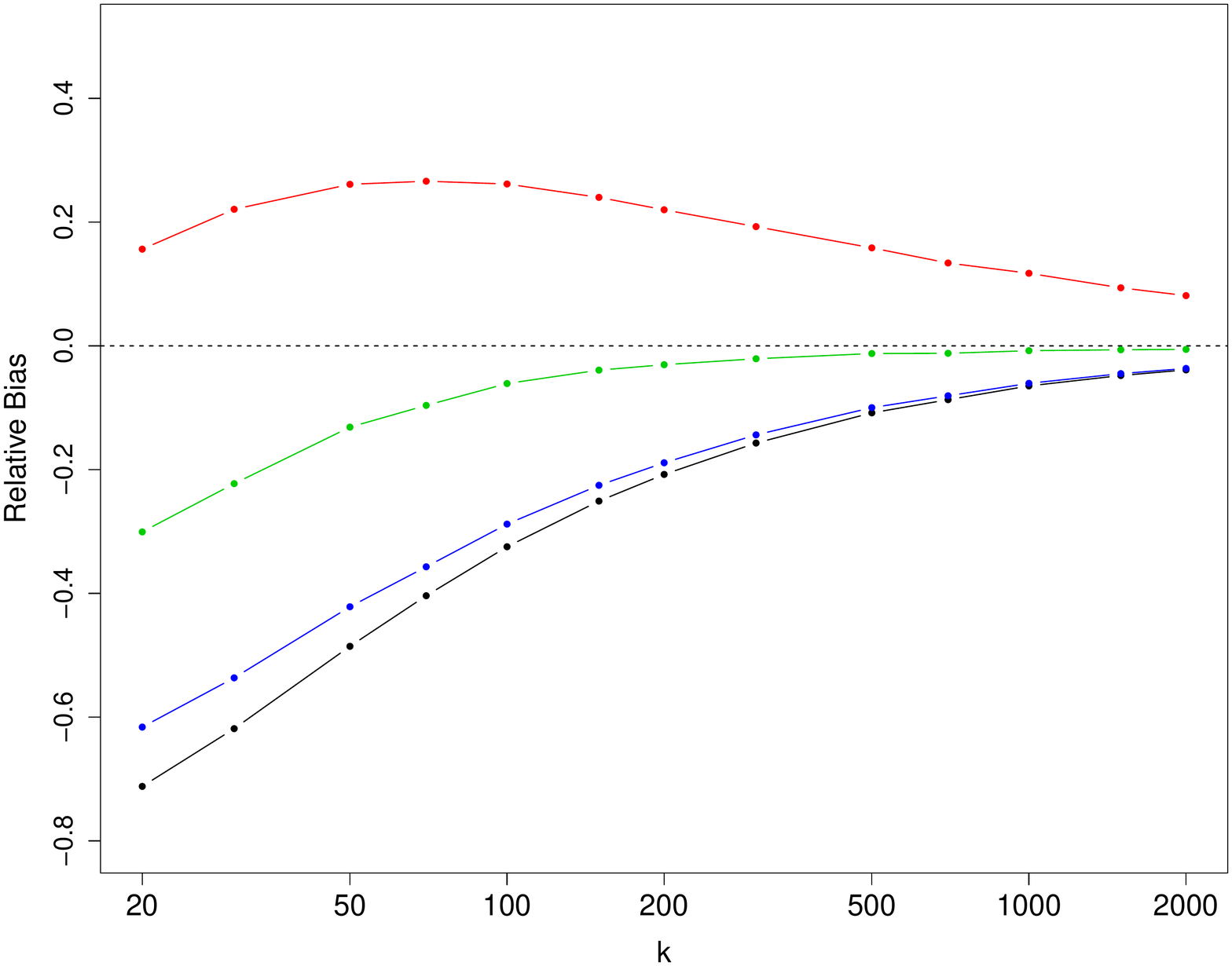}}
		\subfigure{\includegraphics[width=0.32\textwidth]{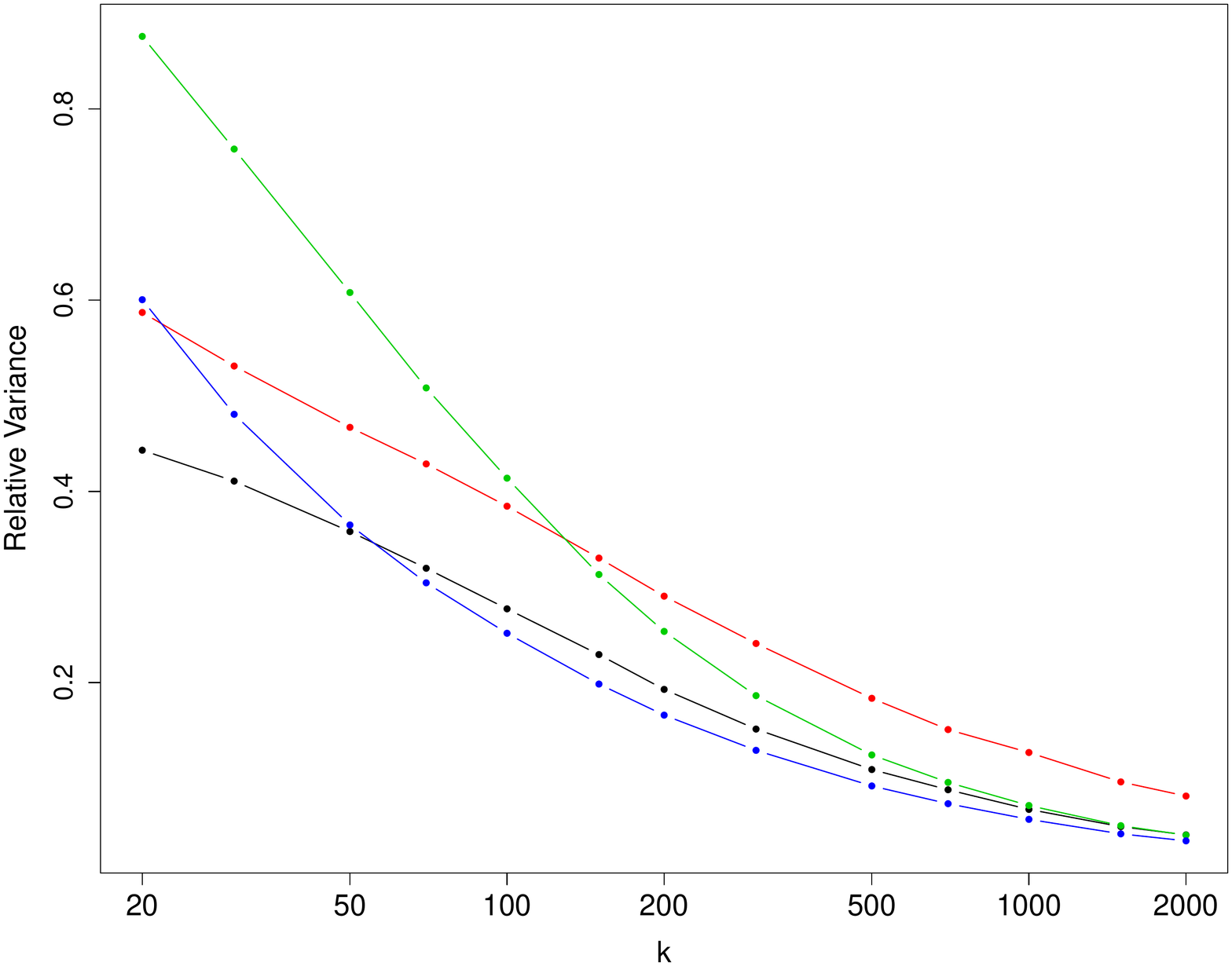}}
		\subfigure{\includegraphics[width=0.32\textwidth]{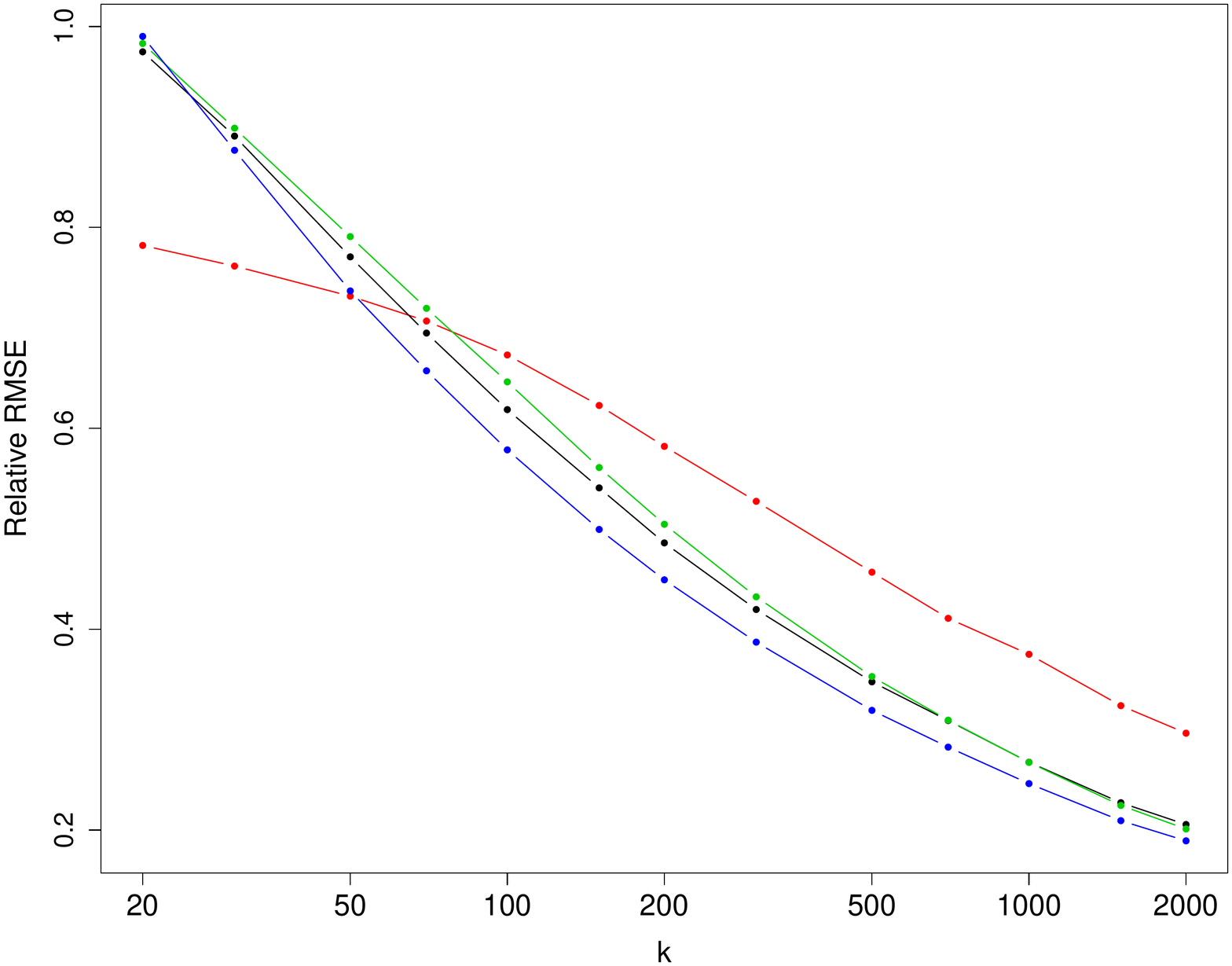}}
		\subfigure{\includegraphics[width=0.32\textwidth]{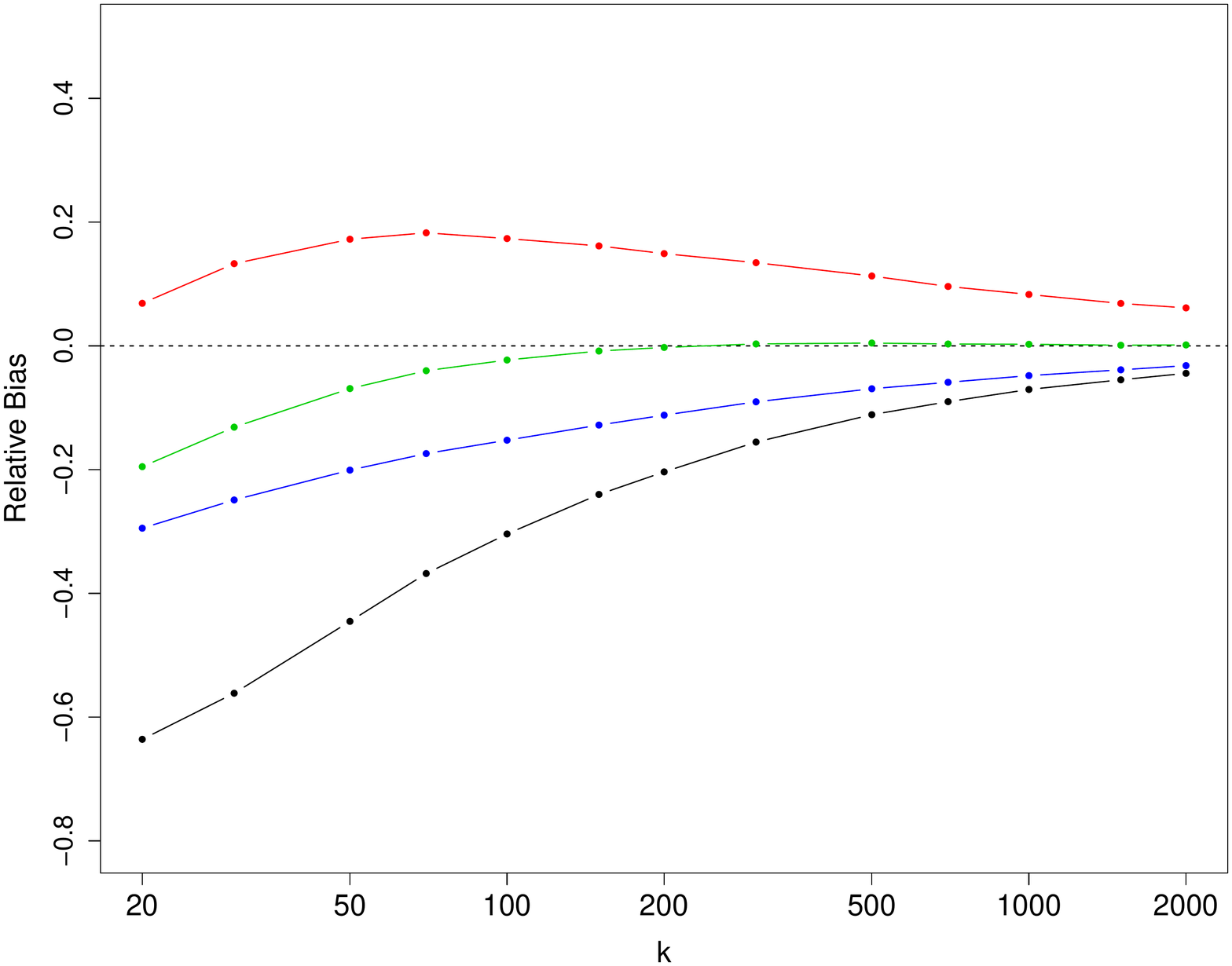}}
		\subfigure{\includegraphics[width=0.32\textwidth]{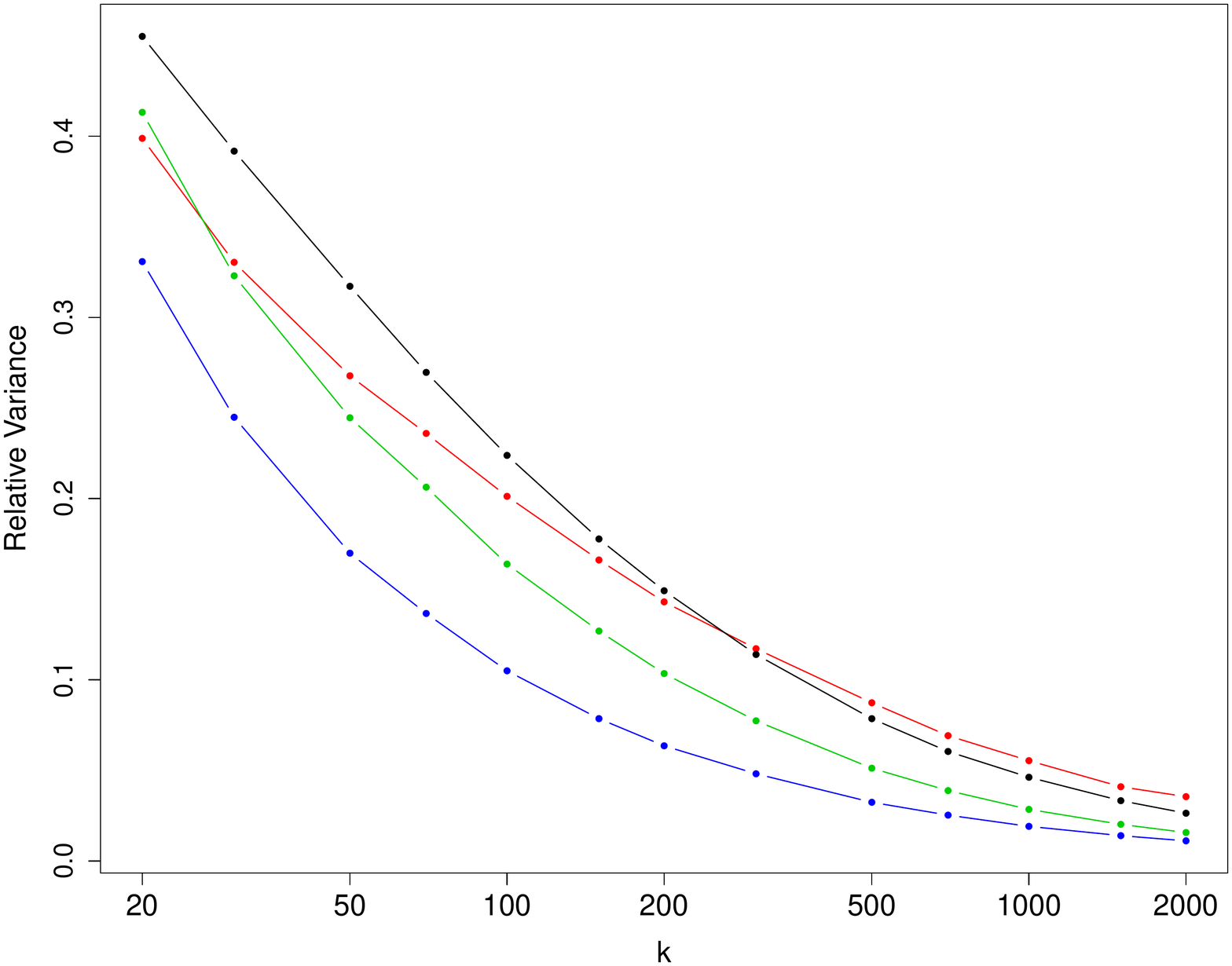}}
		\subfigure{\includegraphics[width=0.32\textwidth]{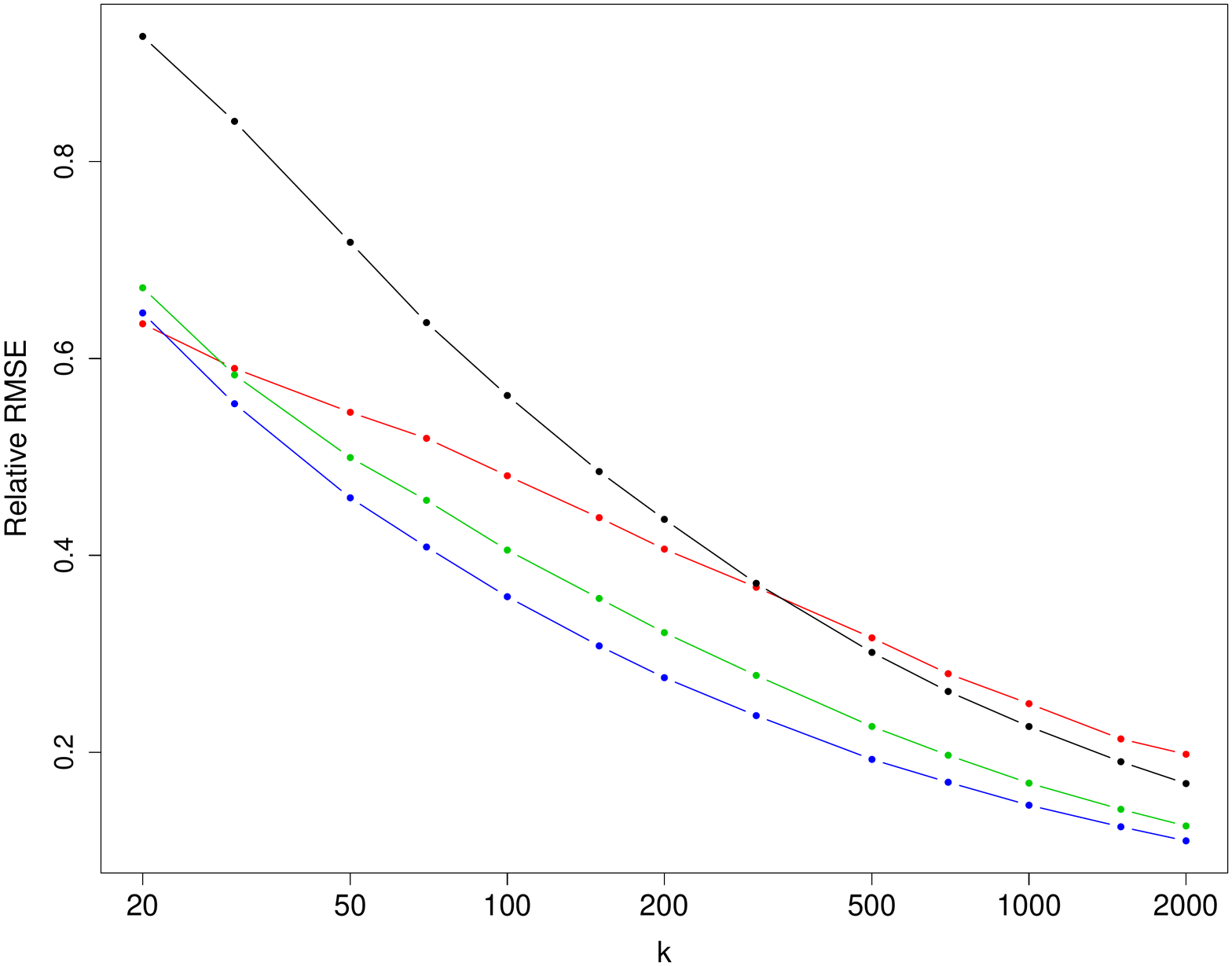}}
		\captionit{Shape parameter estimates when sampling from the GEV distribution with $(\mu, \sigma)=(0, 1)$ using the fixed-threshold stopping rule with threshold $c_k$ and $\xi=0.2$ (top) and $\xi=-0.2$ (bottom) both plotted against $k$. Left: relative bias, centre: relative variance, right: relative RMSE, using: standard likelihood (red), excluding the final observation (black), full conditioning (green) and partial conditioning (blue). Based on $10^5$ replicated samples with the historical data created using approach~\eqref{eqn:initial}.}
		\label{fig:shape}
	\end{center}
\end{figure}

The shape parameter, $\xi$, is important in determining the tail behaviour. Figure~\ref{fig:shape} shows the relative bias, variance and RMSE of each of the estimators when sampling using the fixed-threshold stopping rule for $\xi=0.2$ and $-0.2$ (top and bottom rows respectively). Judged by RRMSE, we find that $\ell_{pc}$ is generally best for moderate to large $k$, with clear benefits for $\xi=-0.2$; however $\ell_{fc}$ has generally quite similar RRMSE and low bias. 
As one would expect the lighter the tail of the distribution, the smaller both the relative variance and, in most cases, the relative bias of the shape parameter estimators resulting in smaller RRMSE. To help understand why these RRMSE results arise we now look at more detail at the performance of the four estimators.


The standard MLE for the shape parameter, $\hat{\xi}_{std}$, is almost always positively biased while $\hat{\xi}_{ex}$ leads to quite large negative bias (with $E(\hat{\xi}_{ex}) < 0.1$ when $\xi=0.2$ and $k<50$ (Figure~\ref{fig:shape})) since we lose information about the upper tail of the underlying distribution. In particular, the fitted distribution typically has a lighter tail and can even have an upper end point which could be less than the excluded observation. Unlike all other estimators considered, the variance of $\hat{\xi}_{ex}$ is not substantially lower when the tail is lighter and so has quite large RRMSE when $\xi=-0.2$. 


The partial conditioning method generally has $\hat{\xi}_{pc}$ lower than the truth however, for moderate $k$, they consistently have low variance relative to the other methods over a range of $\xi$. Therefore, partial conditioning 
provides $\xi$ estimators with the lowest RRMSE for $k>100$.
In contrast, $\ell_{fc}$ leads to very little bias in $\xi$ estimates for $k>100$ but the variance can be large, particularly when $\xi=0.2$ with $k < 100$. 
This is in agreement with \cite{Molenberghs} findings that the full-conditional estimator has poor precision despite it's unbiasedness. However, unlike in \cite{Molenberghs}, we find that, in our context, full conditioning can improve upon the standard estimator especially when the stopping threshold is high (\ie , for large $k$). 



\subsection{Variable-threshold stopping rule}

Properties of the shape parameter estimators under the variable stopping rule are shown in the supplementary material, Figure~\ref{supp:fig:varshape}, for $\xi=0.2$ and $-0.2$. We find that in the variable threshold setting $\hat{\xi}_{std}$ has very low bias (similarly recall in \S\ref{sec:gamma} when sampling from the gamma distribution with this stopping rule we found the standard return-level estimator was unbiased) whereas all other $\xi$ estimators are negatively biased, with $\hat{\xi}_{ex}$ having the largest negative bias out of all the estimators for both values of $\xi$ considered. We find that $\hat{\xi}_{std}$ also has the lowest RRMSE of the estimators. 
Despite $\hat{\xi}_{std}$ performing well under the variable threshold stopping rule, this is not always the case for the $\ell_{std}$ return-level estimators.

\setcounter{section}{0}
\renewcommand{\thesection}{\arabic{section}}
\renewcommand\thefigure{S.\arabic{figure}}    
\setcounter{figure}{0}
\renewcommand\theequation{S.\arabic{equation}}    
\setcounter{equation}{0} 



\renewcommand\thesection{S.\arabic{section}}    
\setcounter{section}{0}
\renewcommand\thefigure{S.\arabic{figure}}    
\setcounter{figure}{0}
\renewcommand\theequation{S.\arabic{equation}}    
\setcounter{equation}{0} 

\title{\Large Supplementary material for `Inference for extreme values under threshold-based stopping rules'}
\date{}
\author{Anna Maria Barlow, Chris Sherlock, Jonathan Tawn}


\maketitle

Here we present extra results from the simulation study for sampling from the GEV distribution and in \S\ref{sec:GPD} discuss our four likelihoods when modelling threshold exceedances.

\section{GEV}

For the fixed-threshold stopping rule the bias, variance and RRMSE results are based on $10^5$ replicated samples and the coverage results on $5000$ replicated samples. The variable-threshold stopping rule requires much more computational effort than the fixed-threshold stopping rule since the parameters must be estimated after each observation in order to calculate the subsequent stopping threshold. The results shown here are based on 10,000 and 20,000 replicated samples for $\xi=0.2$ and $\xi=-0.2$ respectively. There are more simulations for $\xi=-0.2$ as the sample sizes generated, $N$, using the variable-threshold stopping rule are smaller when the distribution has light tails. 

We used the profile-likelihood method to create confidence intervals for the return-levels. Confidence intervals could also be calculated via the delta method or via the bootstrap. In the context of return-level estimation the delta method can produce contradictory confidence intervals (the lower bound for a particular return-level may be lower than that of a return-level with a smaller return period). Bootstrap methods were explored but these were found to have poor coverage and are much more computationally expensive \citep[see][]{thesis}.


The results for the 50 and 1000 year return level estimators follow the same pattern as the 200 year return level estimators shown in the paper but as noted in \S\ref{sec:GEVret} of the paper the bias and variance worsen as the return level estimated becomes more extreme.

\subsection{Other Methods} \label{sec:other}

We also considered a range of alternatives to our two conditioning likelihoods. The two most effective are detailed here.

\begin{figure}[h]
	\begin{center}
		\subfigure{\includegraphics[width=0.32\textwidth]{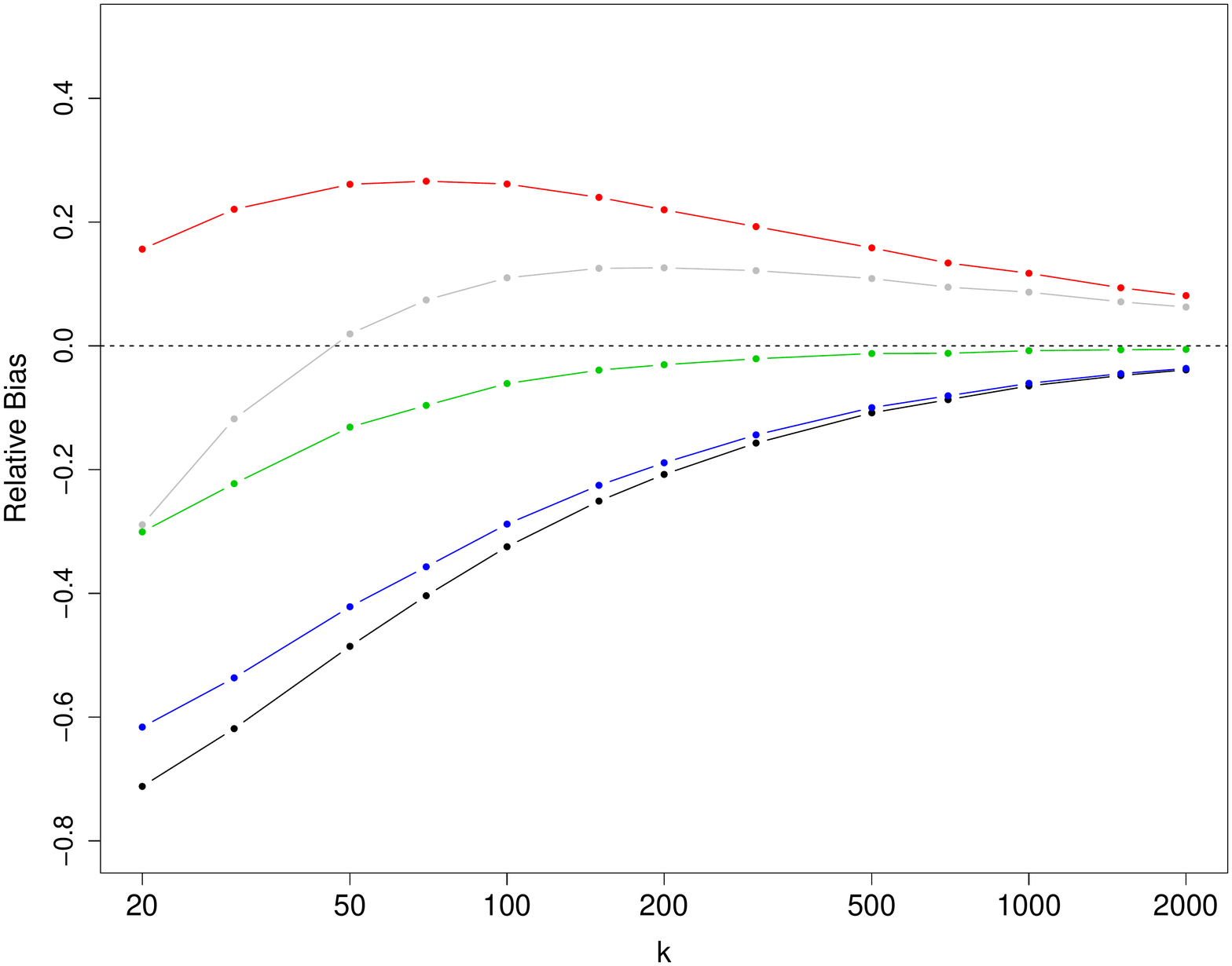}}
		\subfigure{\includegraphics[width=0.32\textwidth]{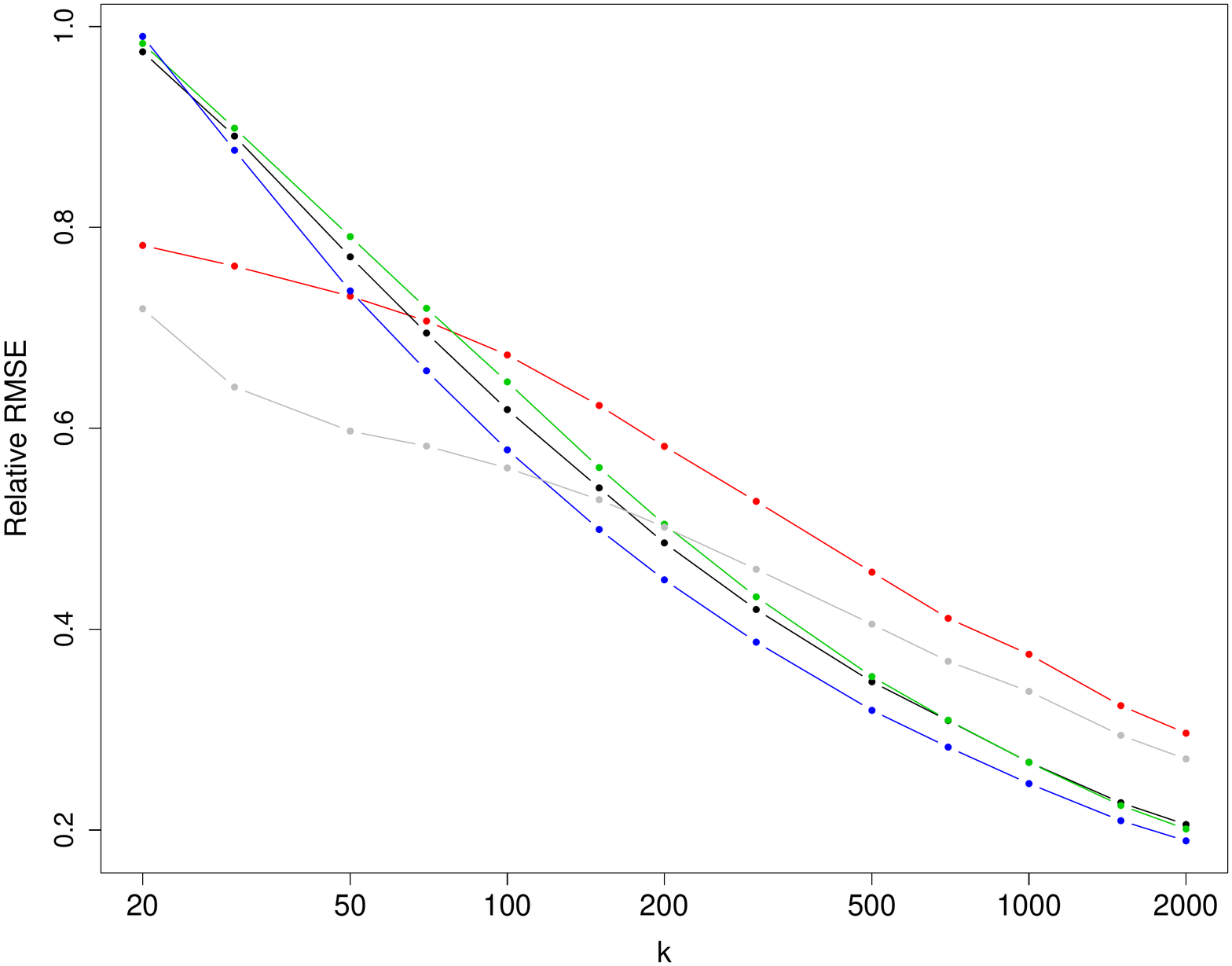}}
		\subfigure{\includegraphics[width=0.32\textwidth]{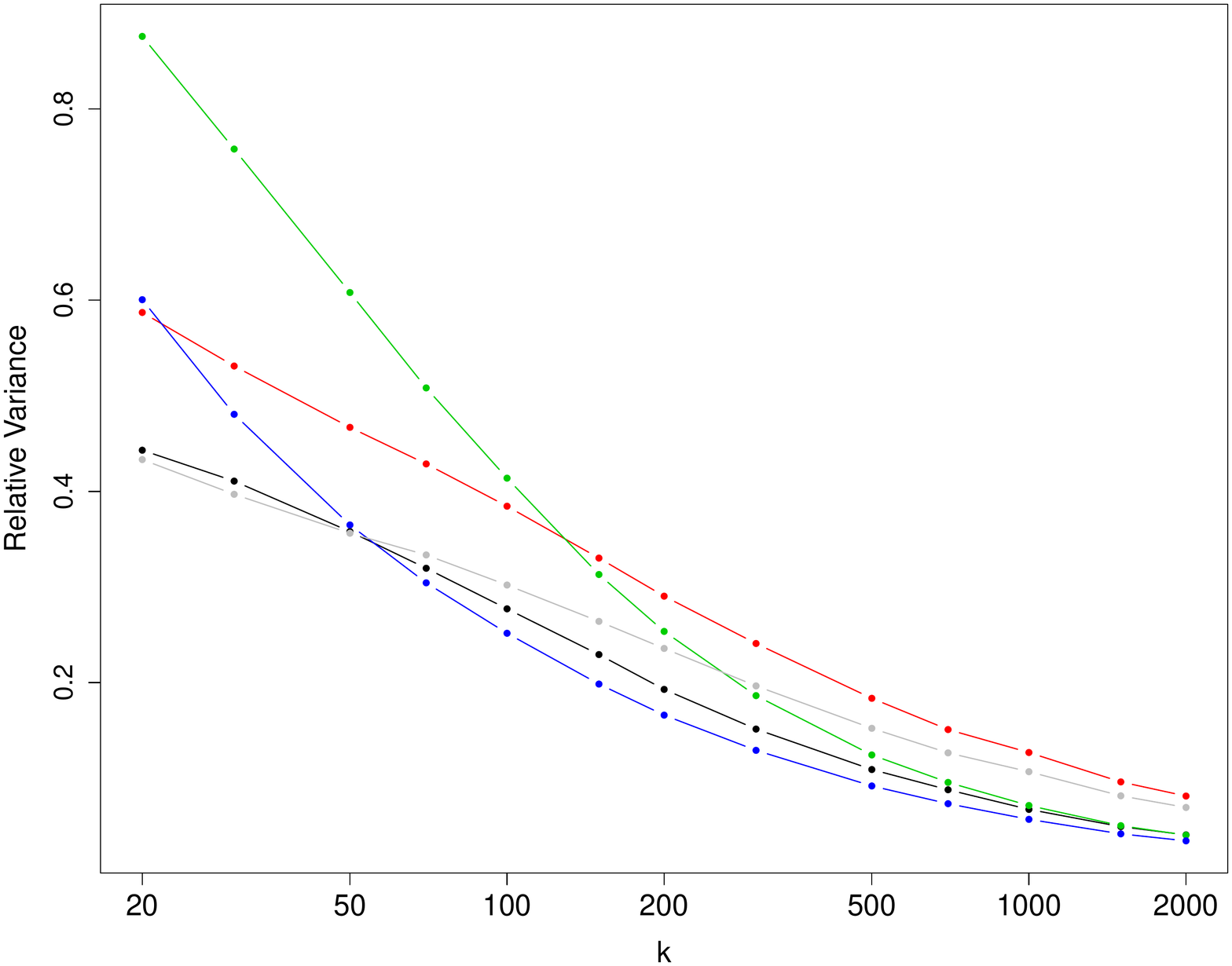}}
		\subfigure{\includegraphics[width=0.32\textwidth]{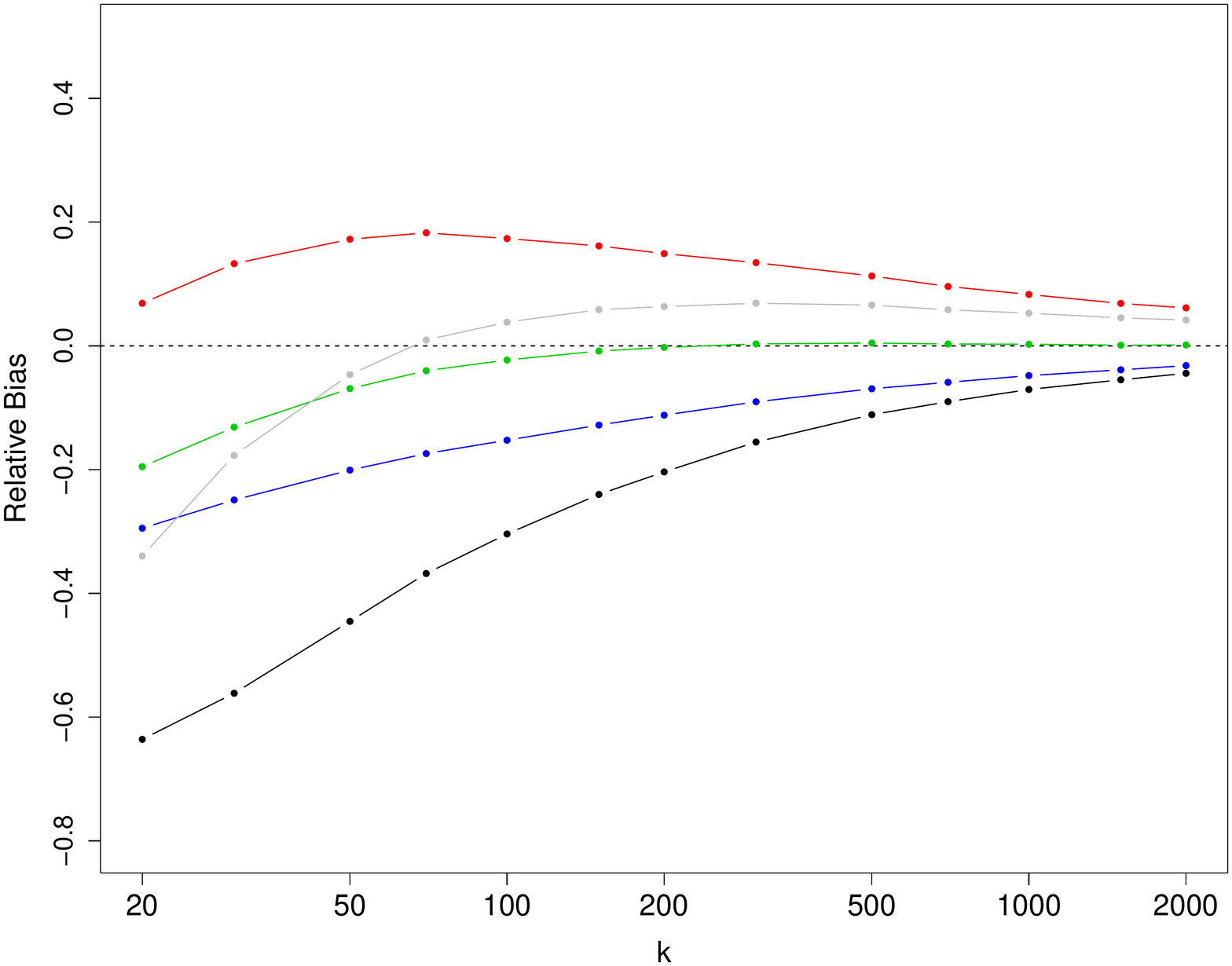}}
		\subfigure{\includegraphics[width=0.32\textwidth]{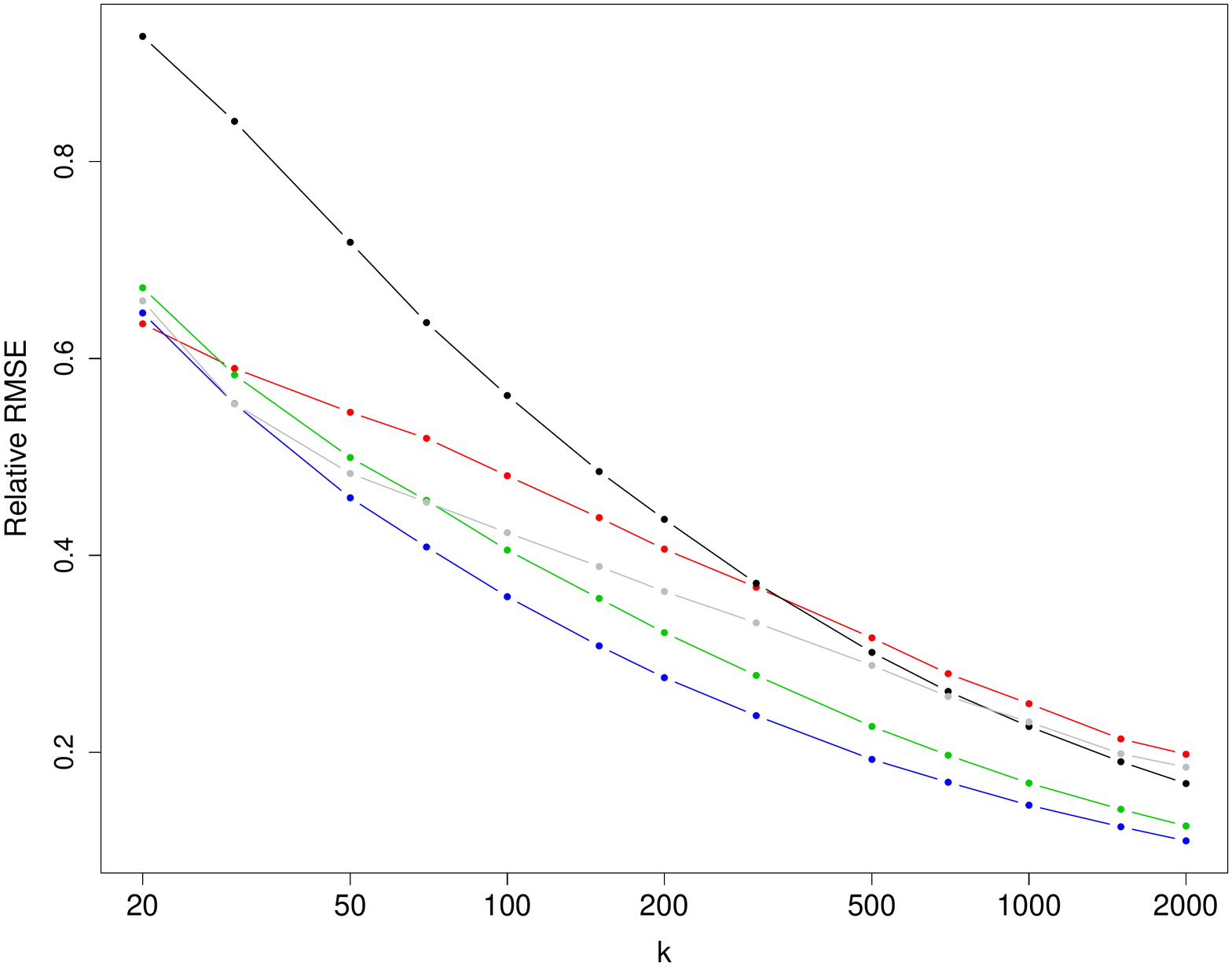}}
		\subfigure{\includegraphics[width=0.32\textwidth]{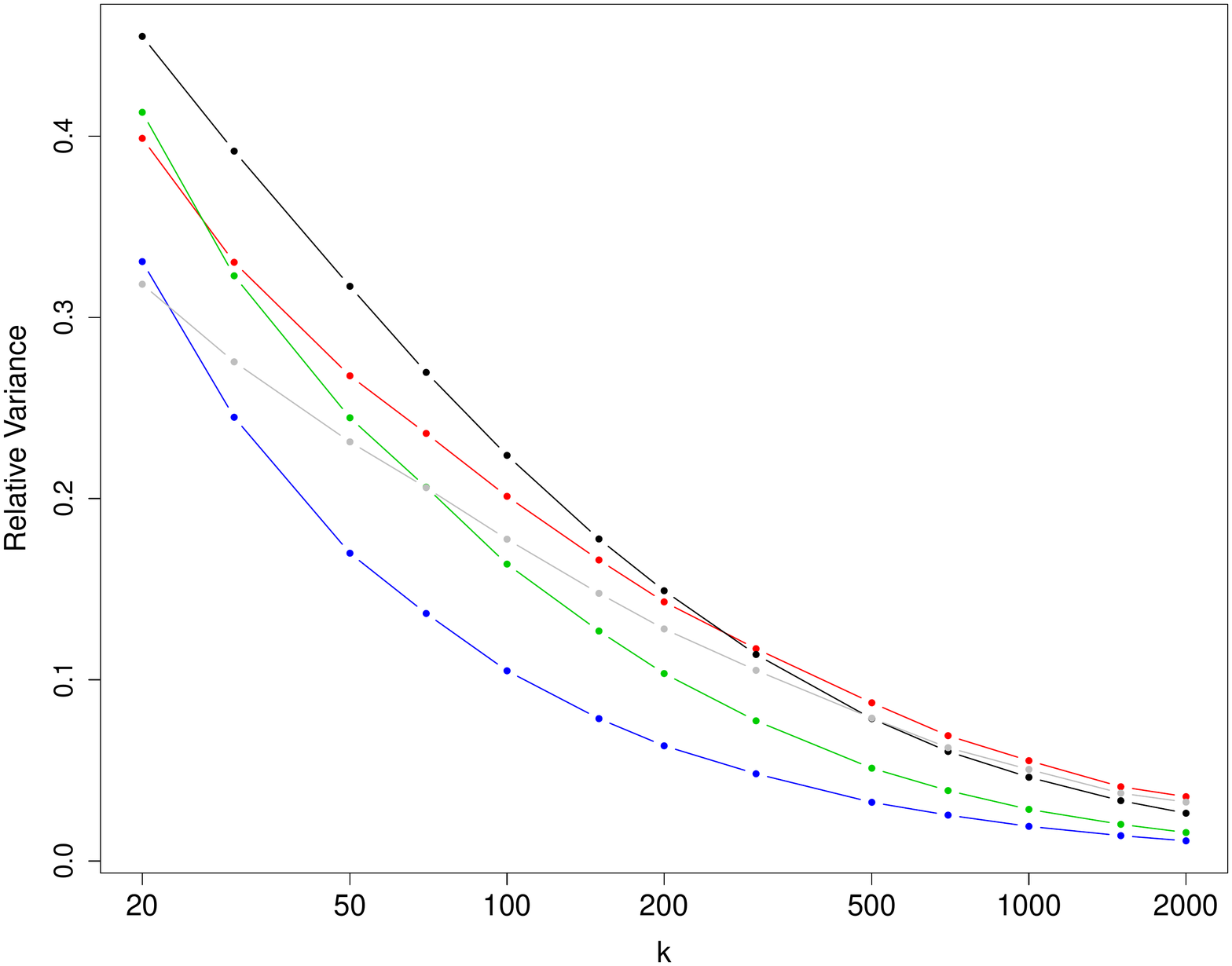}}
		\captionit{Shape parameter estimates when sampling from the GEV distribution with $(\mu, \sigma, \xi)=(0, 1)$ using the fixed-threshold stopping rule with threshold $c_k$ with $\xi=0.2$(top) and $\xi=-0.2$ (bottom) both plotted against $k$. Left: relative bias, centre: relative RMSE, right: relative variance, using: standard likelihood (red), excluding the final observation (black), full conditioning (green), partial conditioning (blue) and truncating (grey). Based on $10^5$ replicated samples with the historical data created using approach~\eqref{eqn:initial} of the paper.}
		\label{fig:sup.shape}
	\end{center}
\end{figure}

\begin{figure}[h]
		\subfigure{\includegraphics[width=0.33\textwidth]{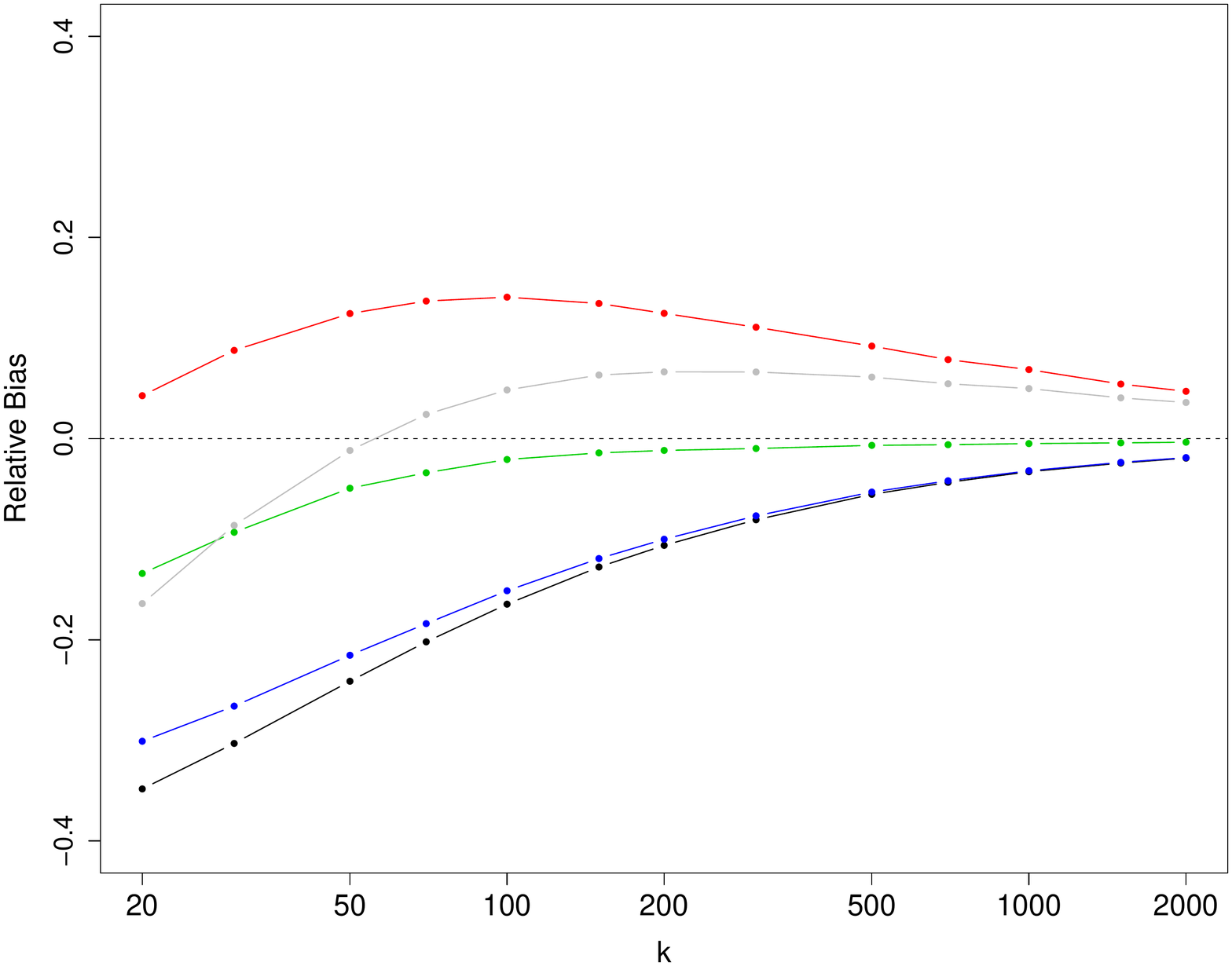}}
		\subfigure{\includegraphics[width=0.33\textwidth]{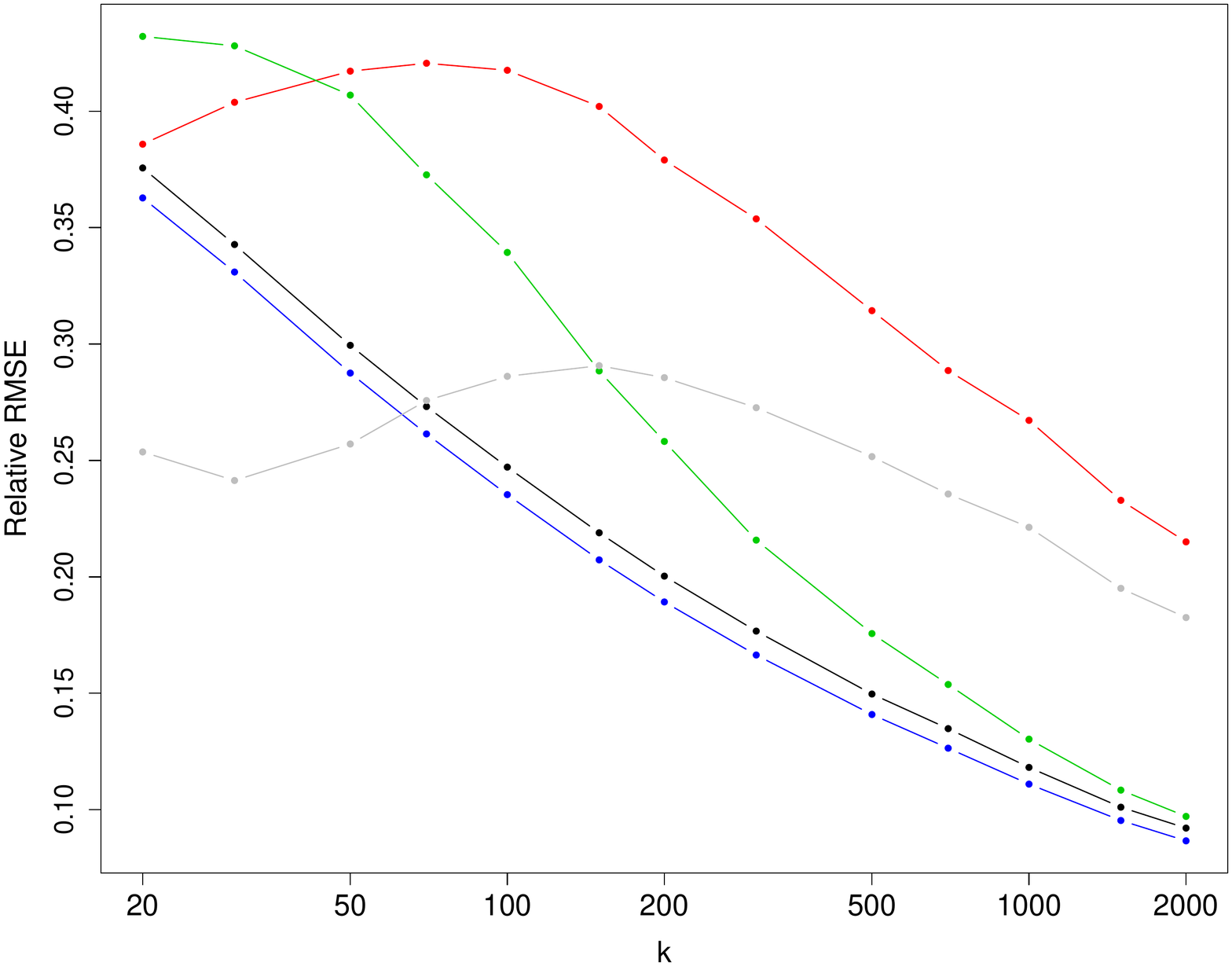}}
		\subfigure{\includegraphics[width=0.33\textwidth]{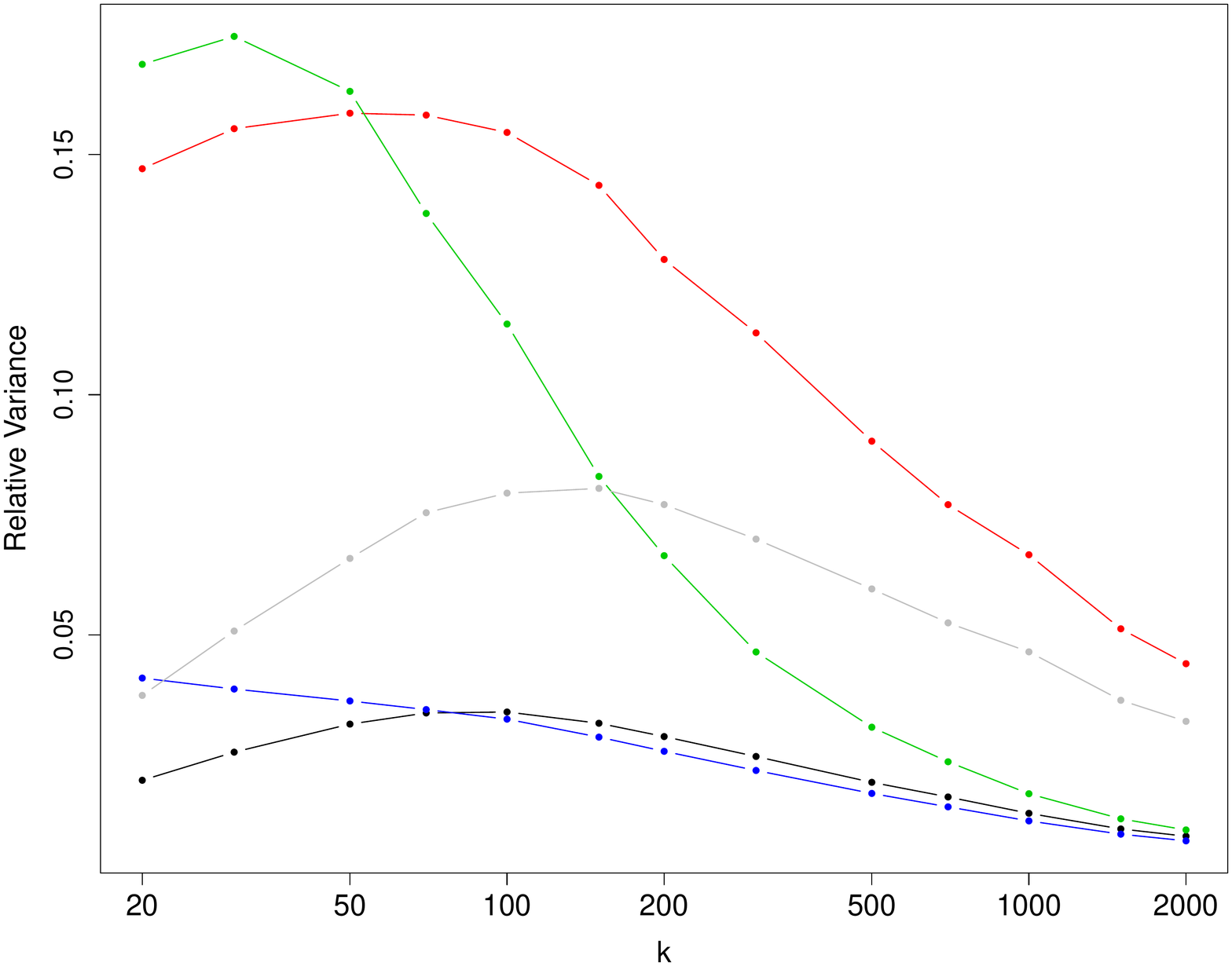}}
		\subfigure{\includegraphics[width=0.33\textwidth]{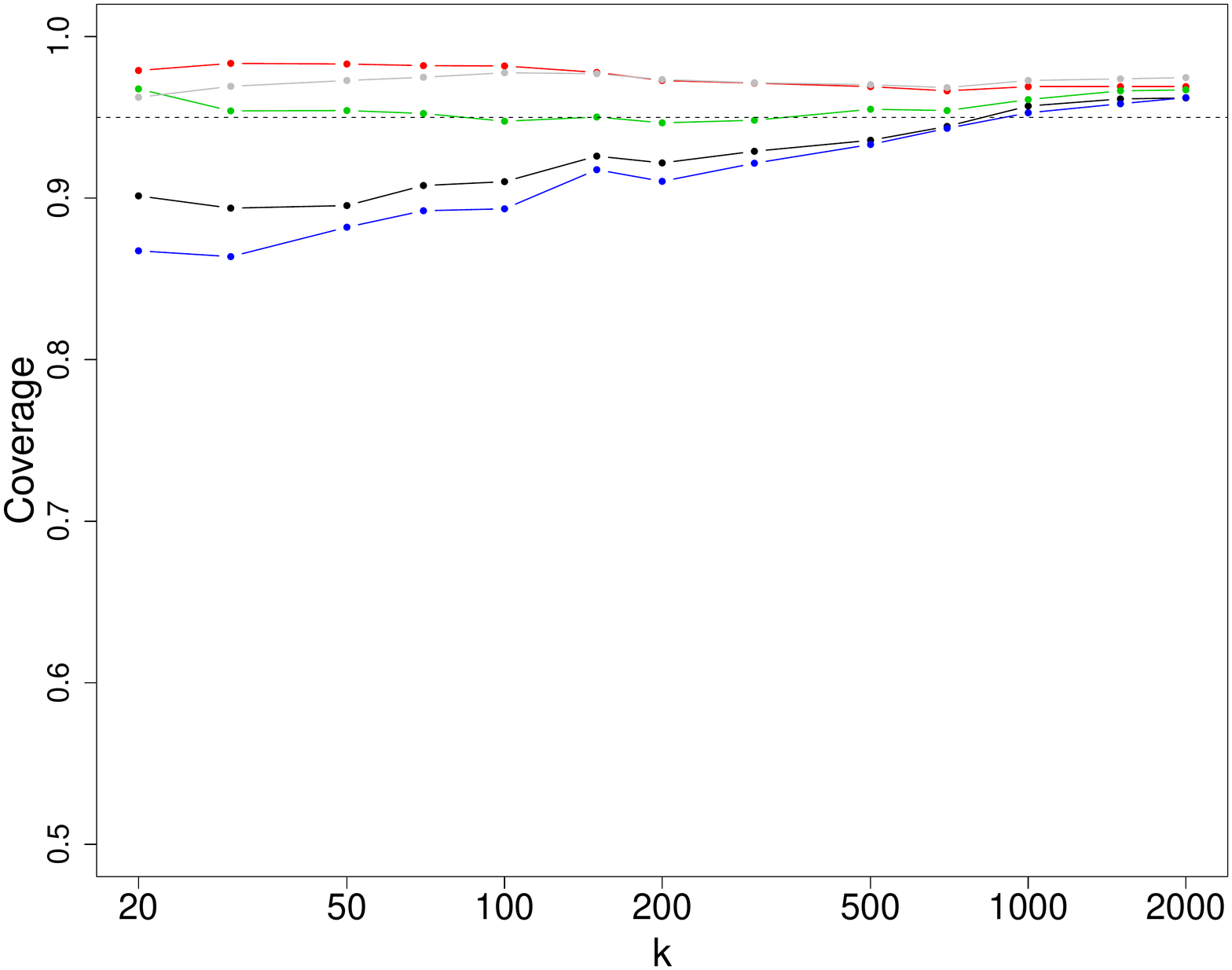}}
		\subfigure{\includegraphics[width=0.33\textwidth]{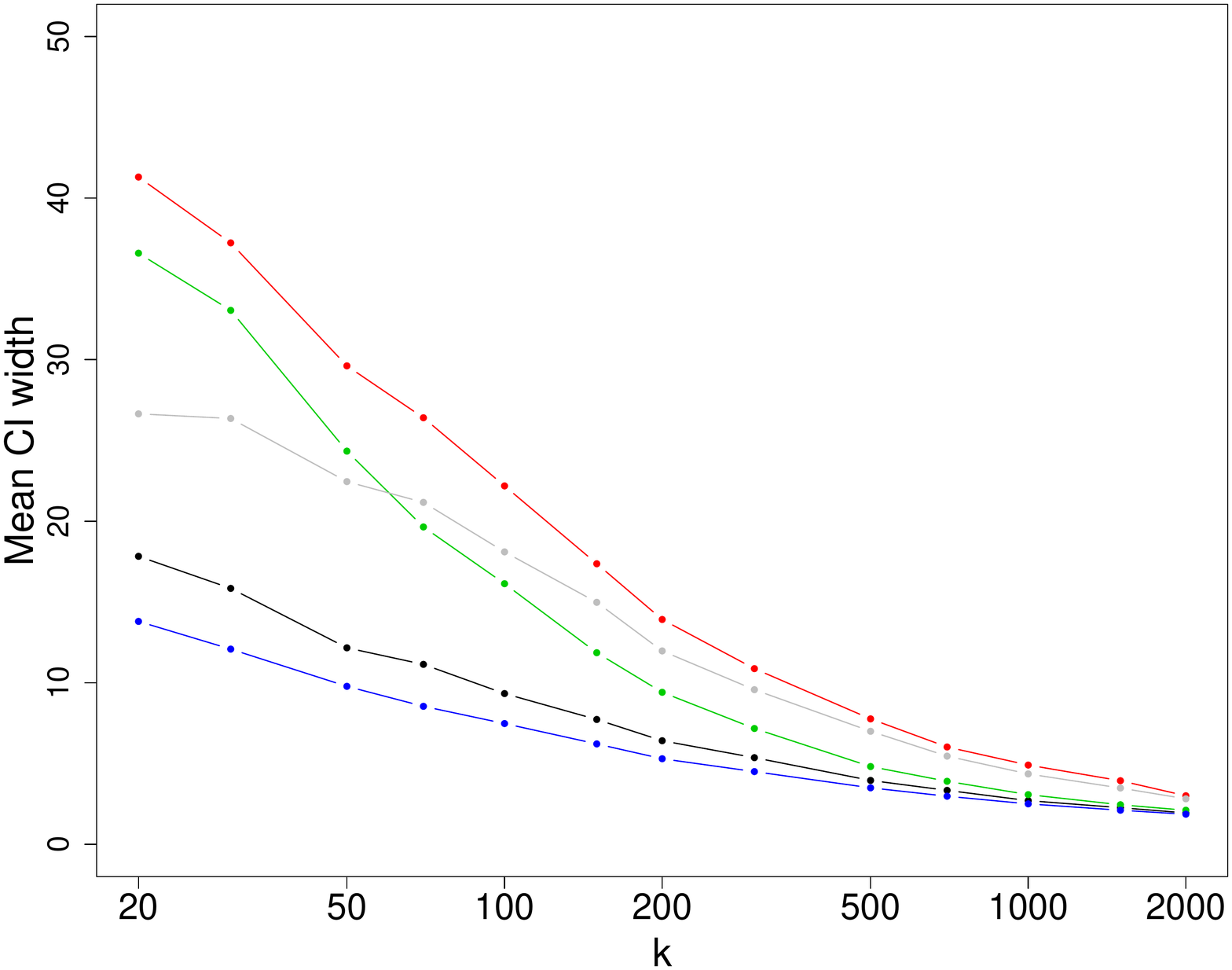}}
		\subfigure{\includegraphics[width=0.33\textwidth]{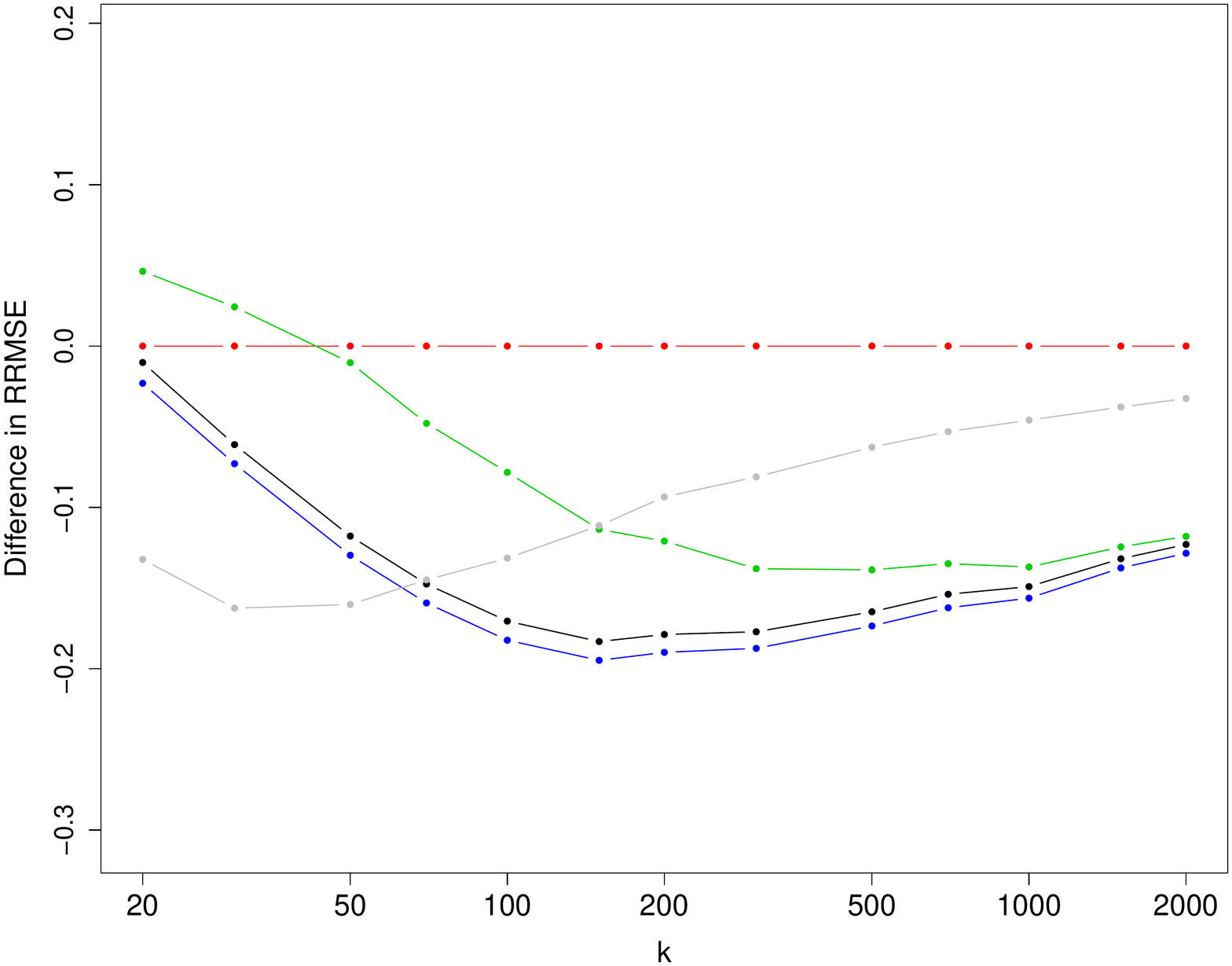}}
		\captionit{50 year return level estimates when sampling from the GEV distribution with $(\mu, \sigma, \xi)=(0, 1, 0.2)$ using the fixed-threshold stopping rule over a range of thresholds. From left to right. Top: relative bias, relative RMSE and relative variance. Bottom: coverage, average CI width, difference in RRMSE to RRMSE of standard estimator. Colour scheme is the same as in Figure \ref{fig:sup.shape}. Based on $10^5$ replicated samples with the historical data created using approach~\eqref{eqn:initial} of the paper. Coverage is based on 5000 replicated samples.}
		\label{fig:sup.200ret}
\end{figure}

\begin{figure}[h]
		\subfigure{\includegraphics[width=0.33\textwidth]{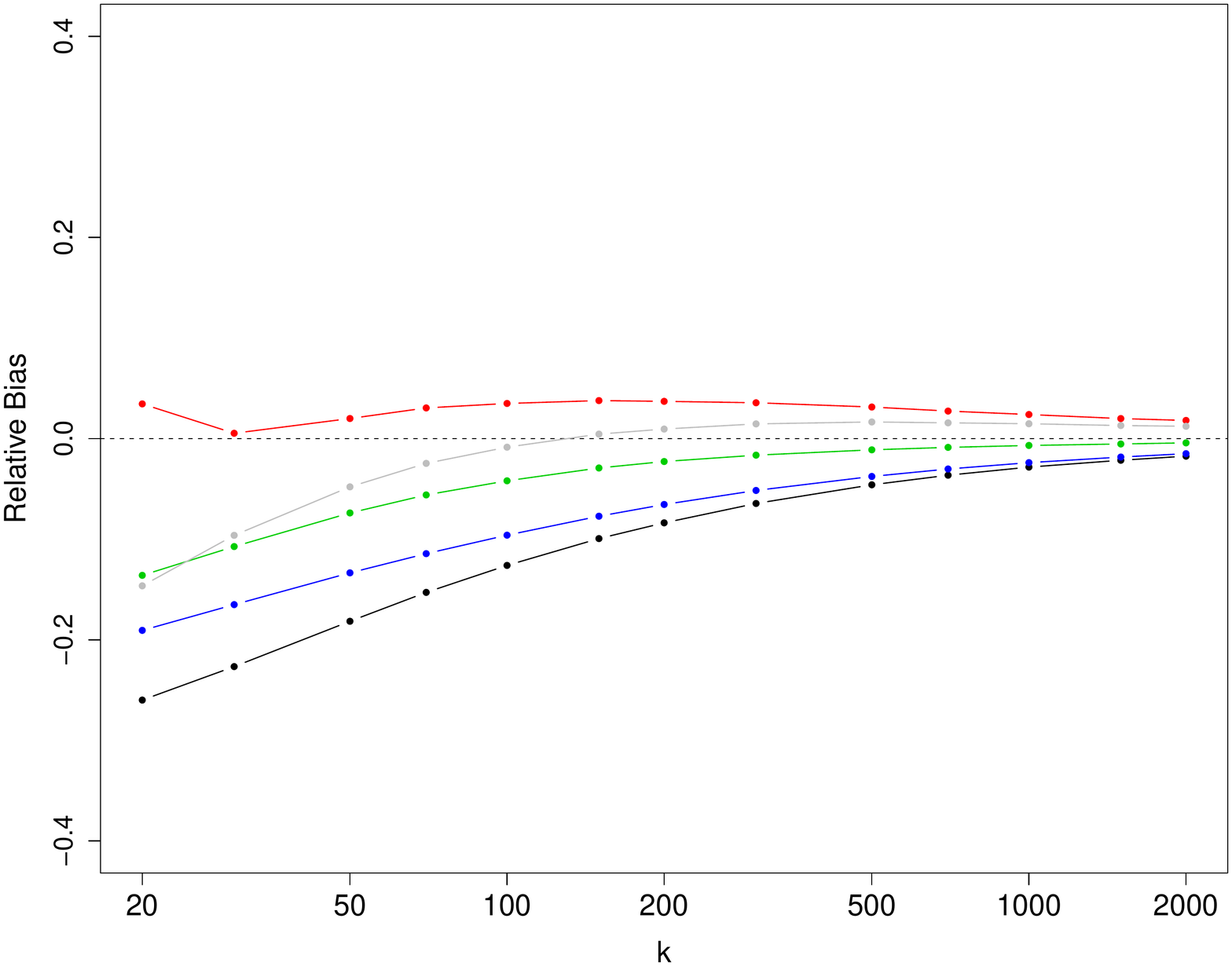}}
		\subfigure{\includegraphics[width=0.33\textwidth]{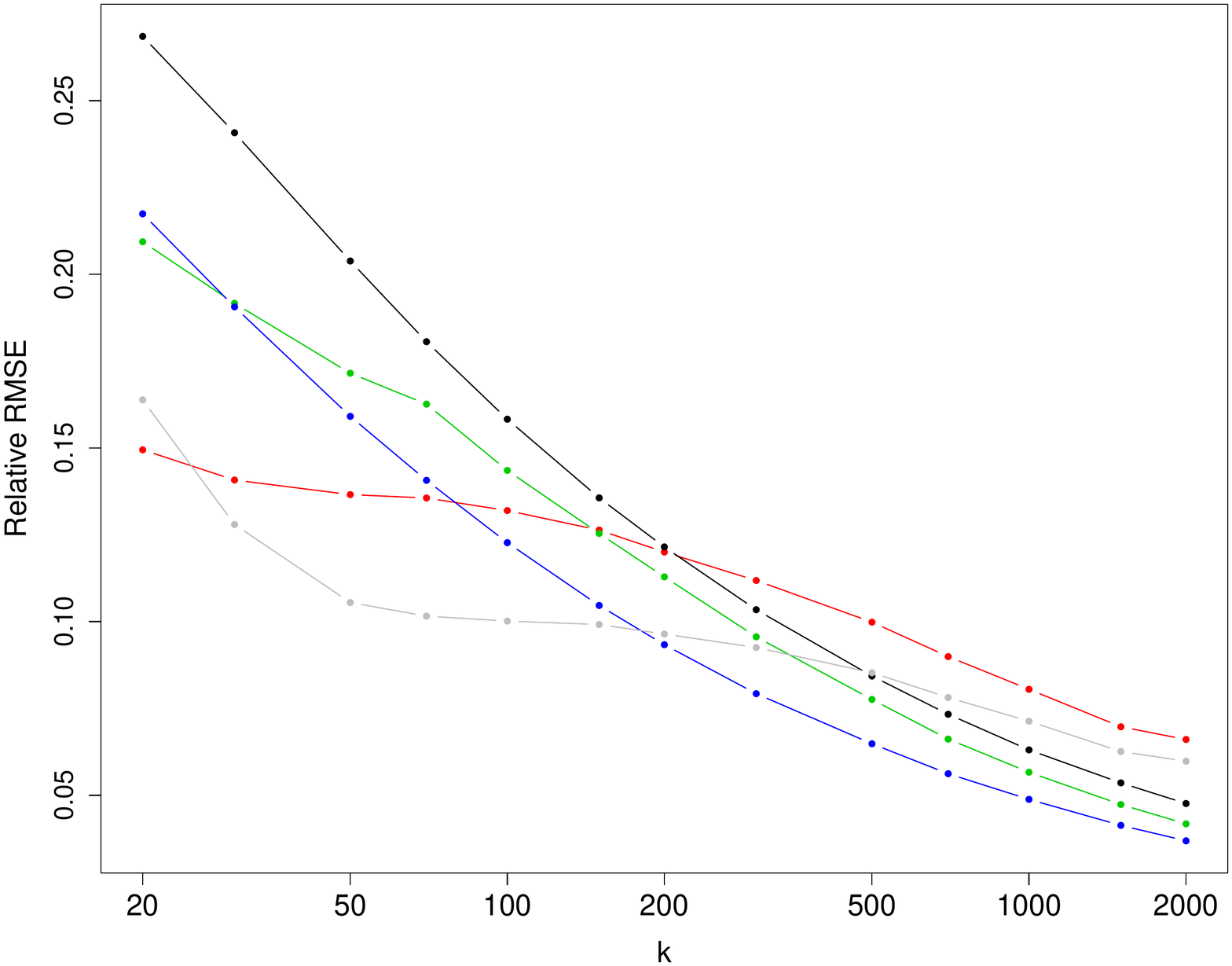}}
		\subfigure{\includegraphics[width=0.33\textwidth]{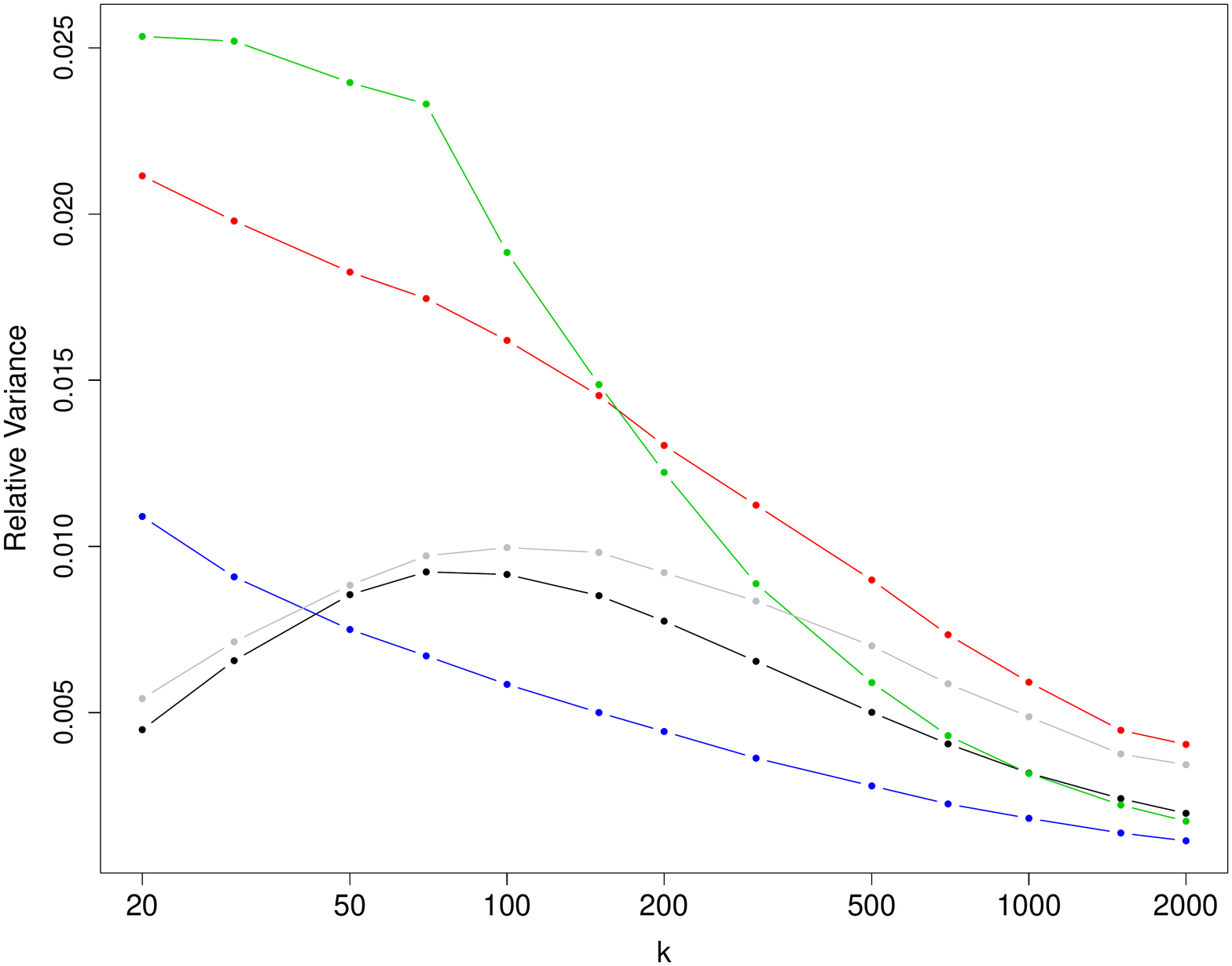}}
		\subfigure{\includegraphics[width=0.33\textwidth]{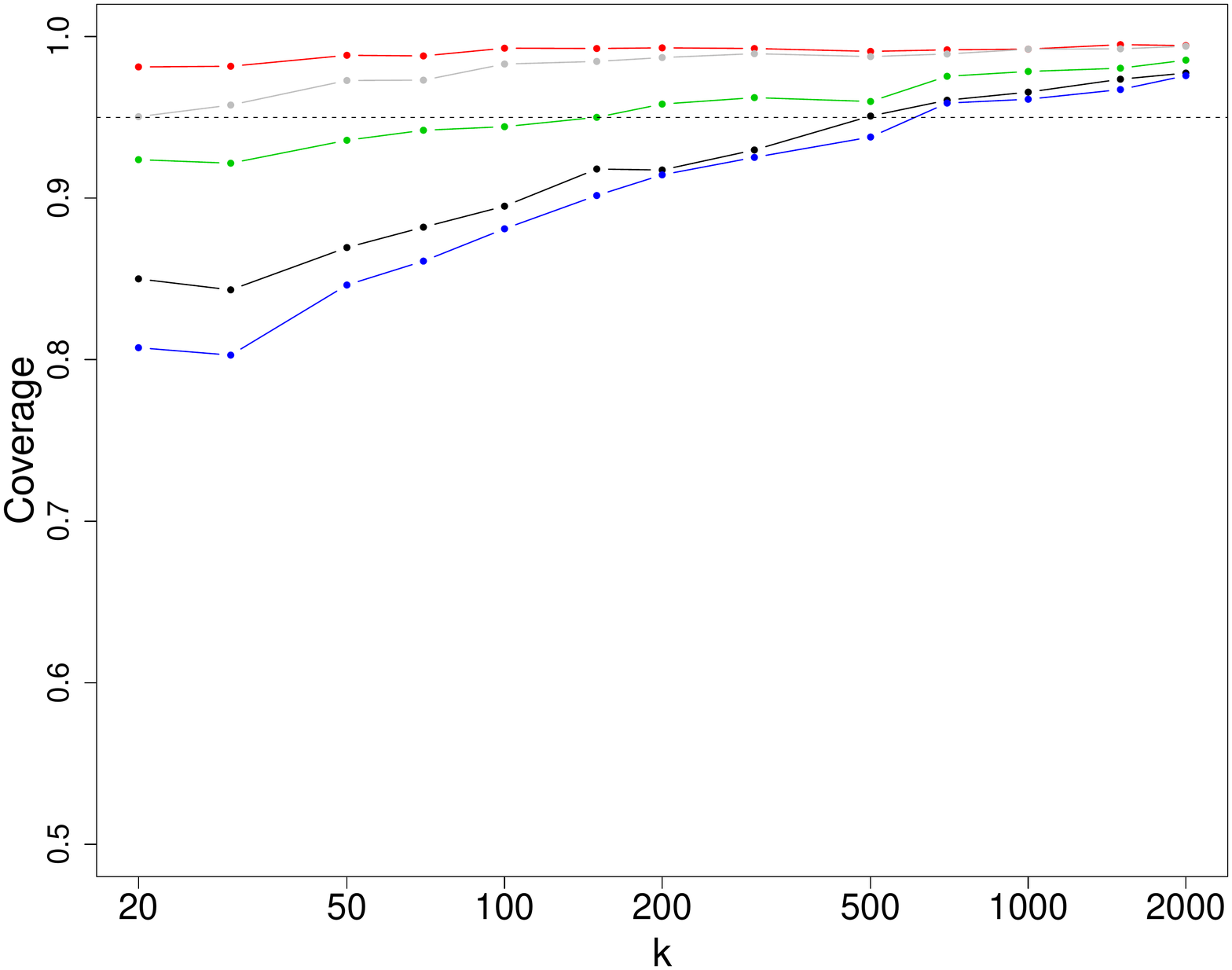}}
		\subfigure{\includegraphics[width=0.33\textwidth]{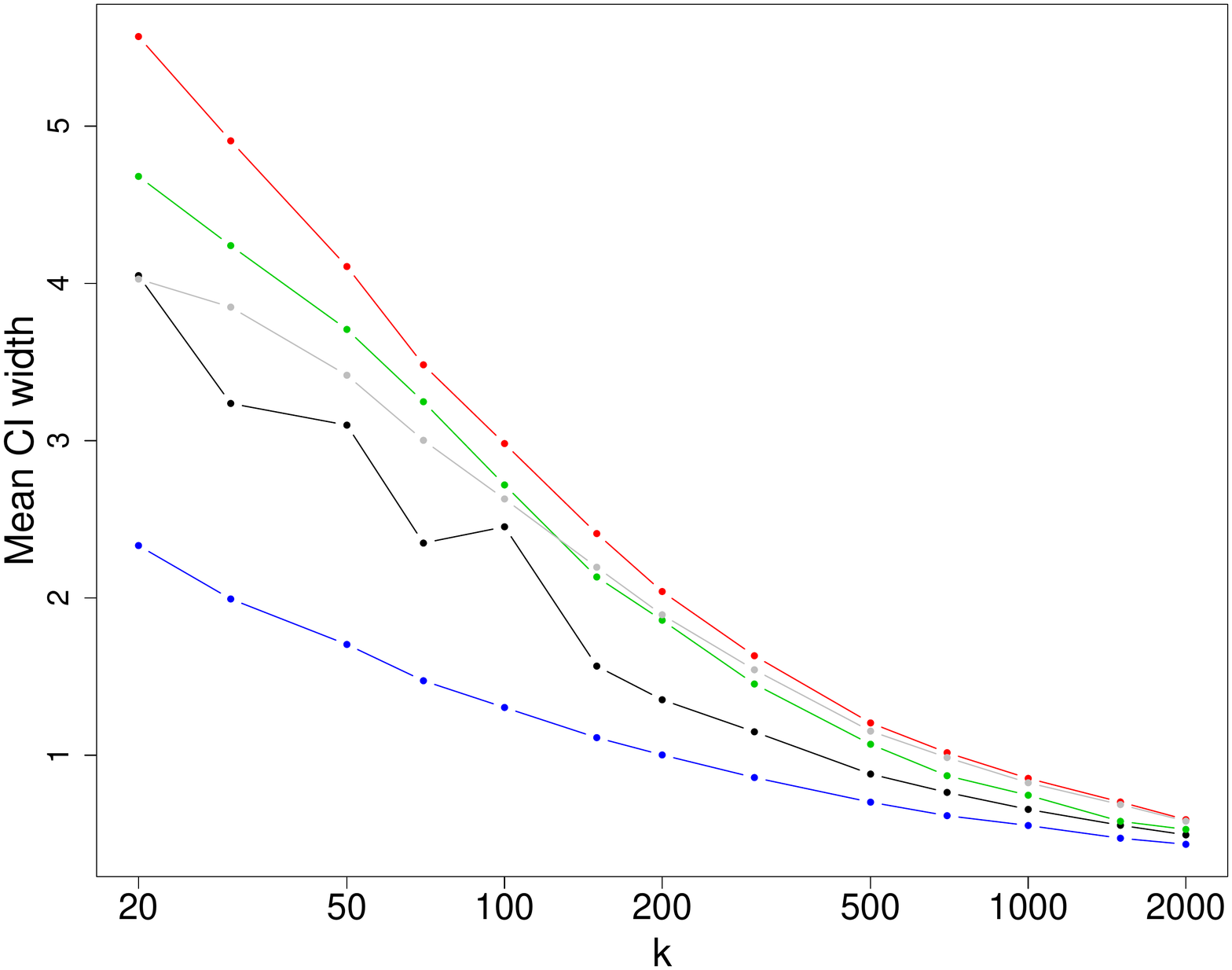}}
		\subfigure{\includegraphics[width=0.33\textwidth]{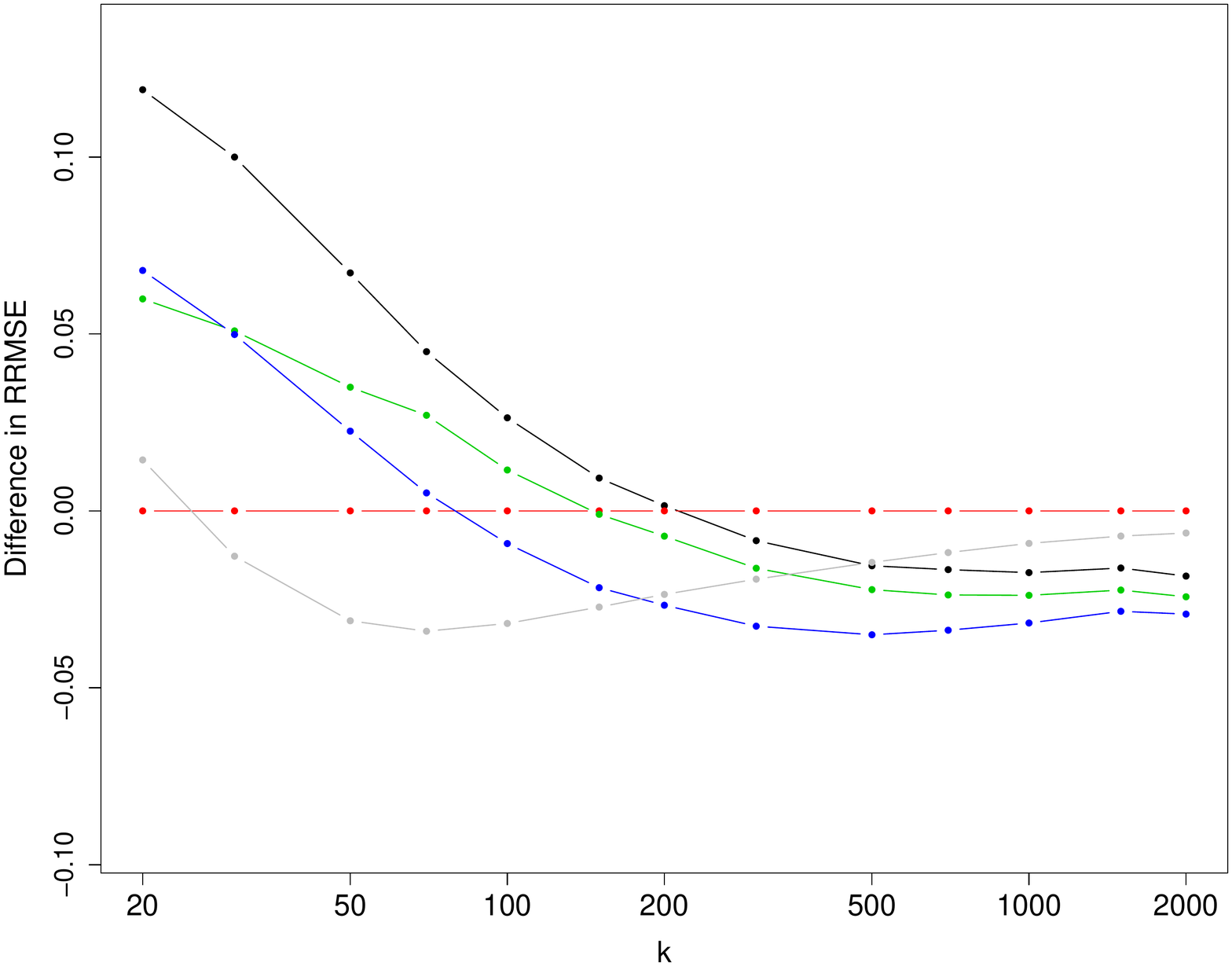}}
		\captionit{50 year return level estimates when sampling from the GEV distribution with $(\mu, \sigma, \xi)=(0, 1, -0.2)$ using the fixed-threshold stopping rule over a range of thresholds. From left to right. Top: relative bias, relative RMSE and relative variance. Bottom: coverage, average CI width, difference in RRMSE to RRMSE of standard estimator. Colour scheme is the same as in Figure \ref{fig:sup.shape}. Based on $10^5$ replicated samples with the historical data created using approach~\eqref{eqn:initial} of the paper. Coverage is based on 5000 replicated samples.}
		\label{fig:sup.200retneg}
\end{figure}

\begin{figure}[h]
		\subfigure{\includegraphics[width=0.33\textwidth]{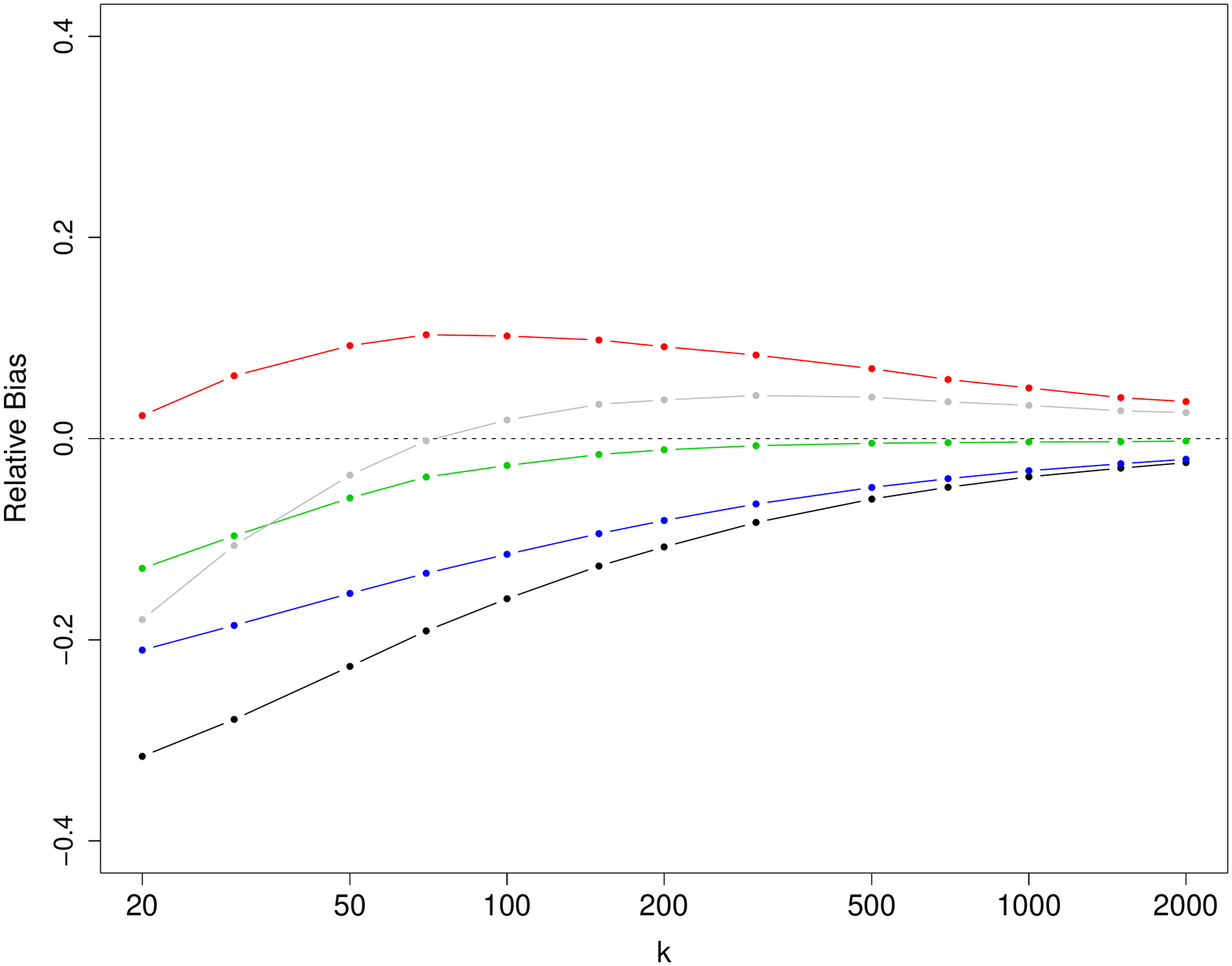}}
		\subfigure{\includegraphics[width=0.33\textwidth]{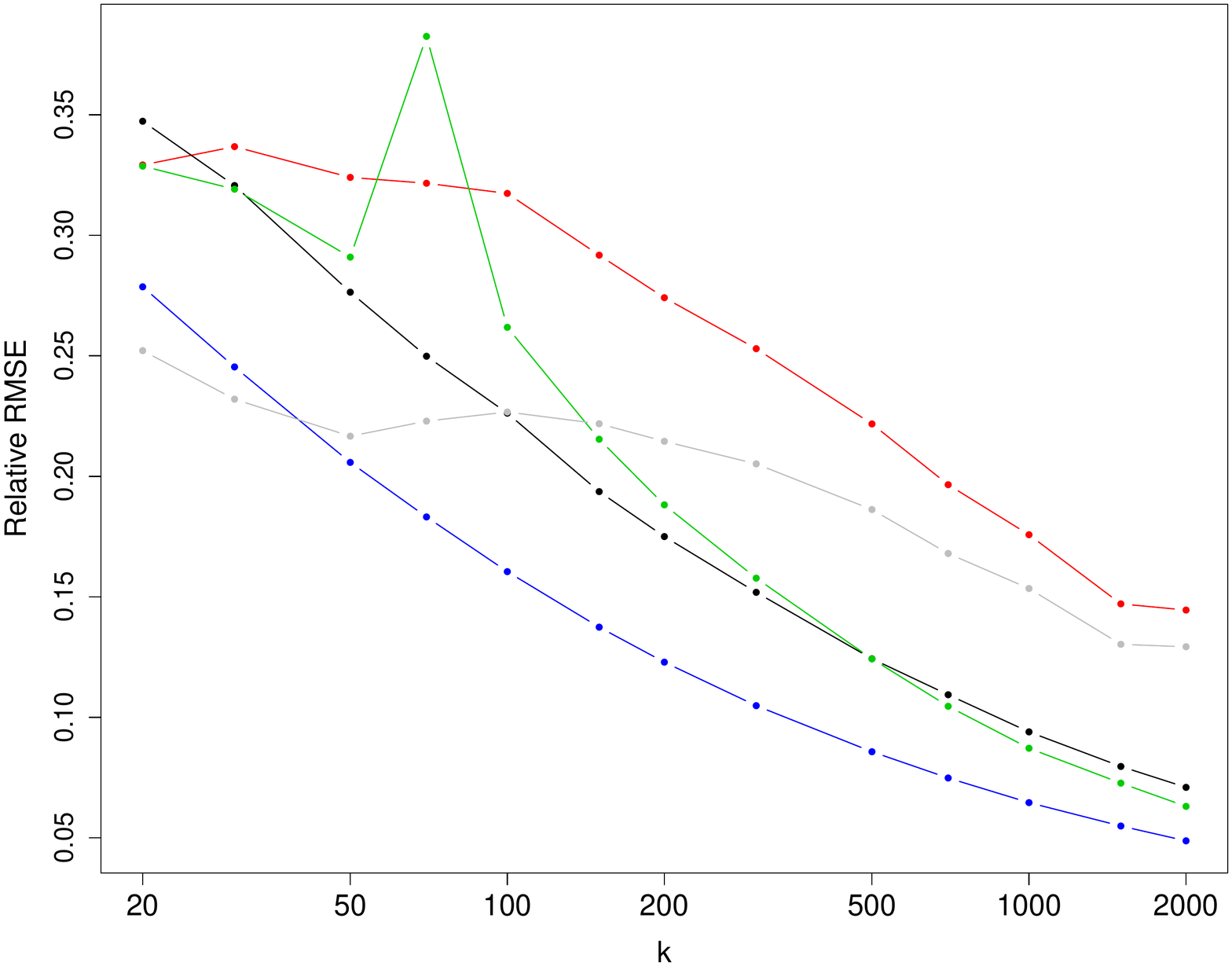}}
		\subfigure{\includegraphics[width=0.33\textwidth]{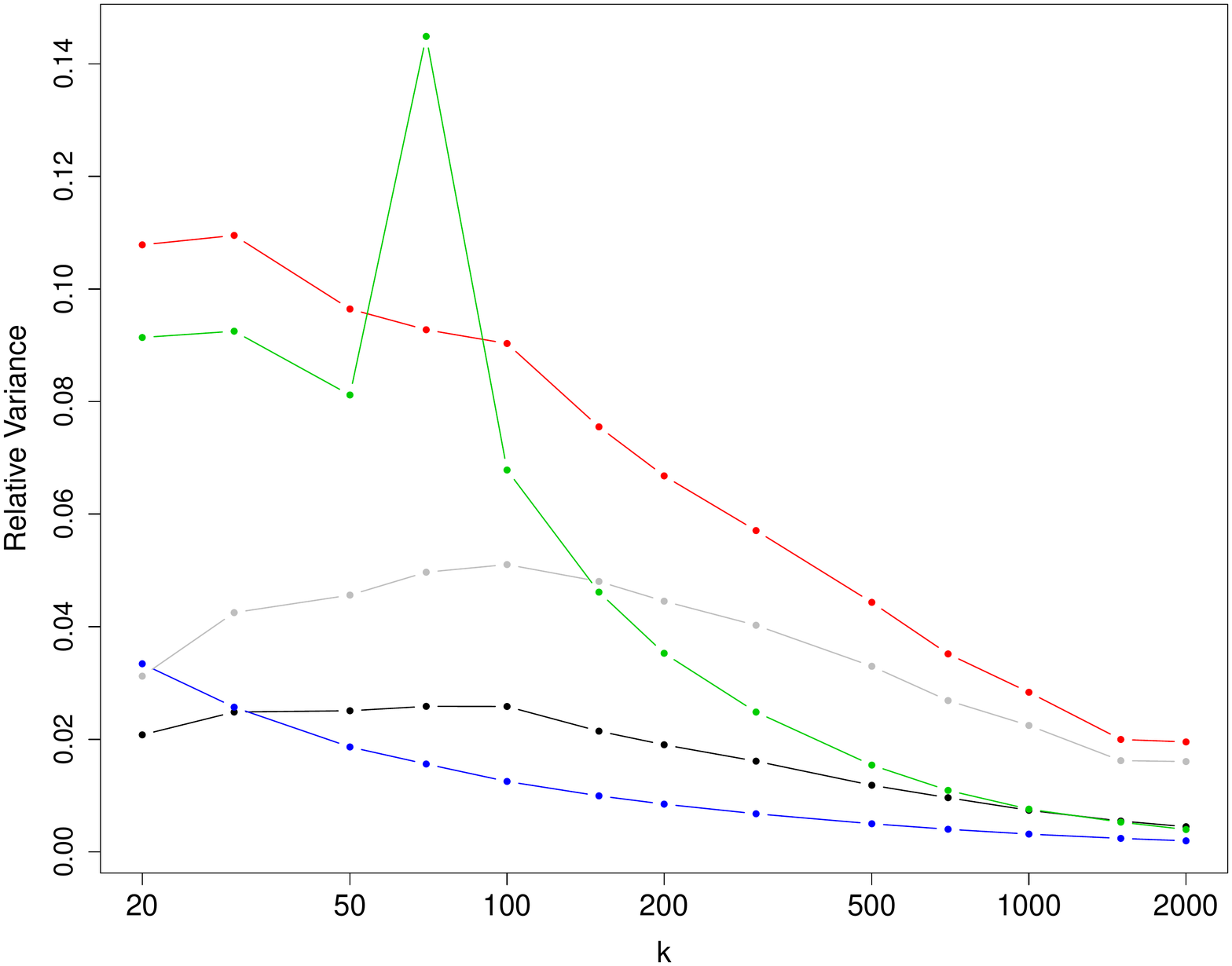}}
		\subfigure{\includegraphics[width=0.33\textwidth]{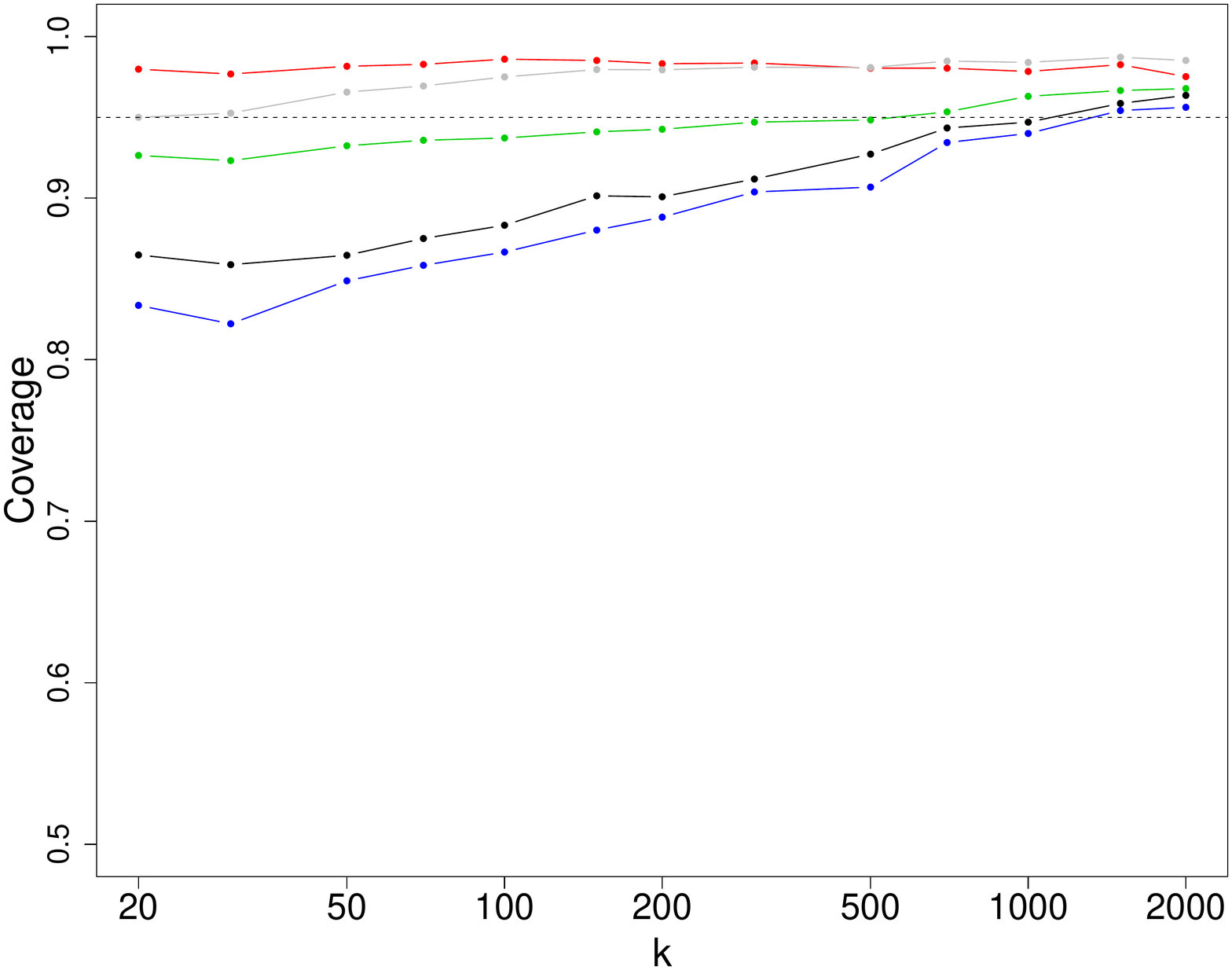}}
		\subfigure{\includegraphics[width=0.33\textwidth]{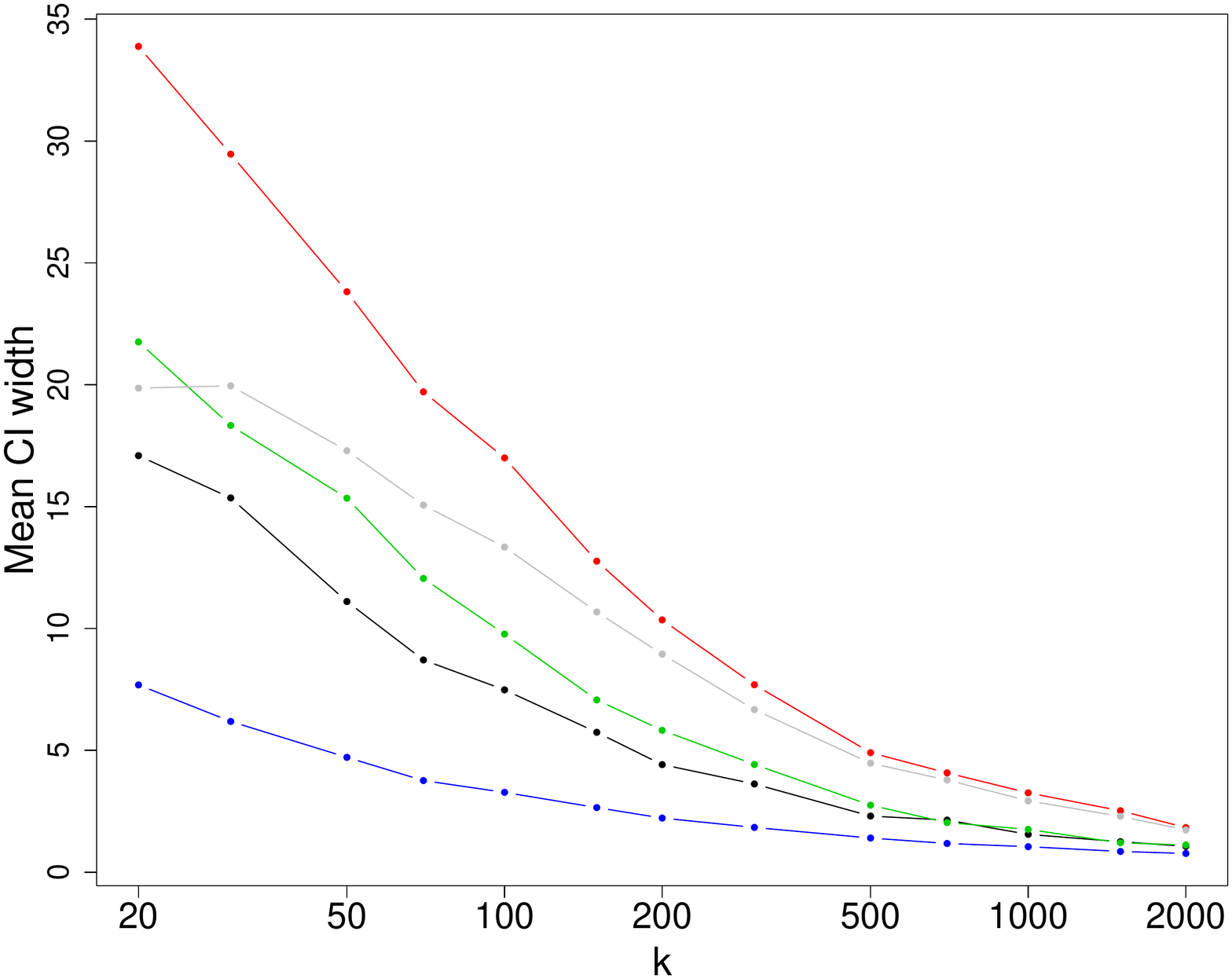}}
		\subfigure{\includegraphics[width=0.33\textwidth]{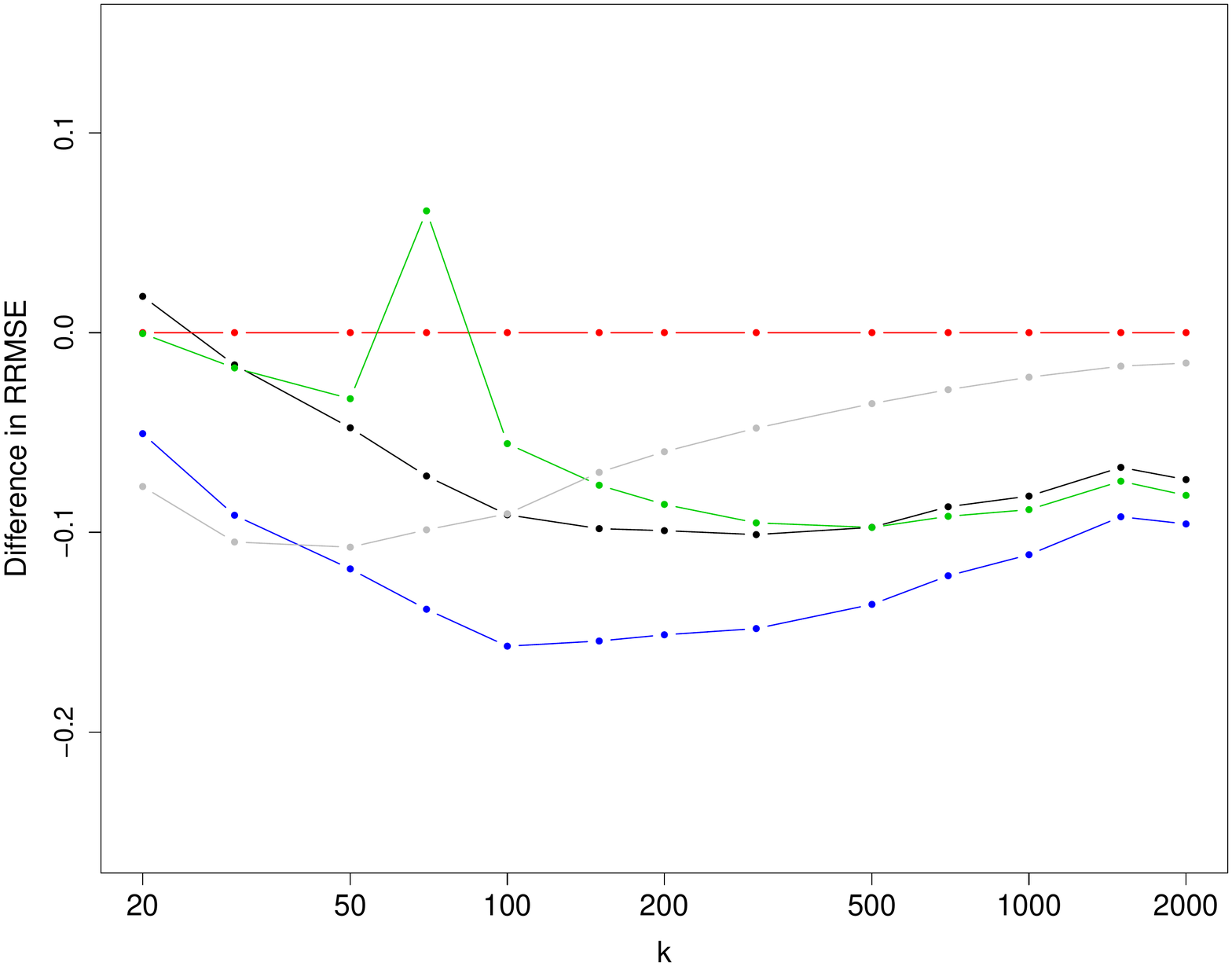}}
		\captionit{1000 year return level estimates when sampling from the GEV distribution with $(\mu, \sigma, \xi)=(0, 1, -0.2)$ using the fixed-threshold stopping rule over a range of thresholds. From left to right. Top: relative bias, relative RMSE and relative variance. Bottom: coverage, average CI width, difference in RRMSE to RRMSE of standard estimator. Colour scheme is the same as in Figure \ref{fig:sup.shape}. Based on $10^5$ replicated samples with the historical data created using approach~\eqref{eqn:initial} of the paper. Coverage is based on 5000 replicated samples.}
		\label{fig:1000retneg}
\end{figure}

\begin{figure}[h]
	\begin{center}
		\subfigure{\includegraphics[width=0.32\textwidth]{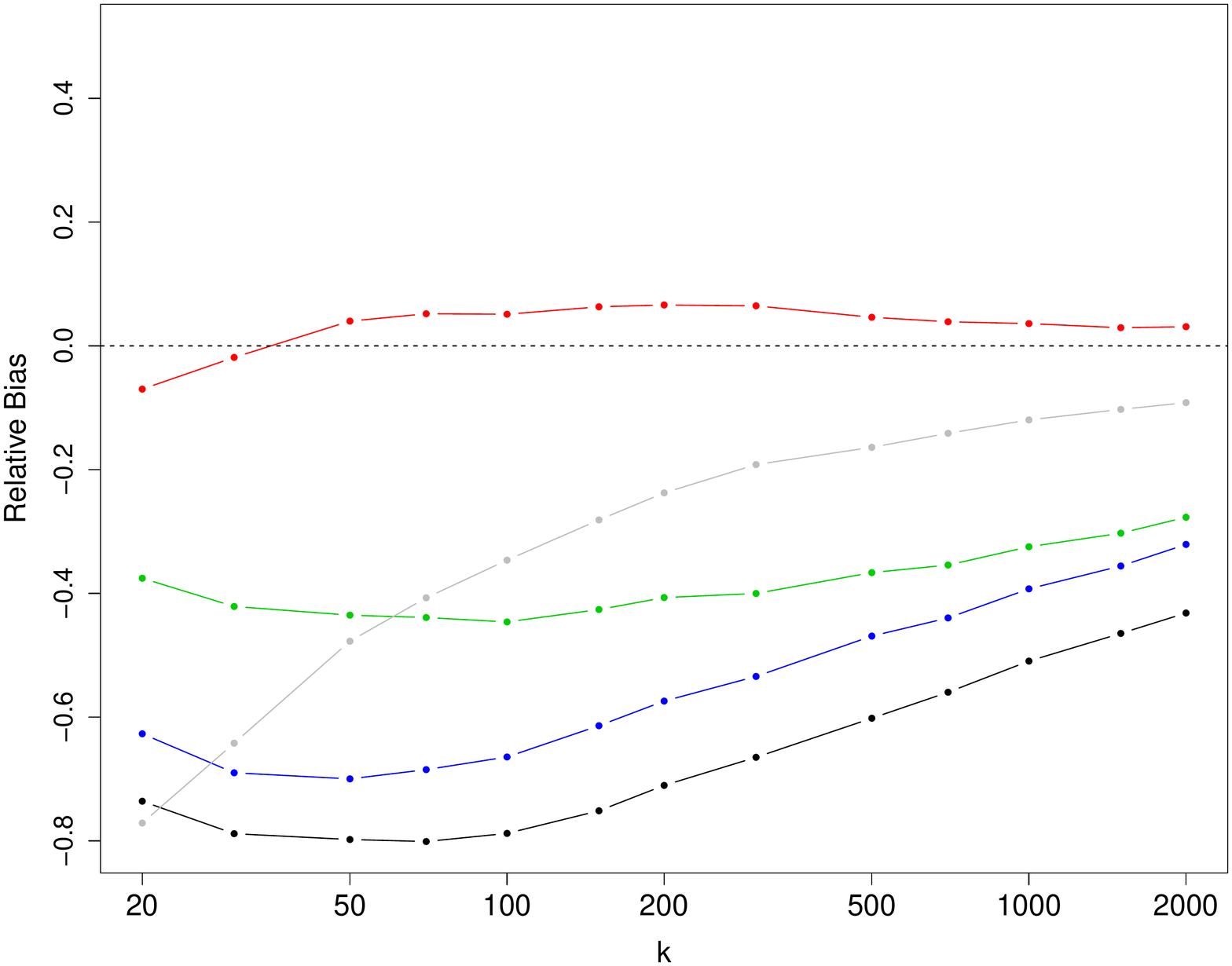}}
		\subfigure{\includegraphics[width=0.32\textwidth]{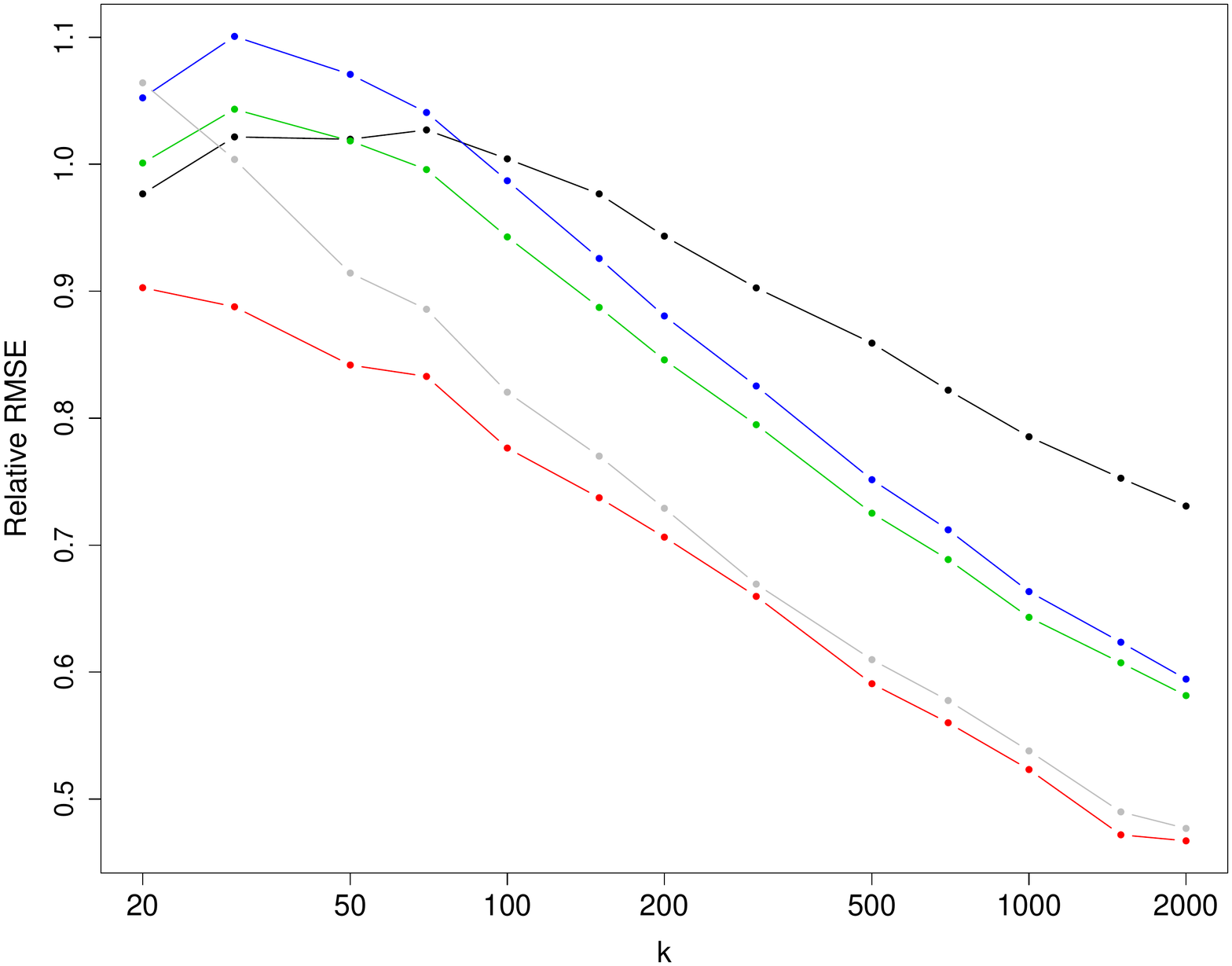}}
		\subfigure{\includegraphics[width=0.32\textwidth]{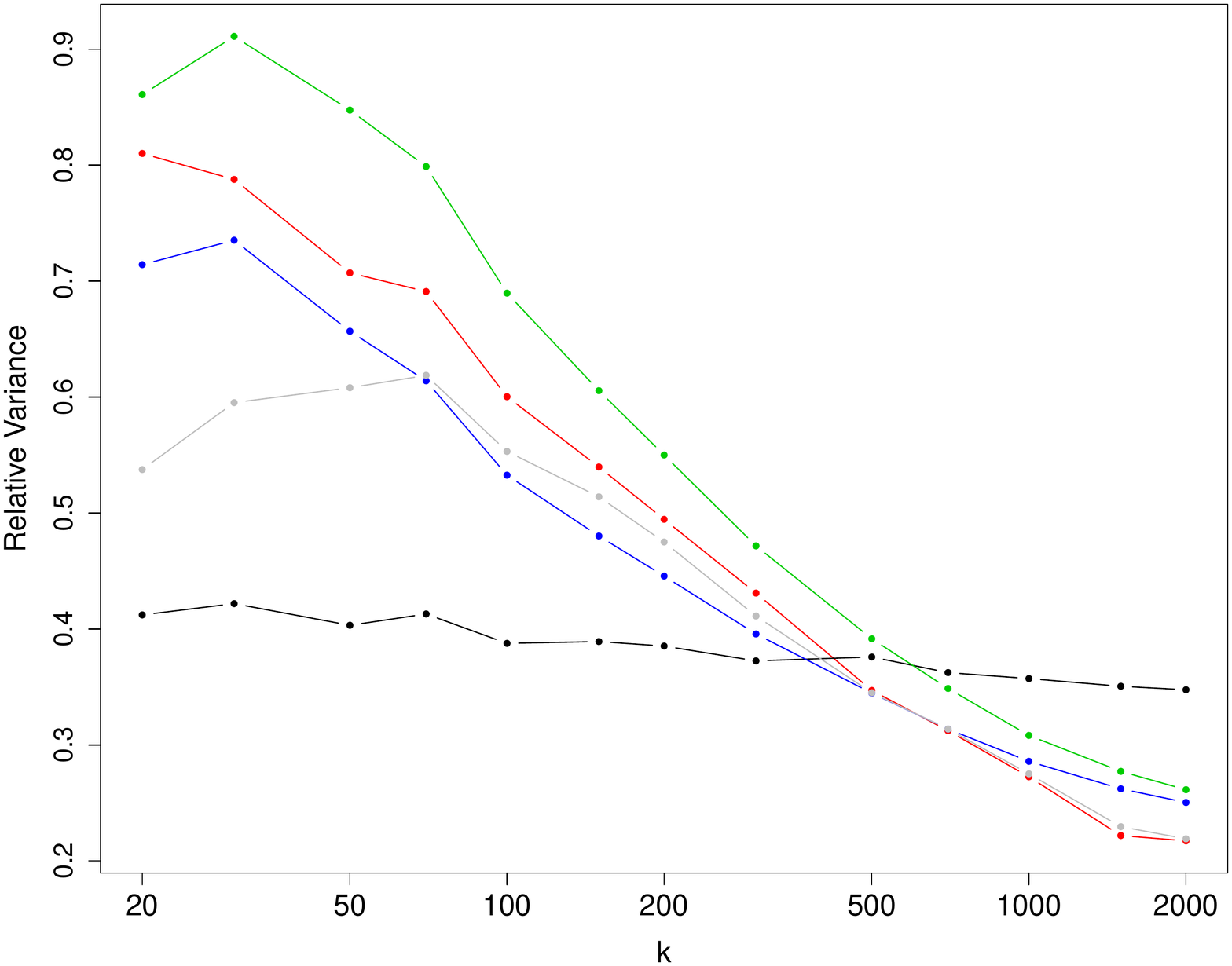}}
		\subfigure{\includegraphics[width=0.32\textwidth]{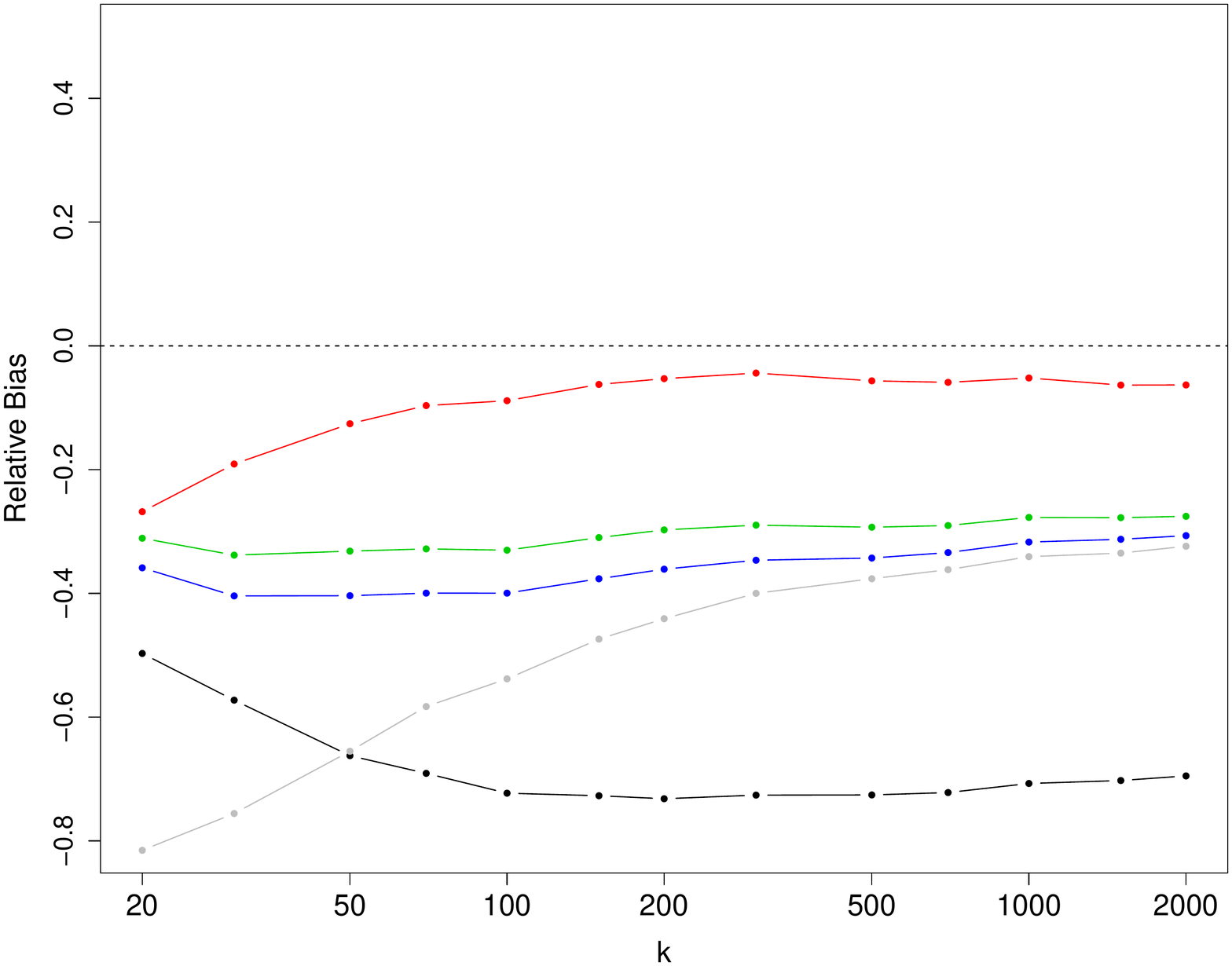}}
		\subfigure{\includegraphics[width=0.32\textwidth]{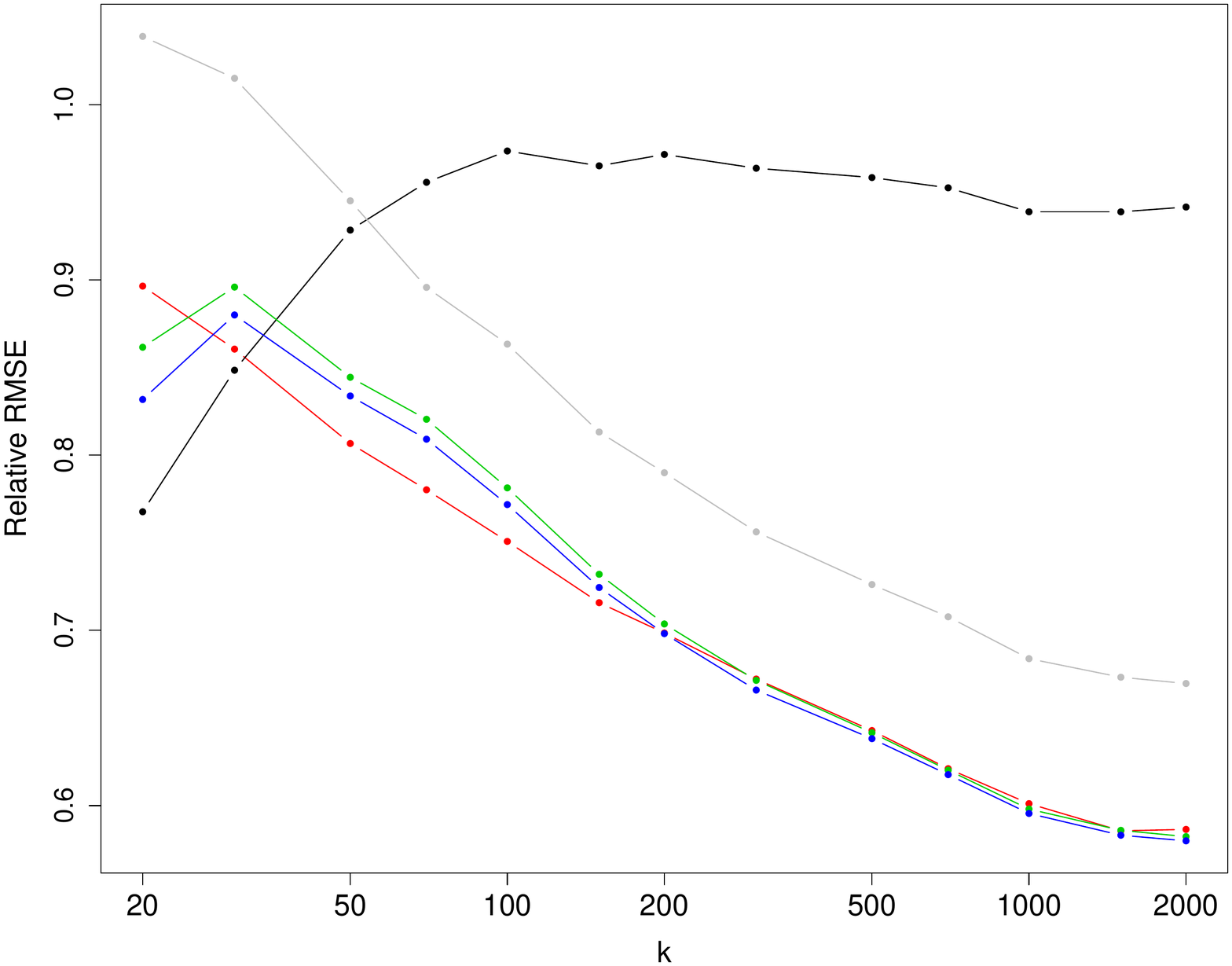}}
		\subfigure{\includegraphics[width=0.32\textwidth]{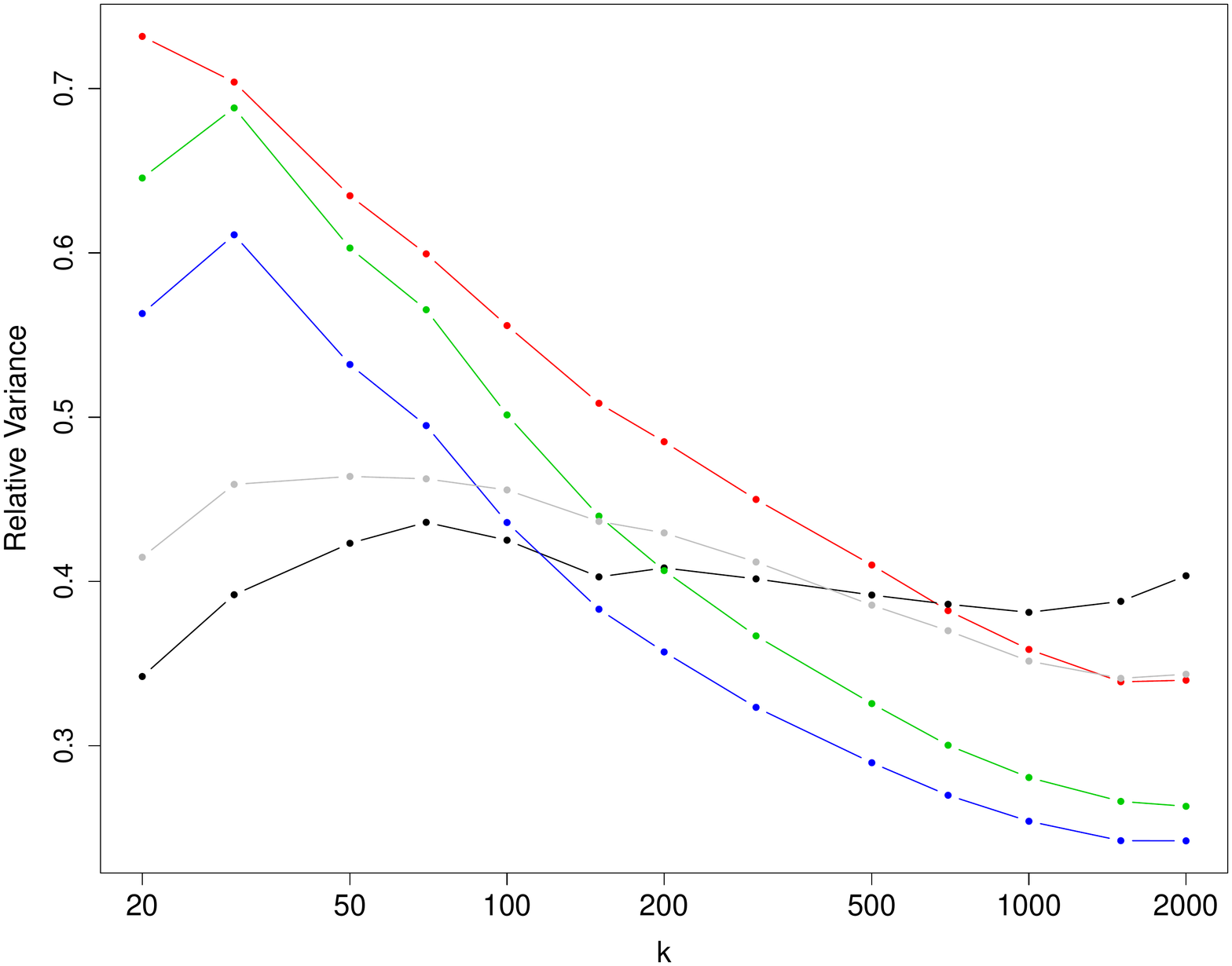}}
		\captionit{Shape parameter estimates when sampling from the GEV distribution with $(\mu, \sigma, \xi)=(0, 1)$ using the variable-threshold stopping rule with $\xi=0.2$(top) and $\xi=-0.2$ (bottom) both plotted against $k$. Left: relative bias, centre: relative RMSE, right: relative variance, using: standard likelihood (red), excluding the final observation (black), full conditioning (green), partial conditioning (blue) and truncating (grey). Based on $10000/20000$ (top/bottom) replicated samples with the historical data created using approach~\eqref{eqn:initial} of the paper.}
		\label{fig:varshape}
	\end{center}
\end{figure}

\begin{figure}[h]
		\subfigure{\includegraphics[width=0.33\textwidth]{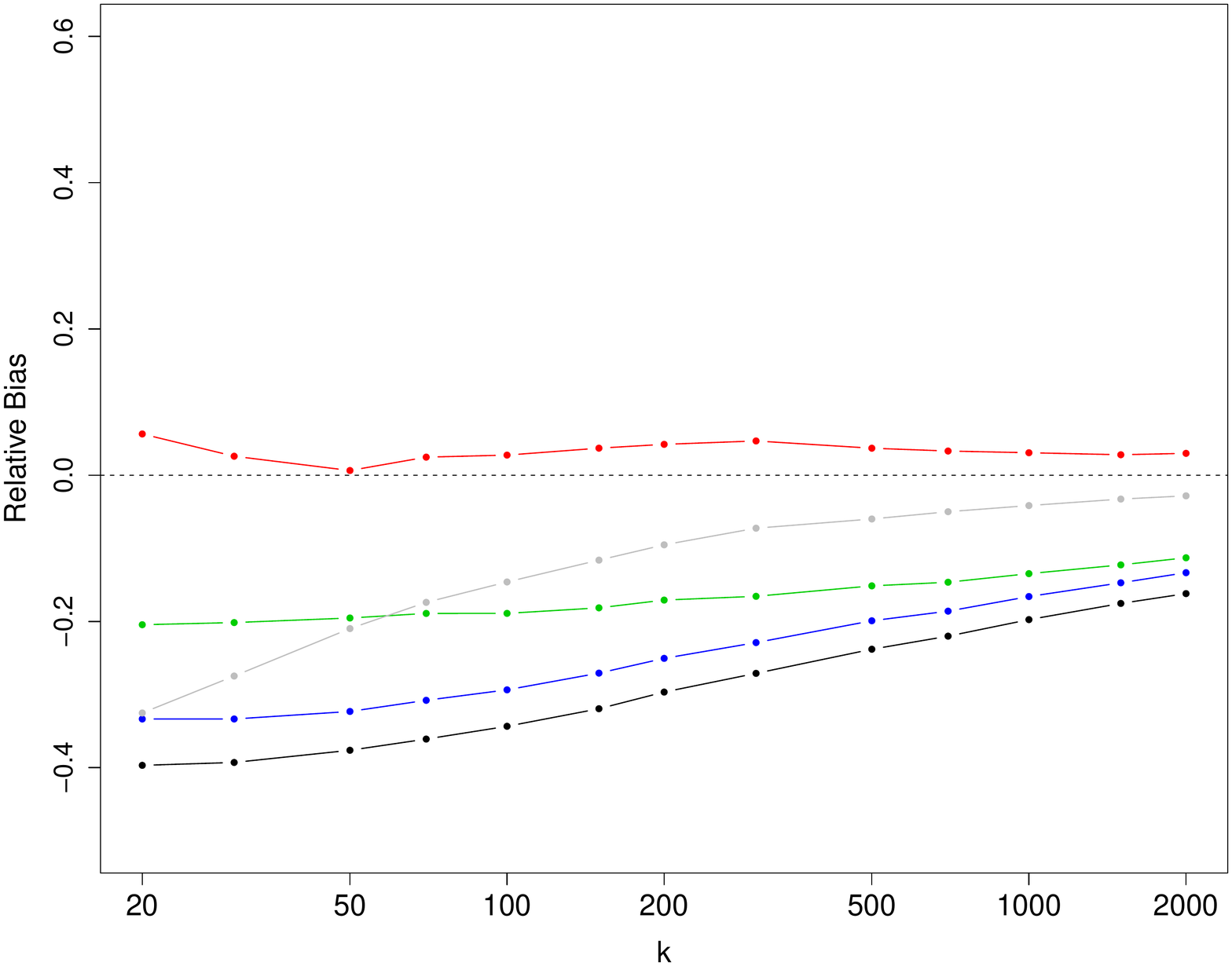}}
		\subfigure{\includegraphics[width=0.33\textwidth]{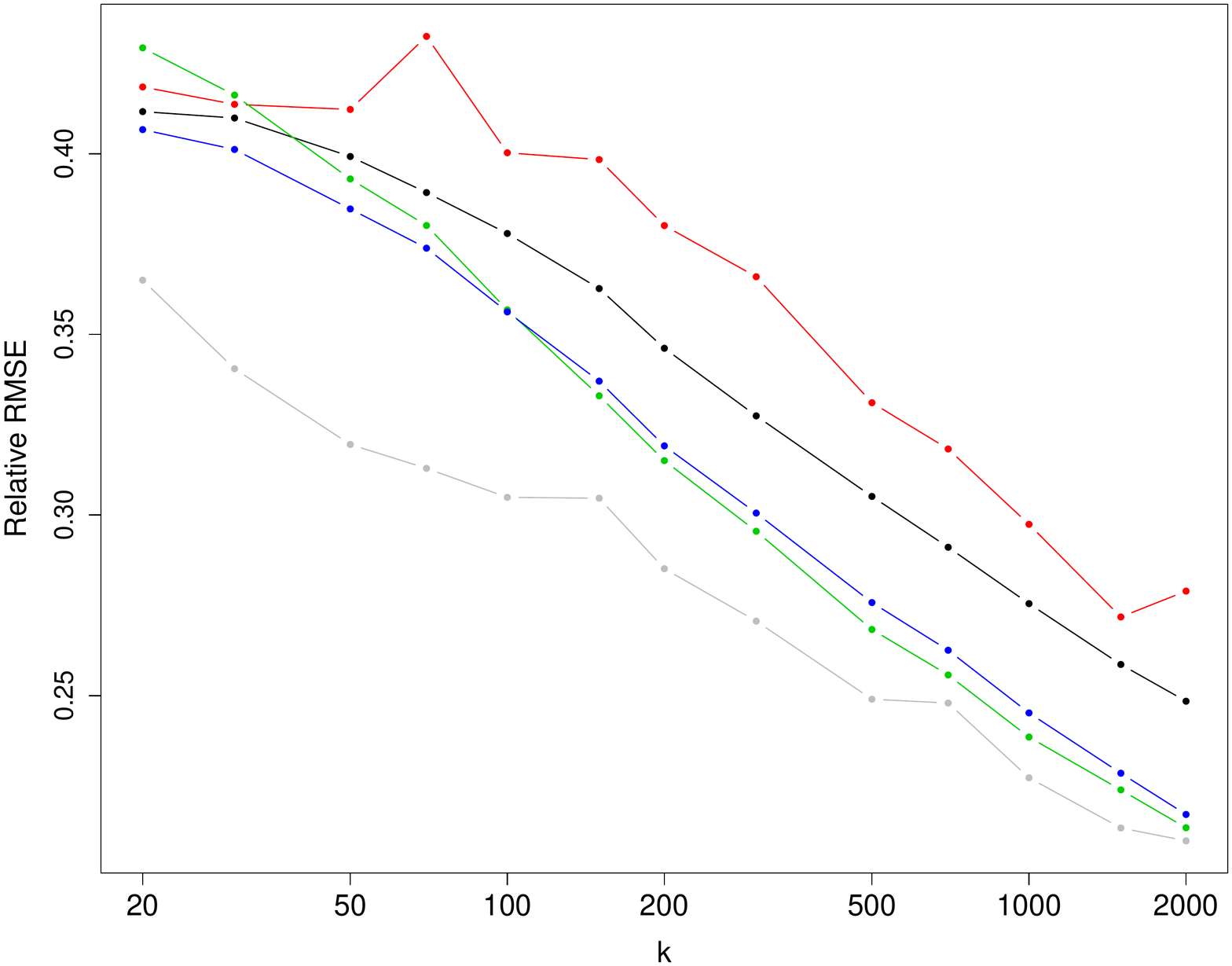}}
		\subfigure{\includegraphics[width=0.33\textwidth]{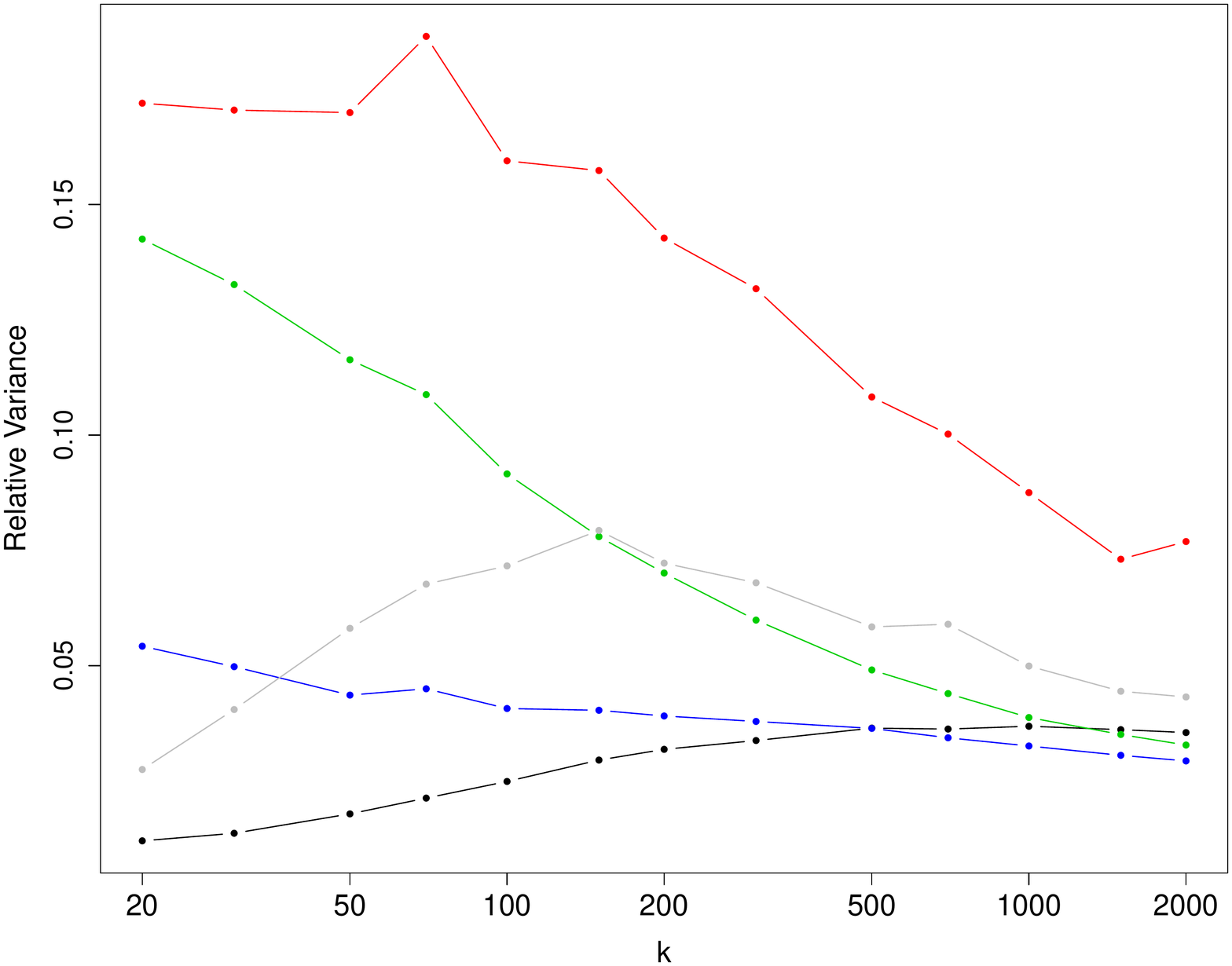}}
		\subfigure{\includegraphics[width=0.33\textwidth]{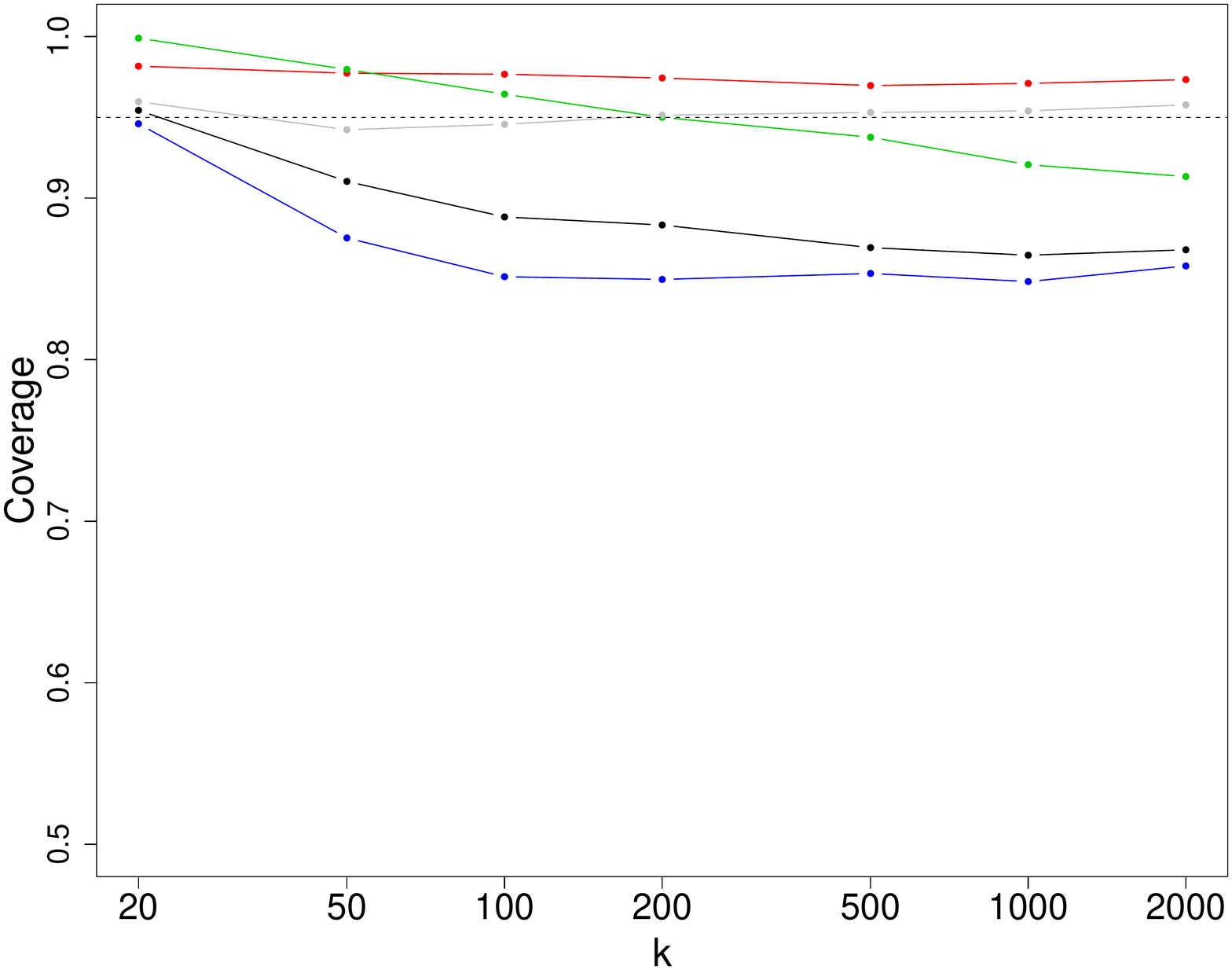}}
		\subfigure{\includegraphics[width=0.33\textwidth]{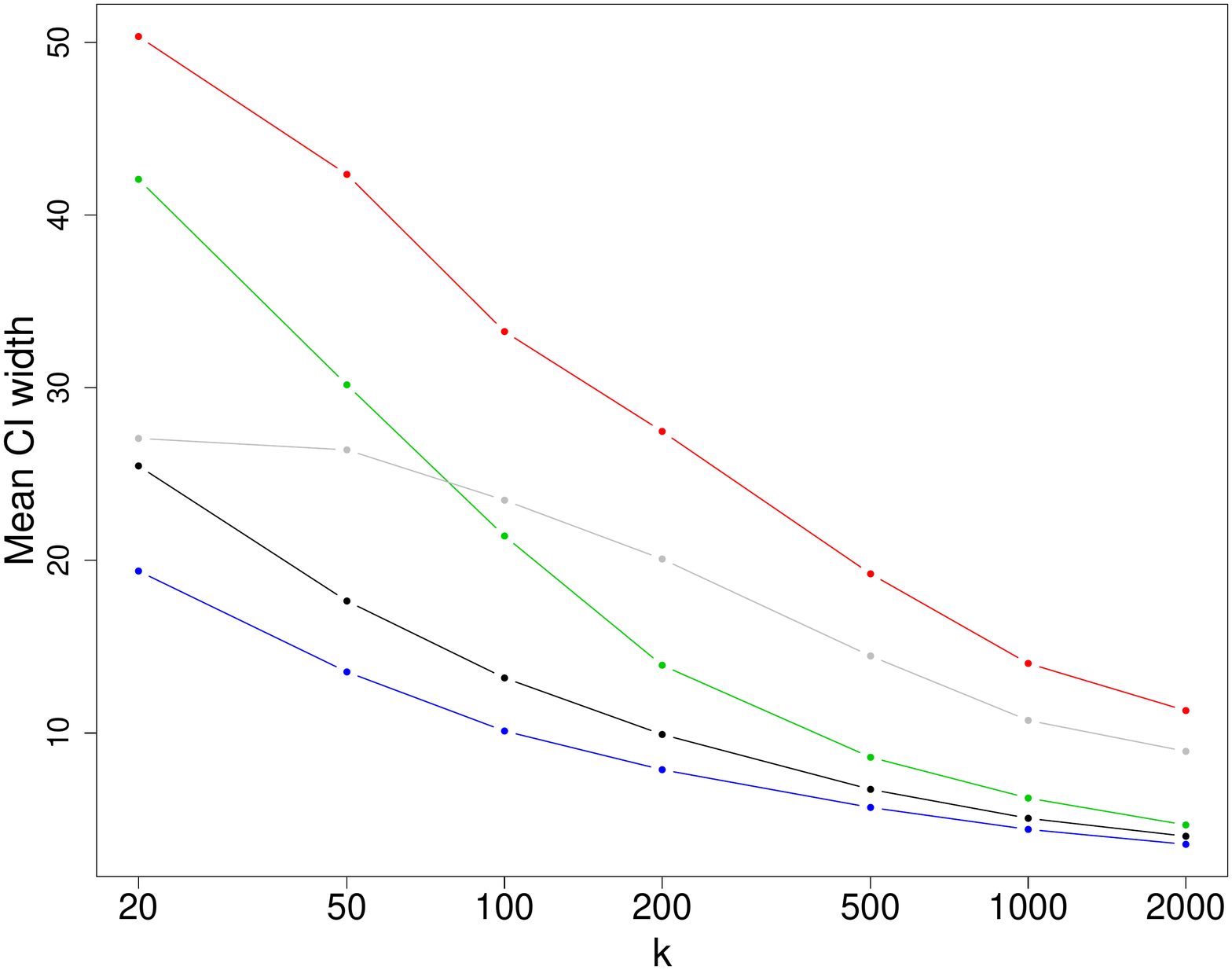}}
		\subfigure{\includegraphics[width=0.33\textwidth]{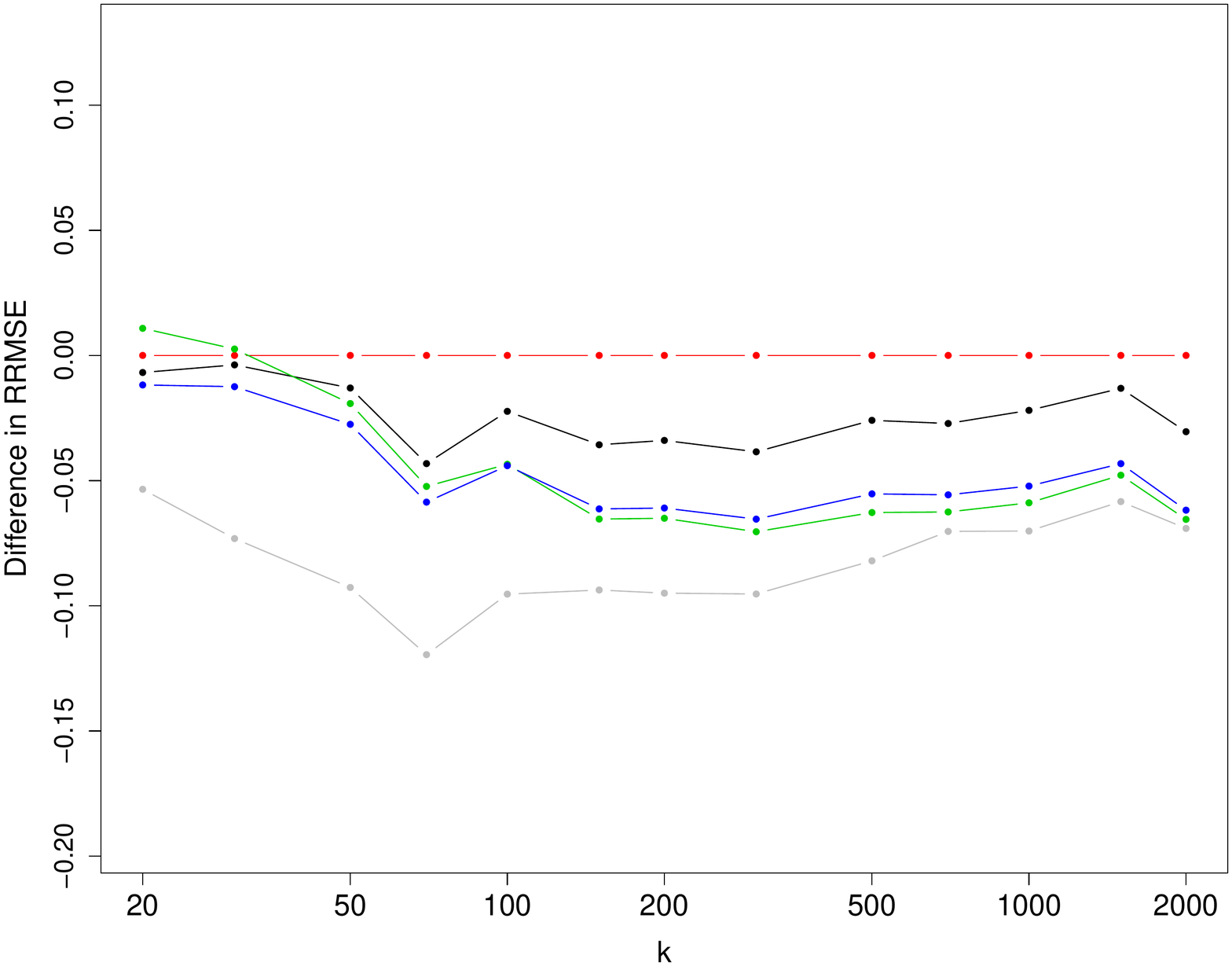}}
		\captionit{50 year return level estimates when sampling from the GEV distribution with $(\mu, \sigma, \xi)=(0, 1, 0.2)$ using the variable-threshold stopping rule over a range of $k$. Based on $10^4$ replicated samples with the historical data created using approach~\eqref{eqn:initial} of the paper. Coverage is based on 3000 replicated samples. See Figure~\ref{fig:sup.200ret} for other associated detail.}
		\label{fig:50retvar}
\end{figure}

\begin{figure}[h]
		\subfigure{\includegraphics[width=0.33\textwidth]{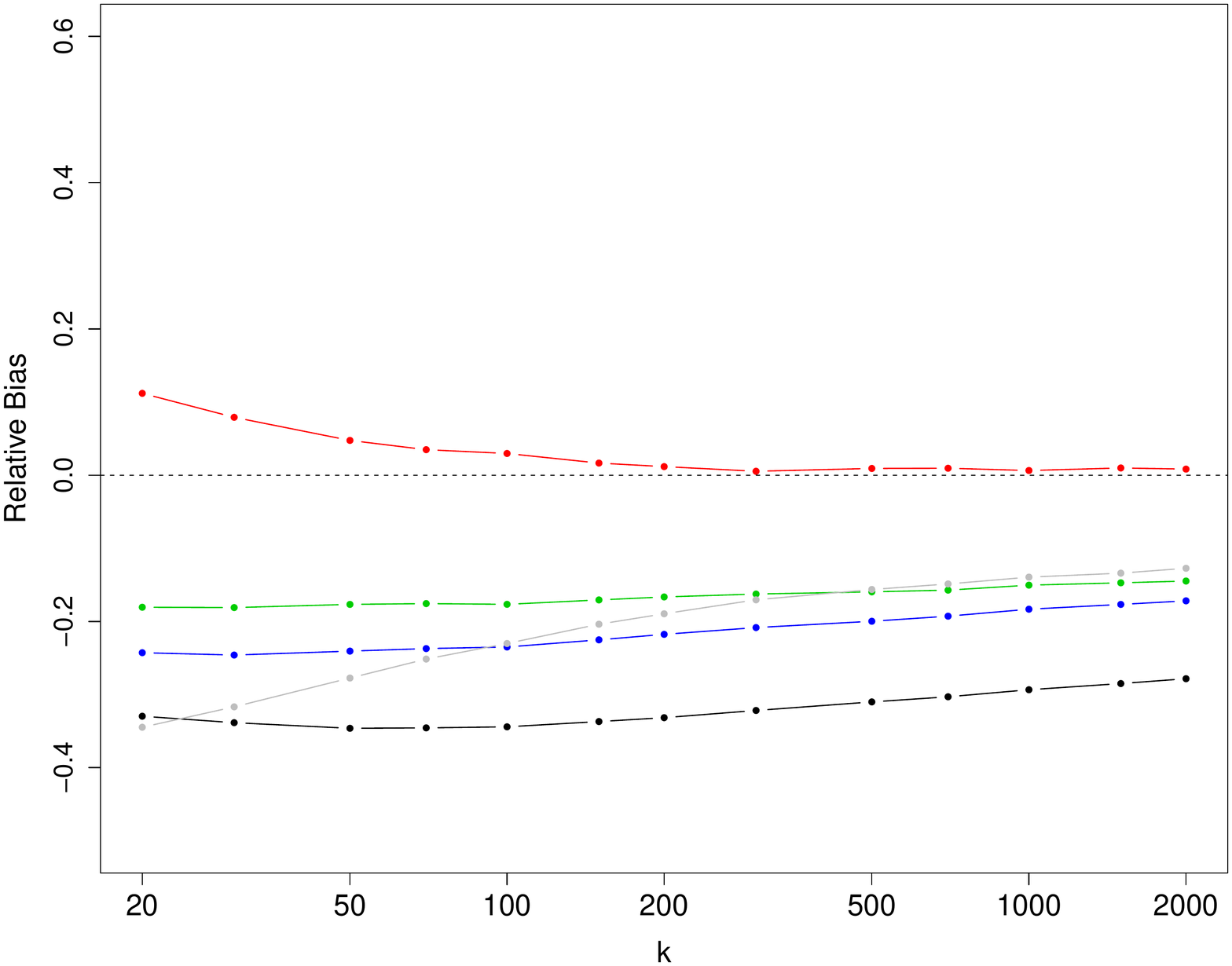}}
		\subfigure{\includegraphics[width=0.33\textwidth]{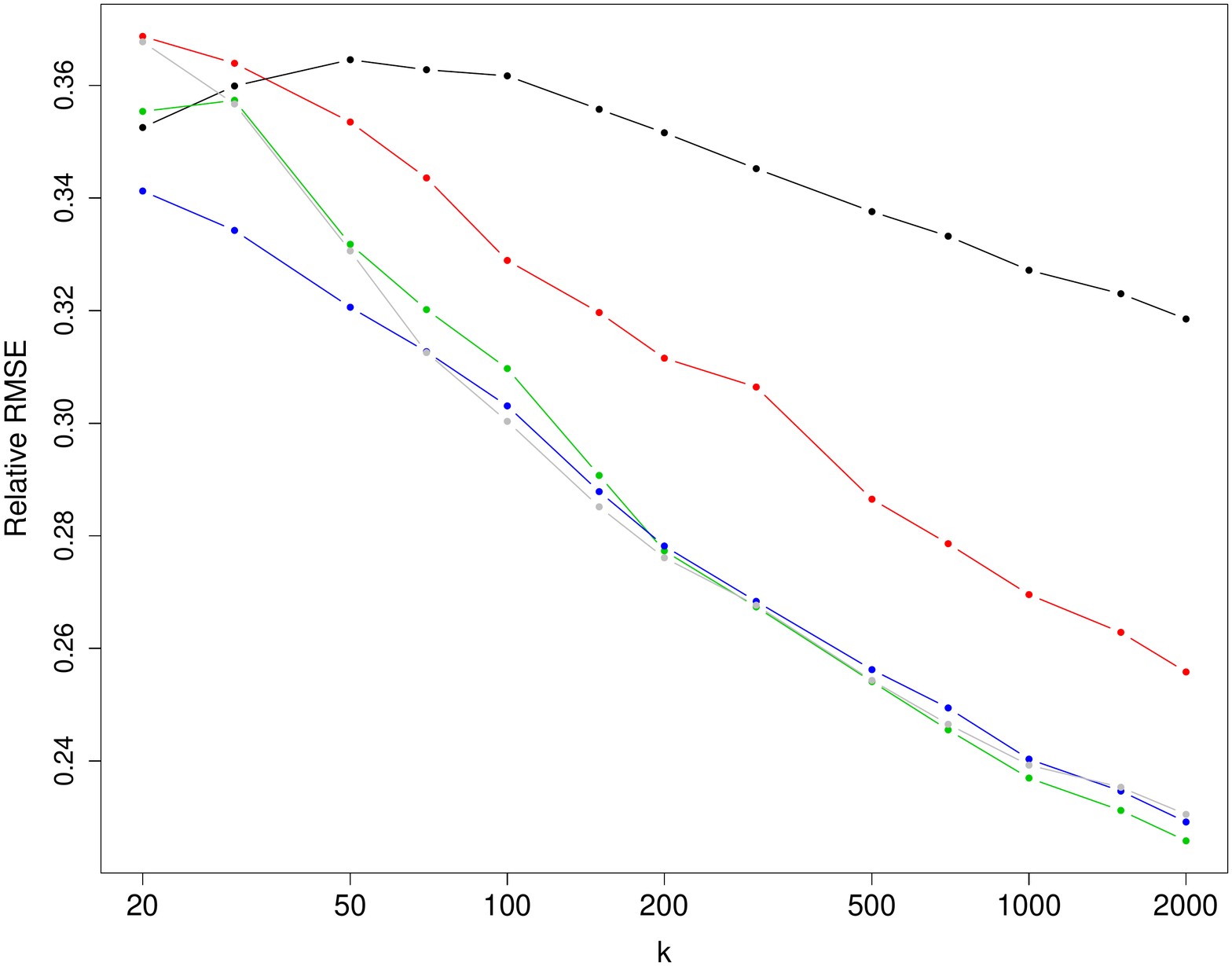}}
		\subfigure{\includegraphics[width=0.33\textwidth]{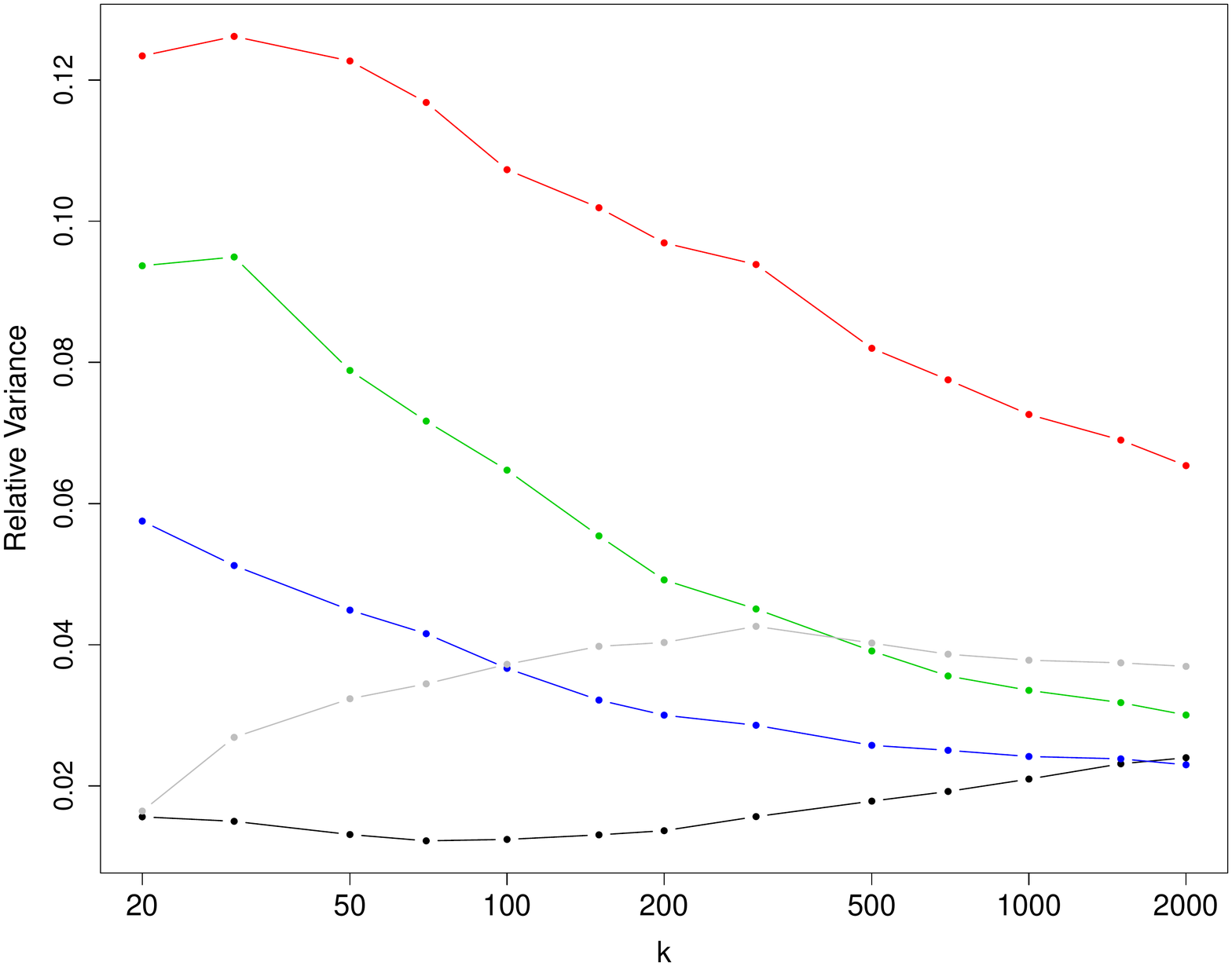}}
		\subfigure{\includegraphics[width=0.33\textwidth]{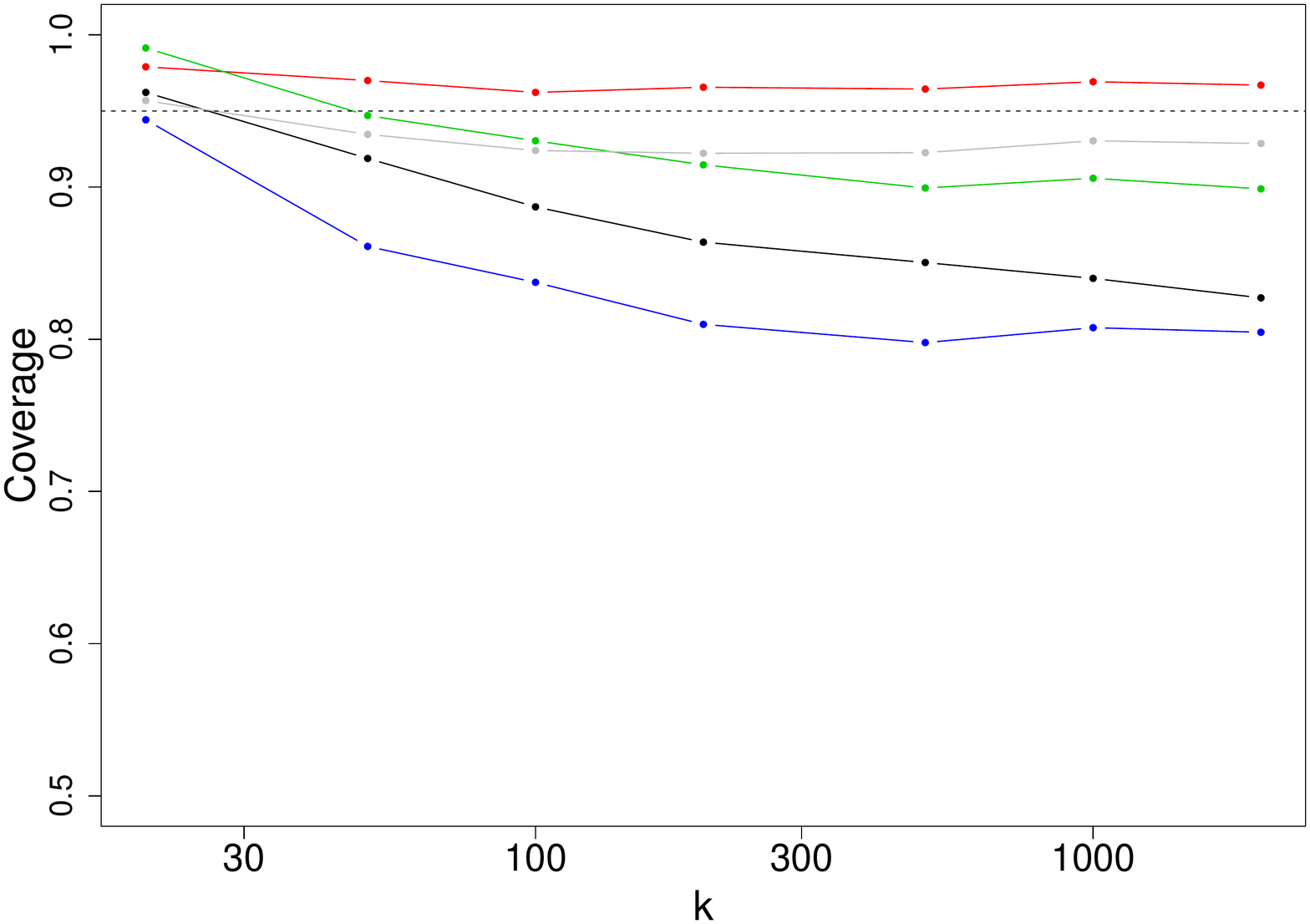}}
		\subfigure{\includegraphics[width=0.33\textwidth]{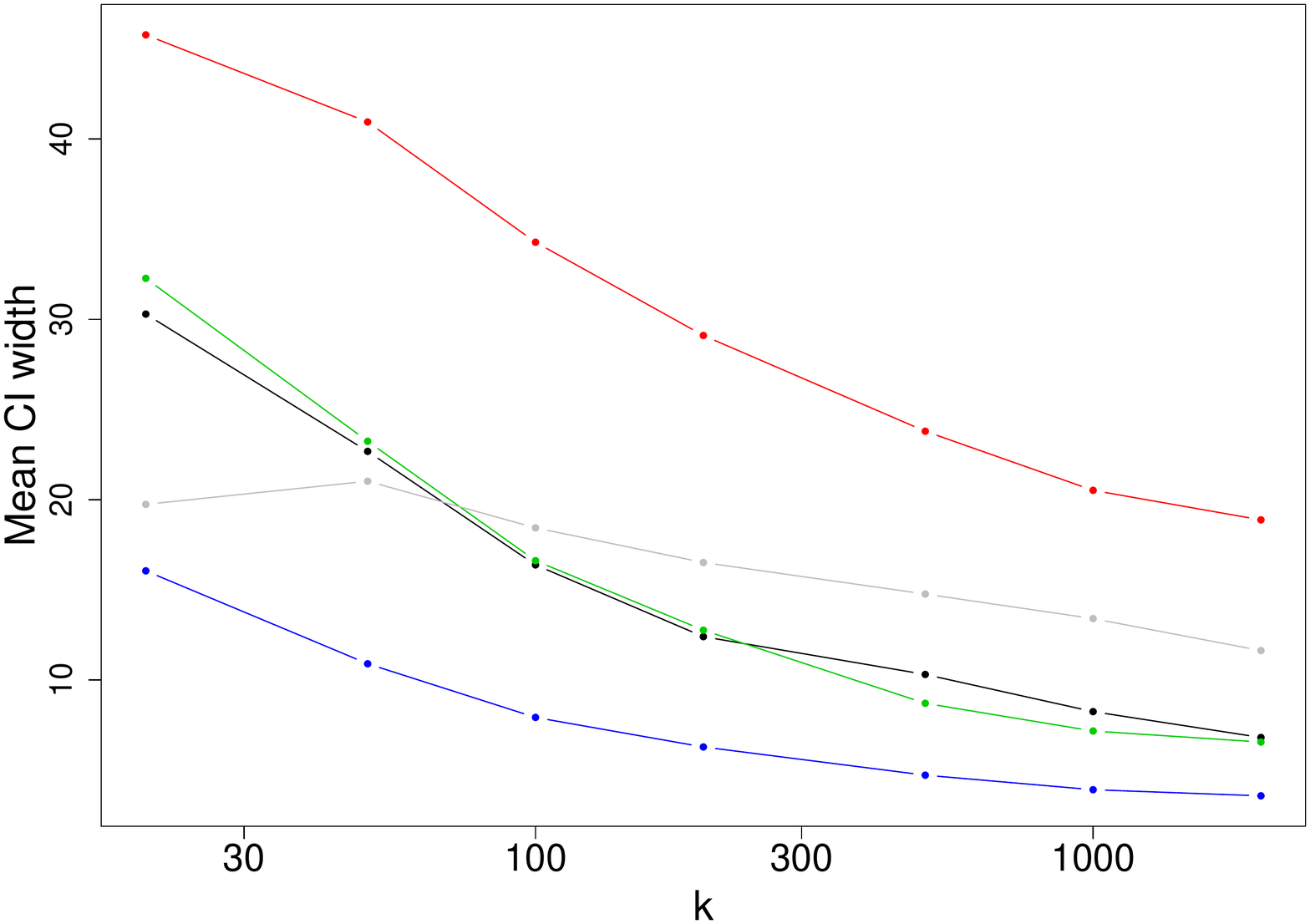}}
		\subfigure{\includegraphics[width=0.33\textwidth]{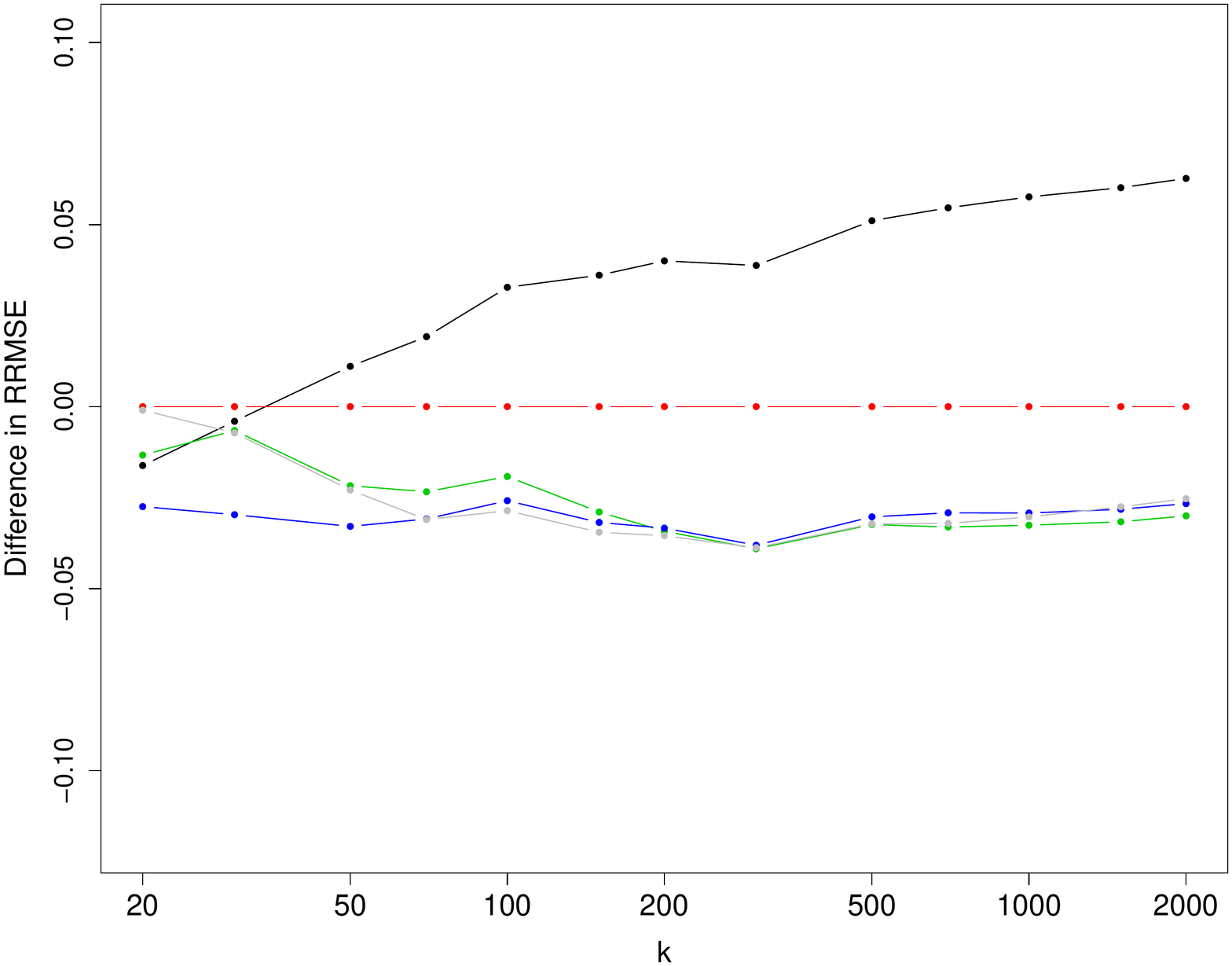}}
		\captionit{1000 year return level estimates when sampling from the GEV distribution with $(\mu, \sigma, \xi)=(0, 1, -0.2)$ using the variable-threshold stopping rule over a range of $k$. Based on $20000$ replicated samples with the historical data created using approach~\eqref{eqn:initial} of the paper. See Figure~\ref{fig:sup.200retneg} for other associated detail.}
		\label{fig:1000retvarneg}
\end{figure}

Firstly, we considered a truncated likelihood: replacing in $\ell_{std}$ the final observation, $x_{n}$, by the stopping threshold, $c_k$. The bias of these truncation estimators will always lie between that of the standard estimator and the exclude estimator. When sampling from the exponential distribution with the fixed-threshold stopping rule the relative bias of the truncation estimator is the sum of the relative biases of the standard and exclude estimators, with the return-level estimators from this method essentially always improving upon the standard method in RMSE. For the variable-threshold stopping rule with $\xi=-0.2$ this estimator performs well however it is outperformed by the conditioning estimators for most cases with positive $\xi$ and the fixed-threshold stopping rule. Figures \ref{fig:sup.shape}-\ref{fig:1000retneg},~\ref{fig:varshape}-\ref{fig:1000retvarneg} show various properties of the truncation estimator (in grey) compared to the those using $\ell_{std}$, $\ell_{ex}$, $\ell_{fc}$ and $\ell_{pc}$. We did not explore this method further as we wanted to obtain an estimator that works well over both stopping rules and for a range of $c_k$ and $k$.

%
%

%
%

Bootstrap bias correction estimators \citep{Efron} were assessed. This method is based on the assumption that the bias of the estimator is approximately the same when the same sampling procedure is used but with the true parameter replaced by an estimate. So one can estimate the bias and correct the standard estimate by taking away this approximated bias. This procedure can be computationally heavy because it involves generating many samples using the stopping rule and evaluating estimates from which to calculate the bias. 
We do not explore them further here as we found that the bootstrap bias corrected return-level estimates 
were generally no better than the partial conditional estimates but required considerable additional computation. 

\section{GPD} \label{sec:GPD}

Consider modelling daily observations above some high threshold, $v$, rather than just modelling the block maxima \citep{Coles}. For suitably large $v$ the exceedances of this threshold are typically assumed to be exactly modelled by their limiting distribution as the threshold tends to the upper end point of the distribution. This limiting distribution is the generalised Pareto distribution (GPD) \citep{DandS} which has distribution function for $x>0$:
	\begin{align}
			F(x) = \PP{X \leq v + x | X > v} = \left\{ \begin{array}{ll} 1 - \left(1+ \frac{\xi x}{\sigma_v}\right)_+^{-\frac{1}{\xi}} \qquad \qquad &\xi \neq 0 \\ 
			1 - \exp\left(-\frac{x}{\sigma_v}\right) &\xi = 0, \end{array} \right.
	\label{eqn:GPD} 
	\end{align}
	
where $\xi$ and $\sigma_v>0$ are the shape and scale parameter respectively. Note that if $\xi$ is zero then the GPD is equivalent to the exponential distribution with rate parameter $\sigma_v^{-1}$. The shape parameter $\xi$ is the same as that under the GEV distribution whereas the scale parameter changes with threshold with $\sigma_v = \sigma + \xi (v - \mu)$ where $(\mu,\sigma,\xi)$ are the associated GEV parameters. For modelling using the GPD we also need to model the rate at which the threshold $v$ is exceeded.

Now consider the estimation of return-levels from a data set of threshold exceedances 
where the sample size is determined by the fixed-threshold stopping rule. A threshold is chosen above which observations are considered to be extreme and the exceedances of this threshold are modelled using the generalised Pareto distribution (GPD). Thus in this setting we have two thresholds: the one above which we fit the GPD, which we denote by $v$, and the fixed stopping threshold $c_k$ which determines the sample size with $c_k > v$.

For comparison with the GEV results, we set $\xi=\pm 0.2$ and $\sigma_v=1+\xi v$ and have an average of 10 exceedances per year. We start using the stopping rule after 10 exceedances have been simulated (\ie ,~the first 10 exceedances are the historical data). When the stopping threshold is exceeded we continue sampling to the end of the current year and take this as the full sample. We omit the final year of observations to calculate the exclude estimator. 

Let $\tau_v$ denote the expected number of exceedances of $v$ in a year, then the $y$-year return-level is:
\begin{align}\label{eqn:GPDret}
	x_y = \left\{ \begin{array}{ll} 
		v + \frac{\sigma_v}{\xi} \left((y\tau_v)^{\xi} - 1 \right) \qquad \qquad &\xi \neq 0 \\
		v + \sigma_v\log{(y\tau_v)} &\xi = 0 \, .
	\end{array}  \right. 
\end{align}

The $y$-year return-level estimate is calculated by substituting the parameter estimates, $(\hat{\xi},\hat{\sigma}_v)$, and $\hat{\tau}_v = {n_v}/{n_y}$ into \eqref{eqn:GPDret}, where $n_v$ is the total number of exceedances and $n_y$ is the number of years.

\begin{figure}
		\subfigure{\includegraphics[width=0.45\textwidth]{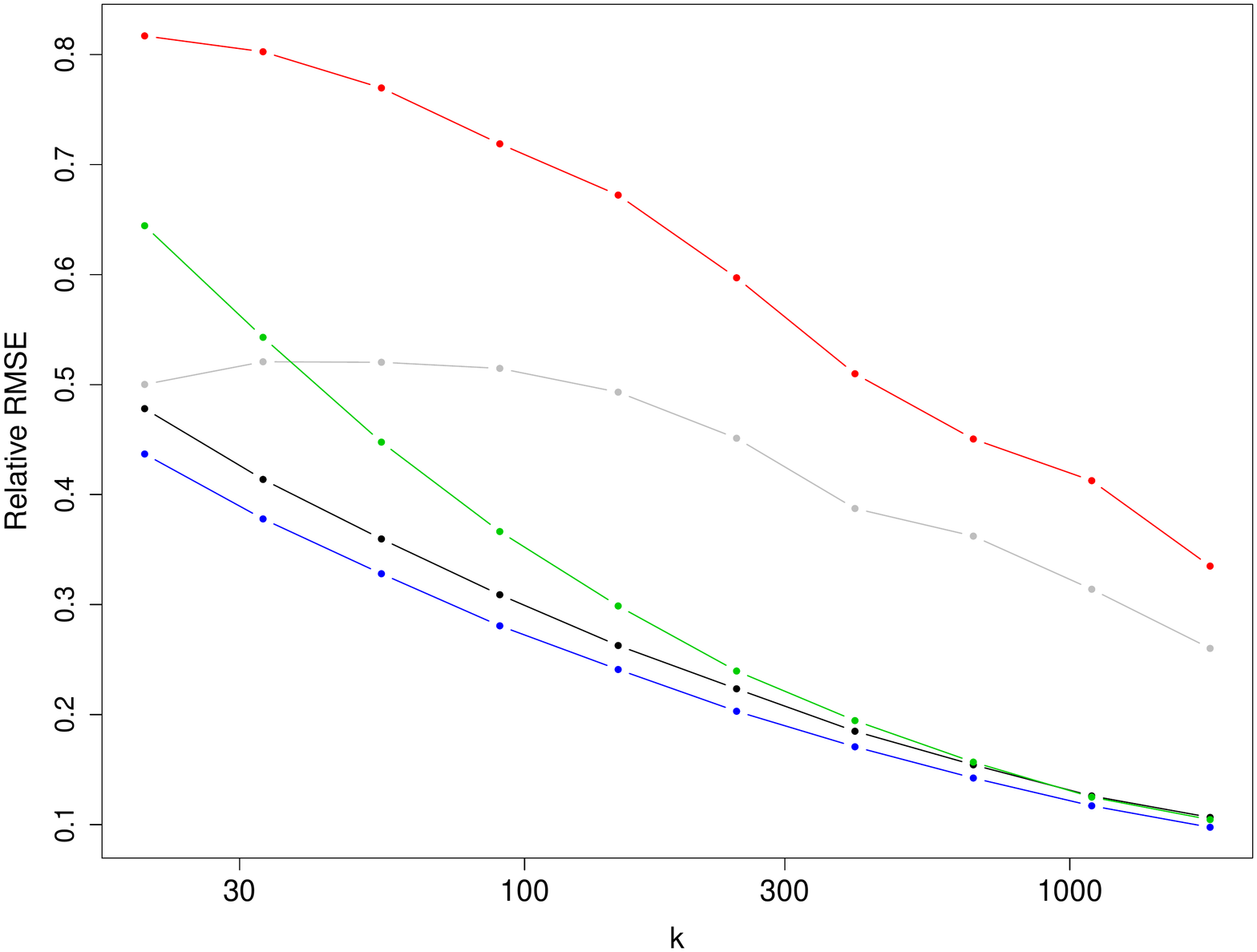}}
		\subfigure{\includegraphics[width=0.45\textwidth]{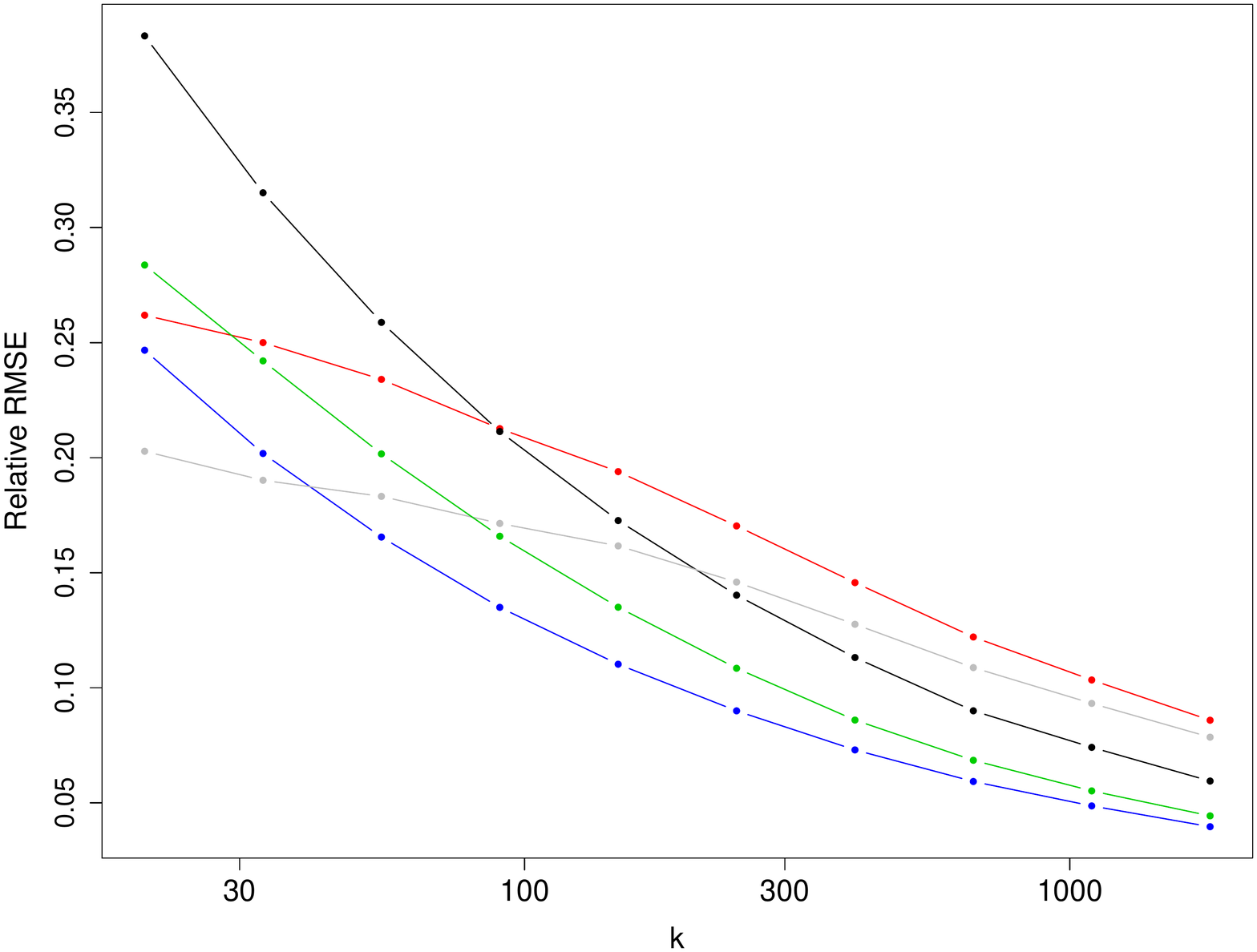}}
		\captionit{Relative RMSE of 200-yr return level estimates based on 100000 samples from the GPD distribution with $\xi=0.2$ (left) and $\xi=-0.2$ (right) using the fixed-threshold stopping rule. Colour scheme is the same as in Figure \ref{fig:sup.shape}.}
		\label{fig:GPD}
\end{figure}

Figure~\ref{fig:GPD} shows the RRMSE of the 200 year return-level estimates for the GPD. 
The results of our GPD simulation are similar to that for the GEV with fixed-threshold stopping rule (\S\ref{sec:fixed} of the paper); the partial conditioning method performs best in terms of RRMSE for estimating return-levels. The main difference to the GEV setting is in estimating $\xi$; we find that $\vartext{\hat{\xi}_{ex}} > \vartext{\hat{\xi}_{std}}$. In the GPD setting the standard likelihood includes multiple data points from the final year (both that which triggered the stopping rule and smaller points) which are informative about the shape parameter. The exclude likelihood does not include this information hence the larger variance of the shape parameter estimates compared to the standard estimates. 





\begin{figure}[h]
		\subfigure{\includegraphics[width=0.33\textwidth]{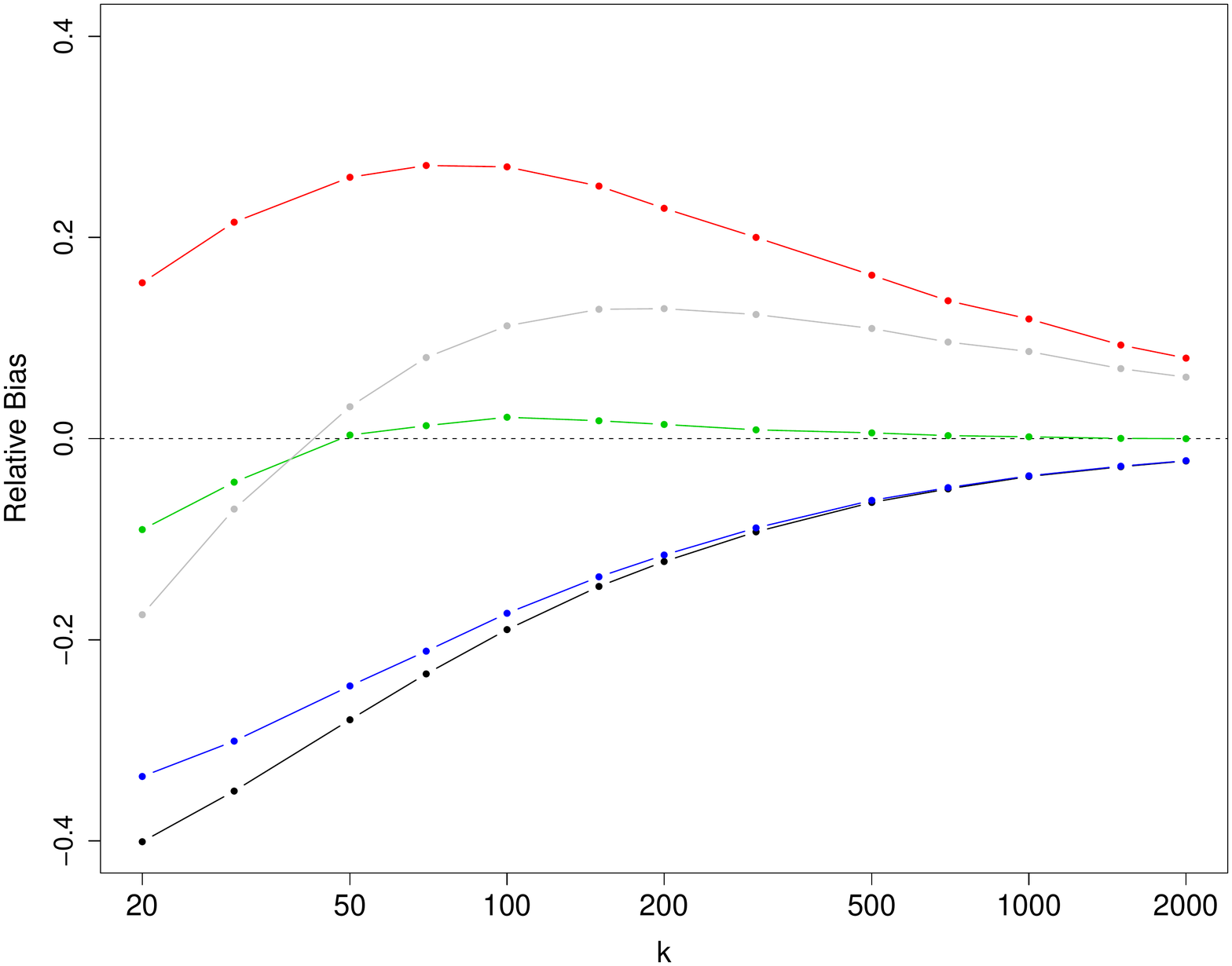}}
		\subfigure{\includegraphics[width=0.33\textwidth]{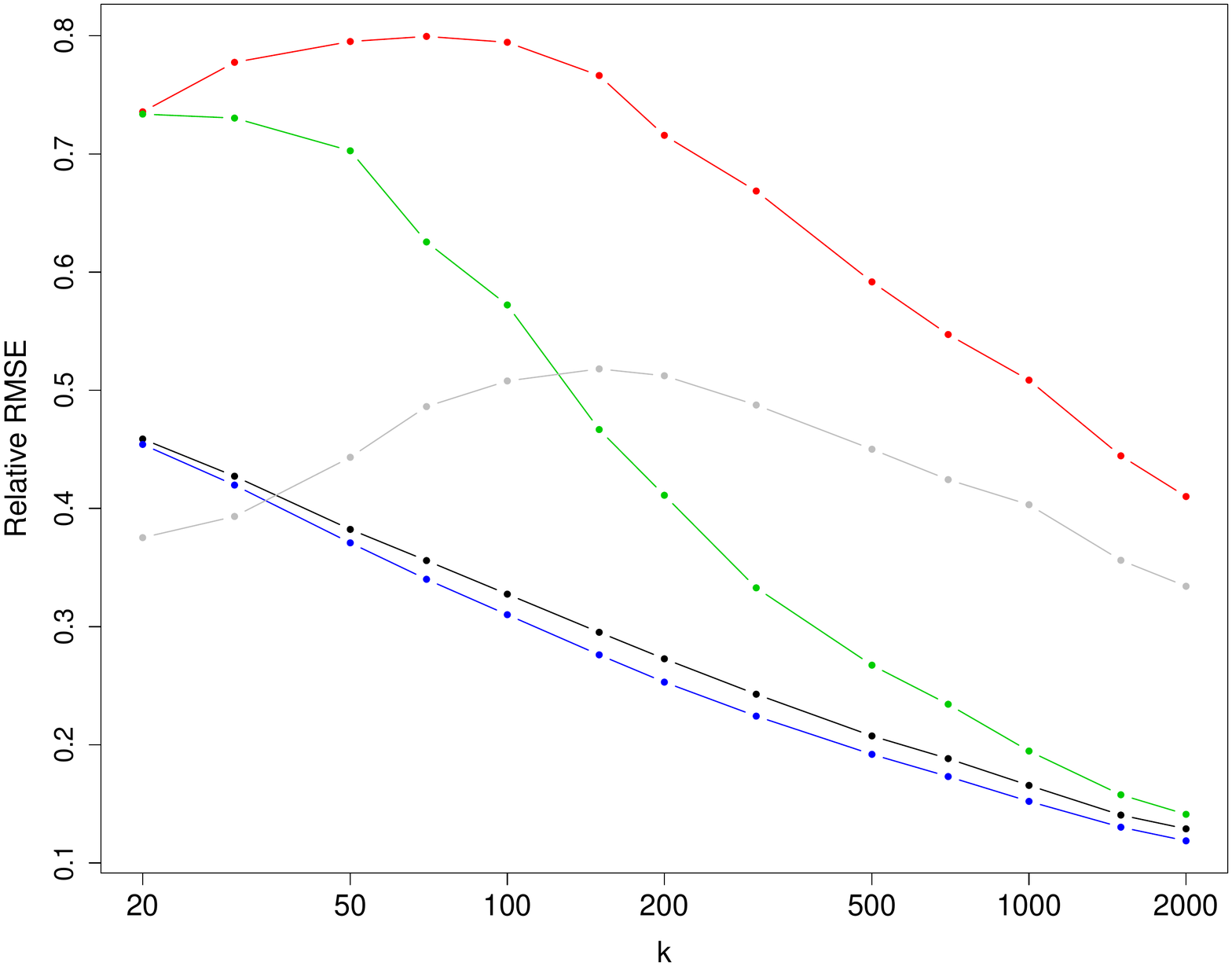}}
		\subfigure{\includegraphics[width=0.33\textwidth]{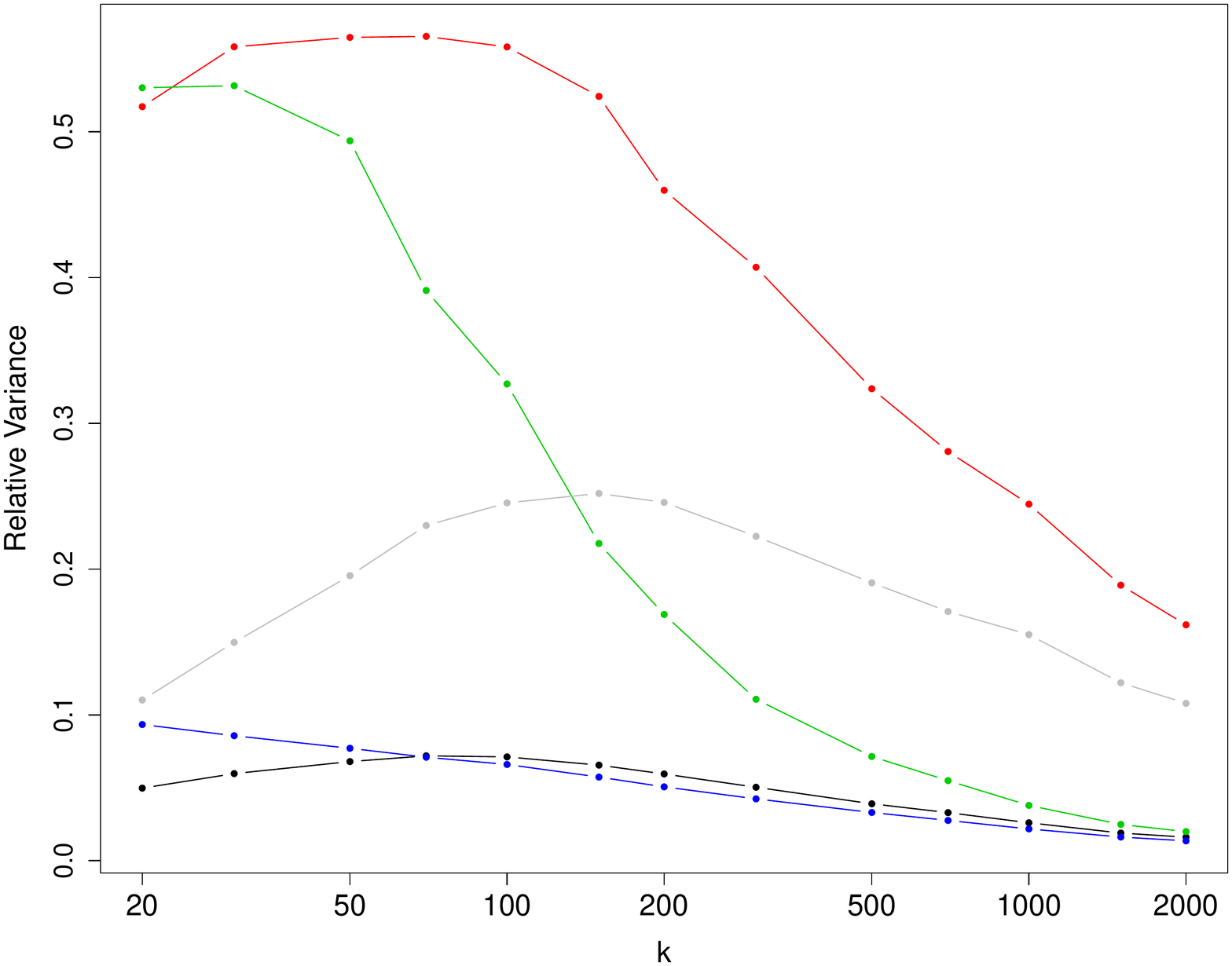}}
		\subfigure{\includegraphics[width=0.33\textwidth]{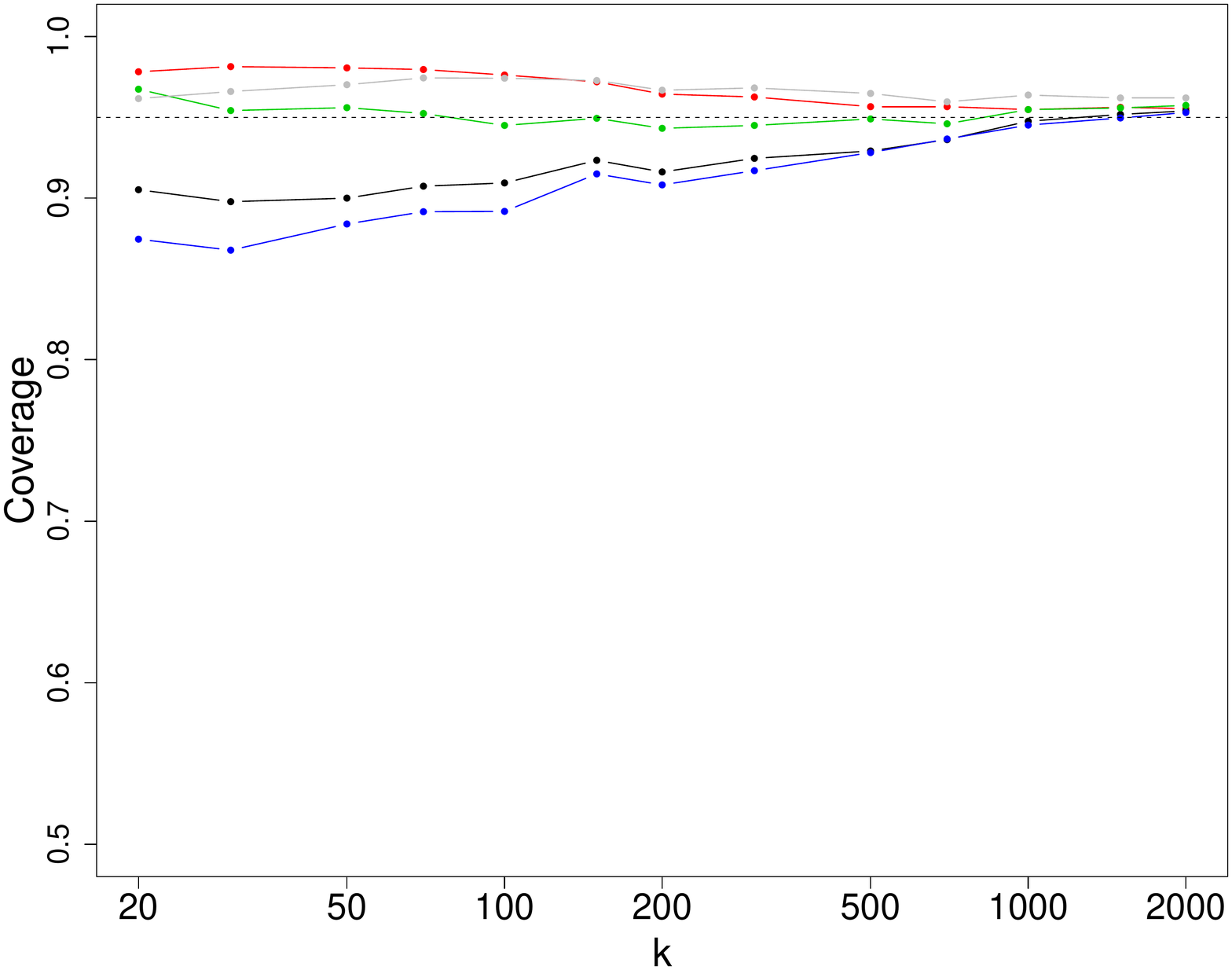}}
		\subfigure{\includegraphics[width=0.33\textwidth]{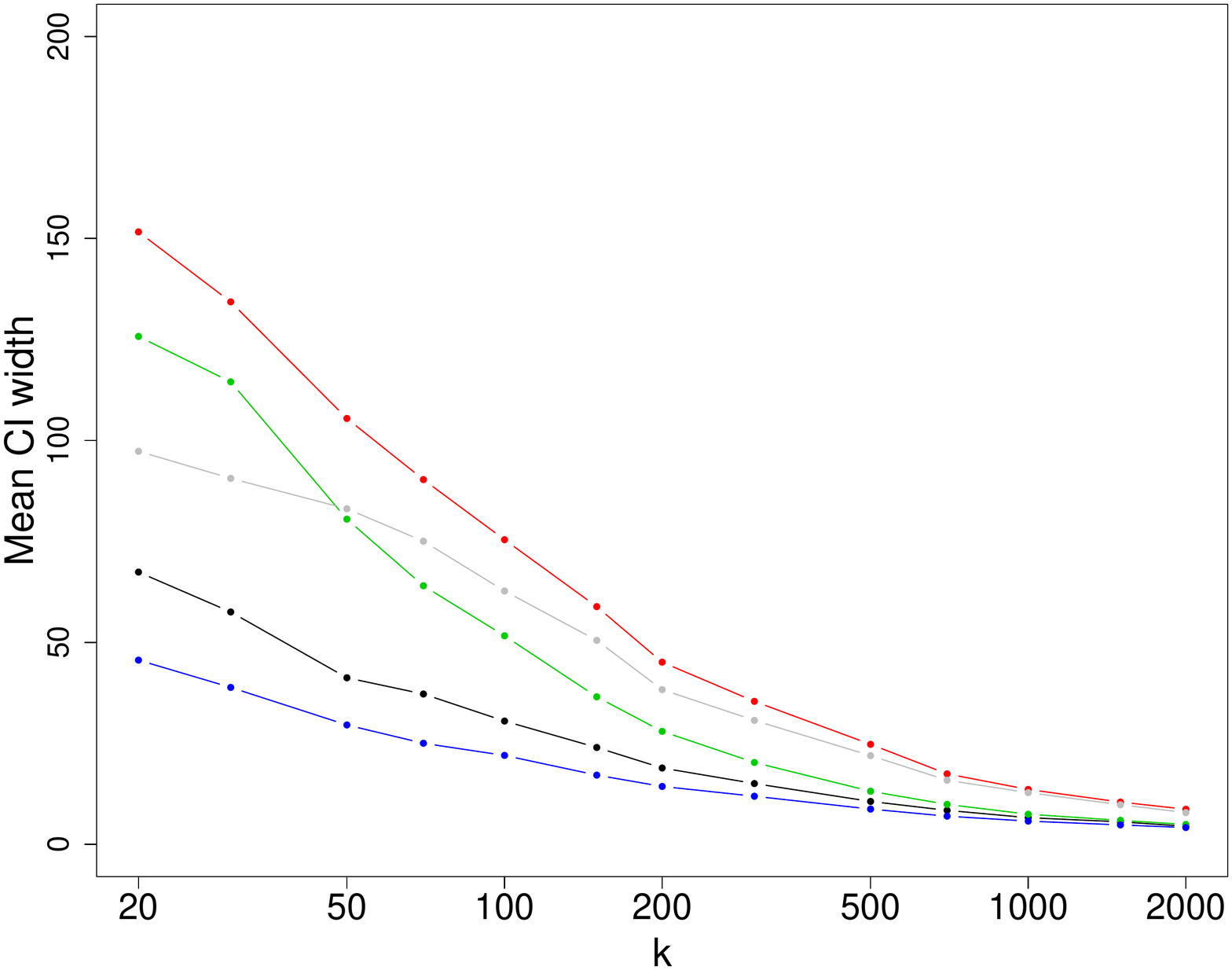}}
		\subfigure{\includegraphics[width=0.33\textwidth]{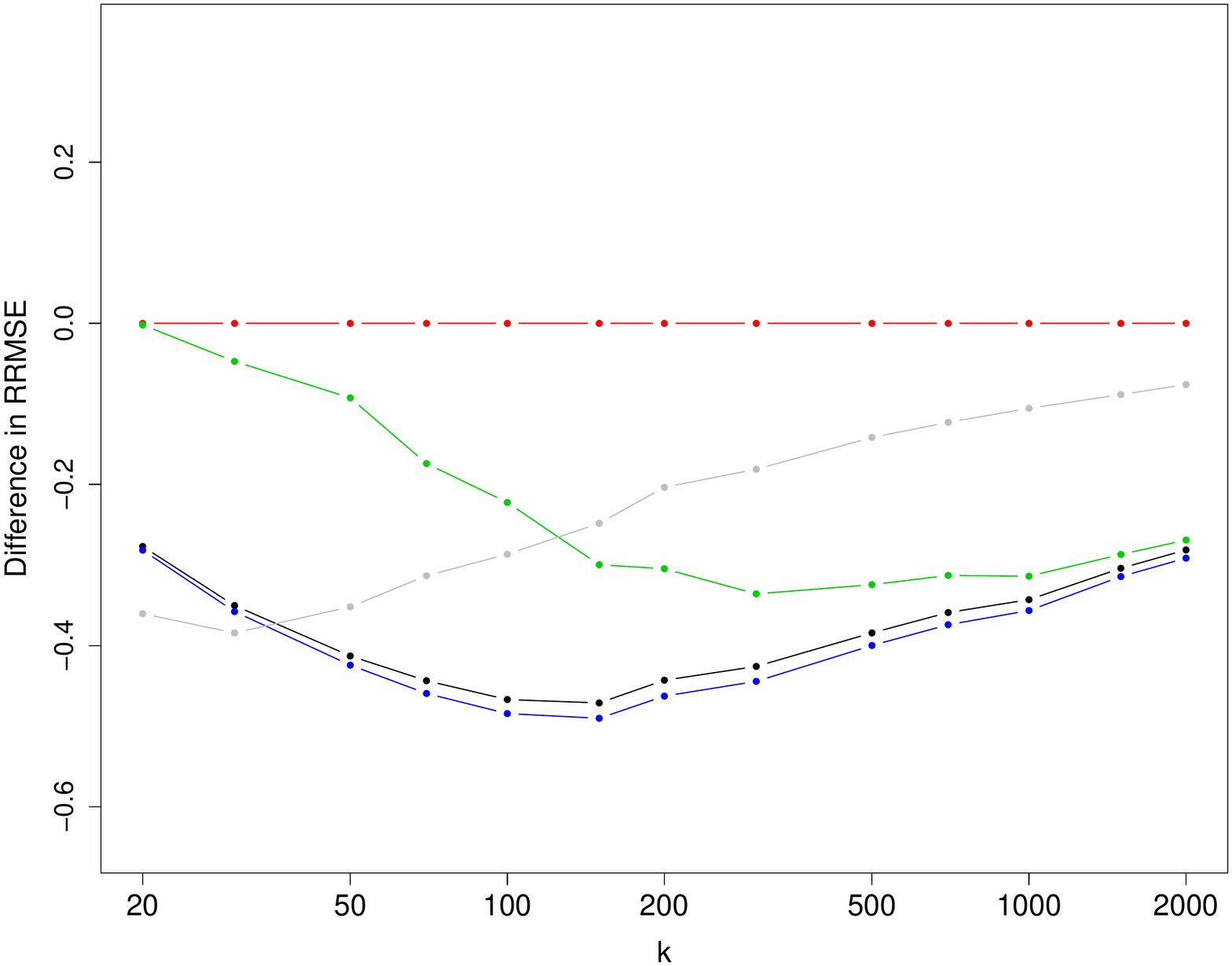}}
		\captionit{200 year return level estimates when sampling from the GEV distribution with $(\mu, \sigma, \xi)=(0, 1, 0.2)$ using the fixed-threshold stopping rule over a range of thresholds. From left to right. Top: relative bias, relative RMSE and relative variance. Bottom: coverage, average CI width, difference in RRMSE to RRMSE of standard estimator. Colour scheme is the same as in Figure \ref{fig:sup.shape}. Based on $10^5$ replicated samples with the historical data created using approach~\eqref{eqn:initial} of the paper. Coverage is based on 5000 replicated samples.}
\end{figure}

\begin{figure}[h]
		\subfigure{\includegraphics[width=0.33\textwidth]{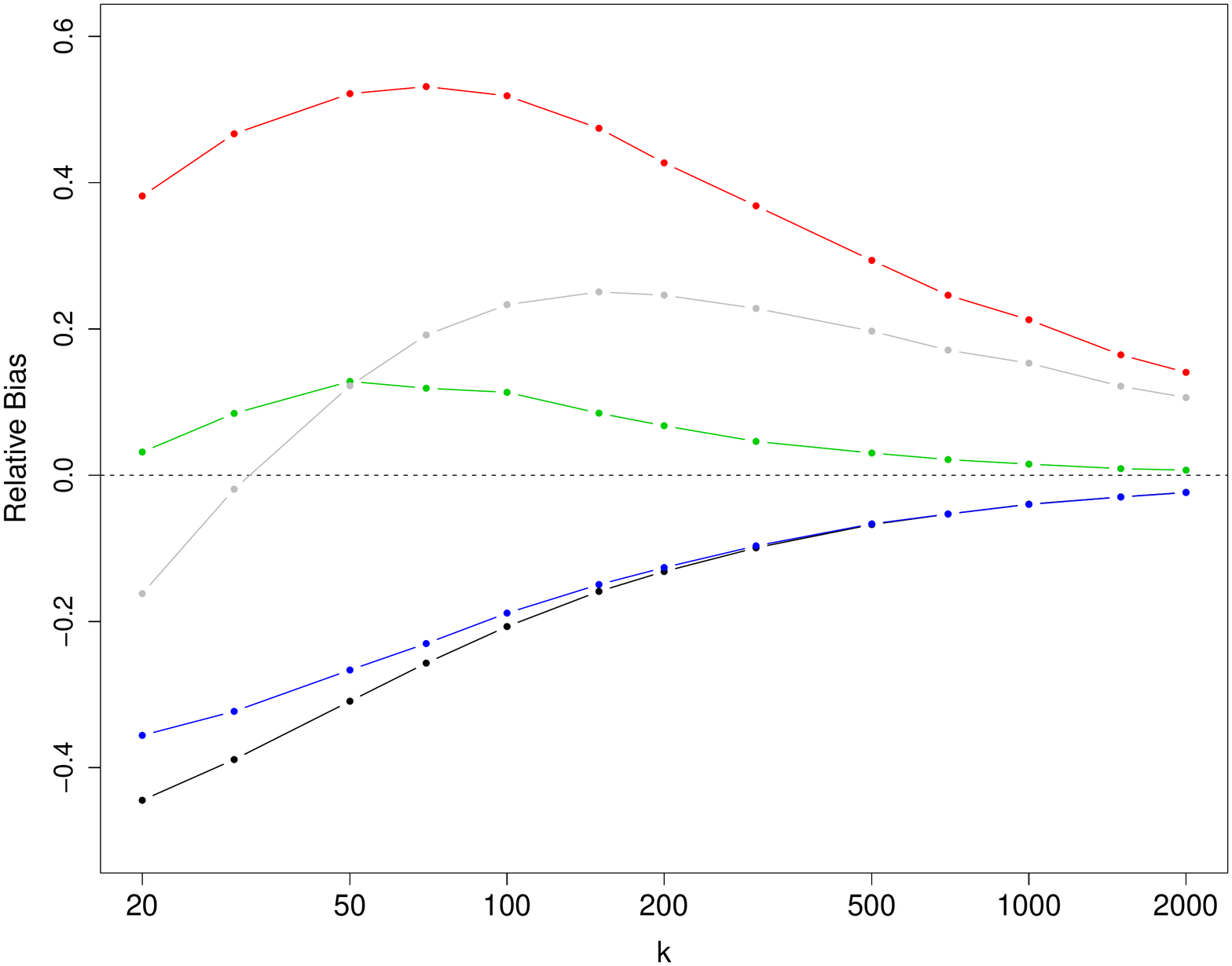}}
		\subfigure{\includegraphics[width=0.33\textwidth]{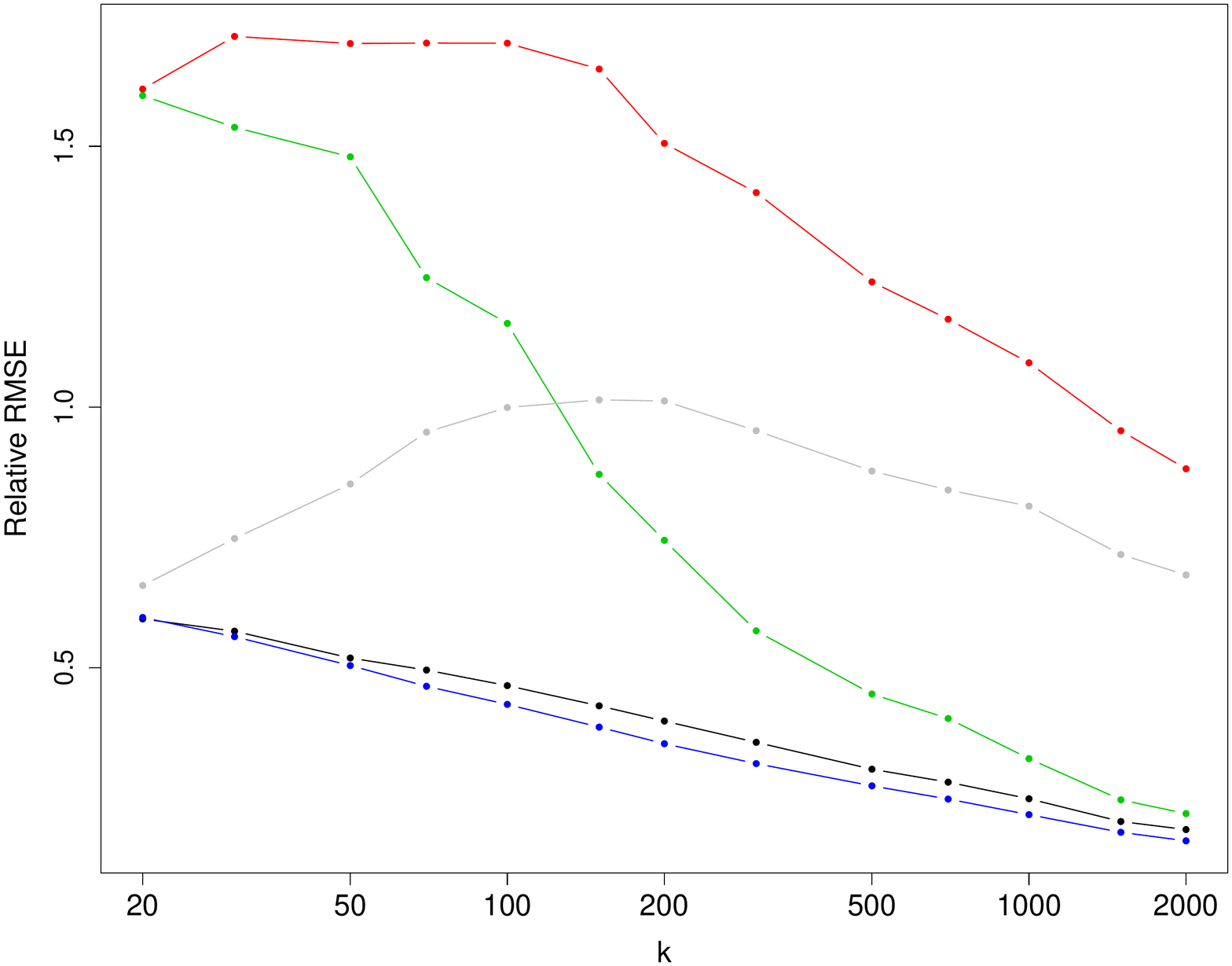}}
		\subfigure{\includegraphics[width=0.33\textwidth]{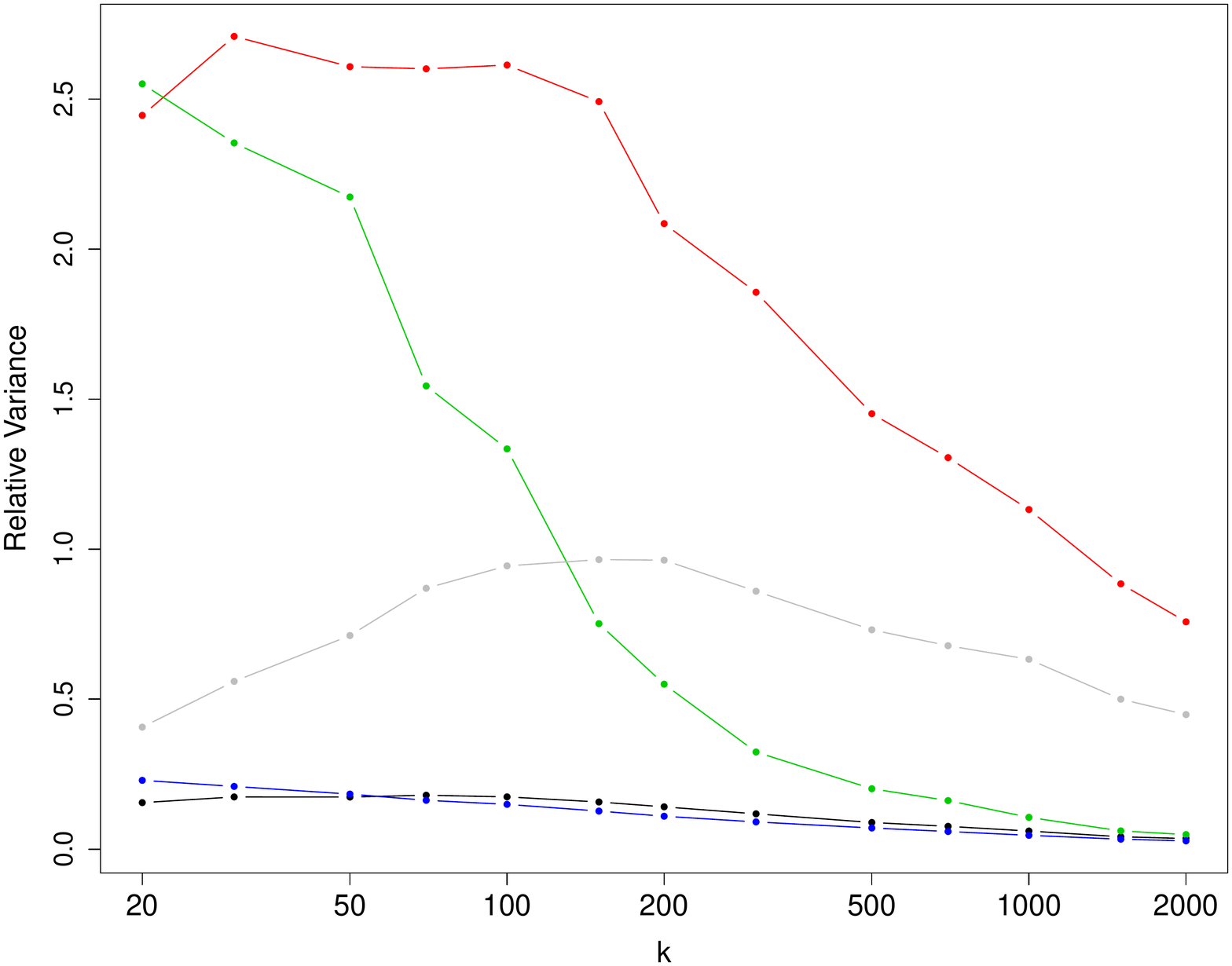}}
		\subfigure{\includegraphics[width=0.33\textwidth]{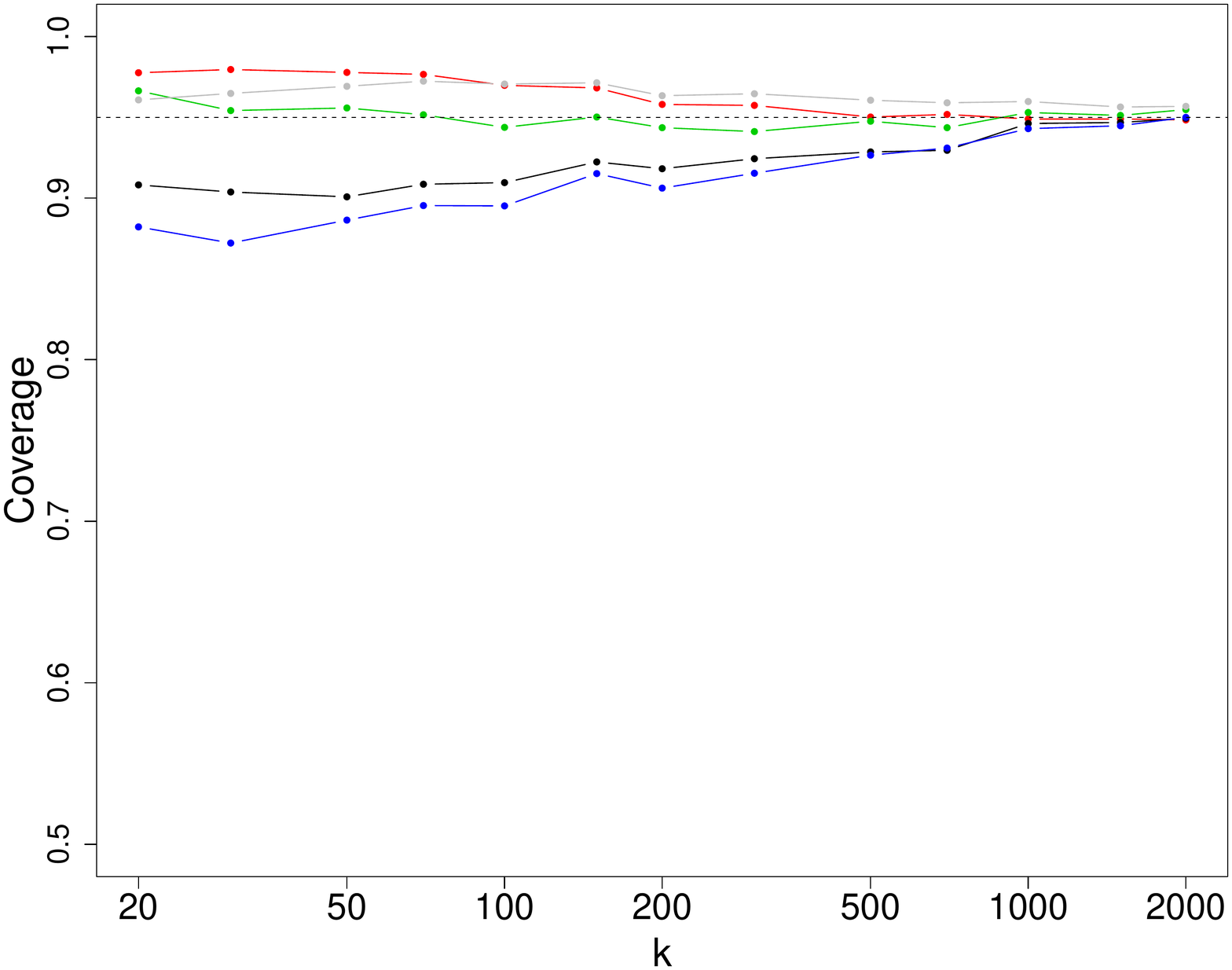}}
		\subfigure{\includegraphics[width=0.33\textwidth]{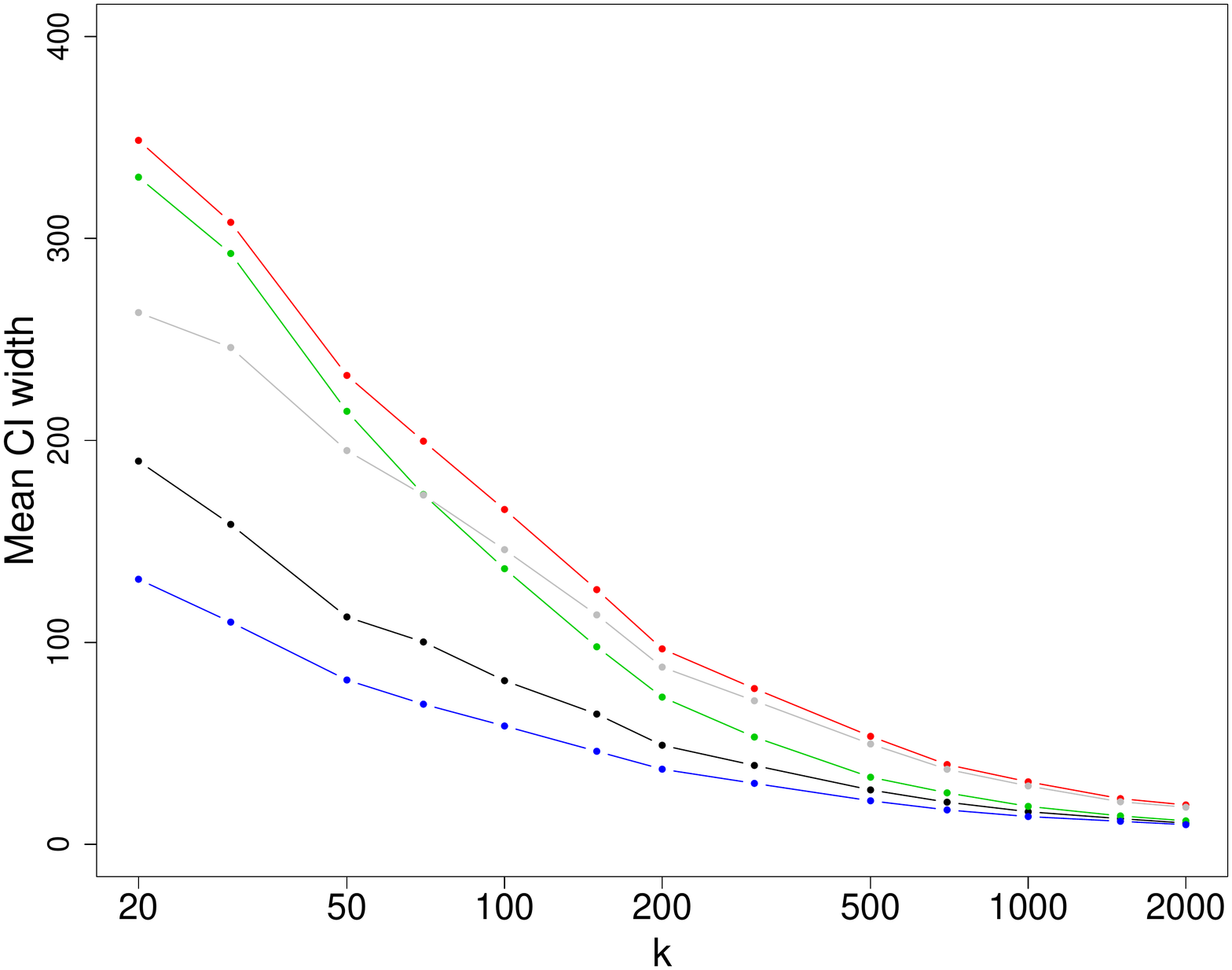}}
		\subfigure{\includegraphics[width=0.33\textwidth]{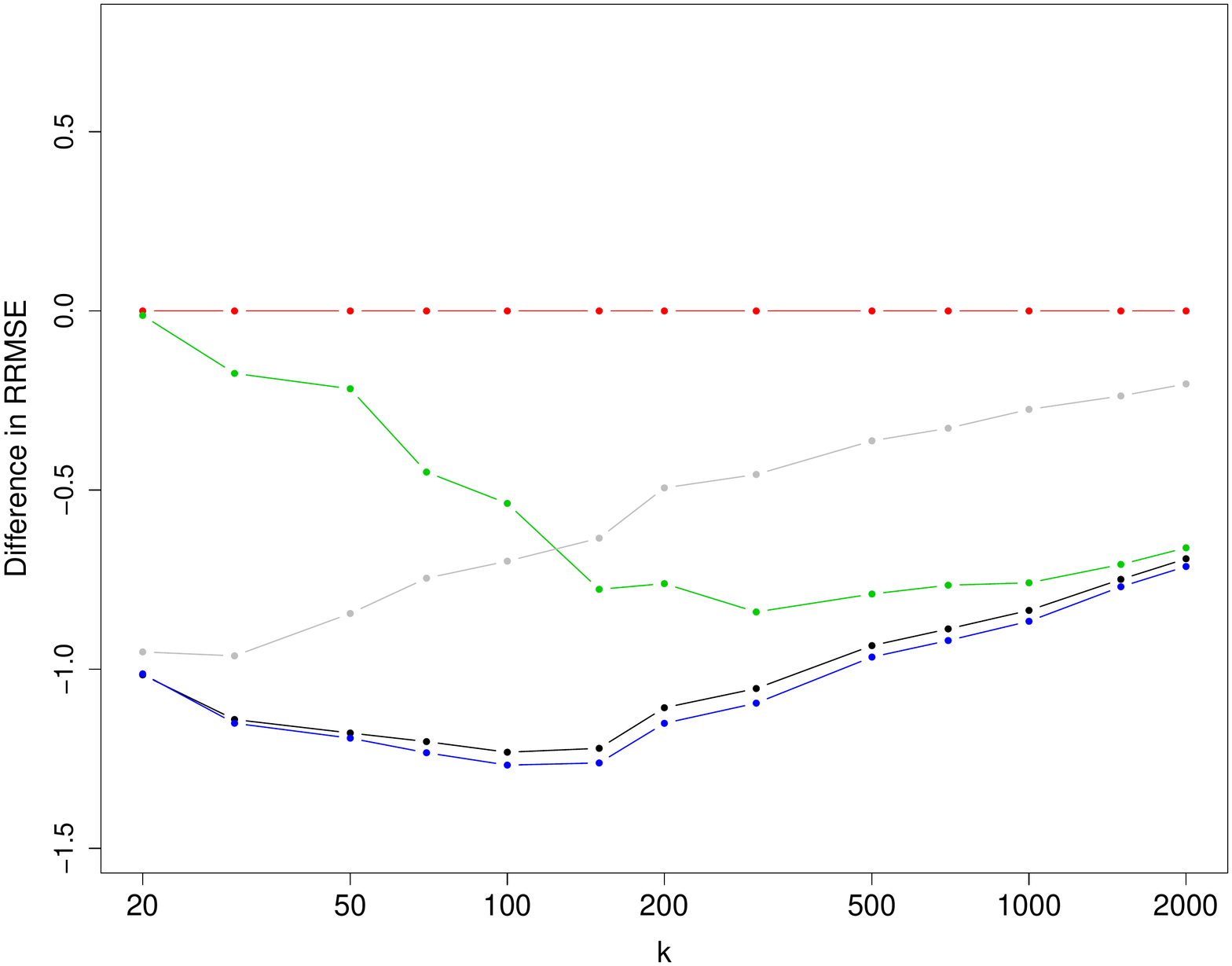}}
		\captionit{1000 year return level estimates when sampling from the GEV distribution with $(\mu, \sigma, \xi)=(0, 1, 0.2)$ using the fixed-threshold stopping rule over a range of thresholds. From left to right. Top: relative bias, relative RMSE and relative variance. Bottom: coverage, average CI width, difference in RRMSE to RRMSE of standard estimator. Colour scheme is the same as in Figure \ref{fig:sup.shape}. Based on $10^5$ replicated samples with the historical data created using approach~\eqref{eqn:initial} of the paper. Coverage is based on 5000 replicated samples.}
\end{figure}

\begin{figure}[h]
		\subfigure{\includegraphics[width=0.33\textwidth]{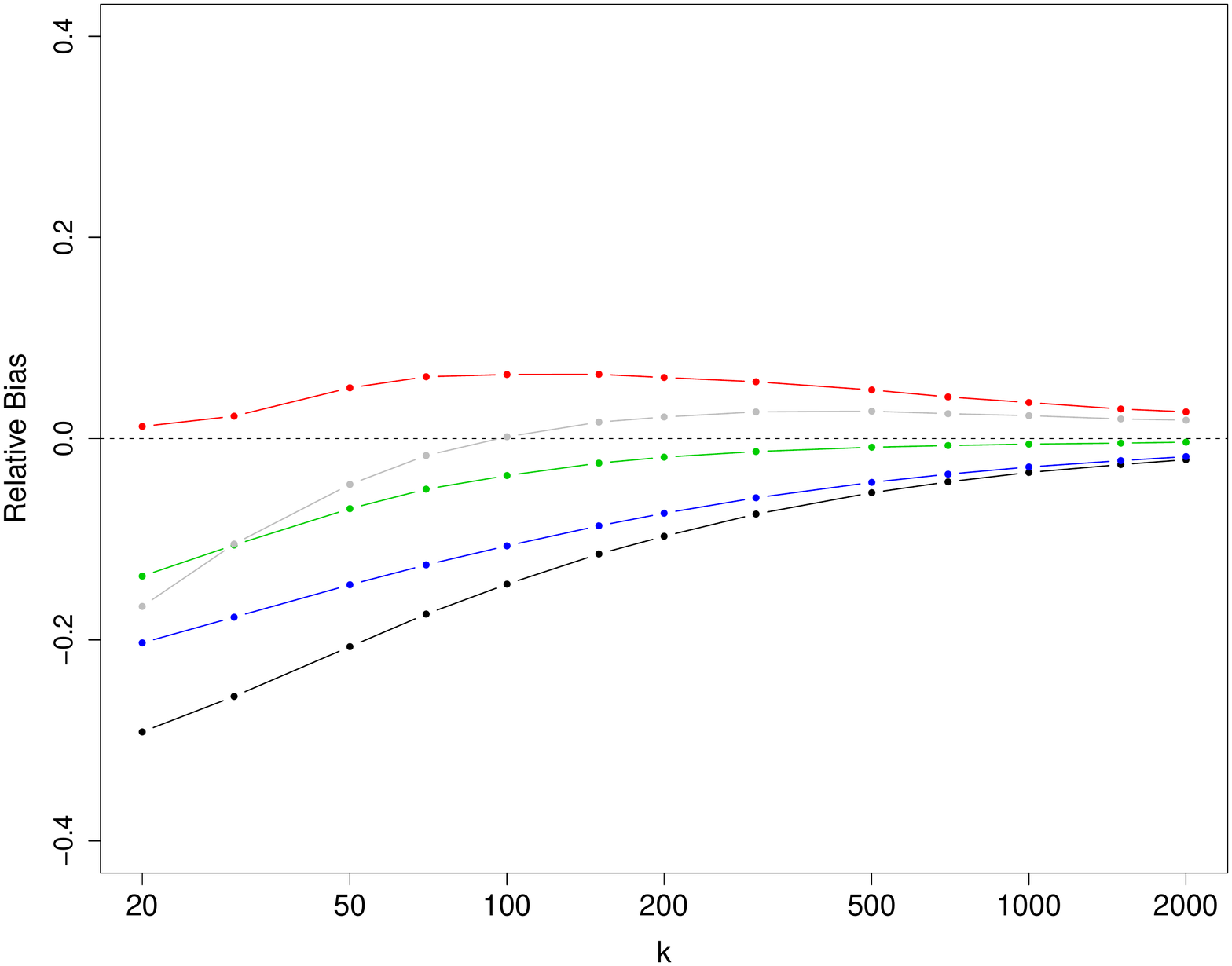}}
		\subfigure{\includegraphics[width=0.33\textwidth]{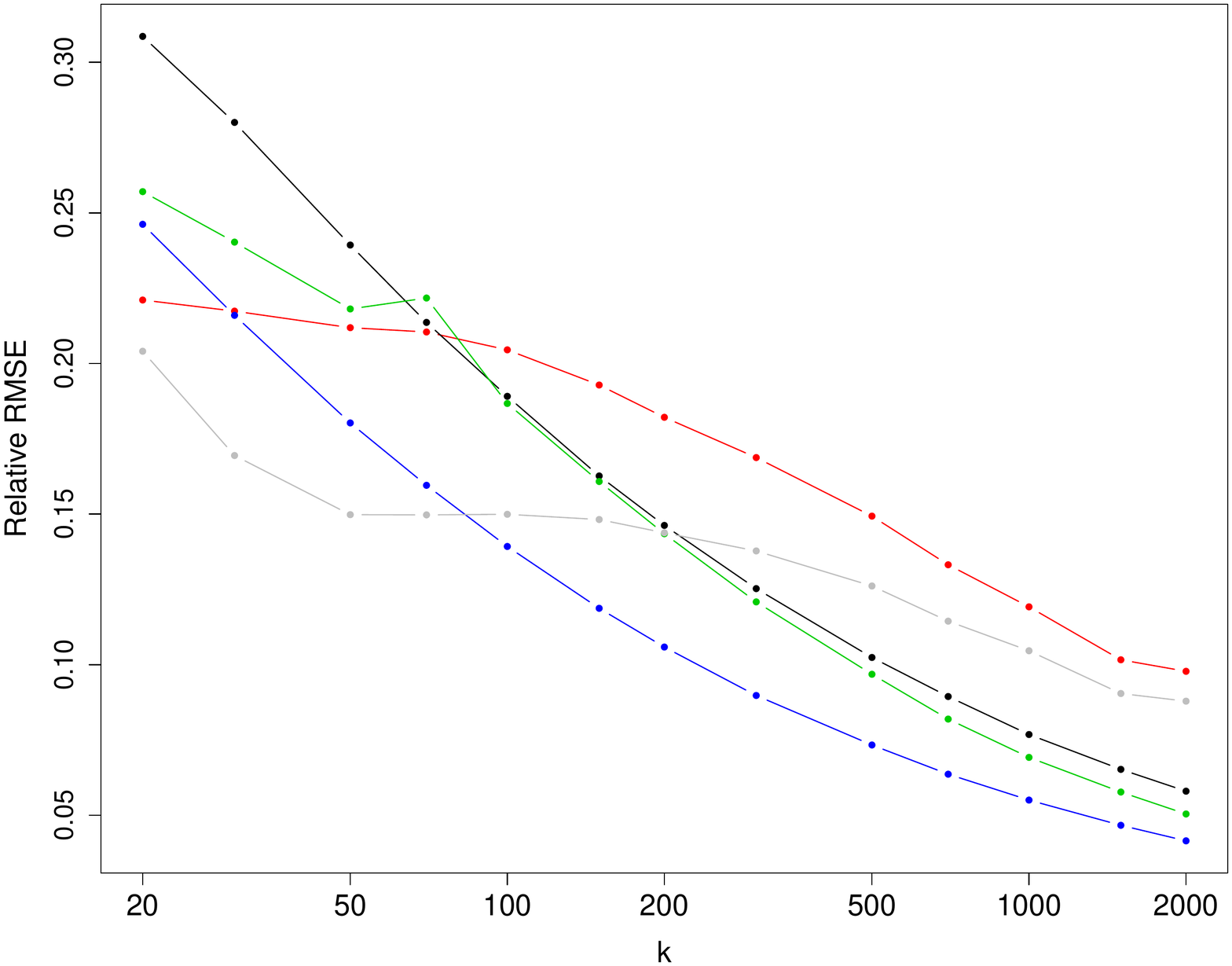}}
		\subfigure{\includegraphics[width=0.33\textwidth]{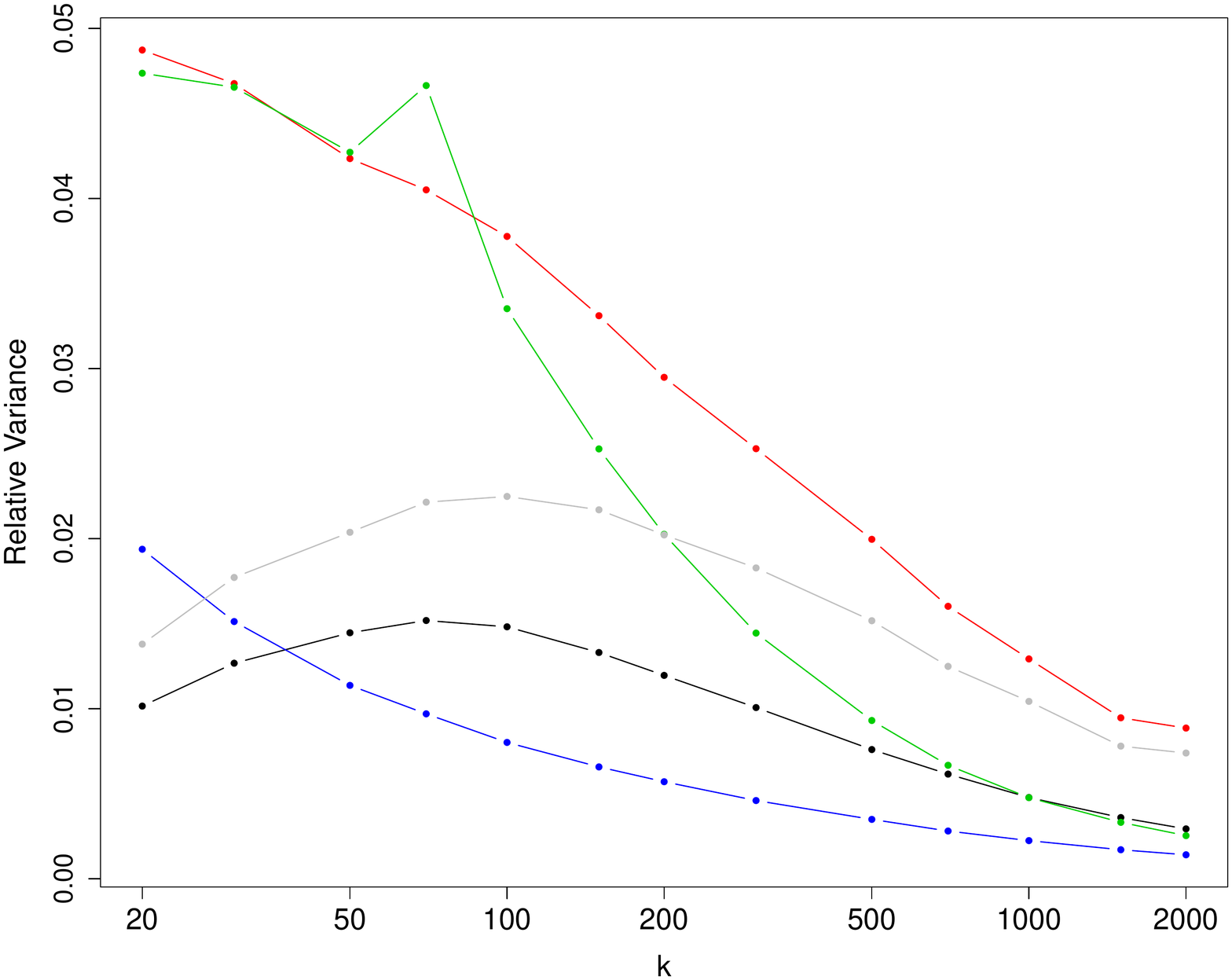}}
		\subfigure{\includegraphics[width=0.33\textwidth]{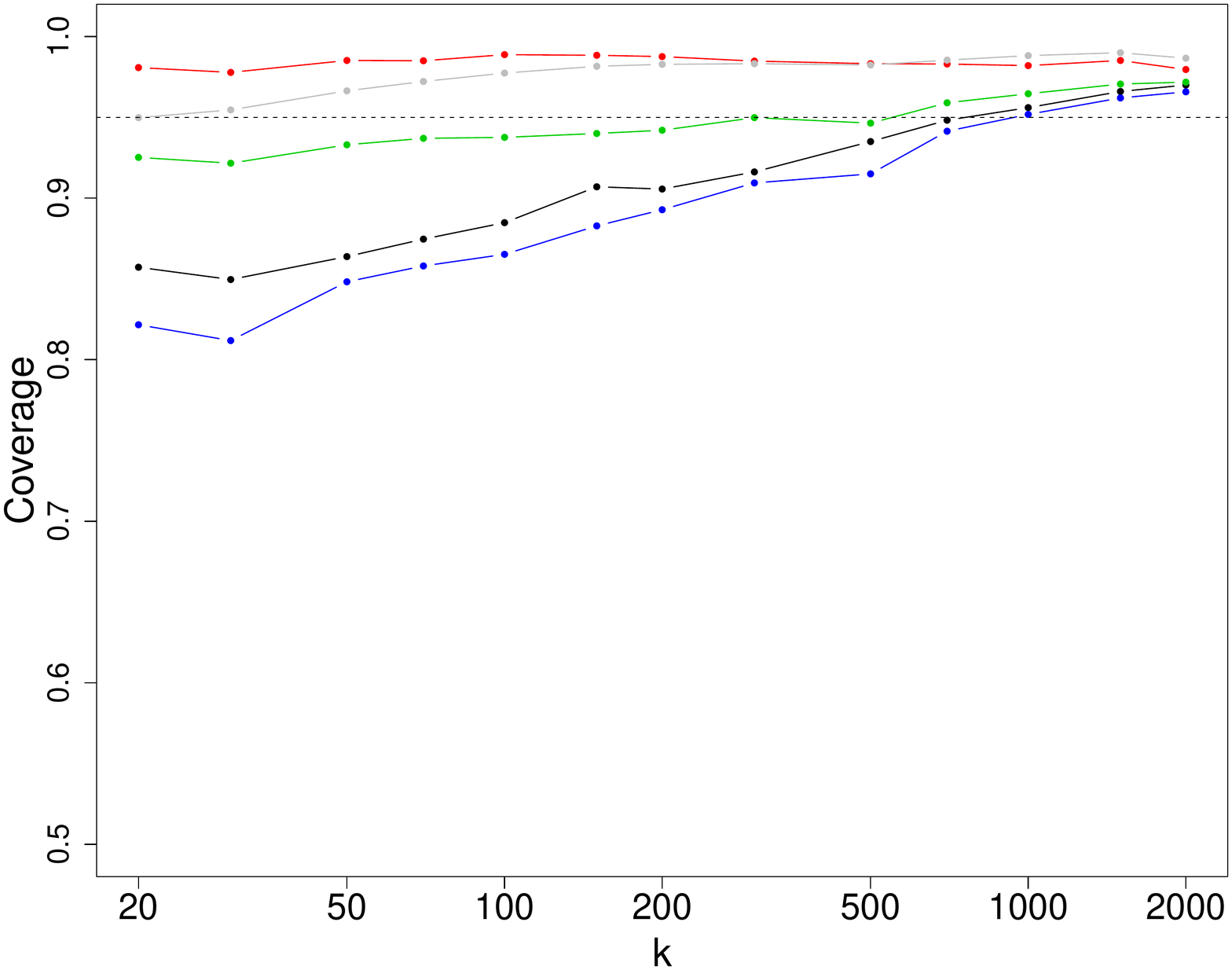}}
		\subfigure{\includegraphics[width=0.33\textwidth]{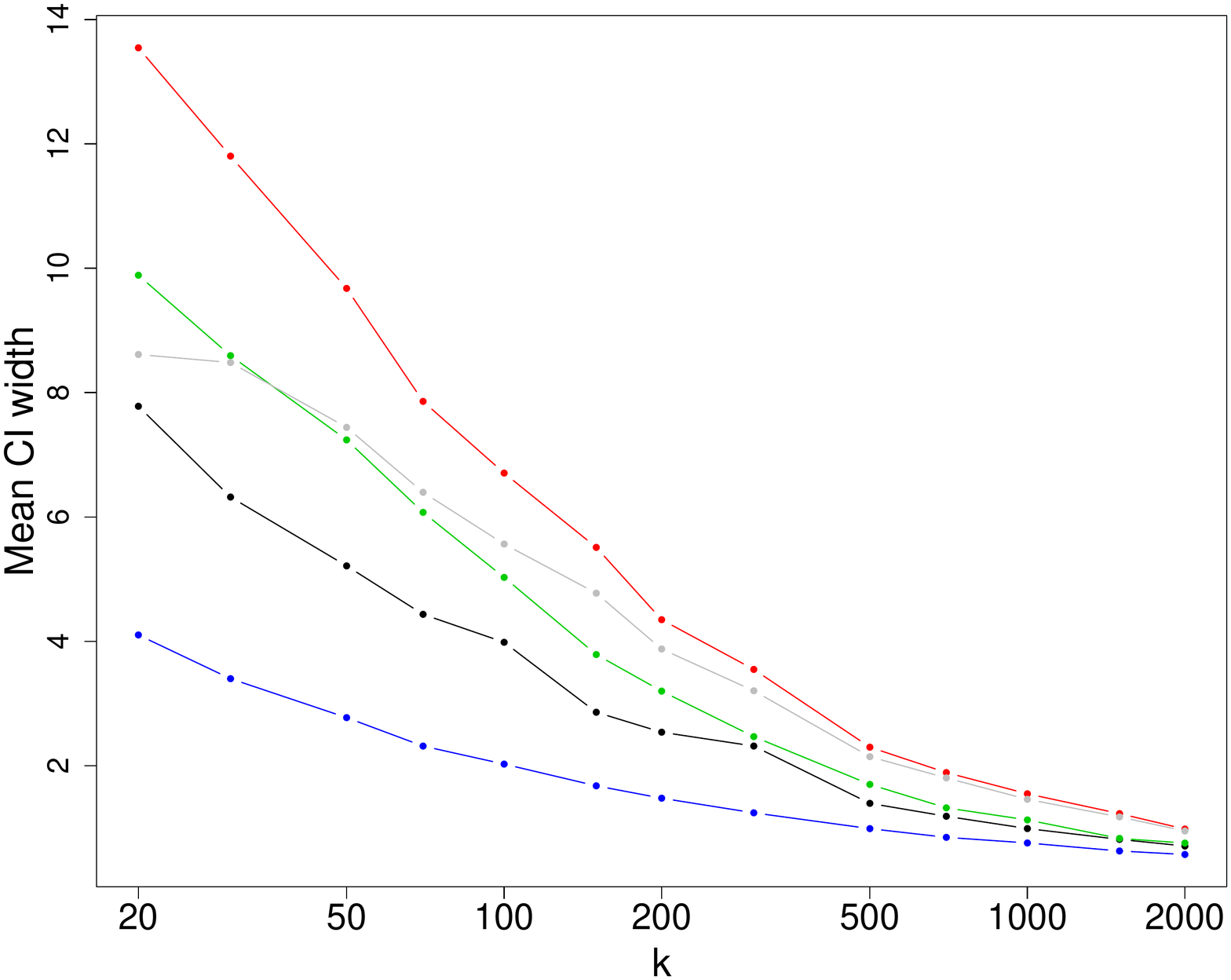}}
		\subfigure{\includegraphics[width=0.33\textwidth]{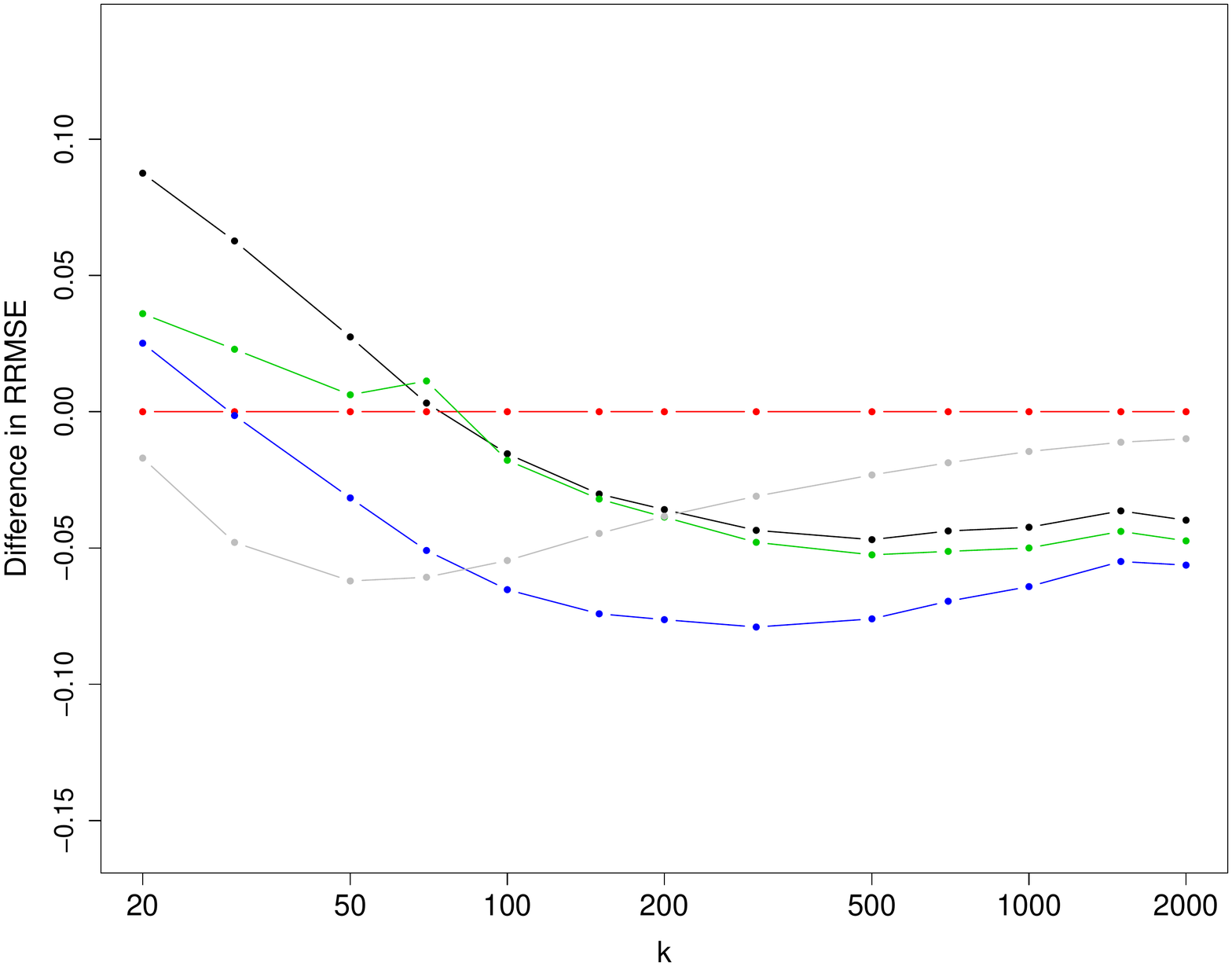}}
		\captionit{200 year return level estimates when sampling from the GEV distribution with $(\mu, \sigma, \xi)=(0, 1, -0.2)$ using the fixed-threshold stopping rule over a range of thresholds. From left to right. Top: relative bias, relative RMSE and relative variance. Bottom: coverage, average CI width, difference in RRMSE to RRMSE of standard estimator. Colour scheme is the same as in Figure \ref{fig:sup.shape}. Based on $10^5$ replicated samples with the historical data created using approach~\eqref{eqn:initial} of the paper. Coverage is based on 5000 replicated samples.}
\end{figure}

\begin{figure}
	\subfigure{\includegraphics[width=0.49\textwidth]{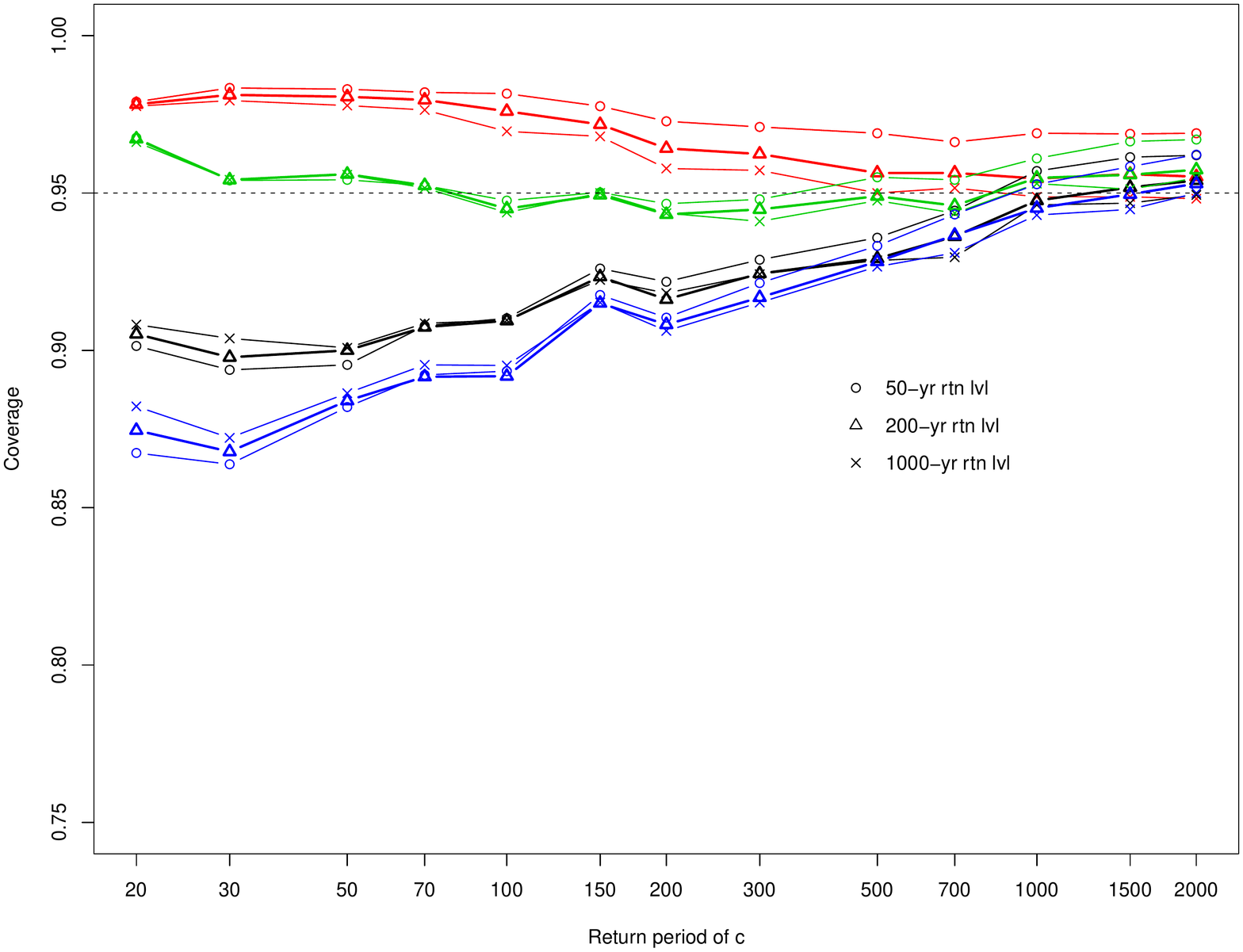}}
	\subfigure{\includegraphics[width=0.49\textwidth]{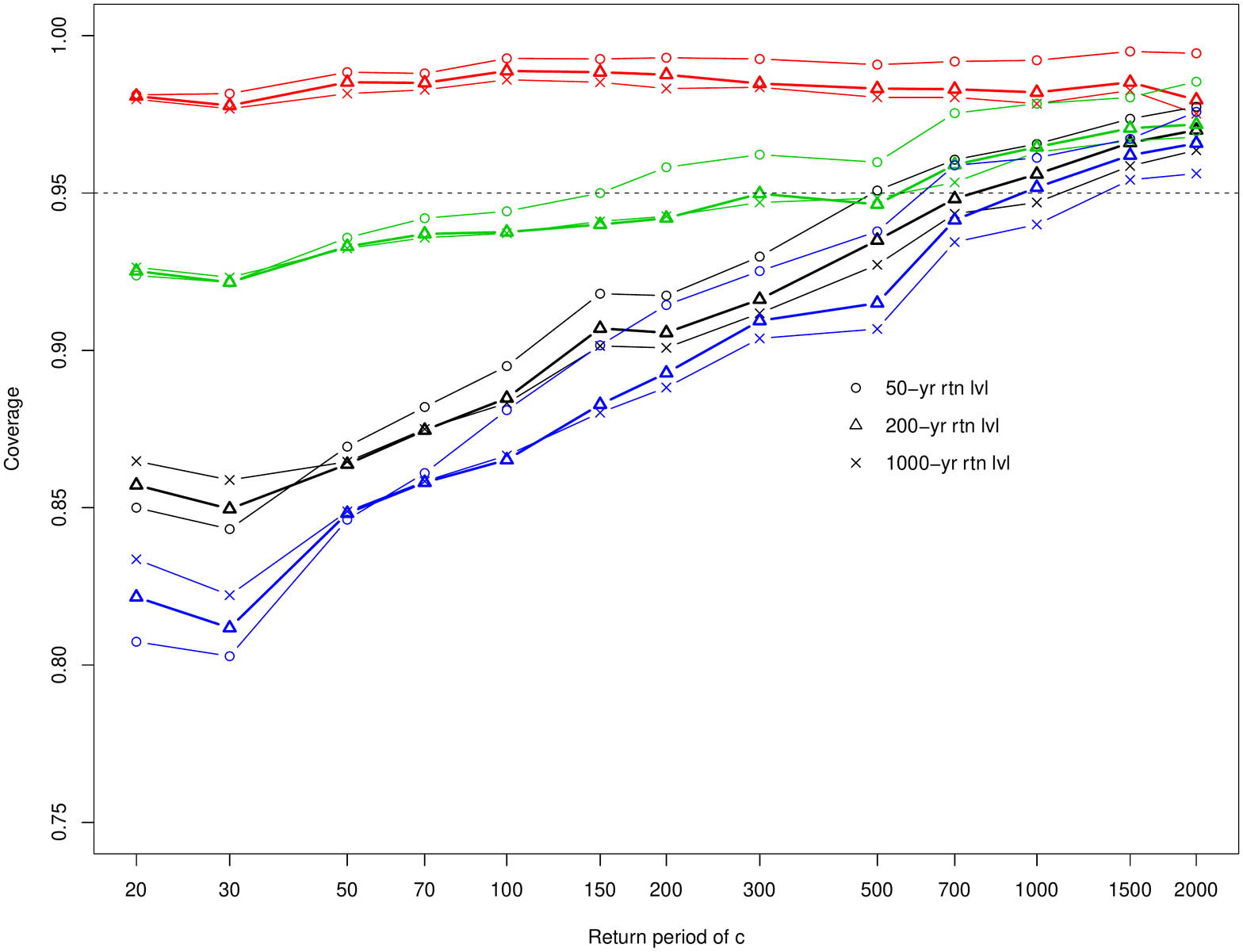}}
	\subfigure{\includegraphics[width=0.49\textwidth]{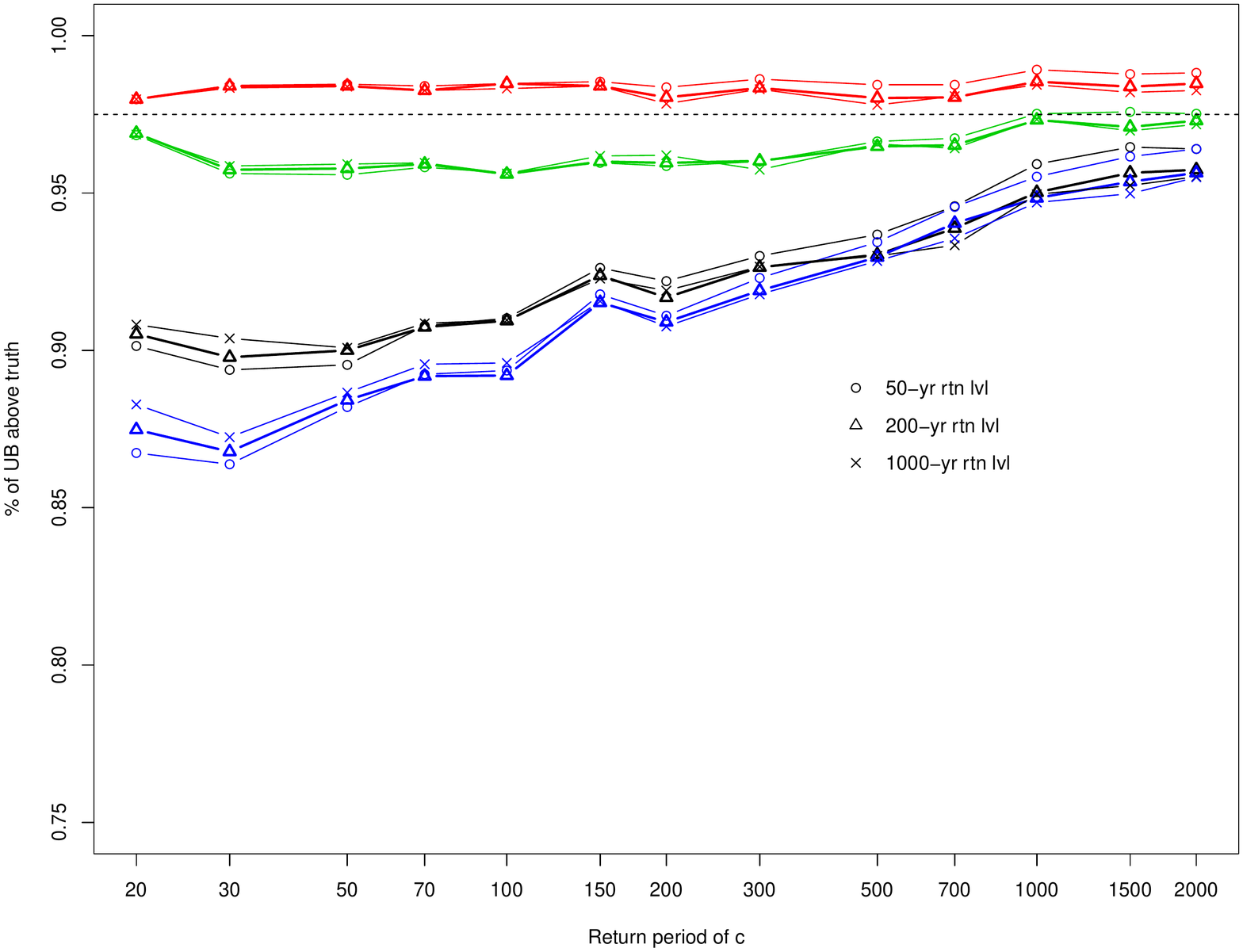}}
	\subfigure{\includegraphics[width=0.49\textwidth]{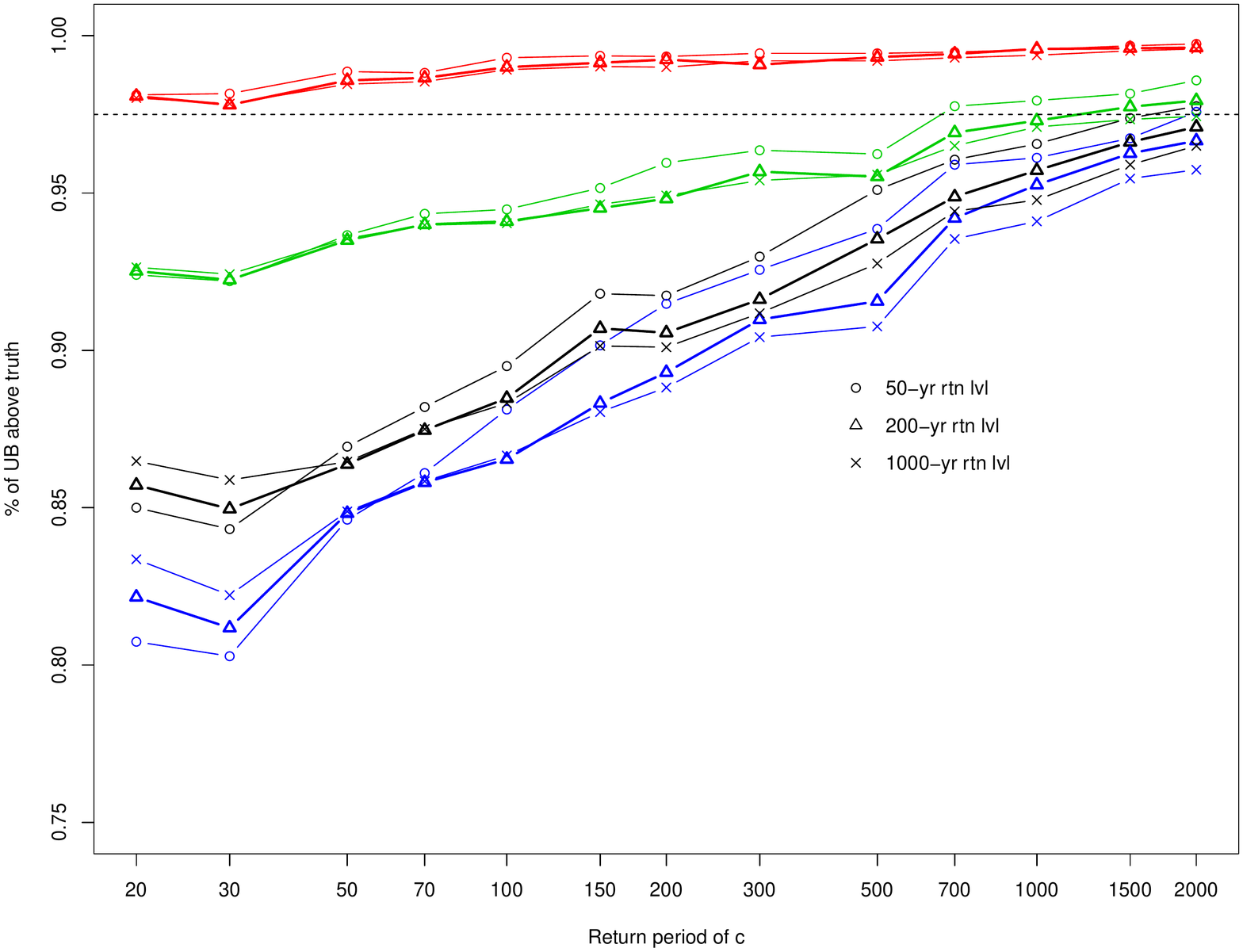}}
	\subfigure{\includegraphics[width=0.49\textwidth]{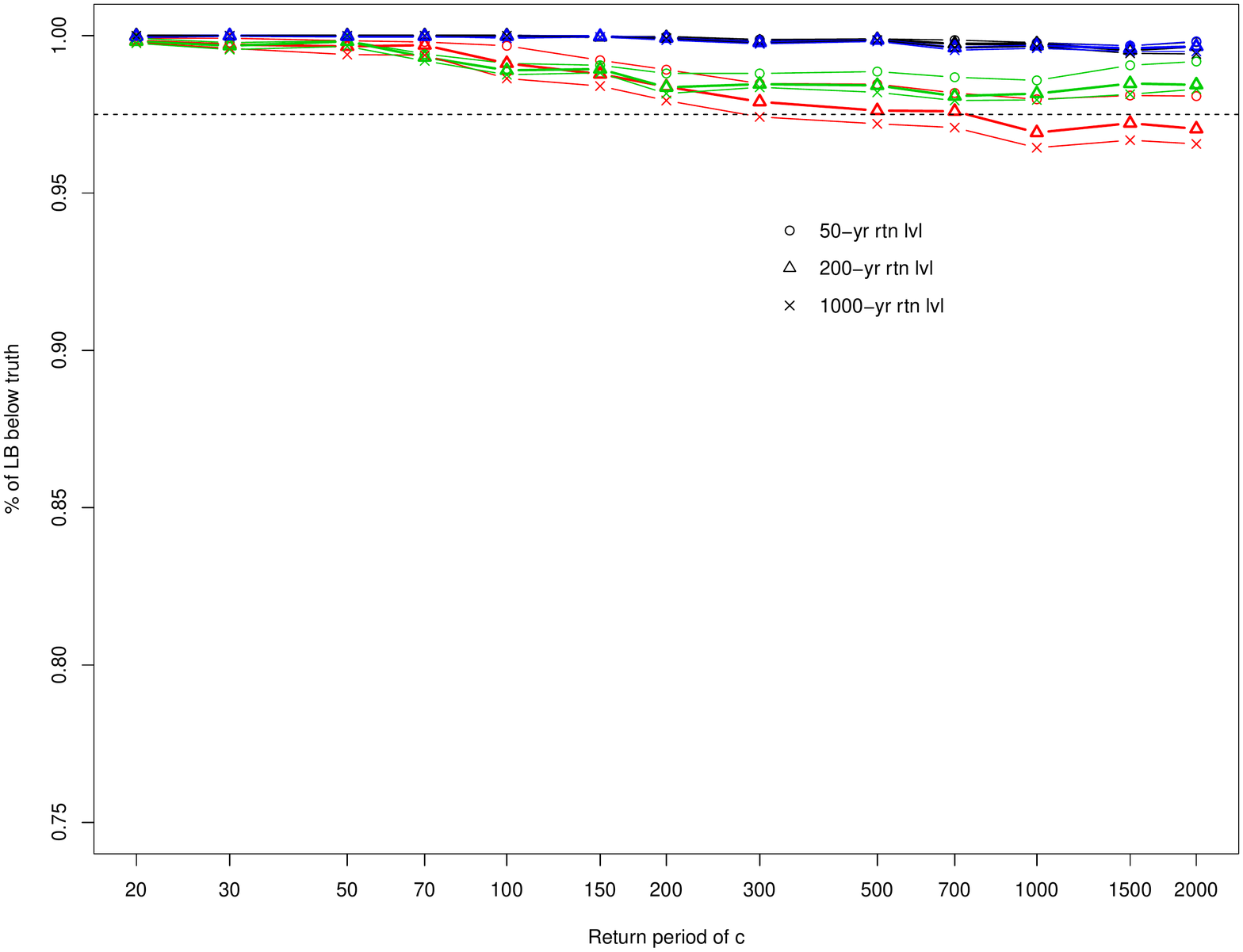}} \hfill
	\subfigure{\includegraphics[width=0.49\textwidth]{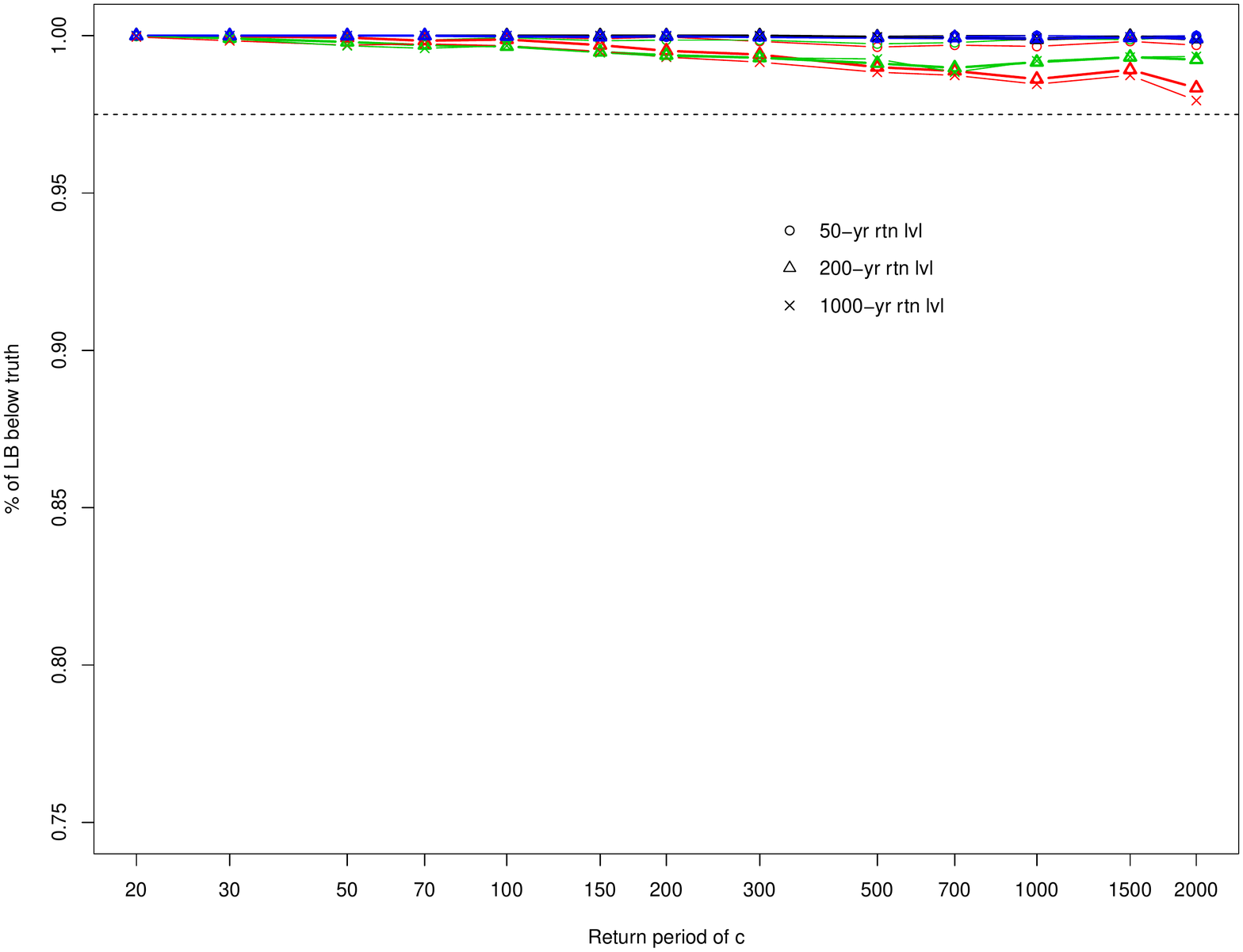}}
	\captionit{Coverage (top), \% of upper bounds greater than $x_y$ (middle), \% lower bounds less than $x_y$ (bottom) of the confidence intervals of $x_y$ for $y=50,200,1000$ when $xi$ is 0.2 (left) and -0.2 (right). Colour scheme is the same as in Figure \ref{fig:sup.shape}. Based on $5000$ replicated samples from the GEV with $(\mu,\sigma)=(0,1)$created using the fixed-threshold stopping rule and historical data created using approach~\eqref{eqn:initial} of the paper.}
\end{figure}

\begin{figure}
	\subfigure{\includegraphics[width=0.49\textwidth]{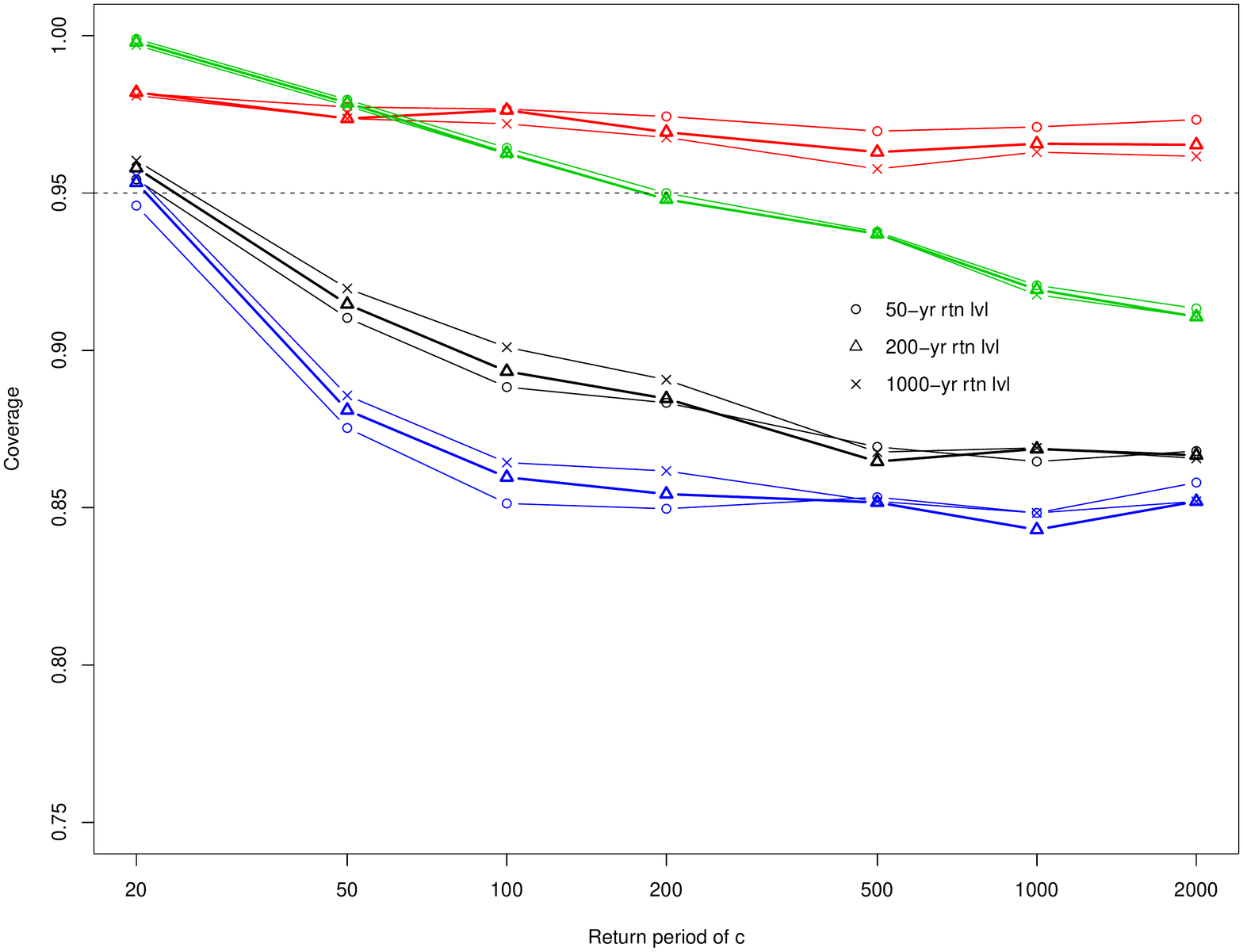}}
	\subfigure{\includegraphics[width=0.49\textwidth]{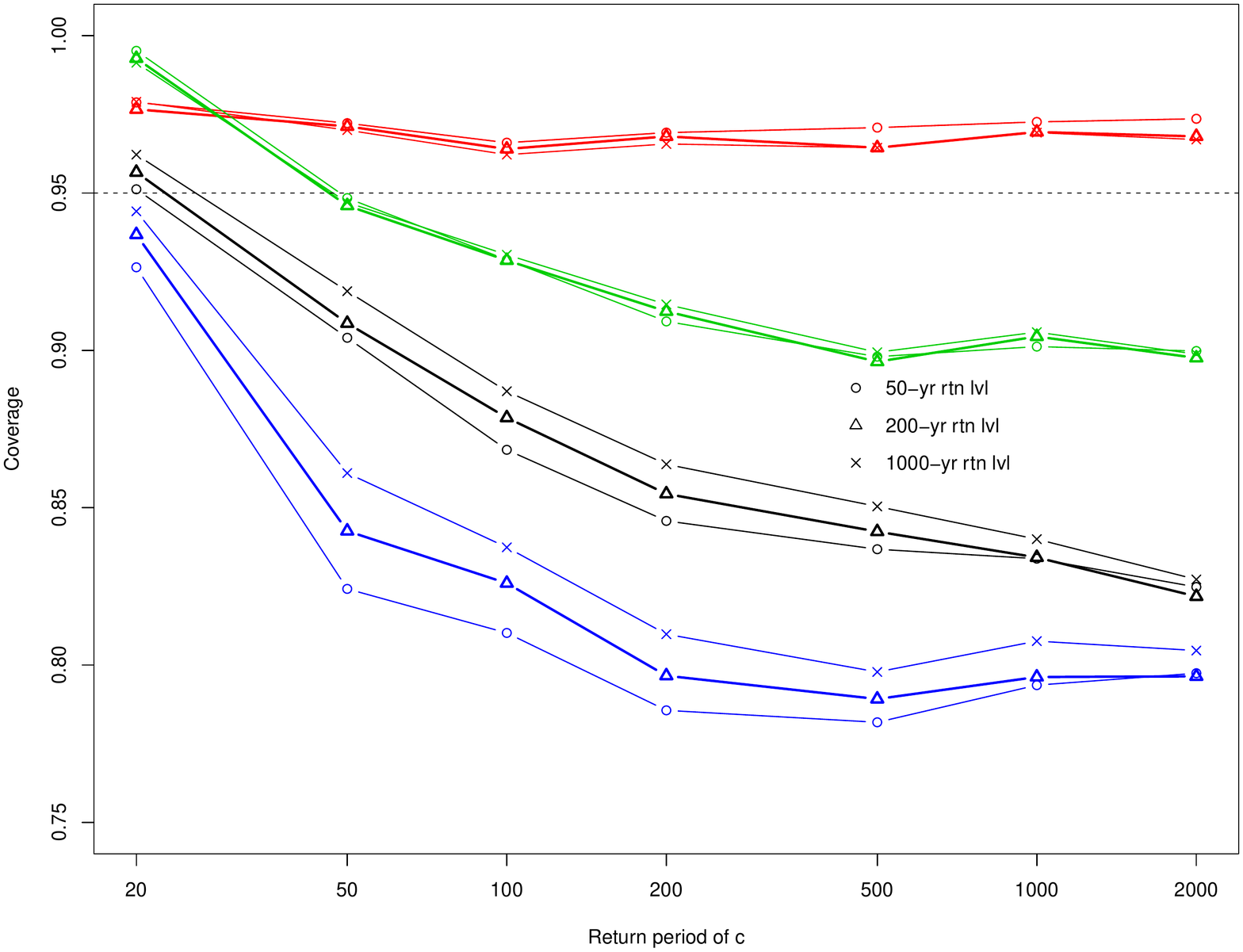}}
	\subfigure{\includegraphics[width=0.49\textwidth]{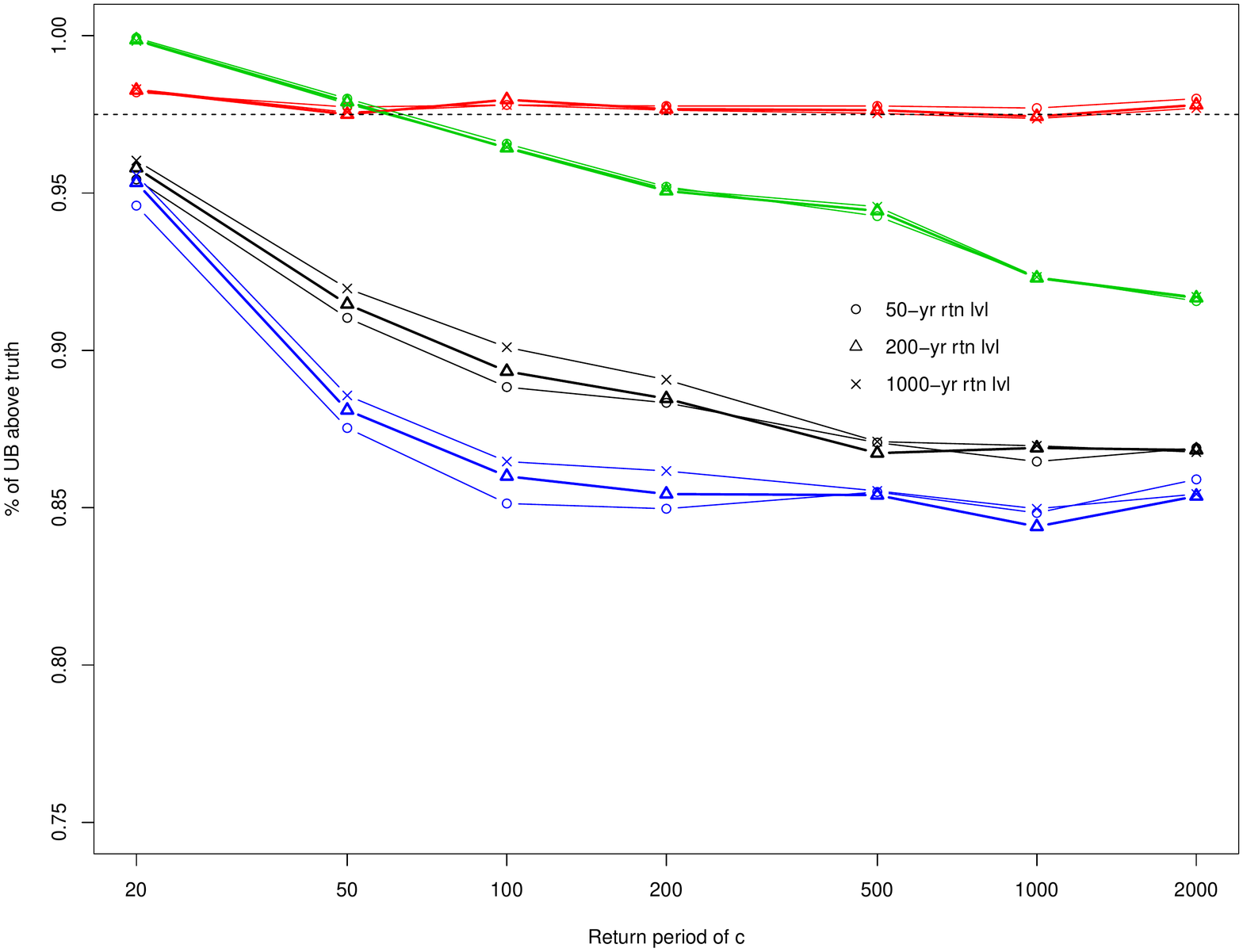}}
	\subfigure{\includegraphics[width=0.49\textwidth]{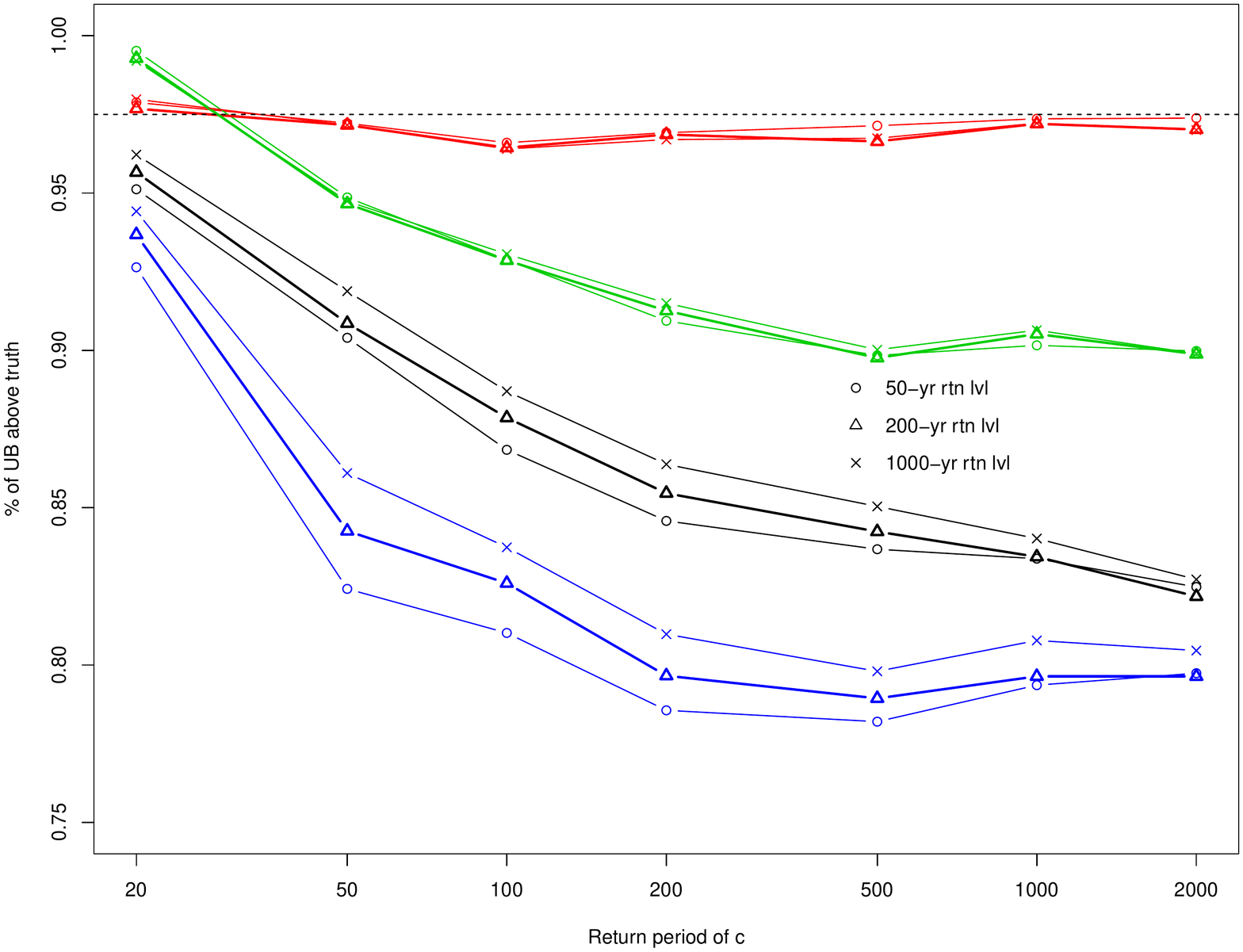}}
	\subfigure{\includegraphics[width=0.49\textwidth]{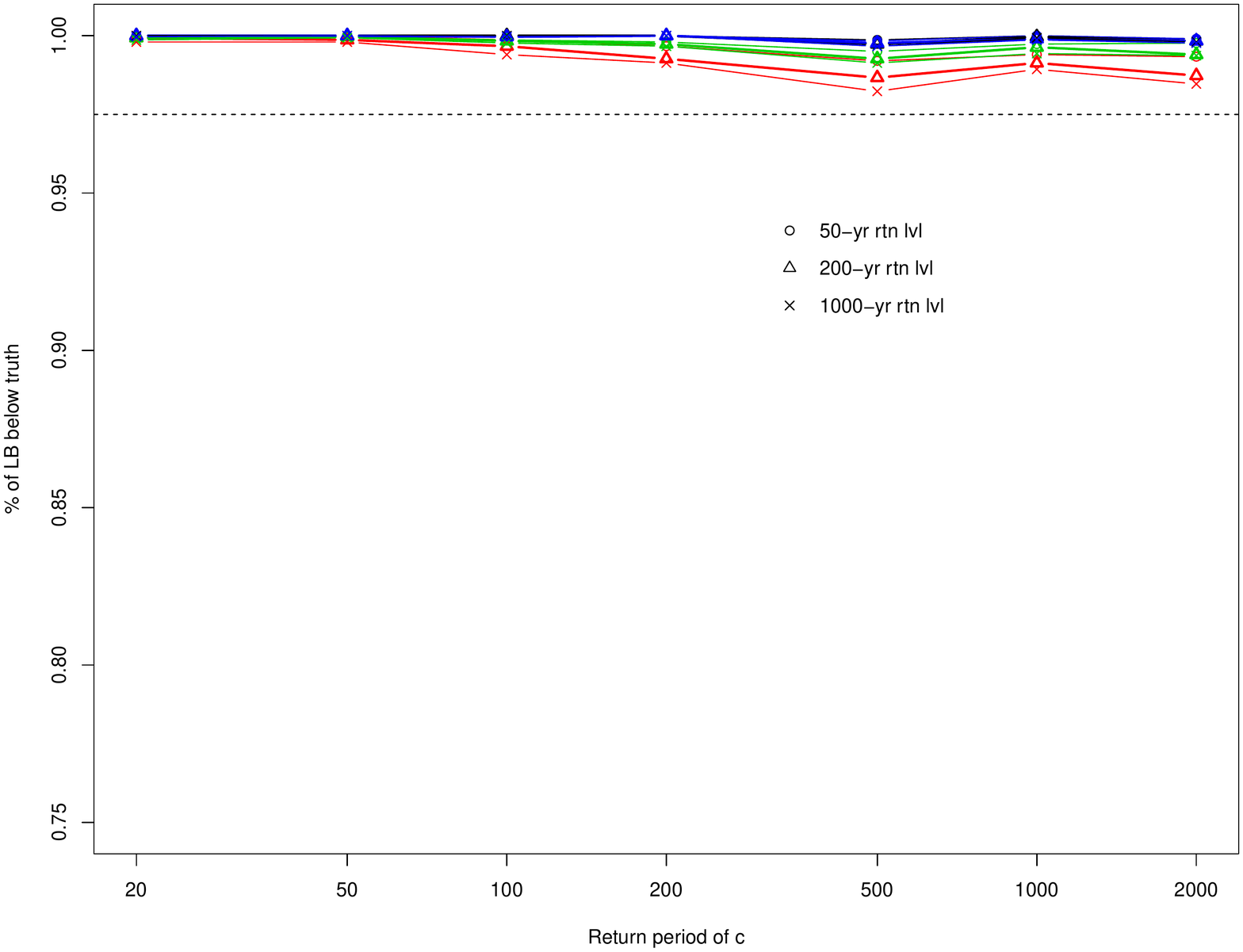}} \hfill
	\subfigure{\includegraphics[width=0.49\textwidth]{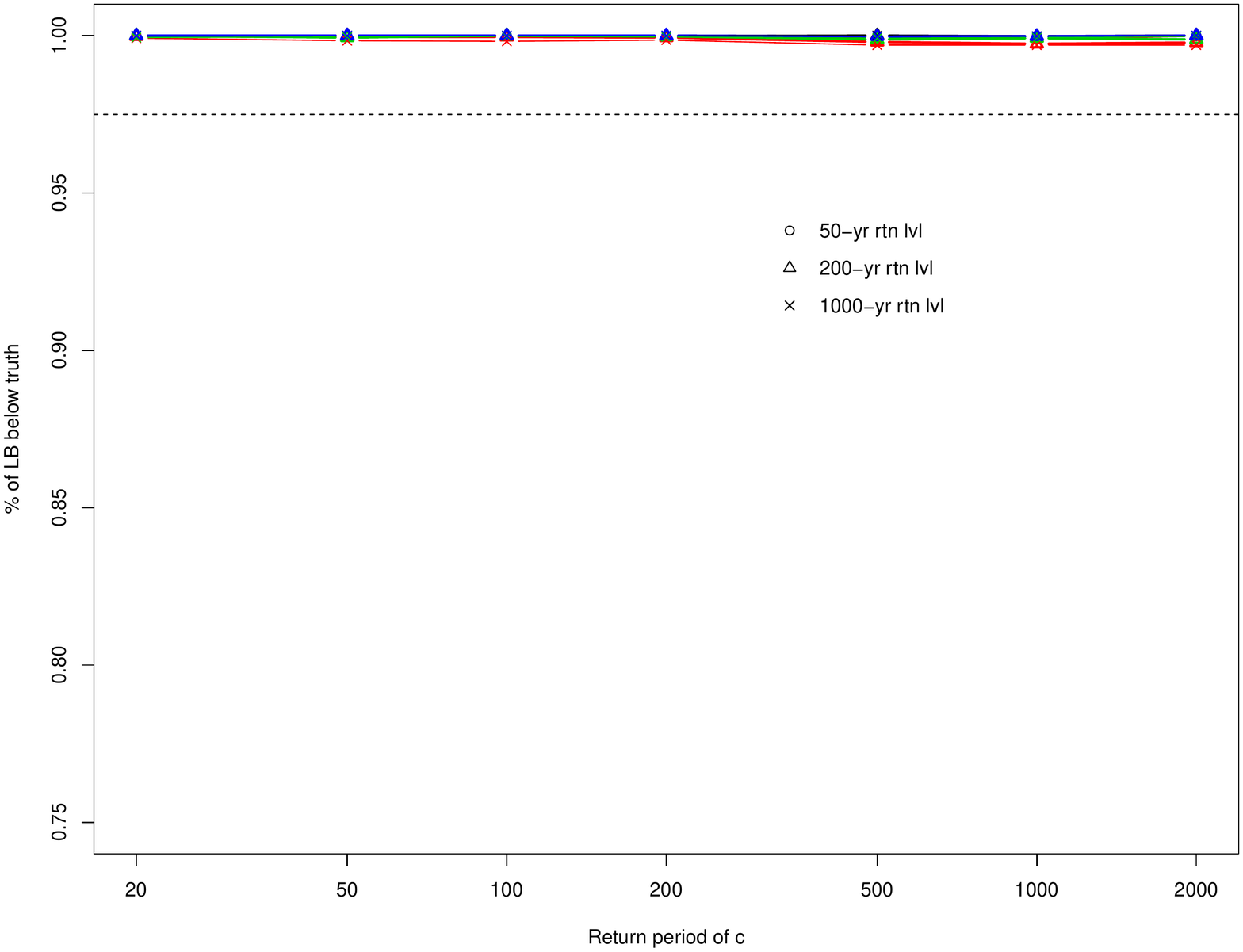}}
	\captionit{Coverage (top), \% of upper bounds greater than $x_y$ (middle), \% lower bounds less than $x_y$ (bottom) of the confidence intervals of $x_y$ for $y=50,200,1000$ when $xi$ is 0.2 (left) and -0.2 (right). Colour scheme is the same as in Figure \ref{fig:sup.shape}. Based on $3000$ replicated samples from the GEV with $(\mu,\sigma)=(0,1)$created using the variable-threshold stopping rule and historical data created using approach~\eqref{eqn:initial} of the paper.}
\end{figure}

\newpage

\begin{figure}
		\subfigure{\includegraphics[width=0.33\textwidth]{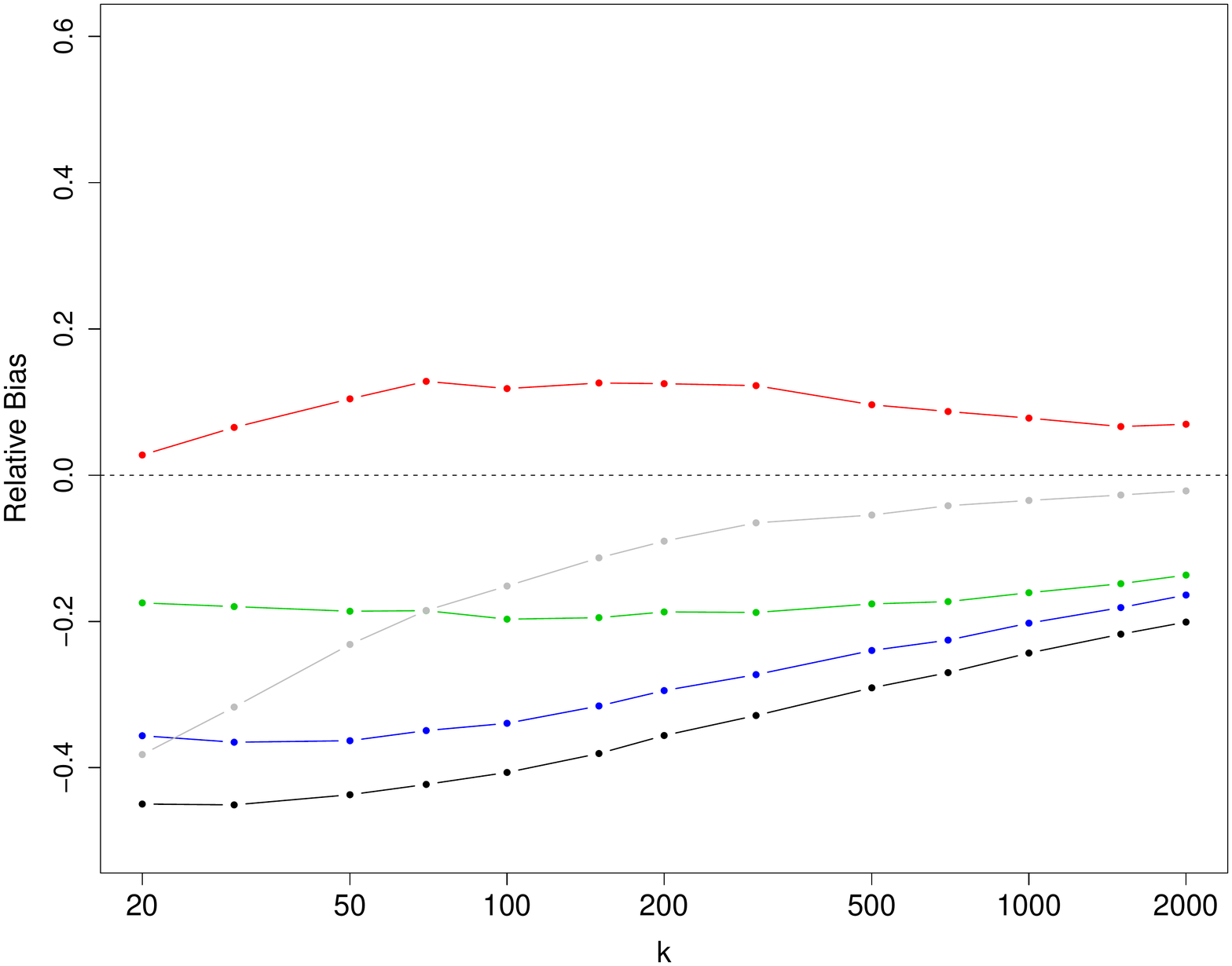}}
		\subfigure{\includegraphics[width=0.33\textwidth]{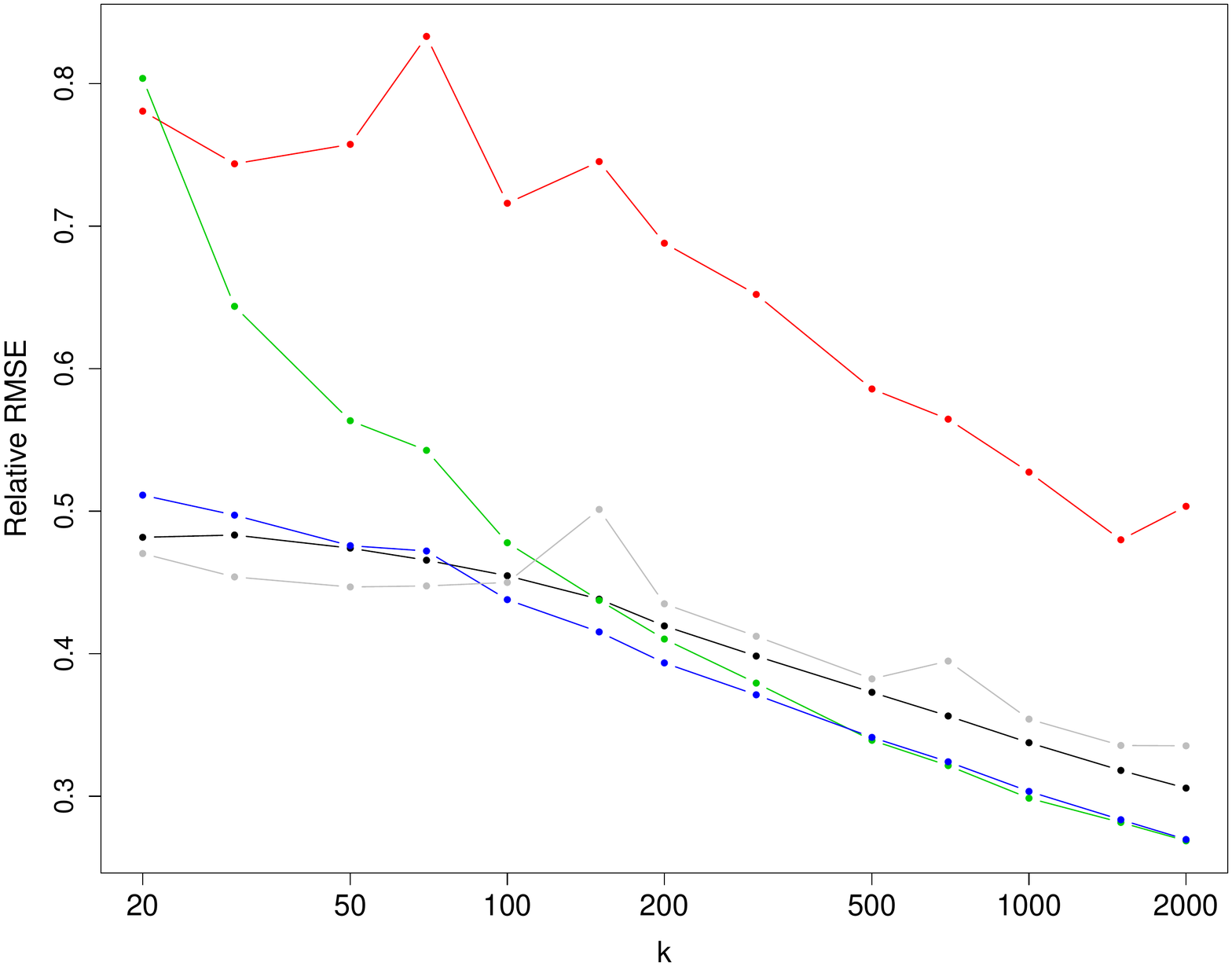}}
		\subfigure{\includegraphics[width=0.33\textwidth]{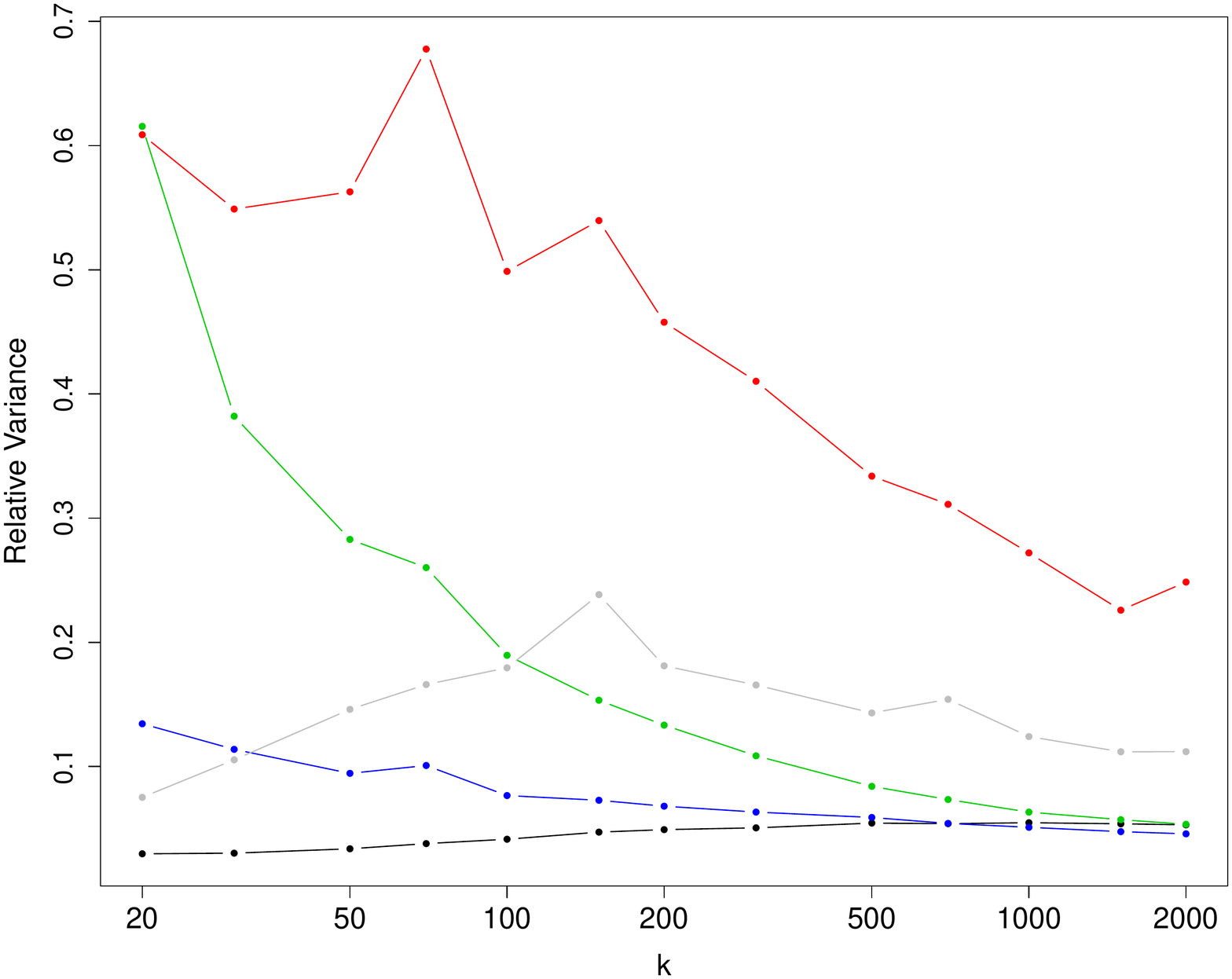}}
		\subfigure{\includegraphics[width=0.33\textwidth]{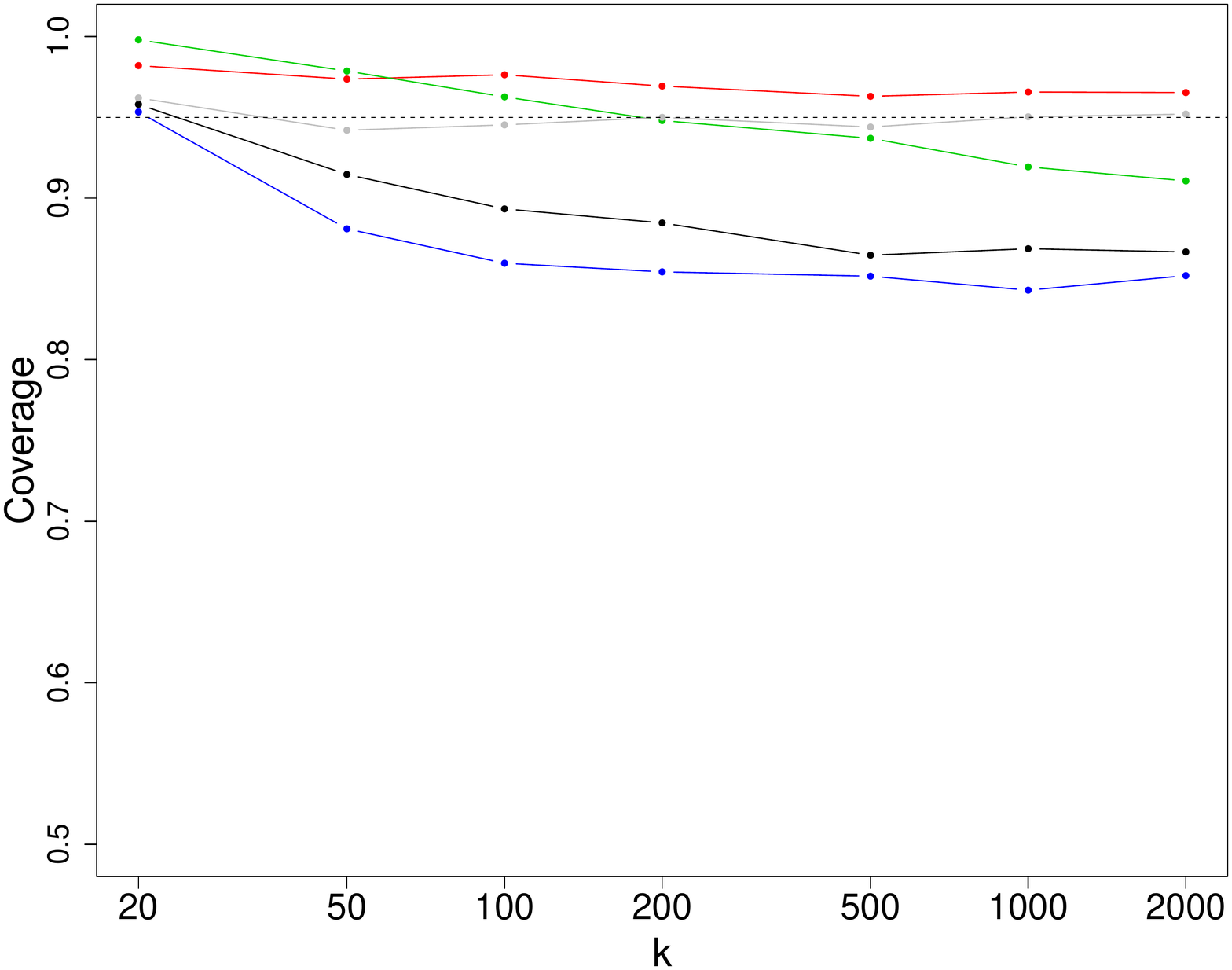}}
		\subfigure{\includegraphics[width=0.33\textwidth]{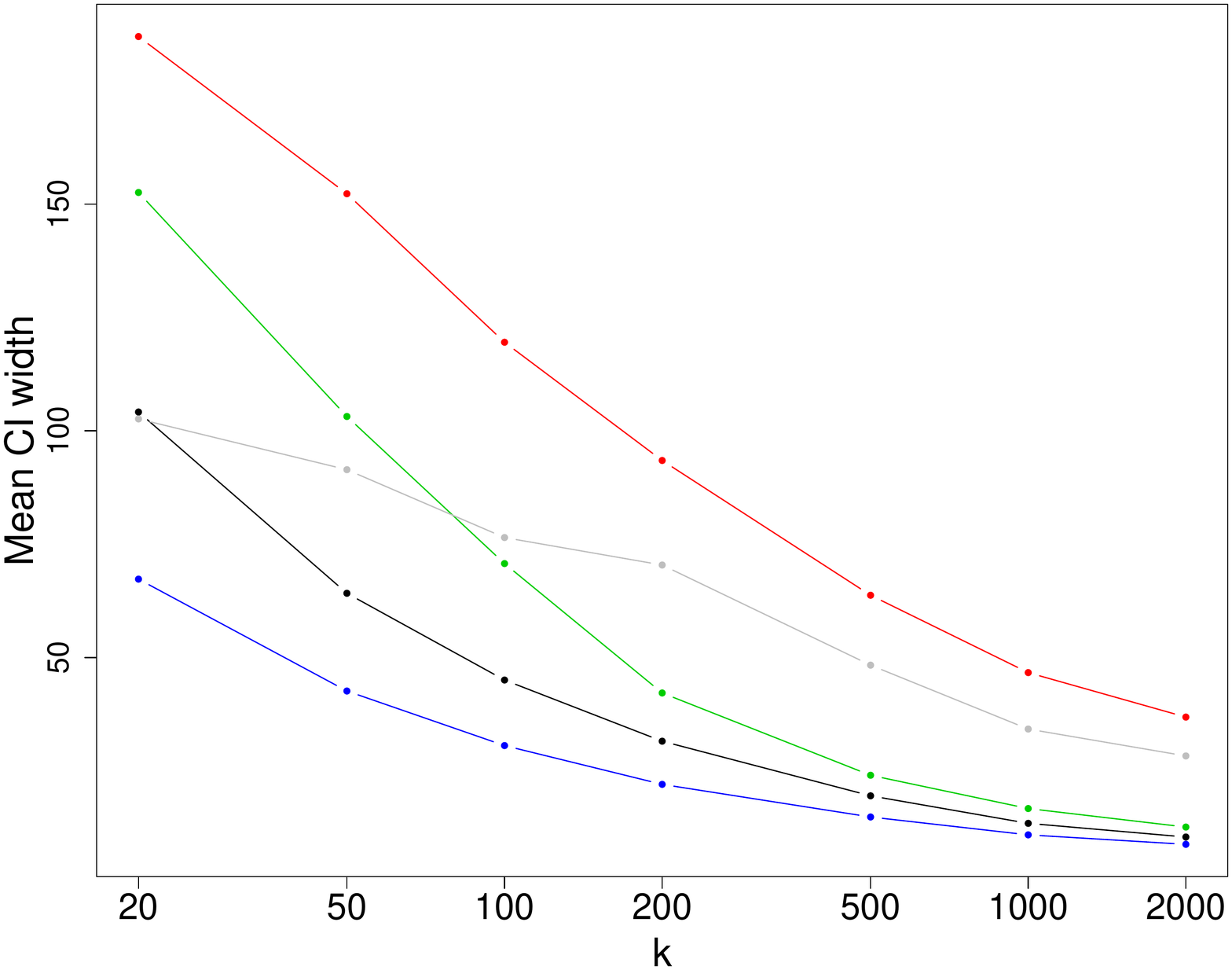}}
		\subfigure{\includegraphics[width=0.33\textwidth]{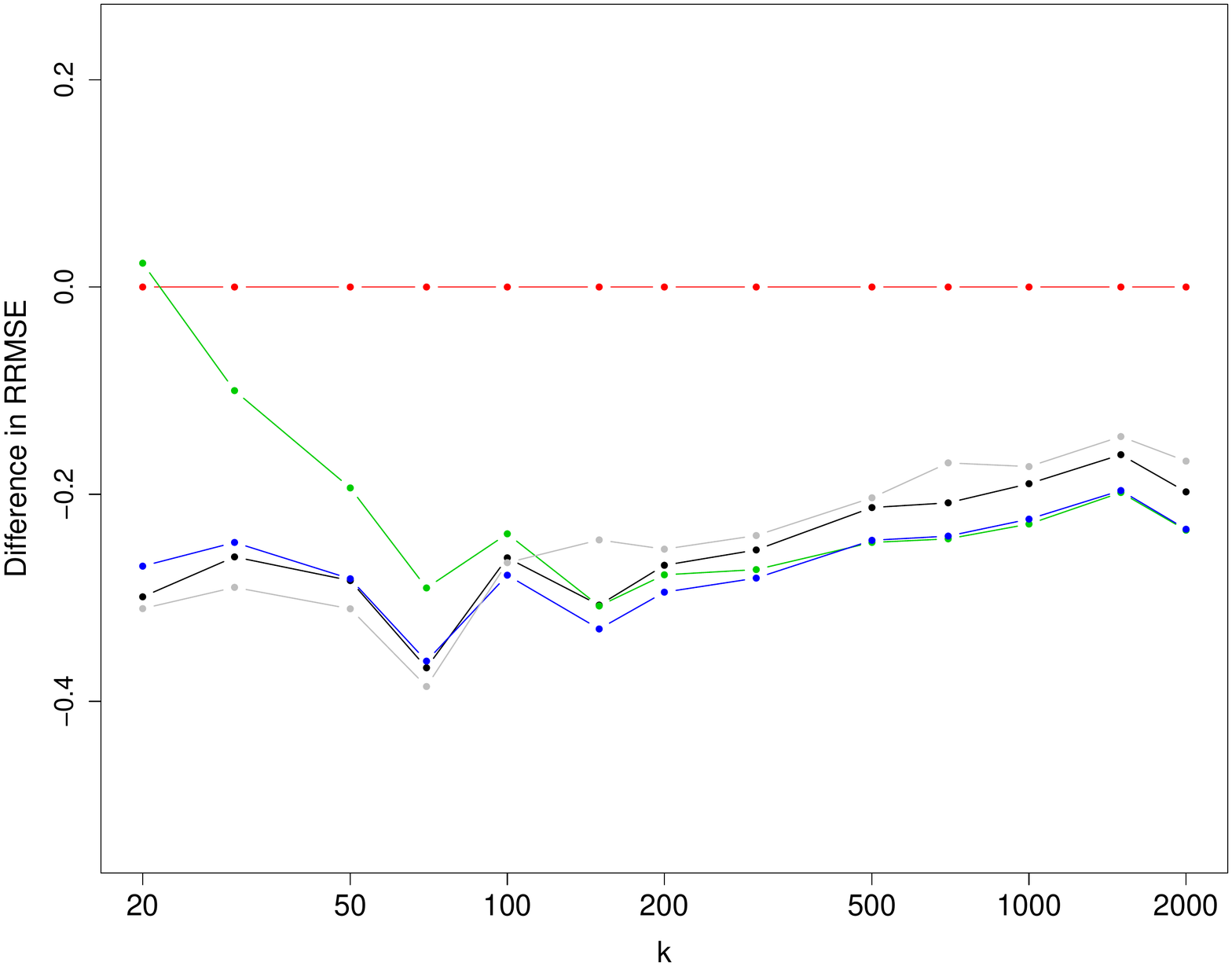}}
		\captionit{200 year return level estimates when sampling from the GEV distribution with $(\mu, \sigma, \xi)=(0, 1, 0.2)$ using the variable-threshold stopping rule over a range of $k$. Based on $10^4$ replicated samples with the historical data created using approach~\eqref{eqn:initial} of the paper. Coverage is based on 3000 replicated samples. See Figure~\ref{fig:sup.200ret} for other associated detail.}
\end{figure}

\begin{figure}
		\subfigure{\includegraphics[width=0.33\textwidth]{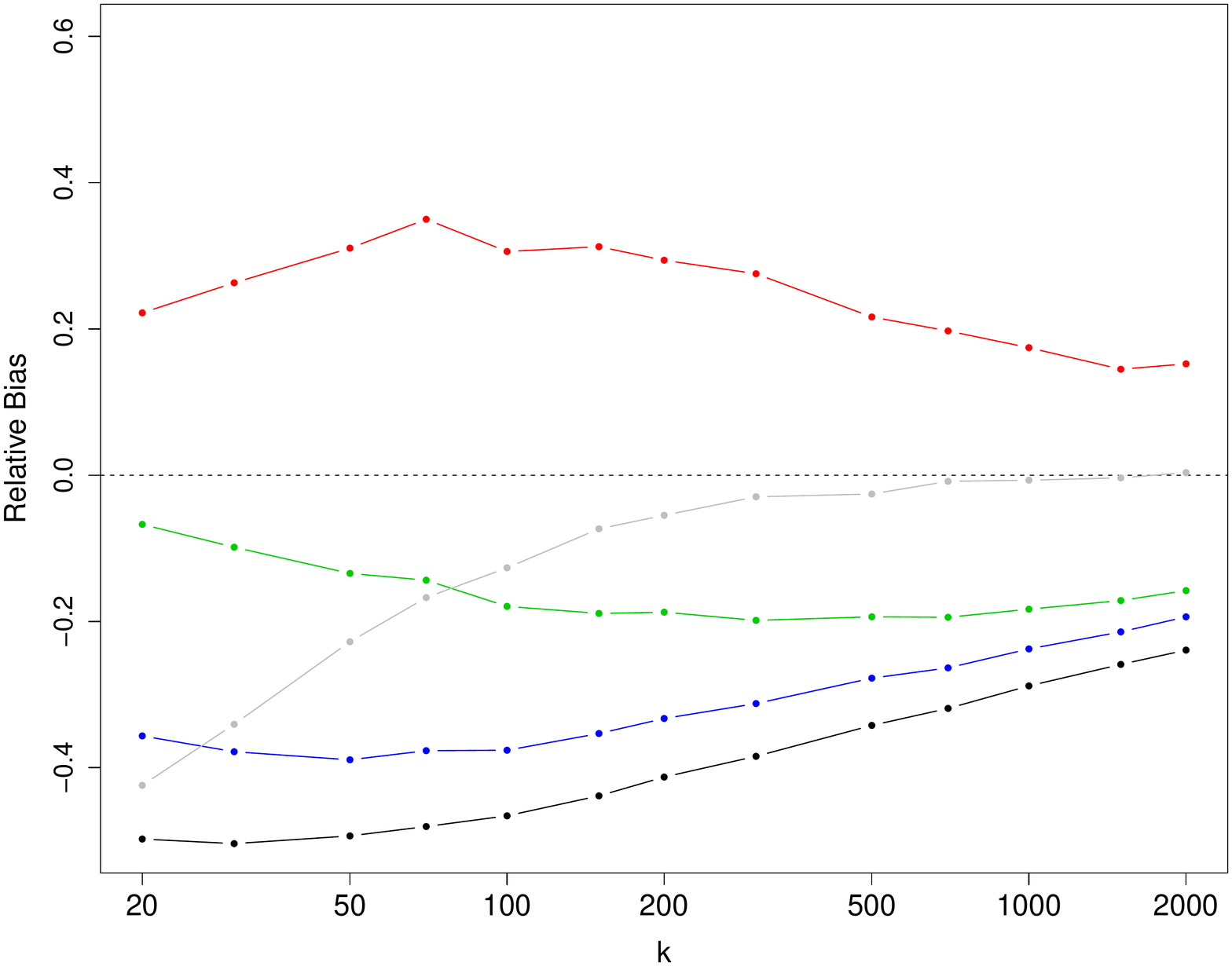}}
		\subfigure{\includegraphics[width=0.33\textwidth]{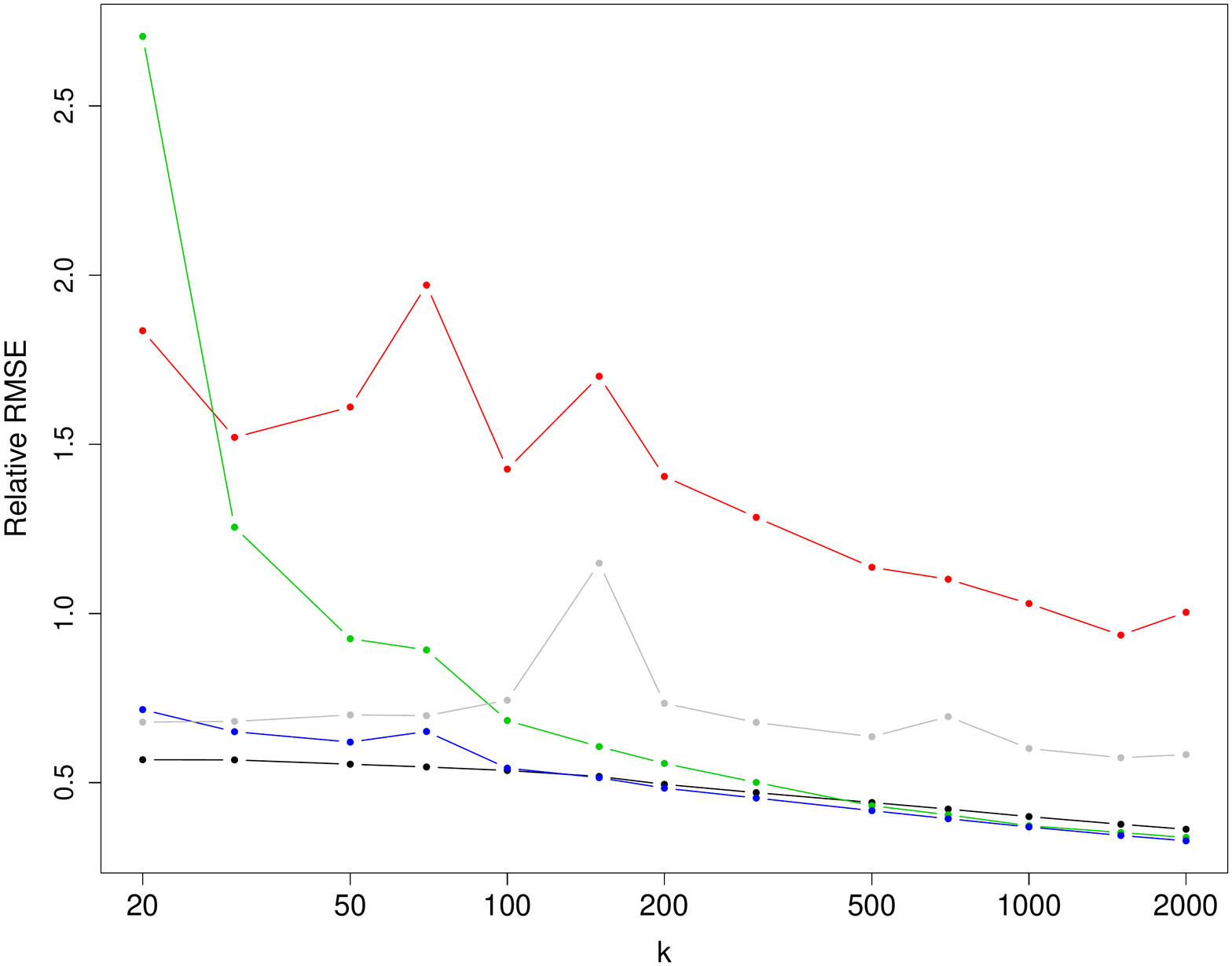}}
		\subfigure{\includegraphics[width=0.33\textwidth]{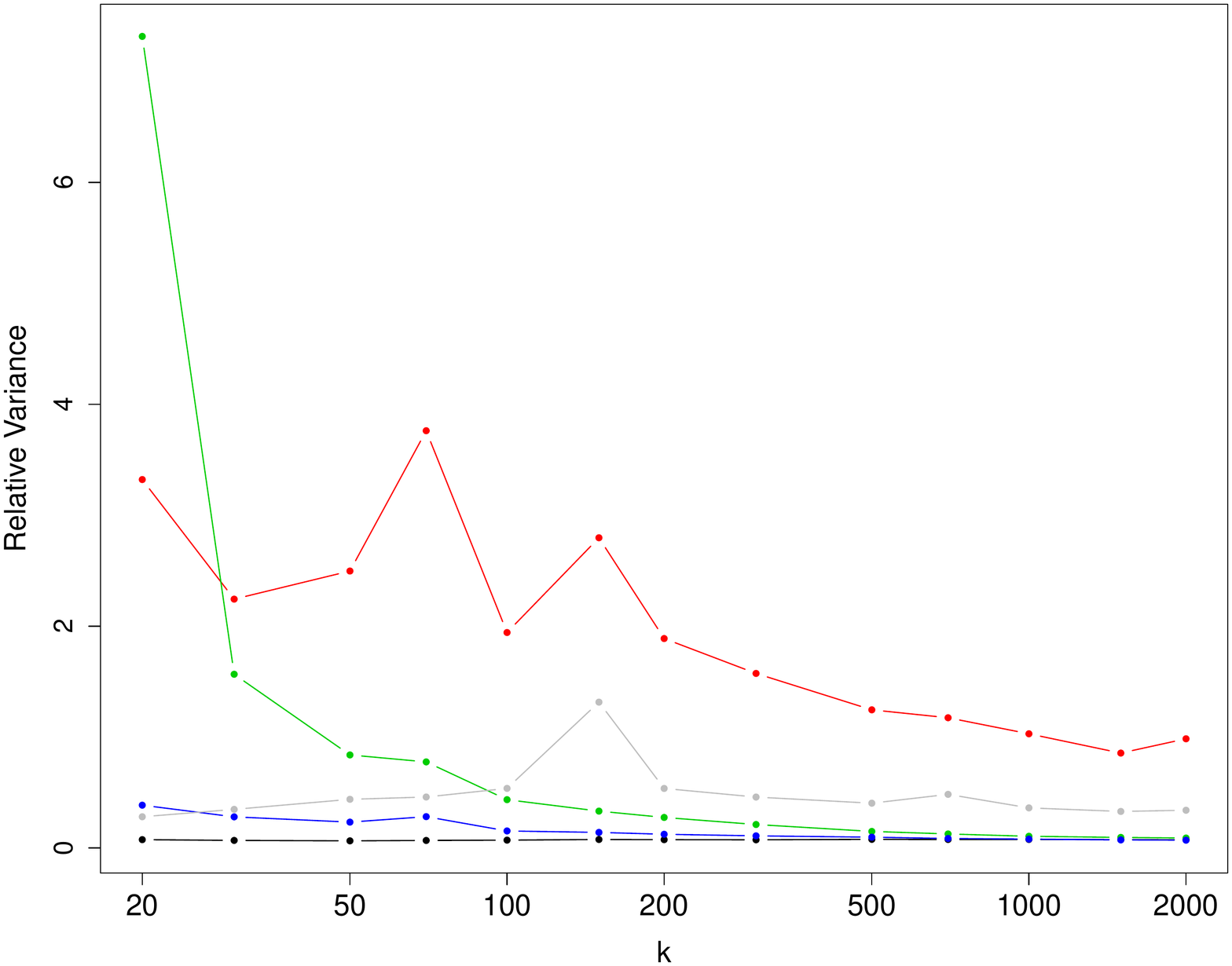}}
		\subfigure{\includegraphics[width=0.33\textwidth]{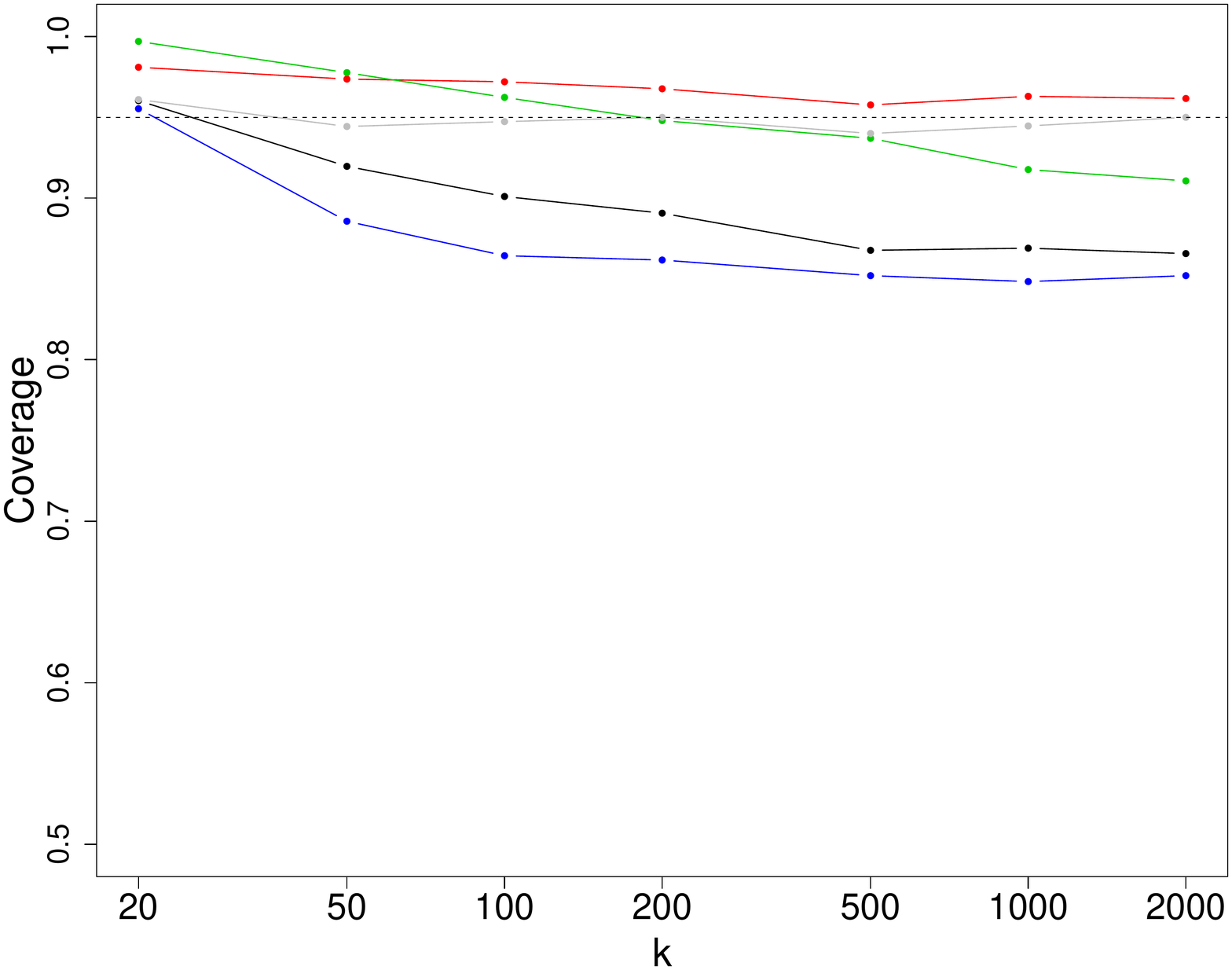}}
		\subfigure{\includegraphics[width=0.33\textwidth]{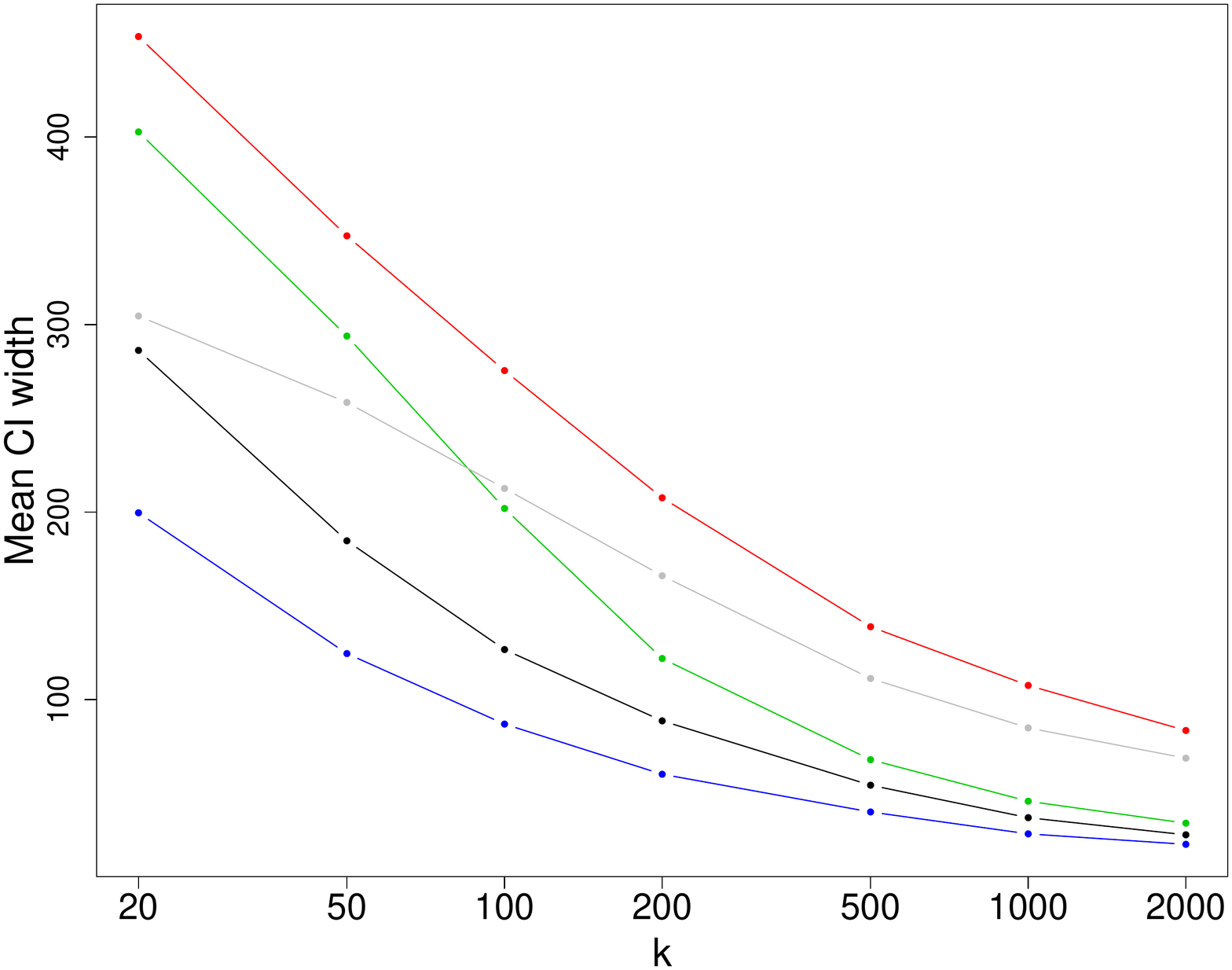}}
		\subfigure{\includegraphics[width=0.33\textwidth]{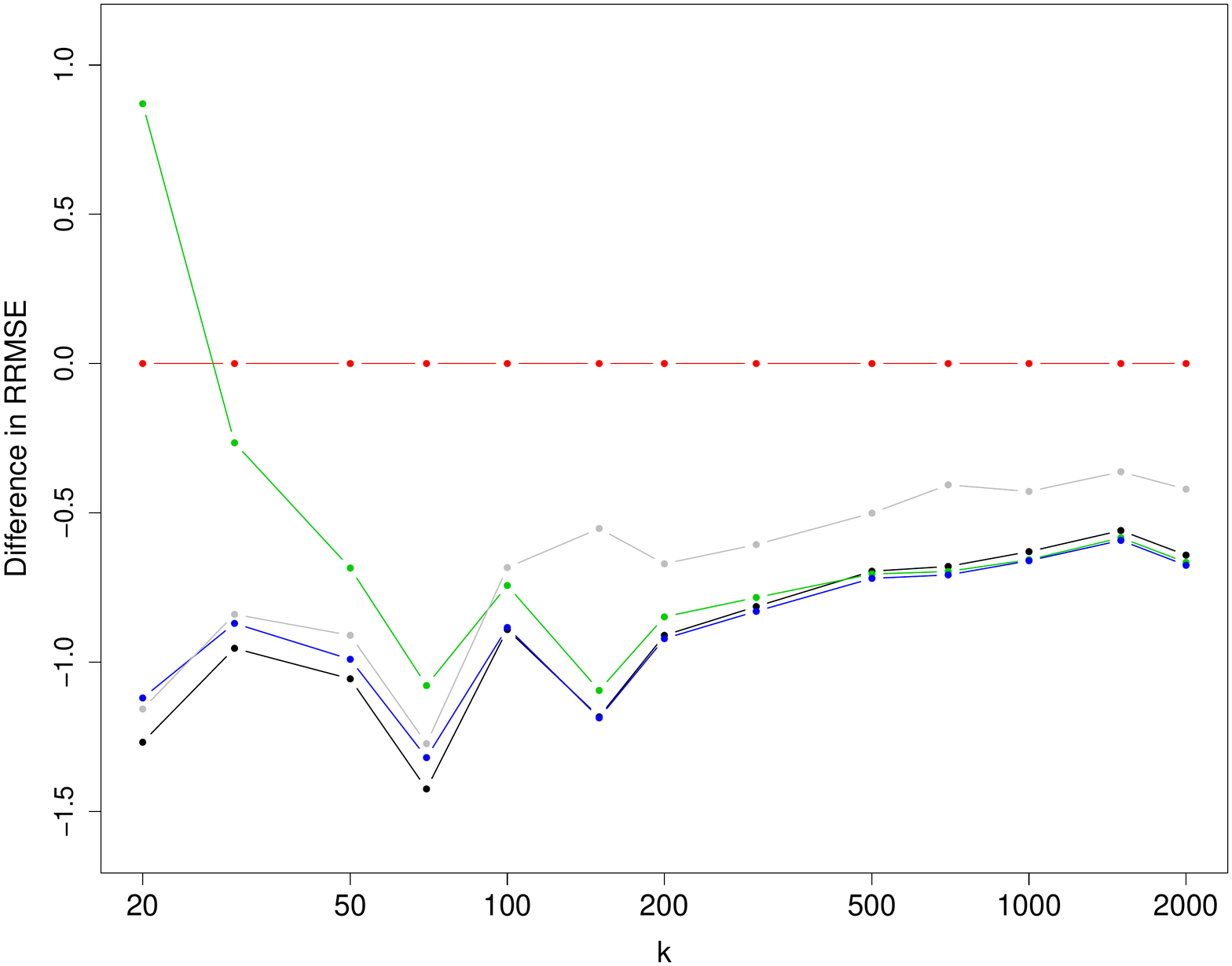}}
		\captionit{1000 year return level estimates when sampling from the GEV distribution with $(\mu, \sigma, \xi)=(0, 1, 0.2)$ using the variable-threshold stopping rule over a range of $k$. Based on $10^4$ replicated samples with the historical data created using approach~\eqref{eqn:initial} of the paper. Coverage is based on 3000 replicated samples. See Figure~\ref{fig:sup.200ret} for other associated detail.}
\end{figure}

\begin{figure}[h]
		\subfigure{\includegraphics[width=0.33\textwidth]{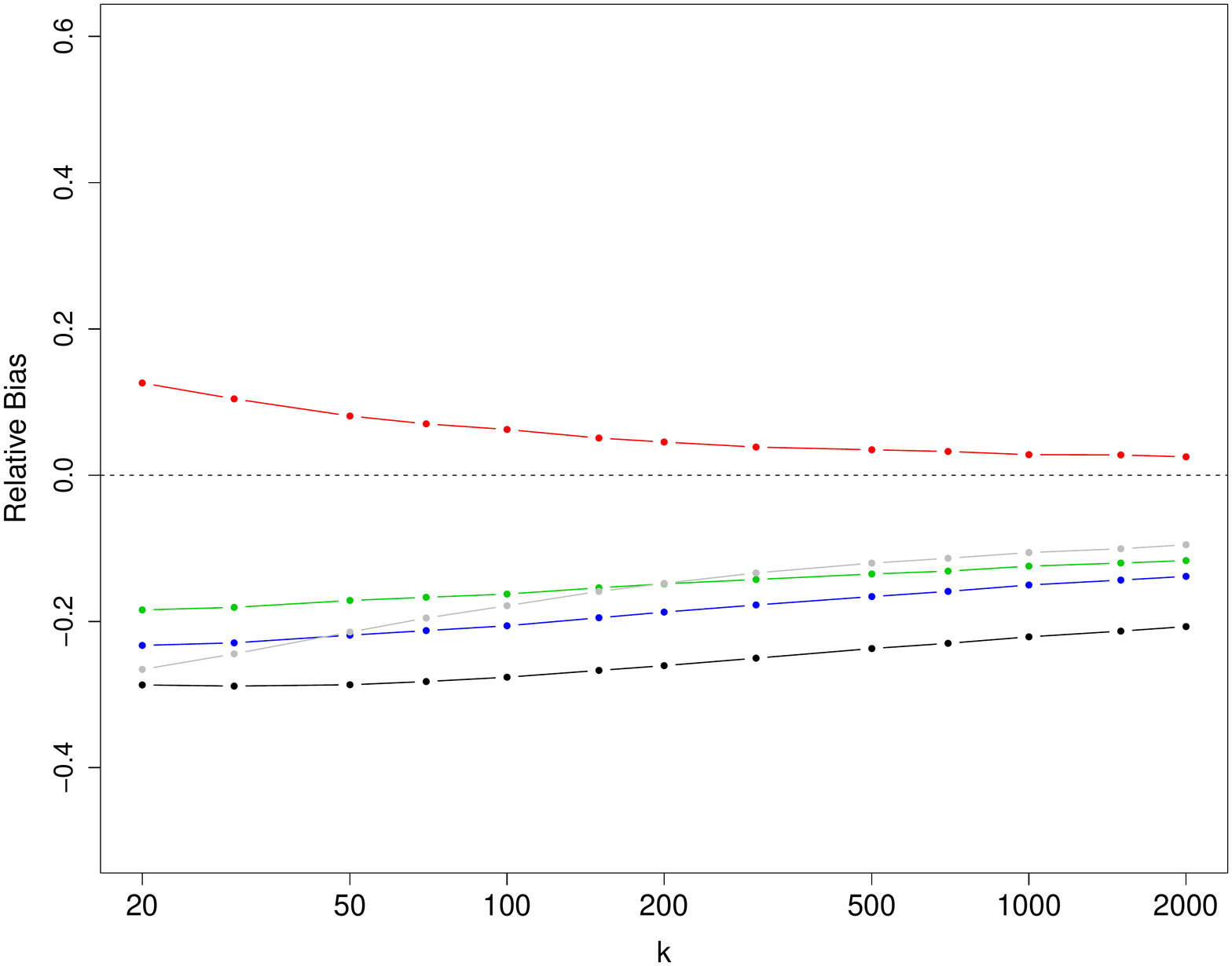}}
		\subfigure{\includegraphics[width=0.33\textwidth]{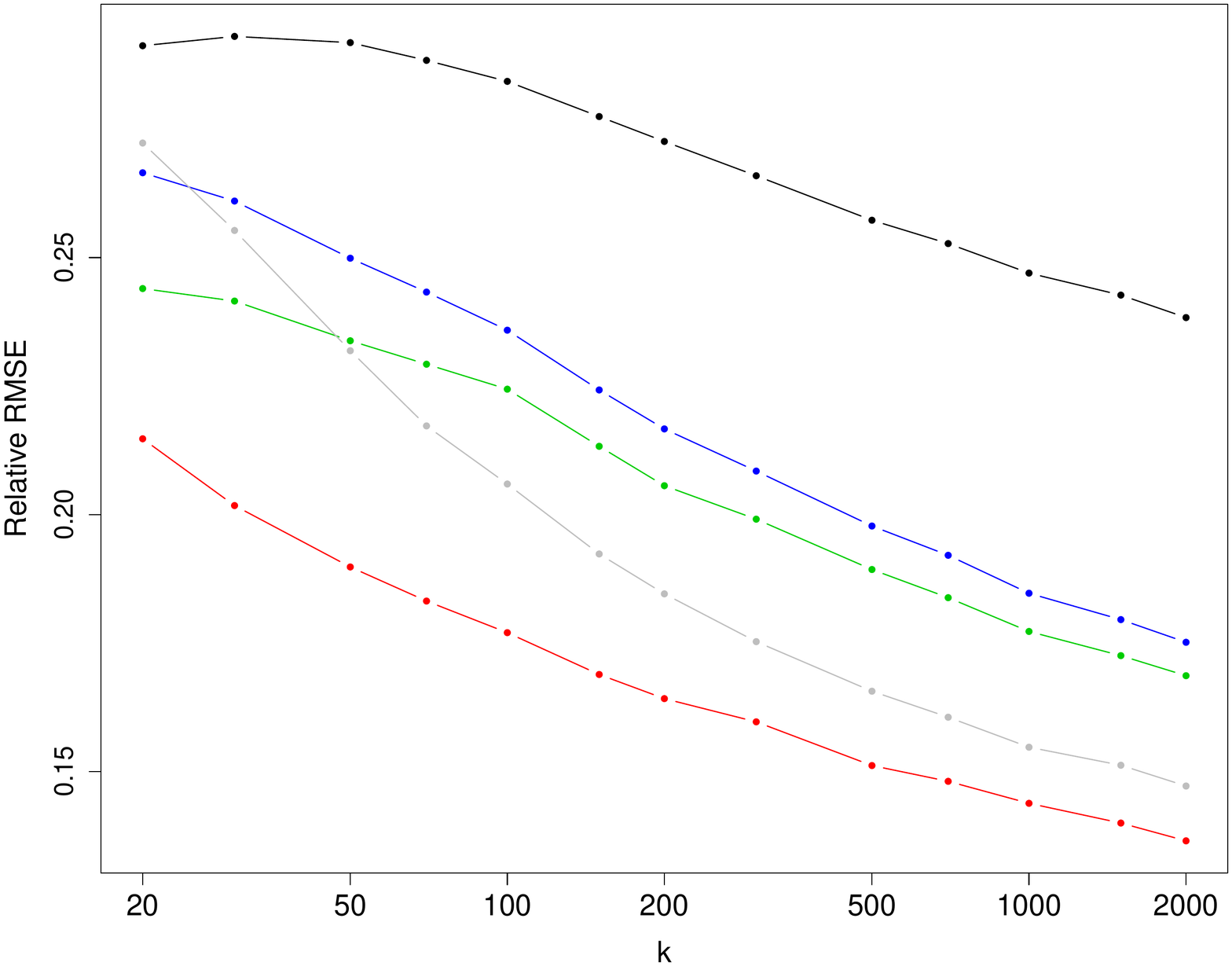}}
		\subfigure{\includegraphics[width=0.33\textwidth]{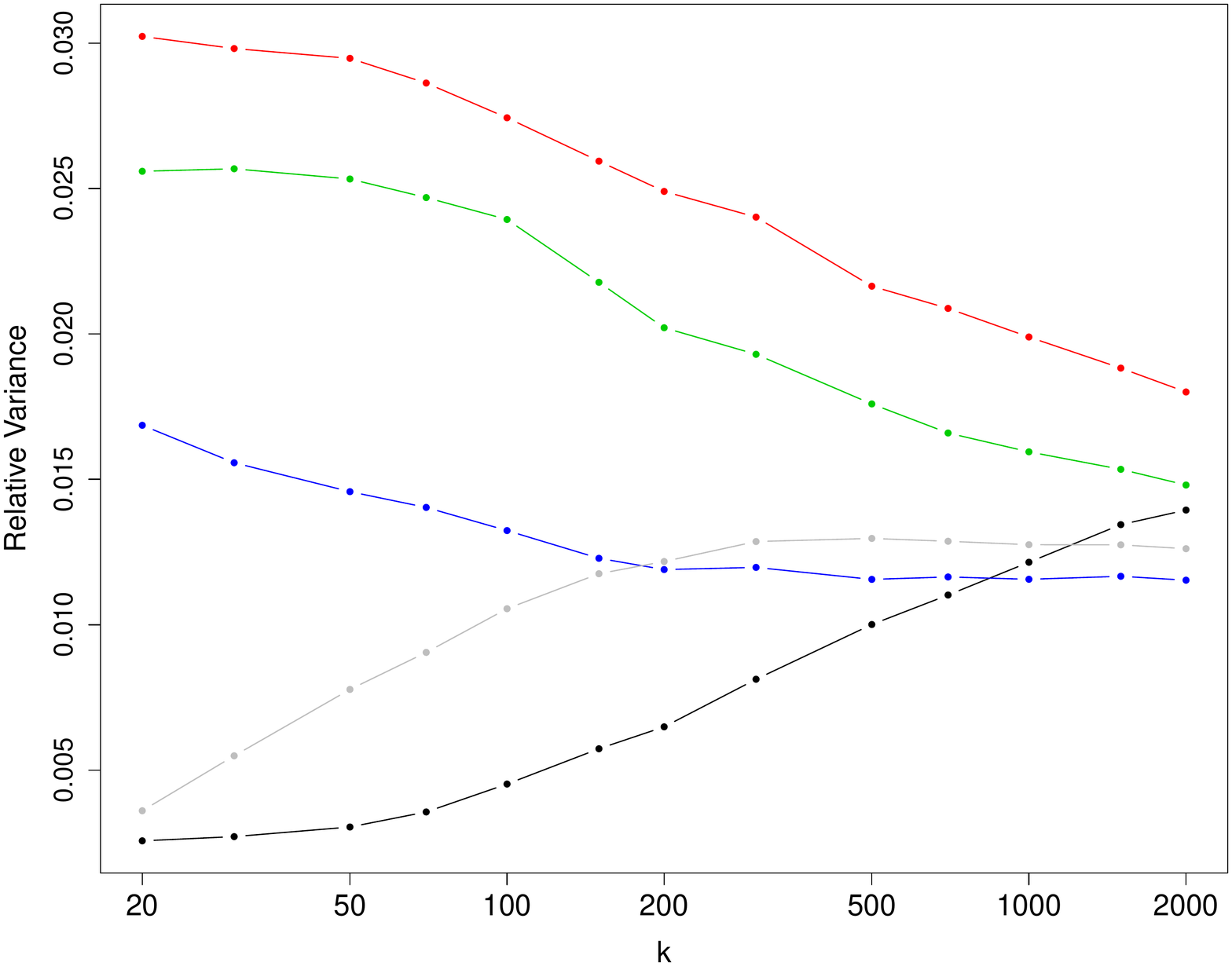}}
		\subfigure{\includegraphics[width=0.33\textwidth]{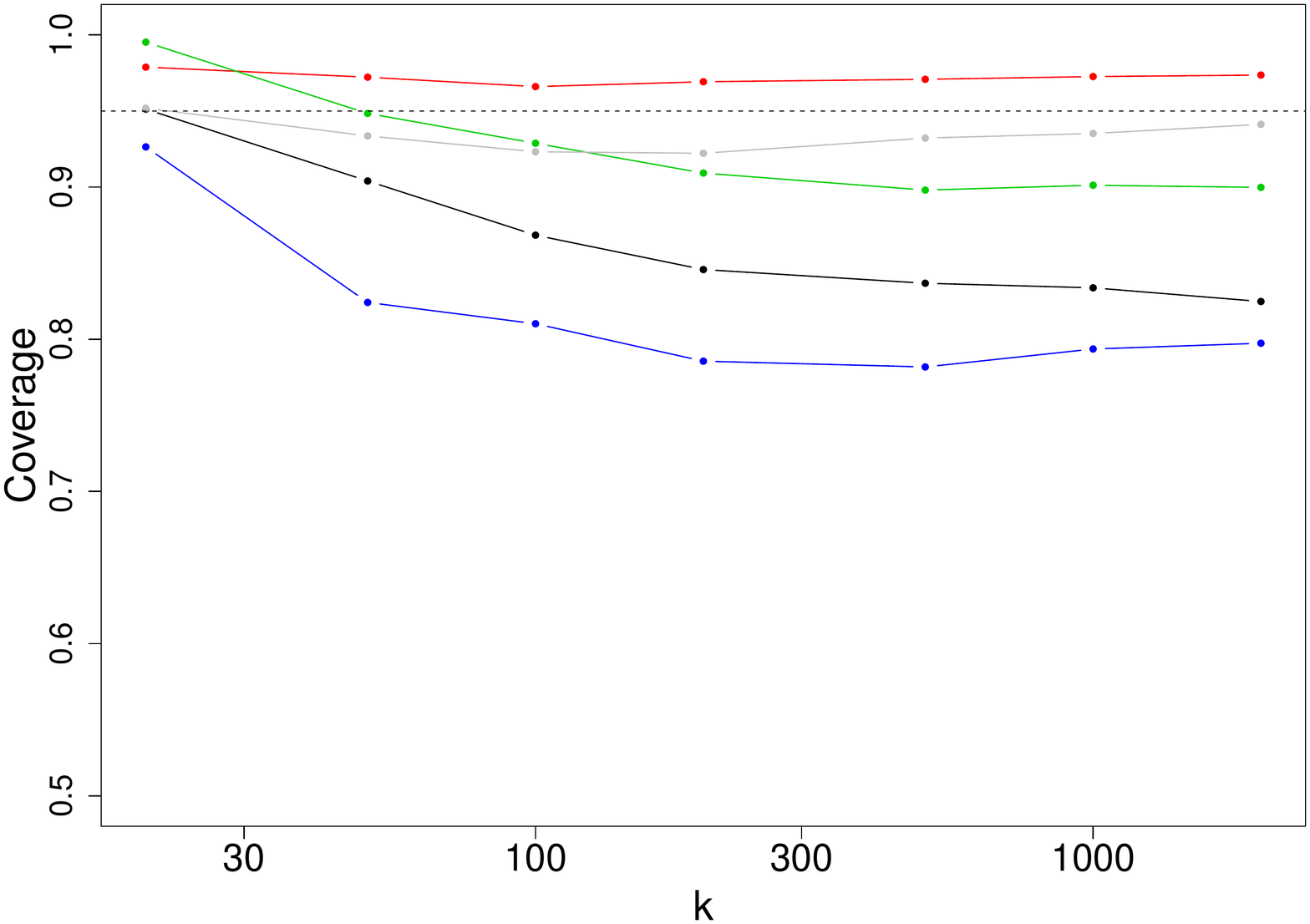}}
		\subfigure{\includegraphics[width=0.33\textwidth]{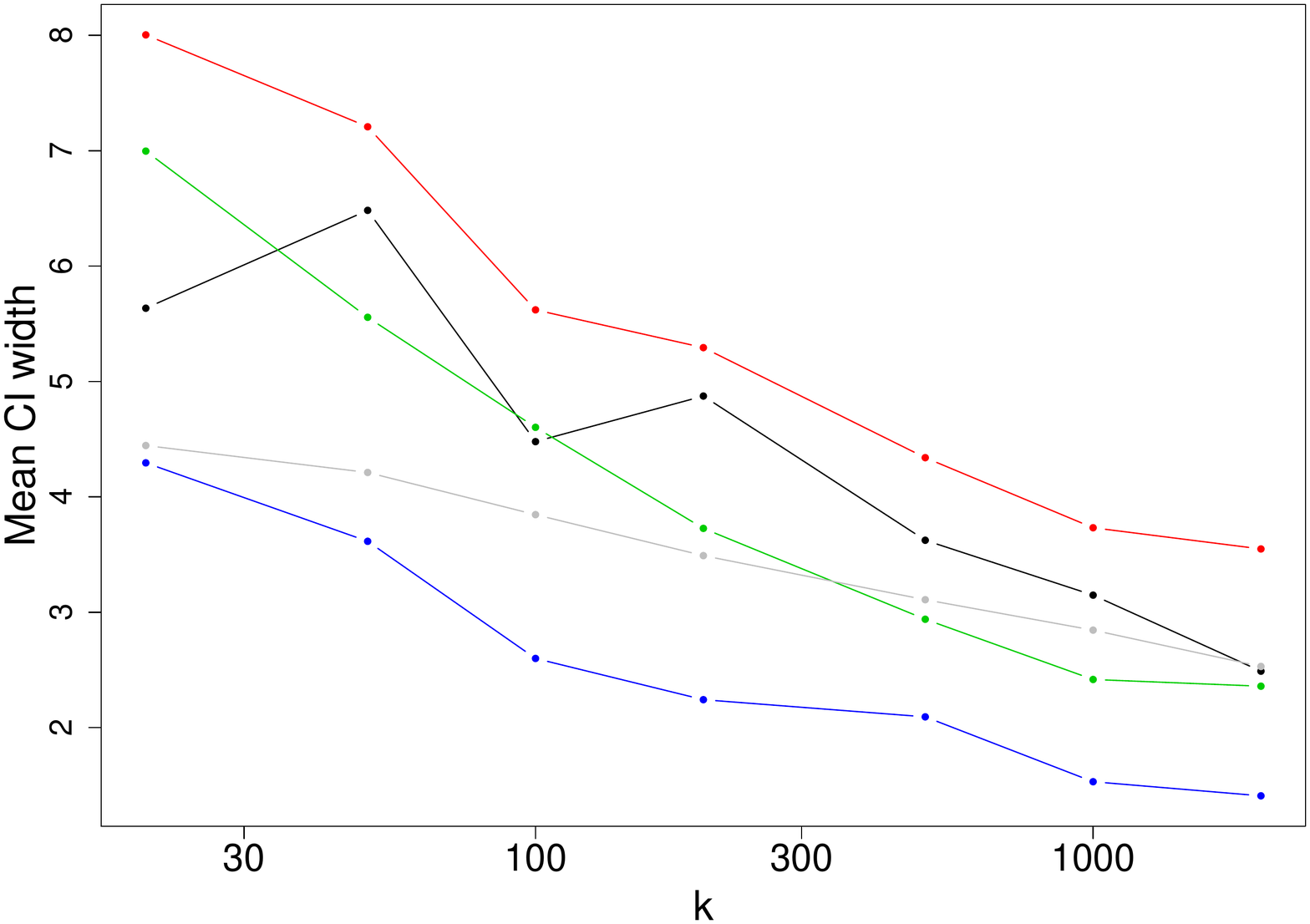}}
		\subfigure{\includegraphics[width=0.33\textwidth]{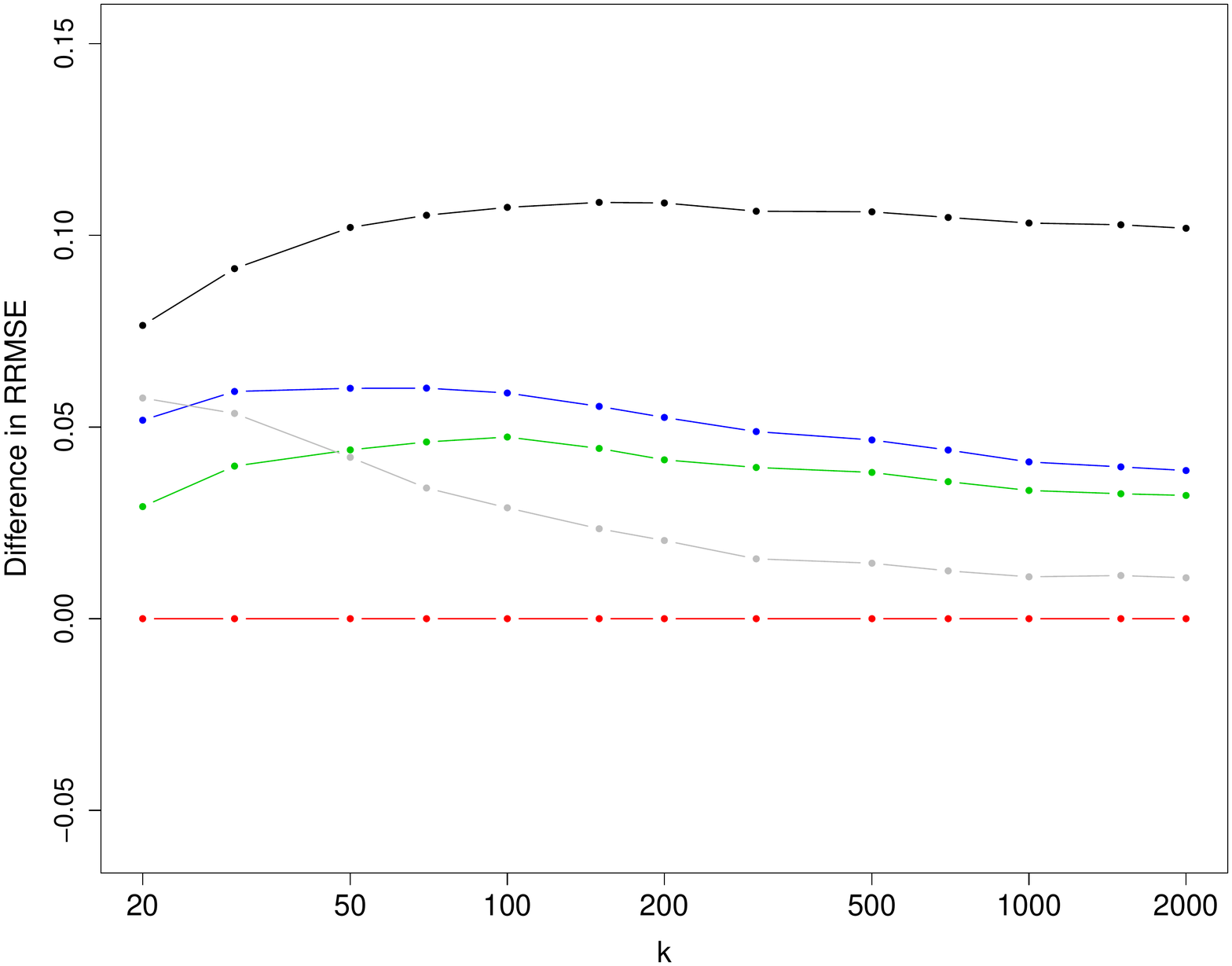}}
		\captionit{50 year return level estimates when sampling from the GEV distribution with $(\mu, \sigma, \xi)=(0, 1, -0.2)$ using the variable-threshold stopping rule over a range of $k$. Based on $20000$ replicated samples with the historical data created using approach~\eqref{eqn:initial} of the paper. See Figure~\ref{fig:sup.200retneg} for other associated detail.}
\end{figure}

\begin{figure}[h]
		\subfigure{\includegraphics[width=0.33\textwidth]{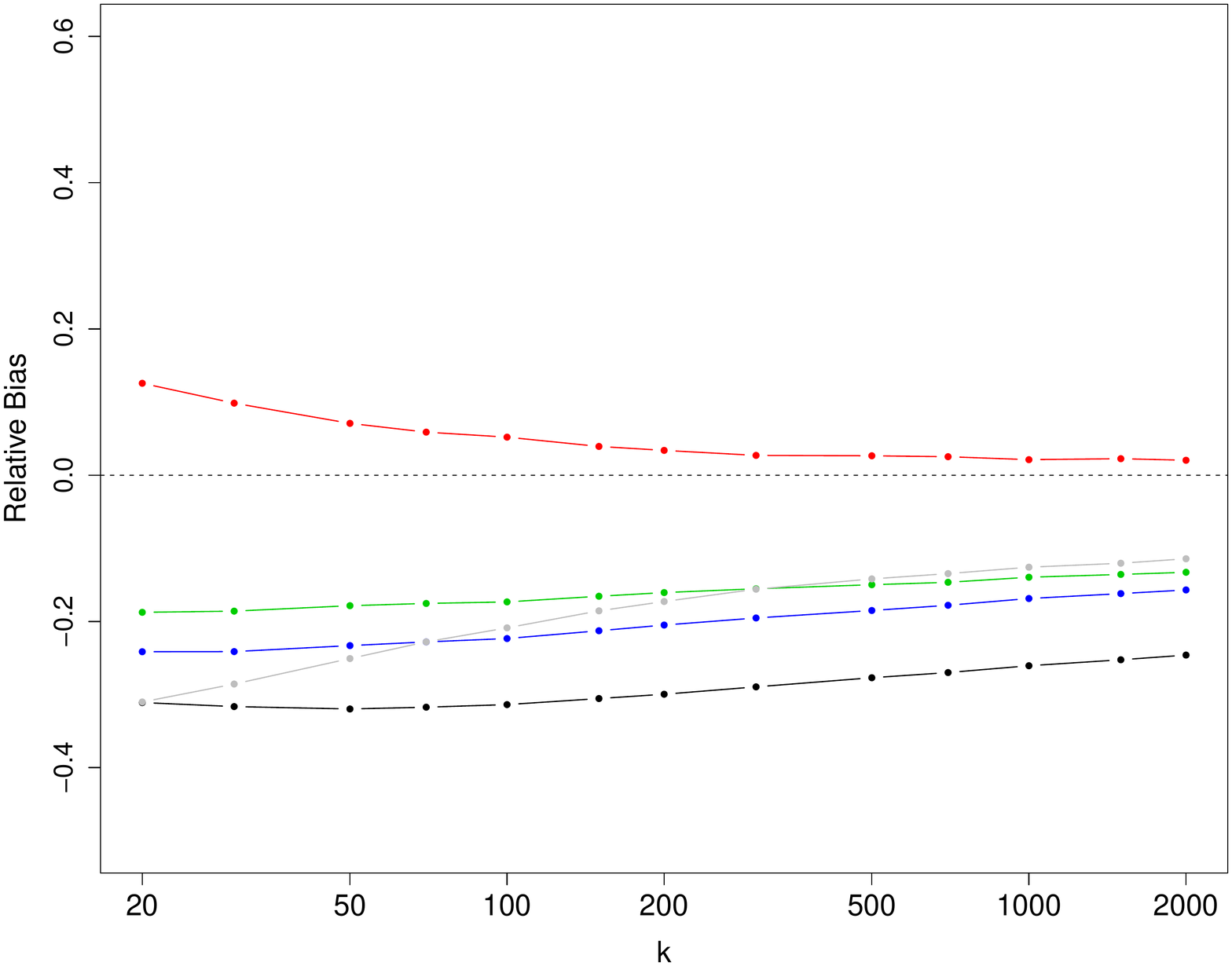}}
		\subfigure{\includegraphics[width=0.33\textwidth]{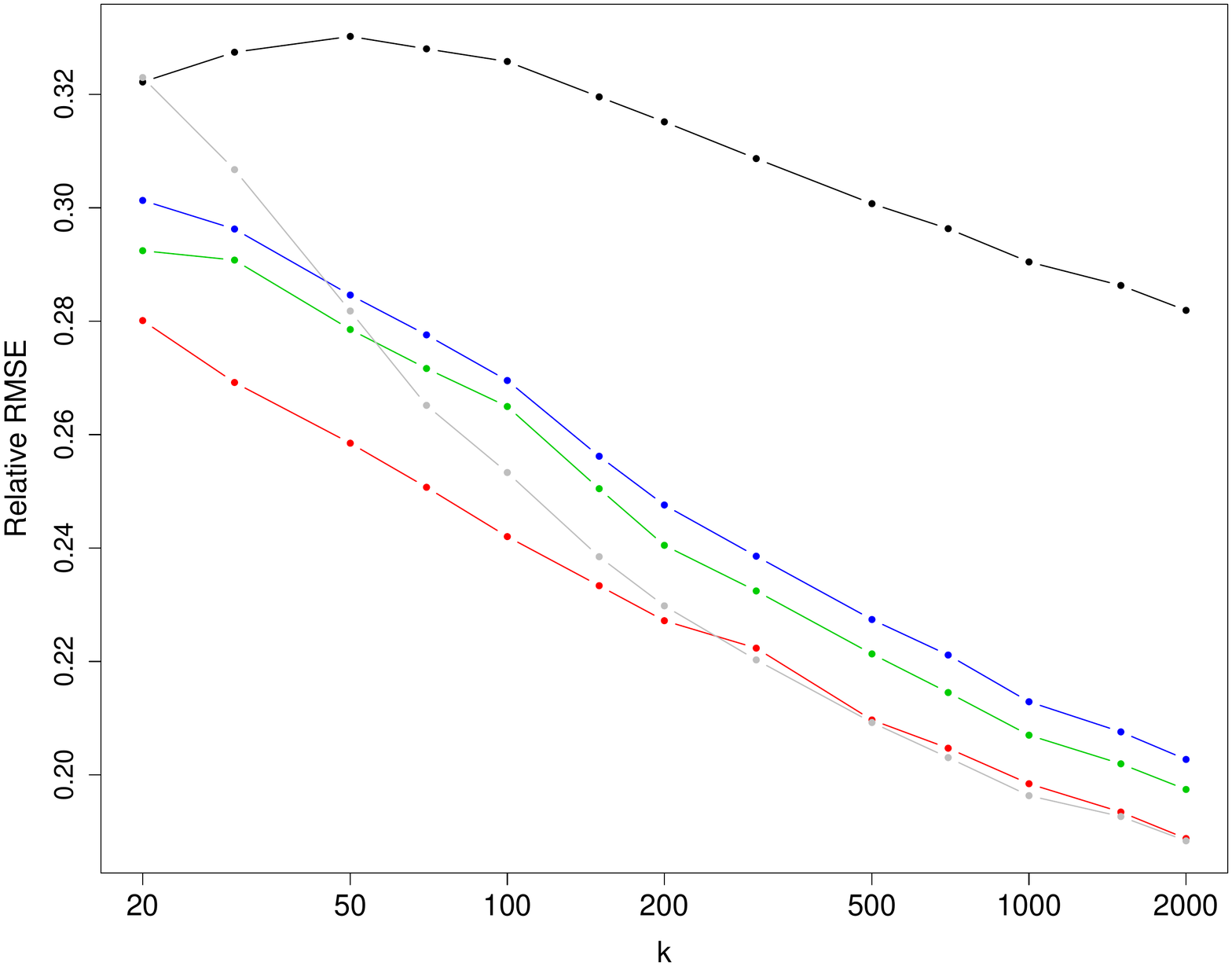}}
		\subfigure{\includegraphics[width=0.33\textwidth]{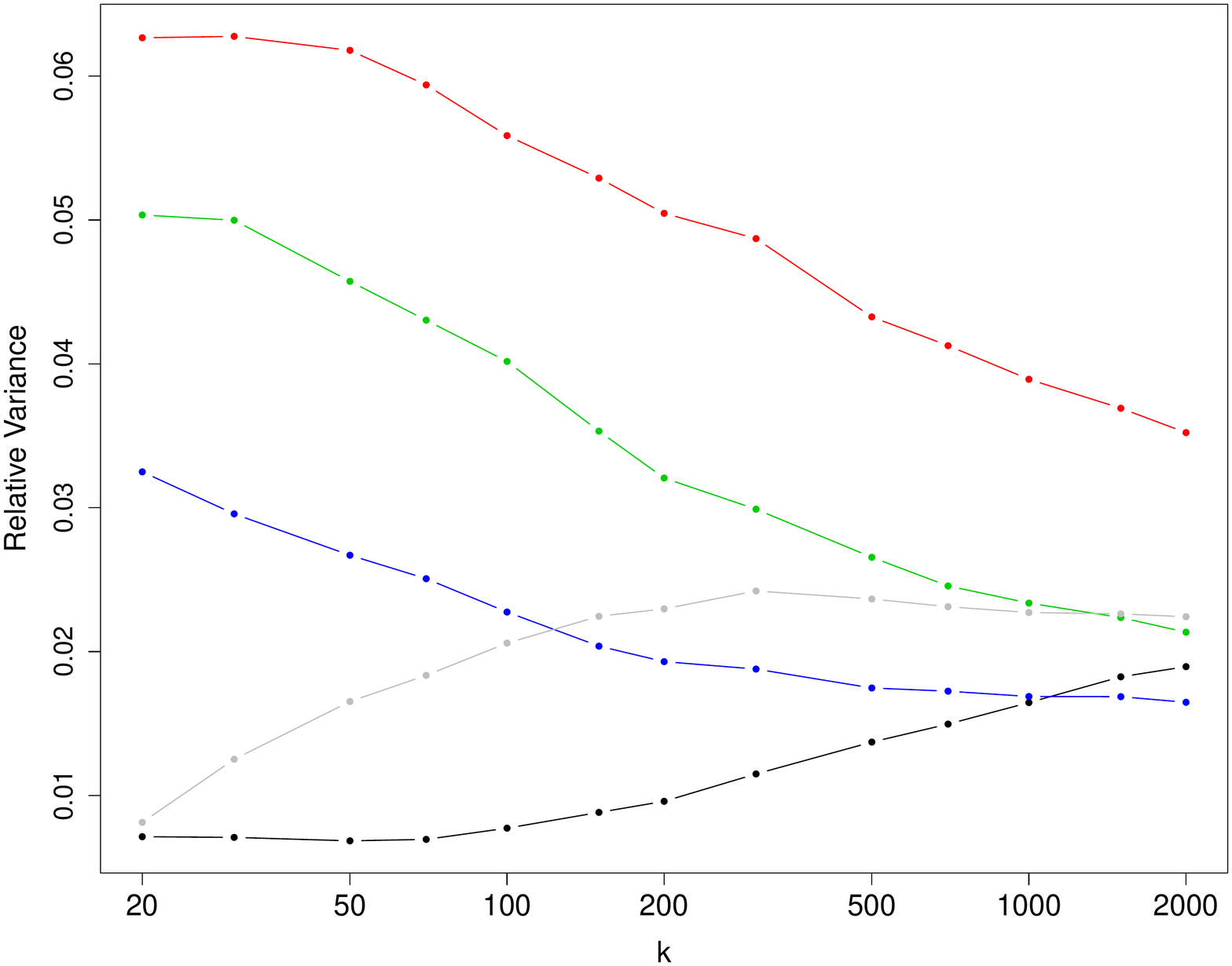}}
		\subfigure{\includegraphics[width=0.33\textwidth]{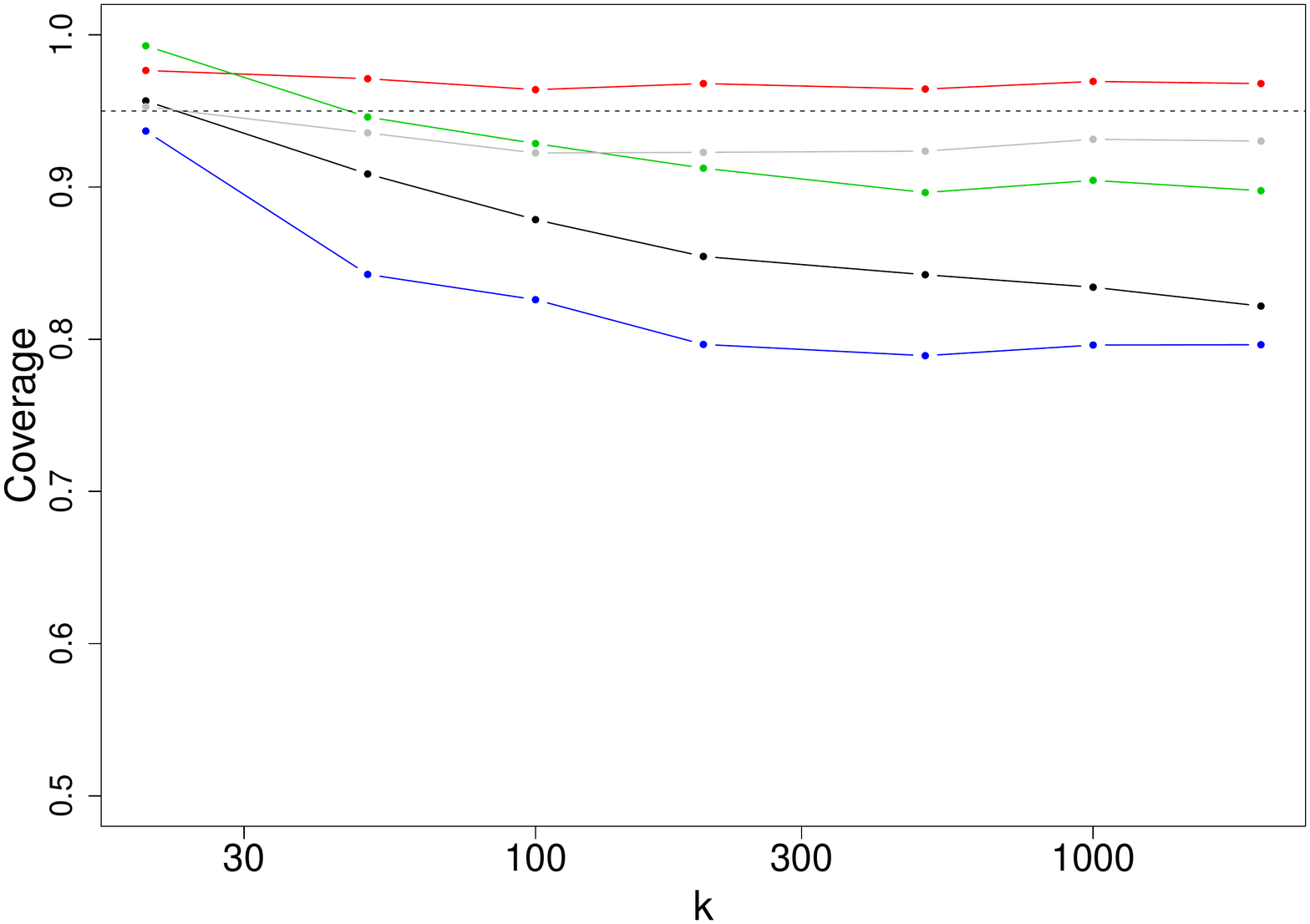}}
		\subfigure{\includegraphics[width=0.33\textwidth]{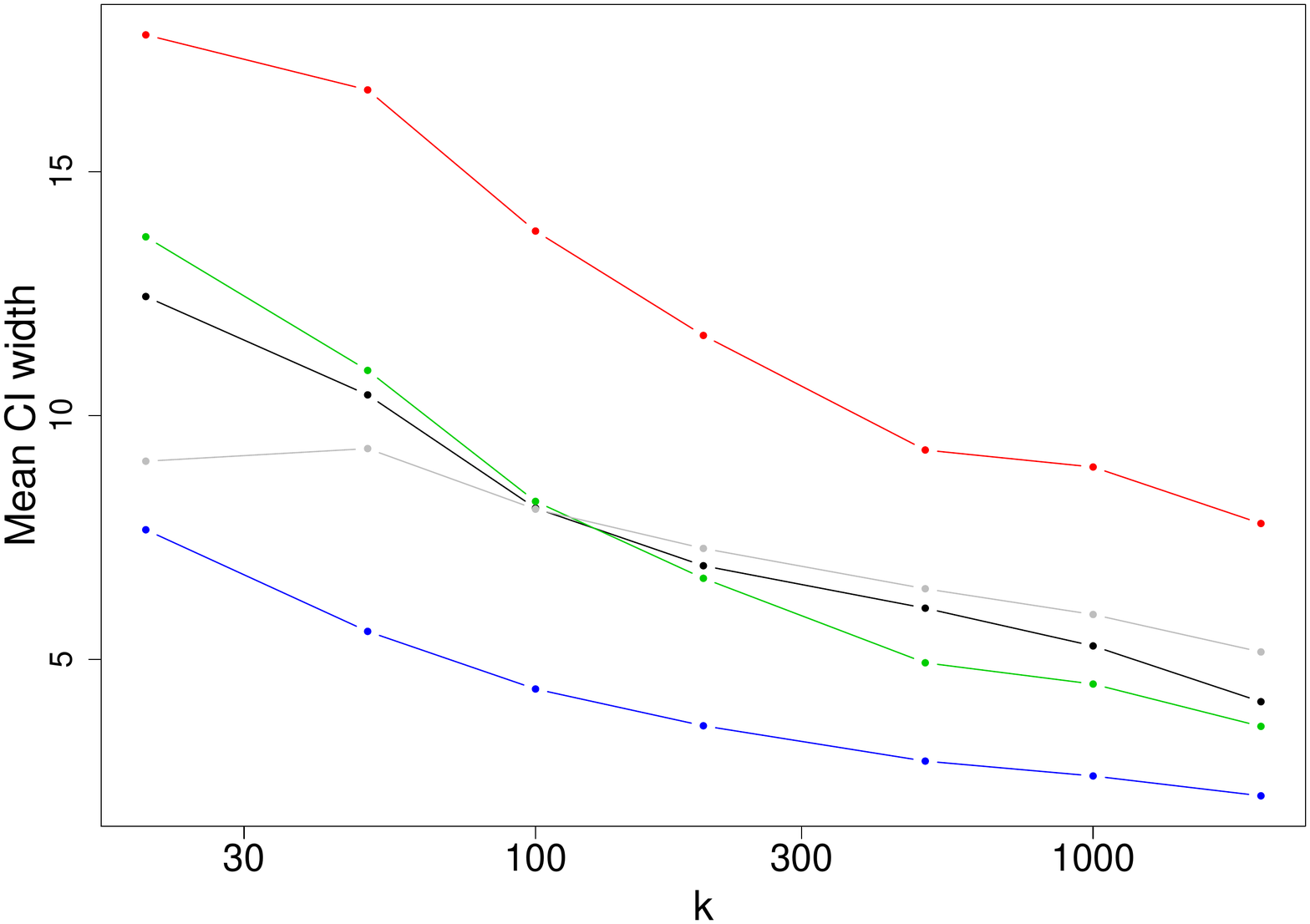}}
		\subfigure{\includegraphics[width=0.33\textwidth]{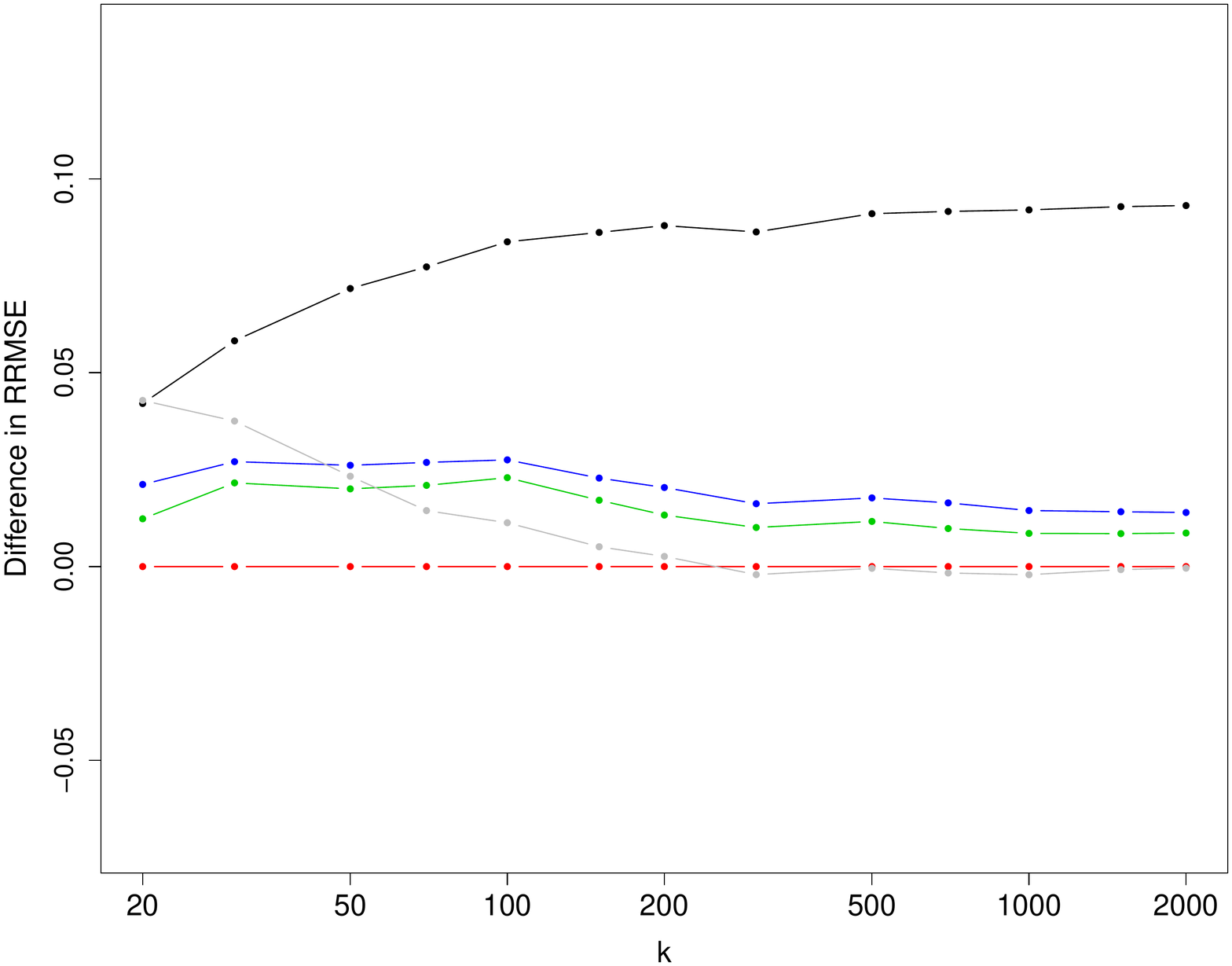}}
		\captionit{200 year return level estimates when sampling from the GEV distribution with $(\mu, \sigma, \xi)=(0, 1, -0.2)$ using the variable-threshold stopping rule over a range of $k$. Based on $20000$ replicated samples with the historical data created using approach~\eqref{eqn:initial} of the paper. See Figure~\ref{fig:sup.200retneg} for other associated detail.}
\end{figure}

\end{document}